\documentclass{aa}  

\usepackage{graphicx}
\usepackage{lscape}
\usepackage{txfonts}
\usepackage{enumitem}
\usepackage{nccmath}
\usepackage[colorlinks=true,linkcolor=red,citecolor=blue]{hyperref}
\begin{document}

\title{Radial metal abundance profiles in the intra-cluster medium of cool-core galaxy clusters, groups, and ellipticals}

\author{F. Mernier\inst{\ref{SRON},\ref{Leiden}} \and J. de Plaa\inst{\ref{SRON}}  \and J. S. Kaastra\inst{\ref{SRON},\ref{Leiden}} Y.-Y. Zhang\inst{\ref{Bonn}}\thanks{This paper is dedicated to the memory of our wonderful colleague Yu-Ying Zhang, who recently passed away.} \and H. Akamatsu\inst{\ref{SRON}} \and L. Gu\inst{\ref{SRON}} \and P. Kosec\inst{\ref{Cambridge}} \and J. Mao\inst{\ref{SRON},\ref{Leiden}} \and C. Pinto\inst{\ref{Cambridge}}  \and T. H. Reiprich\inst{\ref{Bonn}} \and J. S. Sanders\inst{\ref{Munich}} \and A. Simionescu \inst{\ref{Japan}} \and N. Werner\inst{\ref{Budapest},\ref{Brno}}}

\institute{SRON Netherlands Institute for Space Research, Sorbonnelaan 2, 3584 CA Utrecht, The Netherlands\label{SRON} \\ \email{F.Mernier@sron.nl} \and Leiden Observatory, Leiden University, P.O. Box 9513, 2300 RA Leiden,The Netherlands\label{Leiden} \and Argelander-Institut f\"{u}r Astronomie, Auf dem H\"{u}gel 71, D-53121 Bonn, Germany\label{Bonn} \and Institute of Astronomy, Madingley Road, CB3 0HA Cambridge, United Kingdom\label{Cambridge} \and Max-Planck Institut f\"{u}r Extraterrestrische Physik, Giessenbackstrasse 1, D-85748 Garching, Germany\label{Munich} \and Institute of Space and Astronautical Science (ISAS), JAXA, 3-1-1 Yoshinodai, Chuo-ku, Sagamihara, Kanagawa 252-5210, Japan \label{Japan} \and MTA-E\"otv\"os University Lend\"ulet Hot Universe Research Group, P\'azm\'any P\'eter s\'et\'any 1/A, Budapest, 1117, Hungary\label{Budapest} \and Department of Theoretical Physics and Astrophysics, Faculty of Science, Masaryk University, Kotl\'a\v{r}sk\'a 2, Brno, 611 37, Czech Republic\label{Brno}}

\date{Received 16 November 2016 / Accepted 1 March 2017}

\abstract{The hot intra-cluster medium (ICM) permeating galaxy clusters and groups is not pristine, as it has been continuously enriched by metals synthesised in Type Ia (SNIa) and core-collapse (SNcc) supernovae since the major epoch of star formation ($z \simeq $ 2--3). The cluster/group enrichment history and mechanisms responsible for releasing and mixing the metals can be probed via the radial distribution of SNIa and SNcc products within the ICM. In this paper, we use deep \textit{XMM-Newton}/EPIC observations from a sample of 44 nearby cool-core galaxy clusters, groups, and ellipticals (CHEERS) to constrain the average radial O, Mg, Si, S, Ar, Ca, Fe, and Ni abundance profiles. The radial distributions of all these elements, averaged over a large sample for the first time, represent the best constrained profiles available currently. Specific attention is devoted to a proper modelling of the EPIC spectral components, and to other systematic uncertainties that may affect our results. We find an overall decrease of the Fe abundance with radius out to $\sim$$0.9 r_{500}$ and $\sim$$0.6 r_{500}$ for clusters and groups, respectively, in good agreement with predictions from the most recent hydrodynamical simulations. The average radial profiles of all the other elements (X) are also centrally peaked and, when rescaled to their average central X/Fe ratios, follow well the Fe profile out to at least $\sim$0.5$r_{500}$. As predicted by recent simulations, we find that the relative contribution of SNIa (SNcc) to the total ICM enrichment is consistent with being uniform at all radii, both for clusters and groups using two sets of SNIa and SNcc yield models that reproduce the X/Fe abundance pattern in the core well. In addition to implying that the central metal peak is balanced between SNIa and SNcc, our results suggest that the enriching SNIa and SNcc products must share the same origin and that the delay between the bulk of the SNIa and SNcc explosions must be shorter than the timescale necessary to diffuse out the metals.
Finally, we report an apparent abundance drop in the very core of 14 systems ($\sim$32\% of the sample). Possible origins of these drops are discussed.
}


\titlerunning{Radial abundance profiles in the hot intra-cluster medium}
\authorrunning{F. Mernier et al.}

\maketitle

\section{Introduction}\label{sect:intro}

Galaxy clusters and groups are more than a simple collection of galaxies (and dark matter haloes), as they are permeated by large amounts of very hot gas. This intra-cluster medium (ICM) was heated up to $10^7$--$10^8$ K during the gravitational assembly of these systems, and is glowing in the X-ray band, mainly via bremsstrahlung emission, radiative recombination, and line radiation \citep[for a review, see][]{2010A&ARv..18..127B}. Since the first detection of a Fe-K emission feature at $\sim$7 keV in its X-ray spectra \citep{1976MNRAS.175P..29M,1977ApJ...211L..63S}, it is well established that the ICM does not have a primordial origin, but has been enriched with heavy elements, or metals, up to typical values of $\sim$0.5--1 times solar \citep[for reviews, see][]{2008SSRv..134..337W,2013AN....334..416D}. Since the ICM represents about $\sim$80\% of the total baryonic matter in clusters, this means that there is more mass in metals in the ICM than locked in all the cluster galaxies \citep[e.g.][]{2014MNRAS.444.3581R}.

Despite the first detection of several K-shell metal lines with the \textit{Einstein} observatory in the early 1980s
\citep[e.g.][]{1979ApJ...234L..33C,1981ApJ...244L..47M}, before 1993 only the iron (Fe) abundance could be accurately measured in the ICM. After the launch of \textit{ASCA}, abundance studies in clusters could extend (although with a limited accuracy) to oxygen (O), neon (Ne), magnesium (Mg), silicon (Si), sulfur (S), argon (Ar), calcium (Ca), and nickel (Ni), thus opening a new window on the ICM enrichment \citep[e.g.][]{1996ApJ...466..686M,2005ApJ...620..680B}. However, the most spectacular step forward in the field has been achieved by the latest generation of X-ray observatories, i.e. \textit{Chandra}, \textit{XMM-Newton}, and \textit{Suzaku}, which allowed much more accurate abundance measurements of these elements thanks to the significantly improved effective area and spectral resolution of their instruments \citep[e.g.][]{2001A&A...379..107T,2006A&A...452..397D,2006A&A...459..353W}. With excellent \textit{Suzaku} and \textit{XMM-Newton} exposures, the abundance of other elements, such as carbon, nitrogen \citep[e.g.][Mao et al. 2017, to be submitted]{2006A&A...459..353W,2011MNRAS.412L..35S}, or even chromium and manganese \citep{2009ApJ...705L..62T,2016A&A...592A.157M}, could be reasonably constrained as well.

Metals present in the ICM must have been synthesised by stars and supernovae (SNe) explosions, mainly within cluster galaxies. While O, Ne, and Mg are produced almost entirely by core-collapse supernovae (SNcc), the Fe-peak elements mostly originate from Type Ia supernovae (SNIa). Intermediate elements (e.g. Si, S, and Ar) are synthesised by both SNIa and SNcc \citep[for a review, see][]{2013ARA&A..51..457N}. Since the current X-ray missions allow the measurement of the abundance of all these elements with a good level of accuracy in the core of the ICM (i.e. where the overall flux and the metal line emissivities are the highest), several attempts have been made to use these abundances to provide constraints on SNIa and SNcc yield models in individual objects \citep[e.g.][]{2006A&A...449..475W,2006A&A...452..397D,2012ApJ...753...54B} or in samples \citep[e.g.][]{2007A&A...465..345D,2007ApJ...667L..41S,2016A&A...595A.126M}. From these studies, it appears that the typical fraction of SNIa (SNcc) contributing to the enrichment lies within $\sim$20--45\% (55--80\%), depending (mainly) on the selected yield models.

Beyond the overall elemental abundances, witnessing the time-integrated enrichment history in galaxy clusters and groups since the major epoch of star formation \citep[$z \simeq 2$--3; for a review, see][]{2014ARA&A..52..415M} determining the distribution of metals within the ICM is also of crucial importance. Indeed, this metal distribution constitutes a direct signature of, first, the locations and epoch(s) of the enrichment and, second,  the dominant mechanisms transporting the metals into and across the ICM. In turn, these transport mechanisms must also play a fundamental role in governing the thermodynamics of the hot gas. Since the \textit{ASCA} discovery of a strong metallicity gradient in Centaurus \citep{1994MNRAS.269..409A,1994PASJ...46L..55F}, a systematically peaked Fe distribution in cool-core clusters  and groups (i.e. showing a strong ICM temperature decrease towards the centre) has been confirmed by many studies \citep[e.g.][]{1997ApJ...488L.125M,2001ApJ...551..153D,2002ApJ...572..160G,2016A&A...592A..37T}. On the contrary, non-cool-core clusters and groups (i.e. with no central ICM temperature gradient) do not exhibit any clear Fe abundance gradient in their cores \citep{2001ApJ...551..153D}. It is likely that the Fe central excess in cool-core clusters has been produced predominantly by the stellar population of the brightest cluster galaxy (BCG) residing in the centre of the gravitational potential well of the cluster during or after the cluster
assembly \citep{2004A&A...416L..21B,2004A&A...419....7D}. However, this excess is often significantly broader than the light profile of the BCG, suggesting that one or several mechanisms, such as turbulent diffusion \citep{2005MNRAS.359.1041R,2006MNRAS.372.1840R} or active galactic nucleus (AGN) outbursts \citep[e.g.][]{2010ApJ...717..937G}, may efficiently diffuse metals out of the cluster core. Alternatively, the higher concentration of Fe in the core of the ICM may be caused by the release of metals from infalling galaxies via ram-pressure stripping \citep[][]{2006A&A...452..795D} together with galactic winds \citep{2007A&A...466..813K,2009A&A...504..719K}. Other processes, such as galaxy-galaxy interactions, AGN outflows, or an efficient enrichment by intra-cluster stars, may also play a role \citep[for a review, see][]{2008SSRv..134..363S}. In addition to this central excess, there is increasing evidence of a uniform Fe enrichment floor extending out to $r_{200}$\footnote{$r_\Delta$ is defined as the radius within which the mass density corresponds to $\Delta$ times the critical density of the Universe.} and probably beyond \citep{2008PASJ...60S.343F,2013Natur.502..656W,2016A&A...592A..37T}. This suggests an additional early enrichment by promptly exploding SNIa, i.e. having occurred and efficiently diffused before the cluster formation. However, a precise quantification of this uniform level is difficult, since clusters outskirts are very dim and yet poorly understood \citep{2016A&A...586A..32M}.

Whereas the ICM radial distribution of the Fe abundance (rather well constrained thanks to its Fe-K and Fe-L emission complexes, accessible to current X-ray telescopes) has been extensively studied in recent decades, the situation is much less clear for the other elements. Several studies report a rather flat O (and/or Mg) profile, or similarly, an increasing O/Fe (and/or Mg/Fe) ratio towards the outer regions of the cool-core ICM \citep[e.g.][]{2001A&A...379..107T,2003A&A...401..443M,2004A&A...420..135T,2006A&A...459..353W}. As for Fe, there are also indications of a positive and uniform Mg (and other SNcc products) enrichment out to $r_{200}$ \citep{2015ApJ...811L..25S,2016arXiv160903581E}. This apparent flat distribution of SNcc products, contrasting with the enhanced central enrichment from SNIa products, has led to the picture of an early ICM enrichment by SNcc (and prompt SNIa, see above), when galaxies underwent important episodes of star formation. These metals would have mixed efficiently before the cluster assembled, contrary to delayed SNIa enrichment originating from the red and dead BCG. This picture, however, has been questioned by recent observations, suggesting centrally peaked O (and/or Mg) profiles instead \citep[e.g.][]{2007PASJ...59S.327M,2009PASJ...61S.353S,2009A&A...493..409S,2011A&A...528A..60L,2015A&A...575A..37M}. The radial distribution of Si, produced by both SNIa and SNcc, is also unclear, as the Si/Fe profile has been reported to be sometimes flat, sometimes increasing with radius \citep[e.g.][]{2007MNRAS.380.1554R,2011A&A...528A..60L,2011MNRAS.418.2744M,2014ApJ...781...36S}.

In all the studies referred to above, the O, Mg, Si, S, Ar, Ca, and Ni radial abundance profiles have been measured either for individual (mostly cool-core) objects or for very restricted samples ($\le$15 objects). Consequently, in most cases, these profiles suffer from large statistical uncertainties. In parallel, little attention has been drawn to systematic effects that could potentially bias some results. Building average abundance profiles (not only for Fe, but for all the other possible elements mentioned above) over a large sample of cool-core (and, if possible, non-cool-core) systems is clearly needed to clarify the picture of the SNIa and SNcc enrichment history in galaxy clusters and groups. 

In this paper, we use deep \textit{XMM-Newton}/EPIC observations from a sample of 44 nearby cool-core galaxy clusters, groups, and ellipticals to derive the average O, Mg, Si, S, Ar, Ca, Fe, and Ni abundance profiles in the ICM. In order to make our results as robust as possible, specific attention is devoted to understanding all the possible systematic biases and reducing them when possible. This paper is structured as follows. We describe the observations and our data reduction in Sect. \ref{sect:data_reduction}, the adopted spectral modelling in Sect. \ref{sect:spectral_analysis}, and the averaging of the individual profiles over the sample in Sect. \ref{sect:building_radial_profiles}. Our results, and an extensive discussion on the remaining systematic uncertainties, are presented in Sect. \ref{sect:results} and Sect. \ref{sect:systematics}, respectively. We discuss the possible implications of our findings in Sect. \ref{sect:discussion} and conclude in Sect. \ref{sect:conclusion}. Throughout this paper, we adopt the cosmological parameters $H_0 = 70$ km s$^{-1}$ Mpc$^{-1}$, $\Omega_m = 0.3$, and $\Omega_\Lambda = 0.7$. Unless otherwise stated, the error bars are given at 68\% confidence level, and the abundances are given with respect to the proto-solar abundances of \citet{2009LanB...4B...44L}.


\section{Observations and data preparation}\label{sect:data_reduction}


All the observations considered here are taken from the CHEERS\footnote{CHEmical Enrichment Rgs Sample.} catalogue \citep[][]{dePlaa2017,2016A&A...592A.157M}. This sample, optimised to study chemical enrichment in the ICM, consists of 44 nearby cool-core galaxy clusters, groups, and ellipticals for which the \ion{O}{viii} 1s--2p line at $\sim$19 $\AA$ is detected with >5$\sigma$ in their \textit{XMM-Newton}/RGS spectra. This includes archival \textit{XMM-Newton} data and several recent deep observations that were performed to complete the sample in a consistent way \citep{dePlaa2017}.

We reduce the EPIC MOS\,1, MOS\,2, and pn data using the XMM Science Analysis System (SAS) v14.0 and the calibration files dated by March 2015. The standard pipeline commands \texttt{emproc} and \texttt{epproc} are used to extract the event files from the EPIC MOS and pn data, respectively. We filter each observation from soft-flare events by applying the appropriate good time interval (GTI) files following the 2$\sigma$-clipping criterion \citep{2015A&A...575A..37M}. After filtering, the MOS\,1, MOS\,2, and pn exposure times of the full sample are $\sim$4.5 Ms, $\sim$4.6 Ms, and $\sim$3.7 Ms, respectively \citep[see Table 1 of][]{2016A&A...592A.157M}. Following the usual recommendations, we keep the single-, double- and quadruple-pixel events (\texttt{pattern$\le$12}) in MOS, and we only keep the single-pixel events in pn (\texttt{pattern=0}), since the pn double events may suffer from charge transfer inefficiency\footnote{See the XMM-Newton Current Calibration File Release Notes, XMM-CCF-REL-309 (Smith, Guainazzi \& Saxton 2014).}. In both MOS and pn, only the highest quality events are selected (\texttt{flag=0}). The point sources are detected in four distinct energy bands (0.3--2 keV, 2--4.5 keV, 4.5--7.5 keV, and 7.5--12 keV) using the task \texttt{edetect\_chain} and further rechecked by eye. We discard these point sources from the rest of the analysis, by excising a circular region of 10$''$ of radius around their surface brightness peak. This radius is found to be the best compromise between minimising the fraction of contaminating photons from point sources and maximising the fraction of the ICM photons considered in our spectra \citep{2015A&A...575A..37M}. In some specific cases, however, photons from very bright point sources may leak beyond 10$''$, and consequently we adopt a larger excision radius.

In each dataset, we extract the MOS\,1, MOS\,2, and pn spectra of eight concentric annuli of fixed angular size (0$'$--0.5$'$, 0.5$'$--1$'$, 1$'$--2$'$, 2$'$--3$'$, 3$'$--4$'$, 4$'$--6$'$, 6$'$--9$'$, and 9$'$--12$'$), all centred on the X-ray peak emission seen on the EPIC surface brightness images. The redistribution matrix file (RMF) and the ancillary response file (ARF) of each spectrum are produced via the \texttt{rmfgen} and \texttt{arfgen} SAS tasks, respectively.


\section{Spectral modelling}\label{sect:spectral_analysis}


The spectral analysis is performed using the SPEX\footnote{https://www.sron.nl/astrophysics-spex} package \citep{1996uxsa.conf..411K}, version 2.05. Following the method described in \citet{2016A&A...592A.157M}, we start by simultaneously fitting the MOS\,1, MOS\,2, and pn spectra of each pointing. When a target includes two separate observations, we fit their spectra simultaneously. Since the large number of fitting parameters does not allow us to fit more than two observations simultaneously, we form pairs of simultaneous fits when an object contains three (or more) observations. We then combine the results of the fitted pairs using a factor of $1/\sigma_i^2$, where $\sigma_i$ is the error on the considered parameter $i$.
We also note that the second EPIC observation of M\,87 (ObsID:0200920101) is strongly affected by pile-up in its core, owing to a sudden activity of the central AGN \citep{2006A&A...459..353W}. Therefore, the radial profiles within 3$'$ are only estimated with the first observation  (ObsID:0114120101).

Because of calibration issues in the soft X-ray band of the CCDs ($\lesssim$0.5 keV) and beyond $\sim$10 keV, we limit our MOS and pn spectral fittings to the 0.5--10 keV and 0.6--10 keV energy bands, respectively. We rearrange the data bins in each spectrum
via the optimal binning method of \citet{2016A&A...587A.151K} to maximise the amount of information provided by the spectra while keeping reasonable constraints on the model parameters.

\subsection{Thermal emission modelling}\label{sect:thermal_mod}

In principle, we can model the ICM emission in SPEX with the (redshifted and absorbed) \texttt{cie} thermal model. This single-temperature model assumes that the plasma is in (or close to) collisional ionisation equilibrium (CIE), which is a reasonable assumption \citep[e.g.][]{1986RvMP...58....1S}. 

Although the \texttt{cie} model may be a good approximation of the emitting ICM in some specific cases (i.e. when the gas is nearly isothermal), the temperature structure within the core of clusters and groups is often complicated and a multi-temperature model is clearly required. In particular, fitting the spectra of a multi-phase plasma with a single-temperature model can dramatically affect the measured Fe abundance, leading to the "Fe-bias" \citep{1994ApJ...427...86B,1998MNRAS.296..977B,2000MNRAS.311..176B} or to the "inverse Fe-bias" \citep{2008ApJ...674..728R,2009A&A...493..409S,2010A&A...522A..34G}. Taking this caveat into account, we model the ICM emission with a \texttt{gdem} model \citep[e.g.][]{2006A&A...452..397D}, which is also available in SPEX. This multi-temperature component models a CIE plasma following a Gaussian-shaped temperature distribution,
\begin{equation} \label{eq:gdem}
Y(x) = \frac{Y_0}{\sigma_{T} \sqrt{2 \pi}} \exp \left( \frac{(x-x_\text{mean})^2}{2 \sigma^2_{T}} \right),
\end{equation} 
where $x=\log(kT)$, $x_\text{mean}=\log(kT_\text{mean})$, $kT_\text{mean}$ is the mean temperature of the distribution, $\sigma_T$ is the width of the distribution, and $Y_0$ is the total integrated emission measure. The other parameters are similar as in the \texttt{cie} model. By definition, a \texttt{gdem} model with $\sigma_T=0$ reproduces a \texttt{cie} (i.e. single-temperature) model.
The free parameters of the \texttt{gdem} model are the normalisation (or emission measure) $Y_0 = \int n_e n_\ion{H}{} dV$, the temperature parameters $kT_\text{mean}$ and $\sigma_T$, and the abundances of O, Ne, Mg, Si, S, Ar, Ca, Fe, and Ni \citep[given with respect to the proto-solar table of][see Sect. \ref{sect:intro}]{2009LanB...4B...44L}. Because these analyses are out of the scope of this paper, we devote the radial analyses of the temperatures, emission measures, and subsequent densities and entropies for a future work. The abundances of the Z$\le$7 elements are fixed to the proto-solar unity, while the remaining abundances are fixed to the Fe value. As mentioned by \citet{2008A&A...487..461L}, constraining the free abundance parameters to positive values only (for obvious physical reasons) may result in a statistical bias when averaging out the profiles. Therefore, we allow all the best-fit abundances to take positive and negative values. Following \citet{2016A&A...592A.157M}, the measured O abundances have been corrected from updated parametrisation of the radiative recombination rates \citep[see also][]{dePlaa2017}. Since Ne abundances measured with EPIC are highly unreliable (because the main Ne emission feature is entirely blended with the Fe-L complex at EPIC spectral resolution), we do not consider them in the rest of the paper.

The absorption of the ICM photons by neutral interstellar matter is reproduced by a \texttt{hot} model, where the temperature parameter is fixed to 0.5 eV (see the SPEX manual). Because adopting the column densities of \citet{2013MNRAS.431..394W} -- taking  both  atomic and molecular hydrogen into account -- sometimes leads to poor spectral fits, we perform a grid search of the best-fit $N_\ion{H}{}$ parameter within the limits
\begin{equation}\label{eq:N_H}
N_\ion{H}{i} - 5\times 10^{19} \text{ cm}^{-2} \le N_\ion{H}{} \le N_\text{H,tot} + 1 \times 10^{20} \text{ cm}^{-2},
\end{equation}
where $N_\ion{H}{i}$ and $N_\text{H,tot}$ are the atomic and total (atomic and molecular) hydrogen column densities, respectively \citep[for further details, see][]{2016A&A...592A.157M}.

\subsection{Background modelling}\label{sect:bg_mod}

Whereas in the core of bright clusters the ICM emission is largely dominant, in cluster outskirts the background plays an important role and sometimes may even dominate. For extended objects, a background subtraction applied to the raw spectra is clearly not advised because a slightly incorrect scaling may lead to dramatic changes in the derived temperatures \citep{2006A&A...452..397D}. In turn, since the metal line emissivities depend on the assumed plasma temperature, this approach may lead to erroneous abundance measurements outside the cluster cores. Moreover, the observed background data (usually obtained from blank-field observations) may significantly vary with time and position on the sky.

Instead, we choose to model the background directly in the spectral fits by adopting the method extensively described in \citet{2015A&A...575A..37M}. The total background emission is decomposed into five components as follows:

\begin{enumerate}
\item The Galactic thermal emission (GTE) is modelled by an absorbed \texttt{cie} component with proto-solar abundances.

\item The local hot bubble (LHB) is modelled by a (unabsorbed) \texttt{cie} component with proto-solar abundances.

\item The unresolved point sources (UPS), whose accumulated flux can account for a significant fraction of the background emission, are modelled by a power law of index $\Gamma_\text{UPS}=1.41$ \citep{2004A&A...419..837D}.

\item The hard particle background (HP, or instrumental background) consists of a continuum and fluorescence lines. The continuum is modelled by a (broken) power law, whose parameters can be constrained using filter wheel closed observations, and the lines are modelled by Gaussian functions. Because this is a particle background, we leave this modelled component unfolded by the effective area of the CCDs.

\item The quiescent soft-protons (SP) may contribute to the total emission, even after filtering of the flaring events. This component is modelled by a power law with an index varying typically within $0.7 \lesssim \Gamma_\text{SP} \lesssim 1.4$. Similarly to the HP background, this component is not folded by the effective area.
\end{enumerate}

The background components have been first derived from spectra covering the total EPIC field of view to obtain good constraints on their parameters. In particular, this approach allows us to determine both the mean temperature of the ICM (which is the dominant emission below $\sim$2 keV) and the slope of the SP component (better visible beyond $\sim$2 keV), while these two parameters are usually degenerate when only analysing one outer annulus. 
In addition to the \texttt{gdem} component, the free parameters of the background components in the fitted annuli are the normalisations of the HP continuum, HP Gaussian lines (because their emissivities vary with time and across the detector), and quiescent SP (beyond 6$'$ only).

\subsection{Local fits}\label{sect:local_fits}

As discussed extensively in \citet{2015A&A...575A..37M} and \citet{2016A&A...592A.157M}, the abundances measured from a fit covering the full EPIC energy band may be significantly biased, especially for deep exposure datasets. In fact, a slightly incorrect calibration in the effective area may result in an incorrect prediction of the local continuum close to an emission line. Since the abundance of an ion is directly related to the measured equivalent width of its corresponding emission lines, a correct estimate of the local continuum level is crucial to derive accurate abundances.

Therefore, in the rest of the analysis, we measure the O, Mg, Si, S, Ca, Ar, and Ni abundances by fitting the EPIC spectra within several narrow energy ranges centred around their K-shell emission lines \citep[hereafter the "local" fits;][]{2016A&A...592A.157M}. The temperature parameters ($kT_\text{mean}$ and $\sigma_T$) are fixed to their values derived from initial fits performed within the broad energy band (hereafter the "global" fits). In order to assess the systematic uncertainties related to remaining cross-calibration issues between the different EPIC detectors (Sect. \ref{sect:MOS-pn_uncertainties}), we perform our local fits in MOS (i.e. the combined MOS\,1+MOS\,2) and pn spectra independently. Finally,  the Fe abundance can be measured in EPIC using both the K-shell lines ($\sim$6.4 keV) and the L-shell line complex ($\sim$0.9--1.2 keV, although not resolved with CCD instruments). For this reason, in the rest of the paper we use the global fits to derive the Fe abundances.


\section{Building average radial profiles}\label{sect:building_radial_profiles}


Following the approach of \citet{2016A&A...592A.157M}, in addition to the full sample we consider further in this paper, we also split the sample into two subsamples, namely the "clusters" (23 objects) and the "groups" (21 objects), for which the mean temperature within 0.05$r_{500}$ is greater or lower than 1.7 keV, respectively (see also Table \ref{table:observations_radii}). One exception is M\,87, an elliptical galaxy with $kT_\text{mean}(0.05r_{500}) = (2.052 \pm 0.002)$ keV, which we treat in the following as part of the "groups" subsample.

\subsection{Exclusion of fitting artefacts}\label{sect:excl_artefacts}

Since little ICM emission is expected at large radii, one may reasonably expect large statistical uncertainties on our derived fitting parameters in the outermost annuli of every observation. In a few specific cases, however, suspiciously small error bars are reported at large radii, often together with unphysical best-fit values. These peculiar measurements are often due to issues in the fitting process, consequently to bad spectral quality together with a number of fitted parameters that is too large. Since these artefact measurements may significantly pollute our average profiles, we prefer to discard them from the analysis and select outer measurements with reasonably large error bars on their parameters only. To be conservative, we choose to exclude systematically the Fe abundance measurements showing error bars smaller than 0.01, 0.02, and 0.03 in their $4'$--$6'$, $6'$--$9'$, and $9'$--$12'$ annuli, respectively. A similar filtering is applied to the other abundances, this time when their measurements show error bars smaller than 0.01, 0.02, 0.05, and 0.07 in their $3'$--$4'$, $4'$--$6'$, $6'$--$9'$, and $9'$--$12'$ annuli, respectively. These discarded artefacts represent a marginal fraction ($\sim$4\%) of all our data. We list the maximum radial extend for each cluster and all the elements considered ($r_\text{out,X}$) in Table \ref{table:observations_radii}. Finally, we exclude further specific measurements either because their spectral quality could simply not provide reliable estimates or because of possible contamination by the AGN emission. These unaccounted annuli are specified in Table \ref{table:stack_filter}.

\begin{table}
\begin{centering}
\caption{List of the specific measurements that were discarded from our analysis.}             
\label{table:stack_filter}
\resizebox{\hsize}{!}{
\begin{tabular}{l c c c}        
\hline \hline                
Name & Discarded & Element(s) & Comments\\    
 & radii &  & \\    

\hline                        
2A\,0335 & $\ge 6'$ & all & Bad quality\\
A\,4038 & $\ge 9'$ & all & Bad quality\\
A\,3526 & $\ge 9'$ & Mg & HP contamination\\
Hydra A & $\ge 6'$ & all & Bad quality\\
M\,84 & $\le 0.5'$ & all & AGN contamination\\
M\,86 & $\ge 6'$ & all & Bad quality\\
M\,87 & $\le 0.5'$ & all & AGN contamination\\
M\,89 & all & Mg, S, Ar, Ca, Ni & Bad quality\\
           & $\le 0.5'$ & Fe, Si & AGN contamination\\
NGC\,4261 & $\le 0.5'$ & all & AGN contamination\\
NGC\,5044  & $\ge 9'$ & all & Bad quality\\
NGC\,5813  & $\le 0.5'$ & all & AGN contamination\\
                      & $\le 6'$ & Mg & Poor fit in the 1--2 keV band\\
NGC\,5846 & $\le 6'$ & Mg & Poor fit in the 1--2 keV band\\

\hline                                   
\end{tabular}}
\par\end{centering}
\end{table}

\subsection{Stacking method}

Since spectral analysis was performed within annuli of fixed angular sizes regardless of the distances or the cosmological redshifts of the sources, care must be taken to build average radial profiles within consistent spatial scales. As commonly used in the literature, we rescale all the annuli in every object in fractions of $r_{500}$. We adopted the values of $r_{500}$, given for each cluster in Table \ref{table:observations_radii}, from \citet{2015A&A...575A..38P} and references therein. Another unit widely used in the literature is $r_{180}$, as it is often considered (close to) the virial radius of relaxed clusters. Nevertheless, the conversion $r_{500} \simeq 0.6 r_{180}$ is quite straightforward \citep[e.g.][]{2013SSRv..177..195R}.

The number and extent of the reference radial bins of the average profiles are selected such that each bin contains approximately 15--25 individual measurements. The maximum extent of our reference profiles corresponds to the maximum extent reached by the most distant observation: i.e. $1.22 r_{500}$ (based on A\,2597) and $0.97 r_{500}$ (based on A\,189) for clusters and groups, respectively (see Table \ref{table:observations_radii}). After this selection, the average profiles for the full sample and the cluster and group subsamples contain 16, 9, and 8 reference radial bins, respectively. The outermost radial bin of the full sample and the cluster and group subsamples contain 17, 16, and 11 individual measurements, which are located within 0.55--1.22 $r_{500}$, 0.5--1.22 $r_{500}$, and 0.26--0.97 $r_{500}$, respectively. Stacking our individual profiles over the reference bins defined above is not trivial, since some measurements may share their radial extent with two adjacent reference bins. To overcome this issue, we employ the method proposed by \citet{2008A&A...487..461L}. The average abundance profile $X_\text{ref}(k)$, as a function of the $k$-th reference radial bin (defined above), is obtained as
\begin{equation}
X_\text{ref}(k) = \bigg( \sum_{j=1}^{N} \sum_{i=1}^{8} w_{i,j,k} \, \frac{\text{X}(i)_j}{\sigma_{\text{X}(i)_j}^2} \bigg) \bigg/ \bigg( \sum_{j=1}^{N} \sum_{i=1}^{8} w_{i,j,k} \, \frac{1}{\sigma_{\text{X}(i)_j}^2} \bigg) \, ,
\end{equation} 
where X$(i)_j$ is the individual abundance measurement of the $j$-th observation at its $i$-th annulus (as defined in Sect. \ref{sect:data_reduction}), $\sigma_{\text{X}(i)_j}$ is its statistical error (and thus $1/\sigma_{\text{X}(i)_j}^2$ weights each annulus with respect to its emission measure), $N$ is the number of observations, depending of the (sub)sample considered, and $w_{i,j,k}$ a weighting factor. This factor, taking values between 0 and 1, represents the linear overlapping geometric area fraction of the $k$-th reference radial bin on the $i$-th annulus (belonging to the $j$-th observation).

\subsection{MOS-pn uncertainties}\label{sect:MOS-pn_uncertainties}

After stacking the measurements as described above, for each element we are left with $X_\text{ref, MOS}(k)$ and $X_\text{ref, pn}(k)$; i.e. an average MOS and pn abundance profile, respectively, except O, which could only be measured with the MOS instruments, and Fe, which we measured in simultaneous EPIC global fits (see Sect. \ref{sect:local_fits}). The average EPIC (i.e. combined MOS+pn) profiles are then computed as follows:
\begin{align}
X_\text{ref, EPIC}(k) = \bigg( \frac{X_\text{ref, MOS}(k)}{\sigma^2_\text{ref, MOS}(k)} + \frac{X_\text{ref, pn}(k)}{\sigma^2_\text{ref, pn}(k)} \bigg)  \nonumber \\
\bigg/ \bigg( \frac{1}{\sigma^2_\text{ref, MOS}(k)} + \frac{1}{\sigma^2_\text{ref, pn}(k)} \bigg) \, ,
\end{align} 
where $\sigma_\text{ref, MOS}(k)$ and $\sigma_\text{ref, pn}(k)$ are the statistical errors of $X_\text{ref, MOS}(k)$ and $X_\text{ref, pn}(k)$, respectively.
As shown in \citet{2016A&A...592A.157M}, abundance estimates using MOS and pn may sometimes be significantly discrepant. Unsurprisingly, we also find MOS-pn discrepancies in some radial bins of our average abundance profiles. We take this systematic effect into account when combining the MOS and pn profiles by increasing the error bars of the EPIC combined measurements until they cover both their MOS and pn counterparts.


\section{Results}\label{sect:results}


\subsection{Fe abundance profile}\label{sect:results_Fe}

\begin{figure}[!]

                \includegraphics[width=0.5\textwidth]{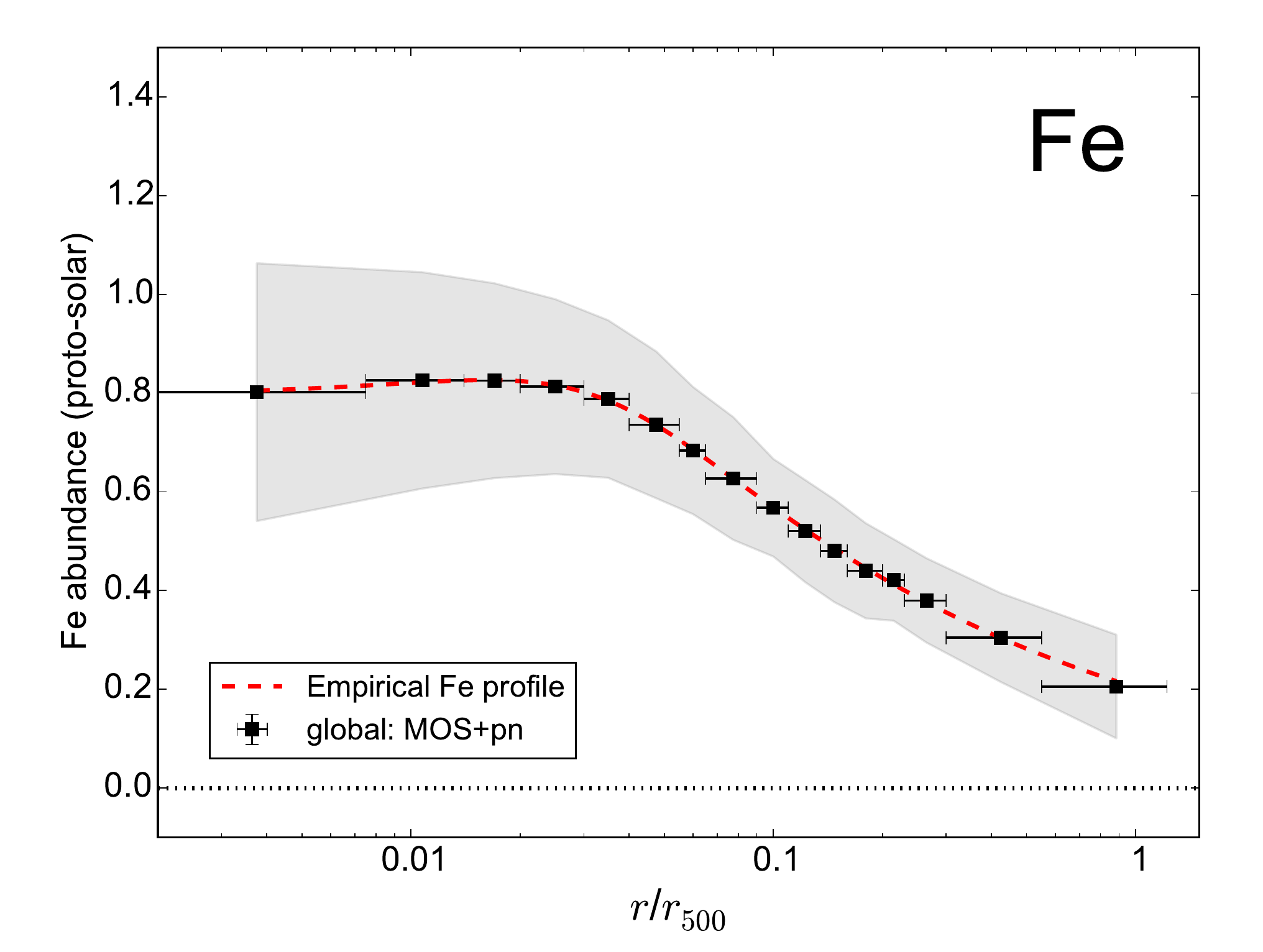}

        \caption{Average radial Fe abundance profile for the full sample. Data points show the average values and their statistical uncertainties ($\sigma_\text{stat}$, barely visible on the plot). The shaded area shows the scatter of the measurements ($\sigma_\text{scatter}$, see text).}
\label{fig:Fe_radial_stacked_profile}
\end{figure}

The average Fe abundance radial profile, measured for the full sample, is shown in Fig. \ref{fig:Fe_radial_stacked_profile}, and the numerical values are detailed in Table \ref{table:Fe_radial_stacked_profile}. The profile shows a clear decreasing trend with radius with a maximum at 0.014--0.02$r_{500}$, and a slight drop below $\sim$0.01$r_{500}$. Such a drop is also observed in the Fe profile of several individual objects (Figs. \ref{fig:Fe_profiles_indiv_clusters} and \ref{fig:Fe_profiles_indiv_groups}) and is discussed in Sect. \ref{sect:Fe_drop}. The very large total exposure time of the sample ($\sim$4.5 Ms) makes the combined statistical uncertainties $\sigma_\text{stat}(k)$ very small --- less than 1\% in the core, up to $\sim$7\% in the outermost radial bin. The scatter of the measurements (grey shaded area in Fig. \ref{fig:Fe_radial_stacked_profile}), expressed as
\begin{align}
\sigma_\text{scatter}(k) = \sqrt{ \sum_{j=1}^{N} \sum_{i=1}^{8} w_{i,j,k} \bigg( \frac{X(i)_j - X_\text{ref}(k)}{\sigma_{\text{X}(i)_j}} \bigg)^2 } \nonumber \\
\Bigg/ \sqrt{ \sum_{j=1}^{N} \sum_{i=1}^{8} w_{i,j,k} \, \frac{1}{\sigma_{\text{X}(i)_j}^2} }
\end{align}
for each $k$-th reference bin, is much larger (up to $\sim$36\% in the innermost bin).

\begin{table}
\begin{centering}
\caption{Average radial Fe abundance profile for the full sample, as shown in Fig. \ref{fig:Fe_radial_stacked_profile}.}             
\label{table:Fe_radial_stacked_profile}
\setlength{\tabcolsep}{12pt}
\scalebox{1}{
\begin{tabular}{r@{ -- }l c c c}        
\hline \hline                
\multicolumn{2}{c}{Radius} & Fe & $\sigma_\text{stat}$ & $\sigma_\text{scatter}$\\    
\multicolumn{2}{c}{($/r_{500}$)} &  &  & \\    

\hline                        
0 & 0.0075 & $0.802$ & $0.005$ & $0.261$\\
0.0075 & 0.014 & $0.826$ & $0.004$ & $0.219$\\
0.014 & 0.02 & $0.825$ & $0.004$ & $0.197$\\
0.02 & 0.03 & $0.813$ & $0.003$ & $0.177$\\
0.03 & 0.04 & $0.788$ & $0.003$ & $0.160$\\
0.04 & 0.055 & $0.736$ & $0.003$ & $0.149$\\
0.055 & 0.065 & $0.684$ & $0.004$ & $0.129$\\
0.065 & 0.09 & $0.627$ & $0.003$ & $0.124$\\
0.09 & 0.11 & $0.568$ & $0.004$ & $0.099$\\
0.11 & 0.135 & $0.520$ & $0.004$ & $0.104$\\
0.135 & 0.16 & $0.480$ & $0.005$ & $0.104$\\
0.16 & 0.2 & $0.440$ & $0.005$ & $0.096$\\
0.2 & 0.23 & $0.421$ & $0.006$ & $0.082$\\
0.23 & 0.3 & $0.380$ & $0.006$ & $0.086$\\
0.3 & 0.55 & $0.304$ & $0.006$ & $0.090$\\
0.55 & 1.22 & $0.205$ & $0.011$ & $0.105$\\

\hline                                   
\end{tabular}}
\par\end{centering}

\end{table}

We parametrise this profile by fitting the empirical function
\begin{equation}
\label{eq:parametrised_profile}
\text{Fe}(r) = A (r-B)^{C} - D \exp \left( -\frac{(r-E)^2}{F} \right) \, ,
\end{equation}
where $r$ is given in units of $r_{500}$, and $A$, $B$, $C$, $D$, $E$, and $F$ are constants to determine. The first term on the right hand side of Eq. (\ref{eq:parametrised_profile}) is a power law that is used to model the decrease beyond $\gtrsim$0.02$r_{500}$. To model the inner metal drop, we subtract a Gaussian (second term) from the power law. The best fit of our empirical distribution is shown in Fig. \ref{fig:Fe_radial_stacked_profile} (red dashed curve) and can be expressed as
\begin{align}
\label{eq:parametrised_profile2}
\text{Fe}(r) = 0.21 (r+0.021)^{-0.48} - 6.54 \exp \left( -\frac{(r+0.0816)^2}{0.0027} \right) \, ,
\end{align}
which provides a reasonable fit to the data ($\chi^2$/d.o.f. = 10.3/9). We also look for possible hints towards a flattening at  the outskirts. When assuming a positive Fe floor in the outskirts (by injecting an additive constant $G$ into Eq. (\ref{eq:parametrised_profile2})), the fit does not improve ($\chi^2$/d.o.f. = 10.3/10, with $G=0.009$) and remains comparable to the former case. Therefore, our data do not allow us to formally confirm the presence of a uniform Fe distribution in the outskirts. The empirical Fe abundance profile of Eq. (\ref{eq:parametrised_profile2}) is compared to the radial profiles of other elements further in our analysis (Sect. \ref{sect:results_abun}).

\begin{figure}[!]
        \centering
                \includegraphics[width=0.49\textwidth]{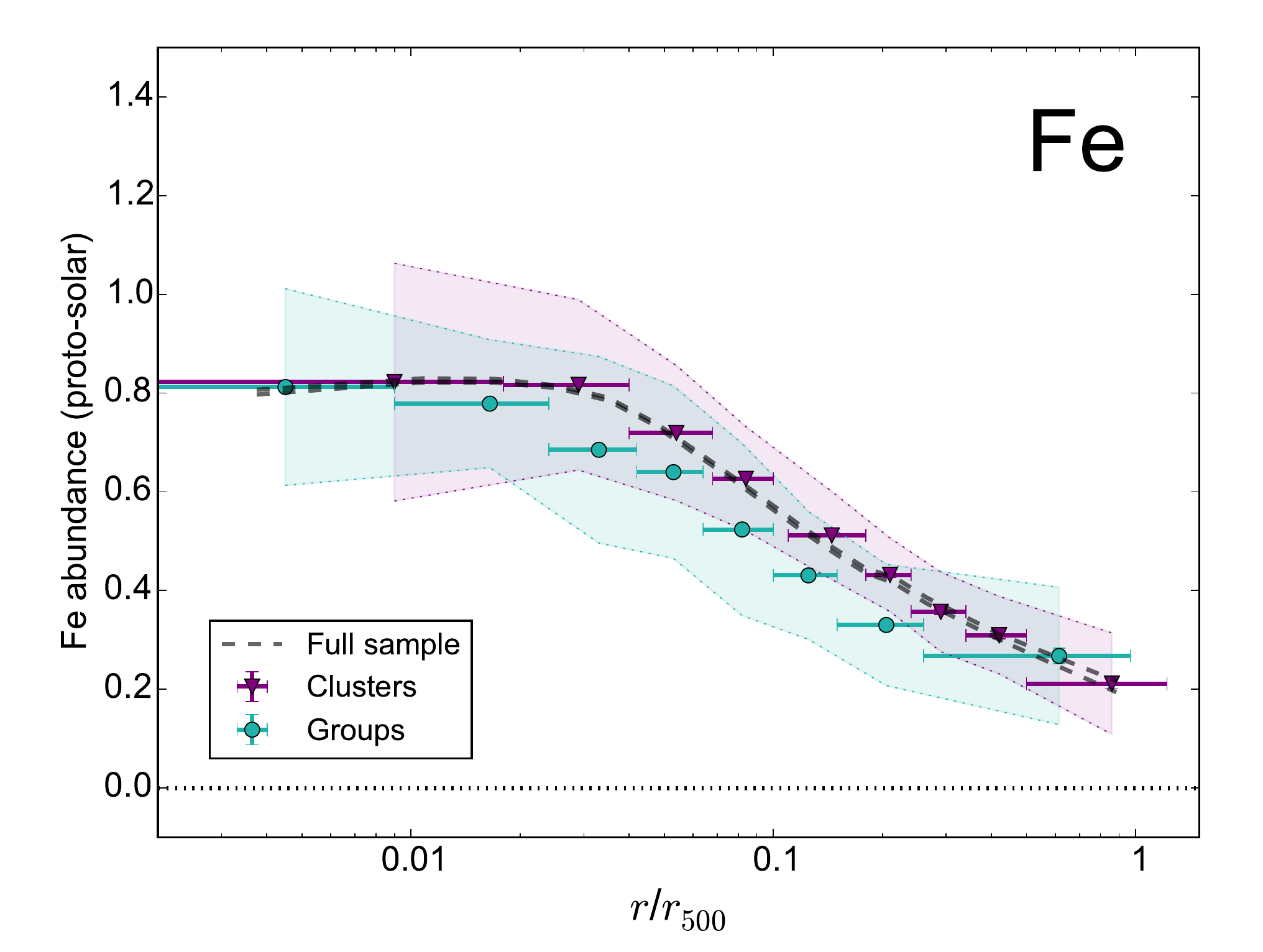}

        \caption{Average Fe profile for clusters (> 1.7 keV, purple) and groups (<1.7 keV, green) within our sample. The corresponding shaded areas show the scatter of the measurements. The two dashed lines indicate the upper and lower statistical error bars of the Fe profile over the full sample (Fig. \ref{fig:Fe_radial_stacked_profile}) without scatter for clarity.}
\label{fig:Fe_radial_stacked_profile_clgr}
\end{figure}

We now compute the average radial Fe abundance profiles separately for the clusters (>1.7 keV) and groups (<1.7 keV) of our sample. The result is shown in Fig. \ref{fig:Fe_radial_stacked_profile_clgr} (where the dashed lines indicate the average profile over the full sample) and Table \ref{table:Fe_radial_stacked_profile_clgr}. The Fe abundance in clusters and groups can be robustly constrained out to $\sim$$0.9 r_{500}$ and $\sim$$0.6 r_{500}$, respectively, and also show a clear decrease with radius. Although both profiles show a similar slope, we note that at each radius, the average Fe abundance for groups is systematically lower than for clusters. The two exceptions are the innermost radial bin (where the cluster and group Fe abundances show consistent values) and the outermost radial bin of these two profiles (where the group Fe abundances appear somewhat higher than in clusters). We discuss this further in Sect. \ref{sect:discussion_clgr}.

\begin{table}
\begin{centering}
\caption{Average radial Fe abundance profile for clusters (>1.7 keV) and groups (<1.7 keV), as shown in Fig. \ref{fig:Fe_radial_stacked_profile_clgr}.}             
\label{table:Fe_radial_stacked_profile_clgr}
\setlength{\tabcolsep}{12pt}
\scalebox{1}{
\begin{tabular}{r@{ -- }l c c c}        
\hline \hline                
\multicolumn{2}{c}{Radius} & Fe & $\sigma_\text{stat}$ & $\sigma_\text{scatter}$\\    
\multicolumn{2}{c}{($/r_{500}$)} &  &  & \\    

\hline                        
\multicolumn{5}{c}{Clusters} \\
\hline
0 & 0.018       & $0.822$ & $0.003$ & $0.241$\\
0.018 & 0.04 & $0.8167$ & $0.0020$ & $0.1725$\\
0.04 & 0.068 & $0.7190$ & $0.0022$ & $0.1369$\\
0.068 & 0.1   & $0.626$ & $0.003$ & $0.106$\\
0.1 & 0.18     & $0.511$ & $0.003$ & $0.089$\\
0.18 & 0.24   & $0.432$ & $0.005$ & $0.075$\\
0.24 & 0.34   & $0.357$ & $0.006$ & $0.081$\\
0.34 & 0.5     & $0.309$ & $0.008$ & $0.079$\\
0.5 & 1.22        & $0.211$ & $0.011$ & $0.102$\\

\hline                        
\multicolumn{5}{c}{Groups} \\
\hline
0 & 0.009         & $0.812$ & $0.009$ & $0.199$\\
0.009 & 0.024 & $0.779$ & $0.005$ & $0.130$\\
0.024 & 0.042 & $0.685$ & $0.007$ & $0.189$\\
0.042 & 0.064 & $0.640$ & $0.009$ & $0.175$\\
0.064 & 0.1     & $0.524$ & $0.007$ & $0.175$\\
0.1 & 0.15       & $0.430$ & $0.007$ & $0.129$\\
0.15 & 0.26     & $0.330$ & $0.010$ & $0.133$\\
0.26 & 0.97       & $0.268$ & $0.016$ & $0.139$\\

\hline                                   
\end{tabular}}
\par\end{centering}

\end{table}

\subsection{Abundance profiles of other elements}\label{sect:results_abun}

While the Fe-L and Fe-K complexes, which are both accessible in the X-ray band, make the Fe abundance rather easy to estimate with a good degree of accuracy, the other elements considered in this paper (O, Mg, Si, S, Ar, Ca, and Ni) can be measured by CCD instruments only via their K-shell main emission lines. Consequently, their radial abundance profiles are in general difficult to constrain in the ICM of individual objects. The deep total exposure of our sample allows us to derive the average radial abundance profiles of elements other than Fe, which we present in this section.

\begin{figure*}[!]
        \centering
                \includegraphics[width=0.42\textwidth]{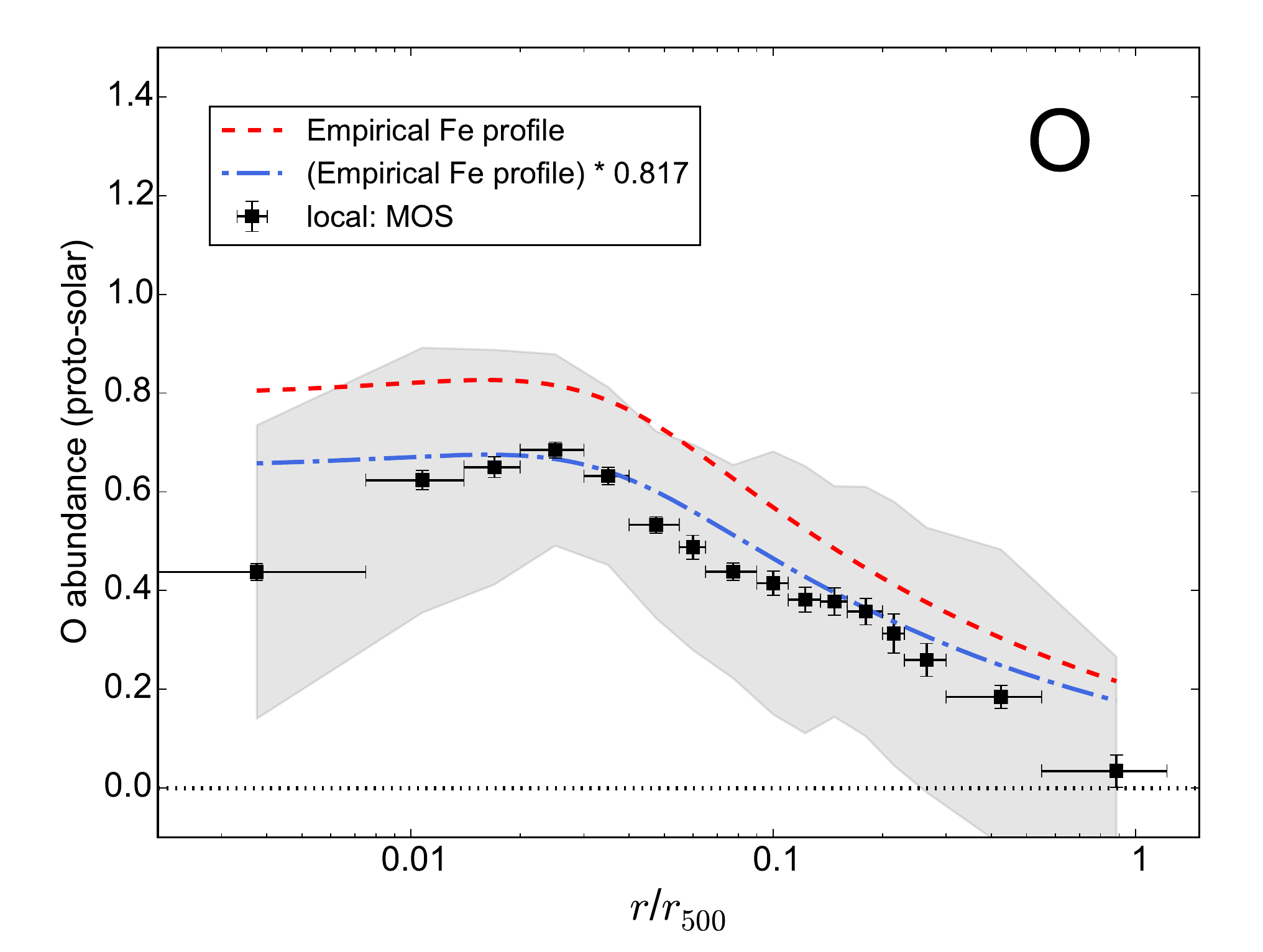}
                \includegraphics[width=0.42\textwidth]{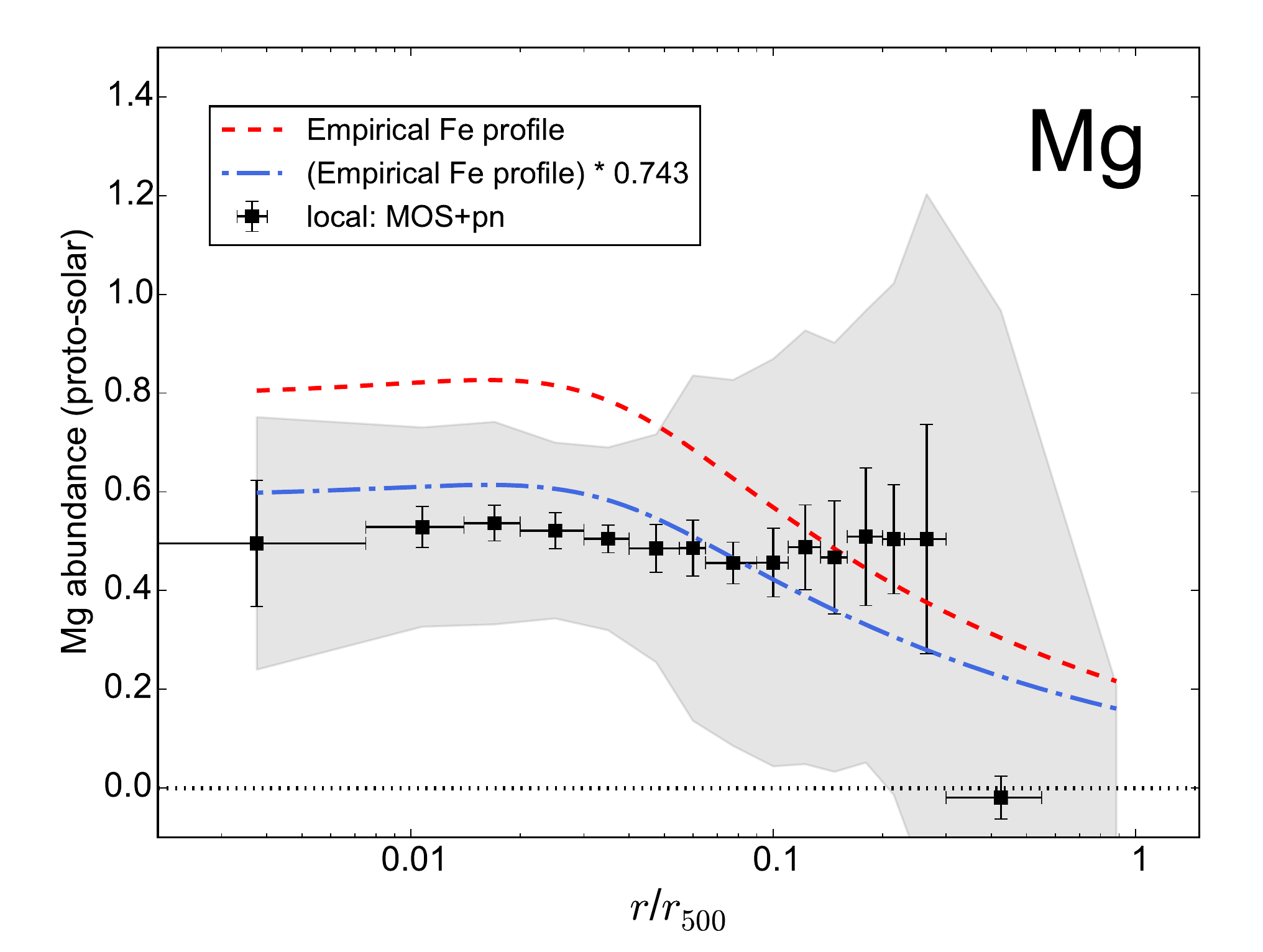}
\\
                \includegraphics[width=0.42\textwidth]{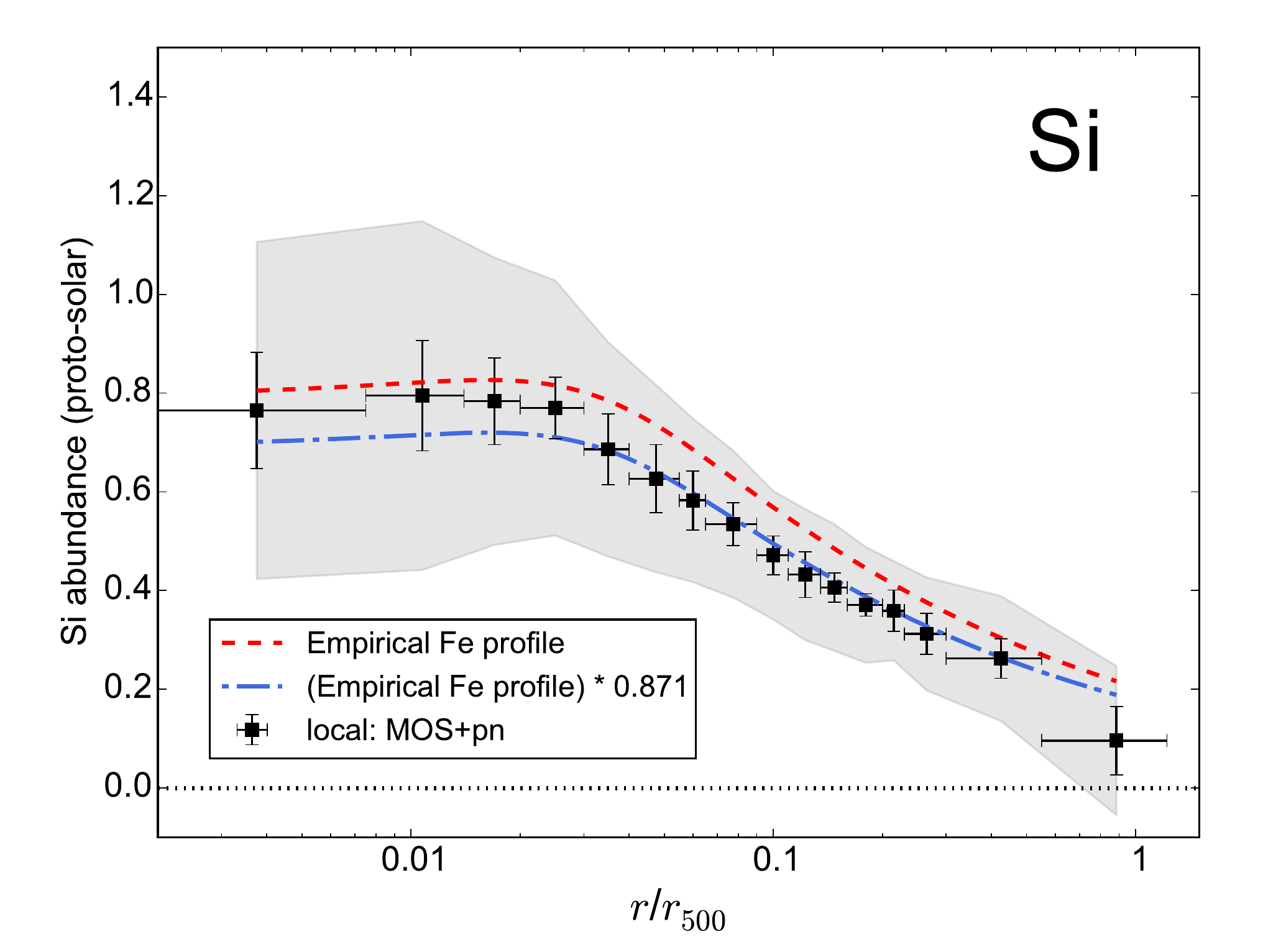}
                \includegraphics[width=0.42\textwidth]{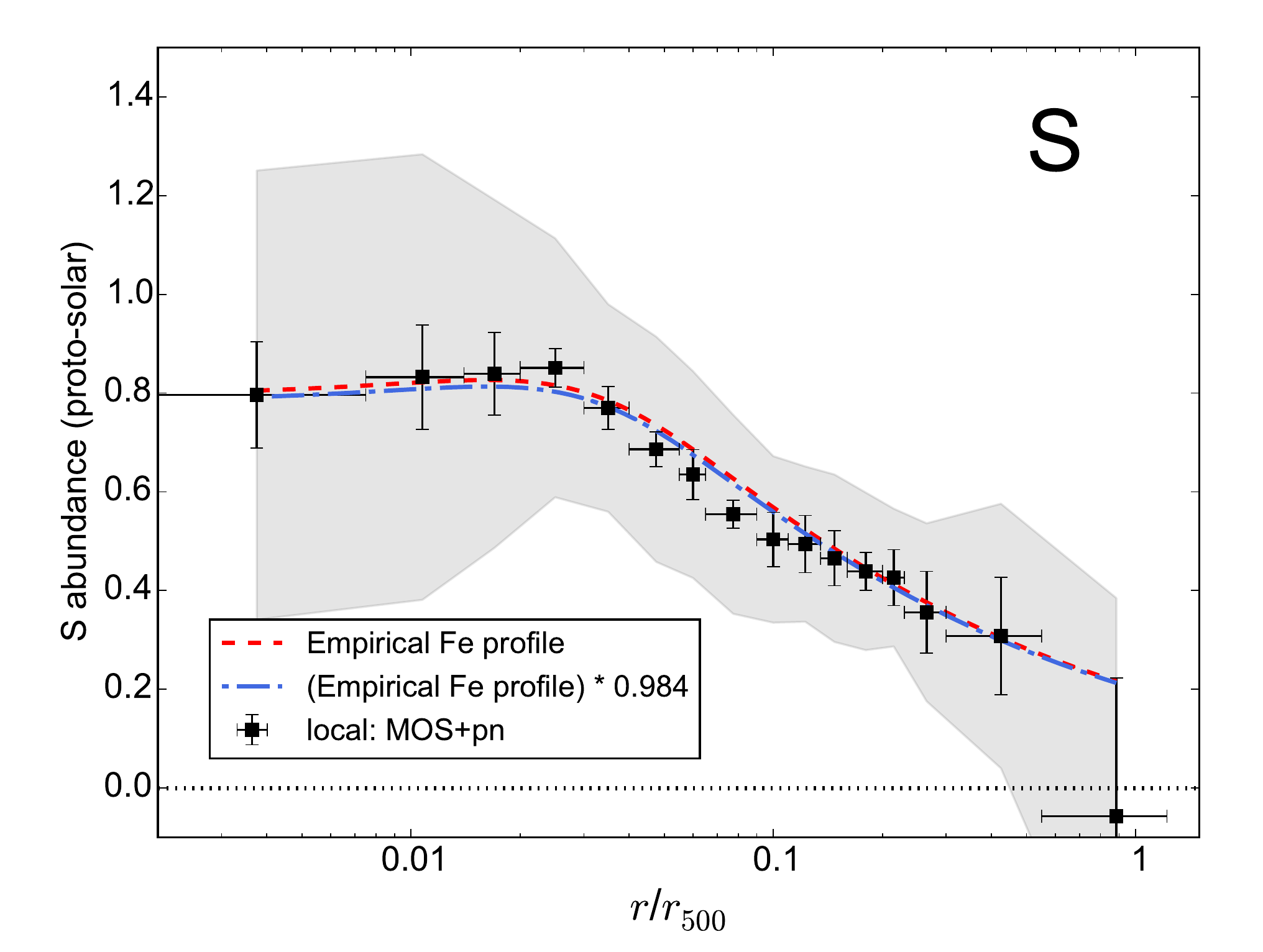}
\\
                \includegraphics[width=0.42\textwidth]{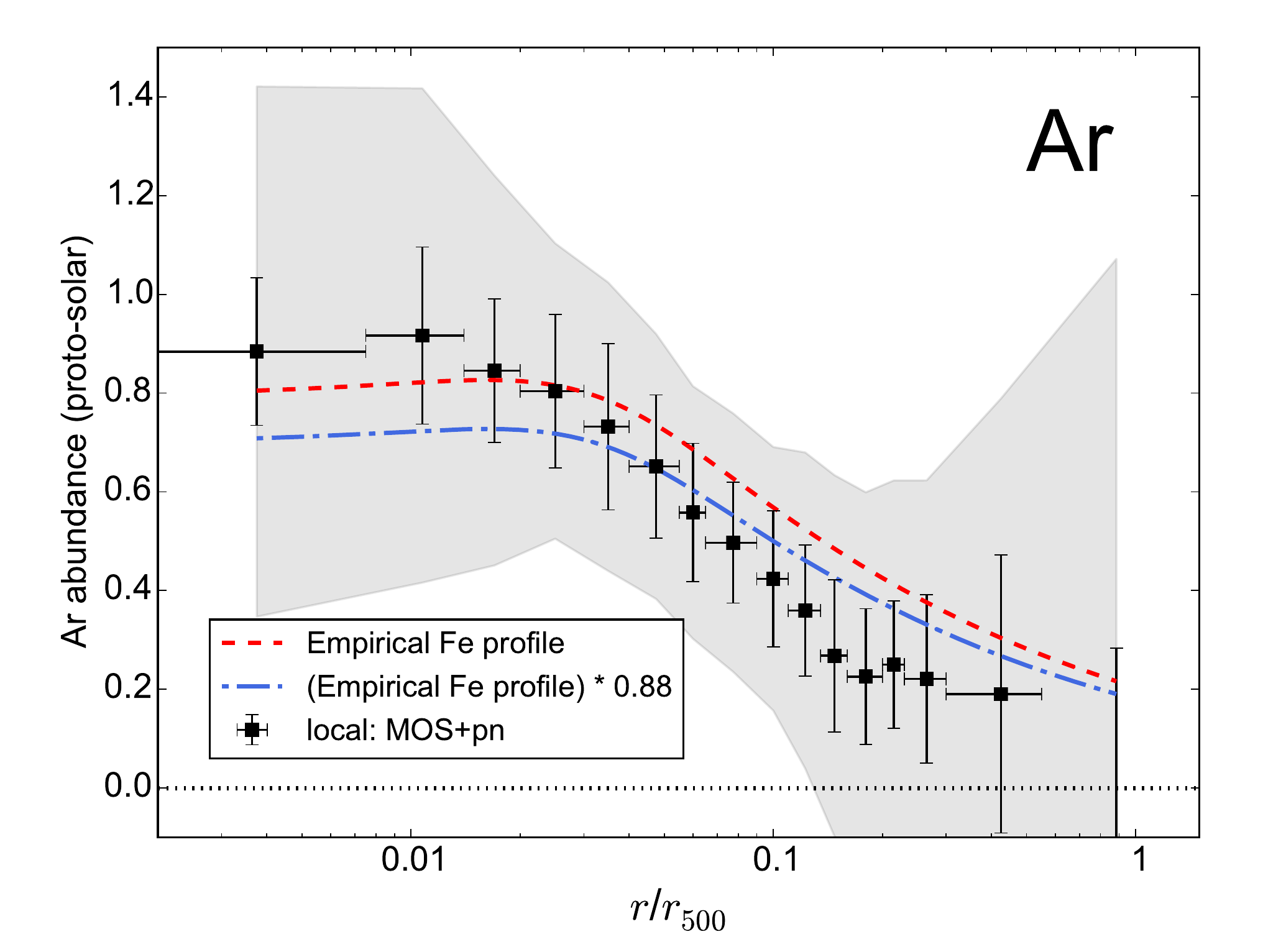}
                \includegraphics[width=0.42\textwidth]{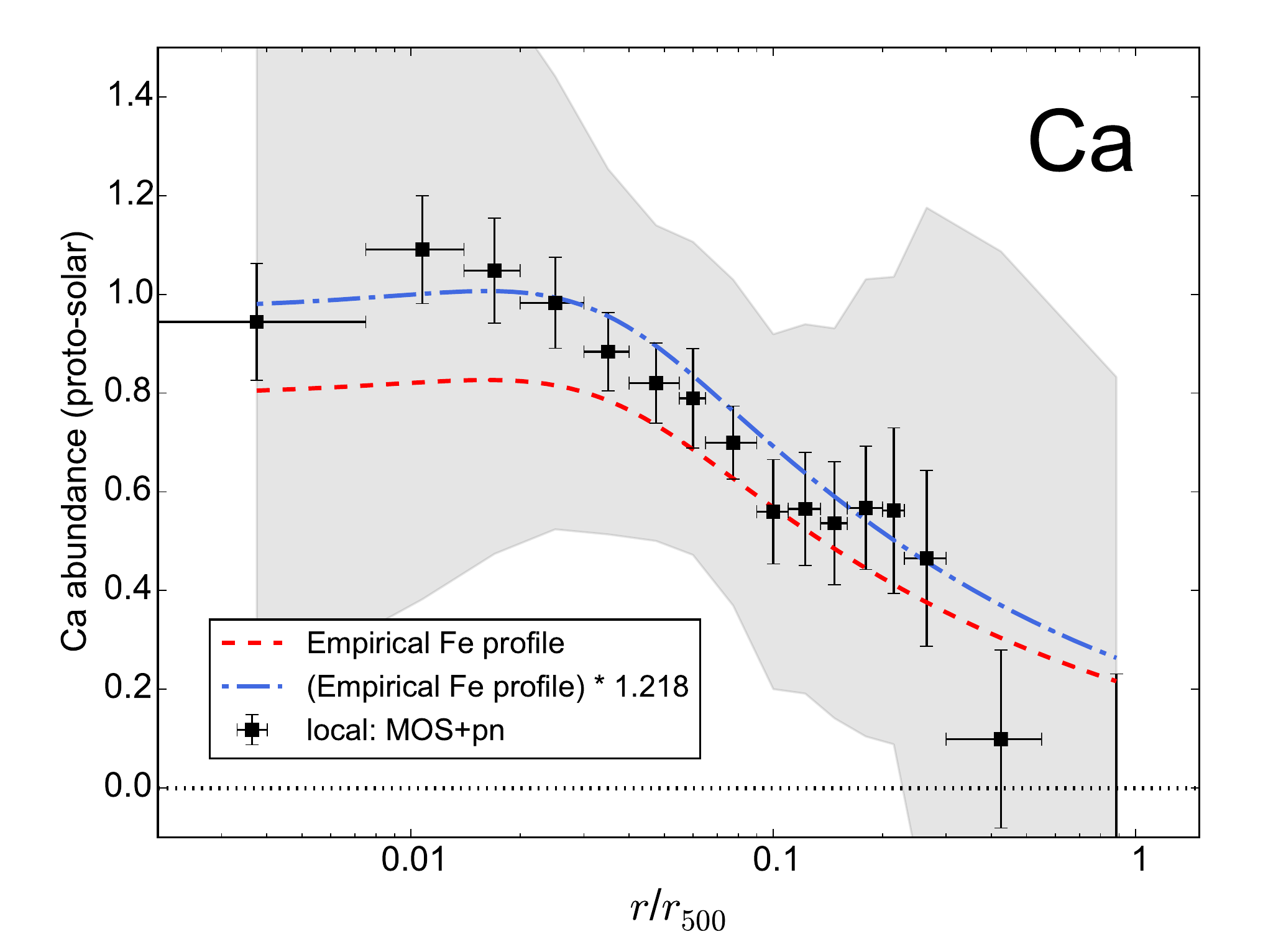}
\\
                \includegraphics[width=0.42\textwidth]{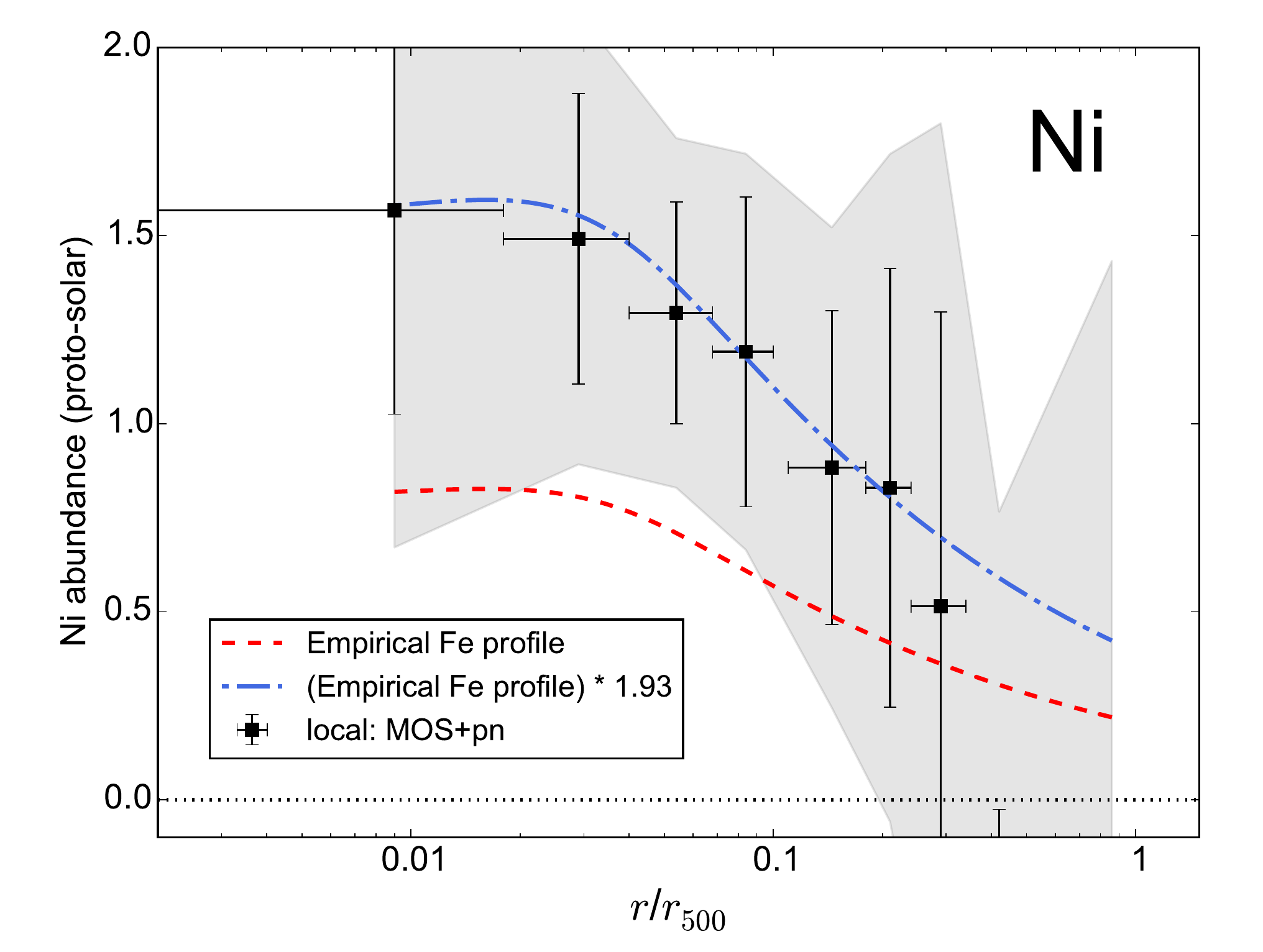}

        \caption{Average radial abundance profiles of all the objects in our sample. The error bars contain the statistical uncertainties and MOS-pn uncertainties (Sect. \ref{sect:MOS-pn_uncertainties}) except for the O abundance profiles, which are only measured with MOS. The corresponding shaded areas show the scatter of the measurements. The Ni profile has only been averaged for clusters (>1.7 keV).}
\label{fig:abundance_profiles}
\end{figure*}

First, and similarly to Fig. \ref{fig:Fe_radial_stacked_profile}, we compute and compare the radial profiles of O, Mg, Si, S, Ar, and Ca, averaged over the full sample. The Ni profile could only be estimated for clusters because the lower temperature of groups and ellipticals prevents a clear detection of the Ni K-shell emission lines. These profiles are shown in Fig. \ref{fig:abundance_profiles} and their numerical values can be found in Table \ref{table:abundance_profiles}. A question of interest is whether these derived profiles follow the shape of the average Fe profile. This can be checked by comparing these radial profiles to the empirical Fe($r$) profile proposed in Eq. (\ref{eq:parametrised_profile2}) and Fig. \ref{fig:Fe_radial_stacked_profile}, shown by the red dashed lines in Fig. \ref{fig:abundance_profiles}. Obviously, the average profile of an element X is not expected to strictly follow the average Fe profile, as the X/Fe ratios may be larger or smaller than unity. A more consistent comparison would be thus to define the empirical X($r$) profiles as 
\begin{equation}
\text{X}(r) = \eta \text{Fe}(r) \, ,
\end{equation}
where $\eta$ is the average X/Fe ratio estimated using our sample, within $0.2 r_{500}$ when possible or $0.05 r_{500}$ otherwise, and tabulated in \citet[][see their Table 2]{2016A&A...592A.157M}. These normalised empirical profiles are shown by the blue dashed lines in Fig. \ref{fig:abundance_profiles} and can be directly compared with our observational data. 

The case of Si is particularly striking, as we find a remarkable agreement (<1$\sigma$) between our measurements and the empirical Si($r$) profile in all the radial bins, except the outermost one (<2$\sigma$). Within $\sim$0.5$r_{500}$, the Ca and Ni profiles follow their empirical counterparts very well (<2$\sigma$).

The O, Mg, and S profiles are somewhat less consistent with their respective X($r$) profiles. The O central drop is significantly more pronounced than the Fe drop, while the Mg profile does not show any clear central drop and appears significantly shallower than expected (blue dashed line). Finally, the S measured profile falls somewhat below the empirical prediction within 0.04--0.1$r_{500}$. However, such discrepancies are almost entirely introduced by a few specific observations. As we show further in Sect. \ref{sect:systematics_weight}, when ignoring (temporarily) these single observations from our sample, a very good agreement is obtained between the data and empirical profiles, both for O, Mg, and S. Moreover, the large plotted error bars at outer radii in the Mg profile are almost entirely due to the MOS-pn discrepancies; while the MOS measurements (located at the lower side of the error bars) follow very well the empirical profile, the pn measurements (located at the upper side of the error bars) increase with radius; this is probably because of contamination of the Mg line with the instrumental Al-K$\alpha$ line (see Sect. \ref{sect:systematics_EPIC} for an extended discussion).
Finally, as we show further in this section, the average O/Fe, Mg/Fe, and S/Fe profiles (compiled from O/Fe and Mg/Fe  measurements of individual observations) show a good agreement with being radially flat. 

The case of Ar is the most interesting one. Despite the large error bars (only covering the MOS-pn discrepancies), the average radial slope of this element appears systematically steeper than its empirical profile. A similar behaviour is found in the average Ar/Fe profile (see further). Unlike the O, Mg, and S profiles, we cannot suppress this overall trend by discarding a few specific objects from the sample (Sect. \ref{sect:systematics_weight}). Although we discuss one possible reason for these differences in Sect. \ref{sect:Fe_drop}, we note that they cannot be confirmed when the scatters are taken into account. 

We also note that in many cases, the average measured abundances in the outermost radial bin are systematically biased low with respect to the empirical prediction. As we show below, this feature is also reported in most of the X/Fe profiles. While at these large distances the scatter is very large and still consistent with the empirical expectations, these values that are systematically lower than expected may emphasise the radial limits beyond which the background uncertainties prevent any robust measurement (see Sect. \ref{sect:systematics_bg}).

\begin{figure*}[!]
        \centering
                \includegraphics[width=0.41\textwidth]{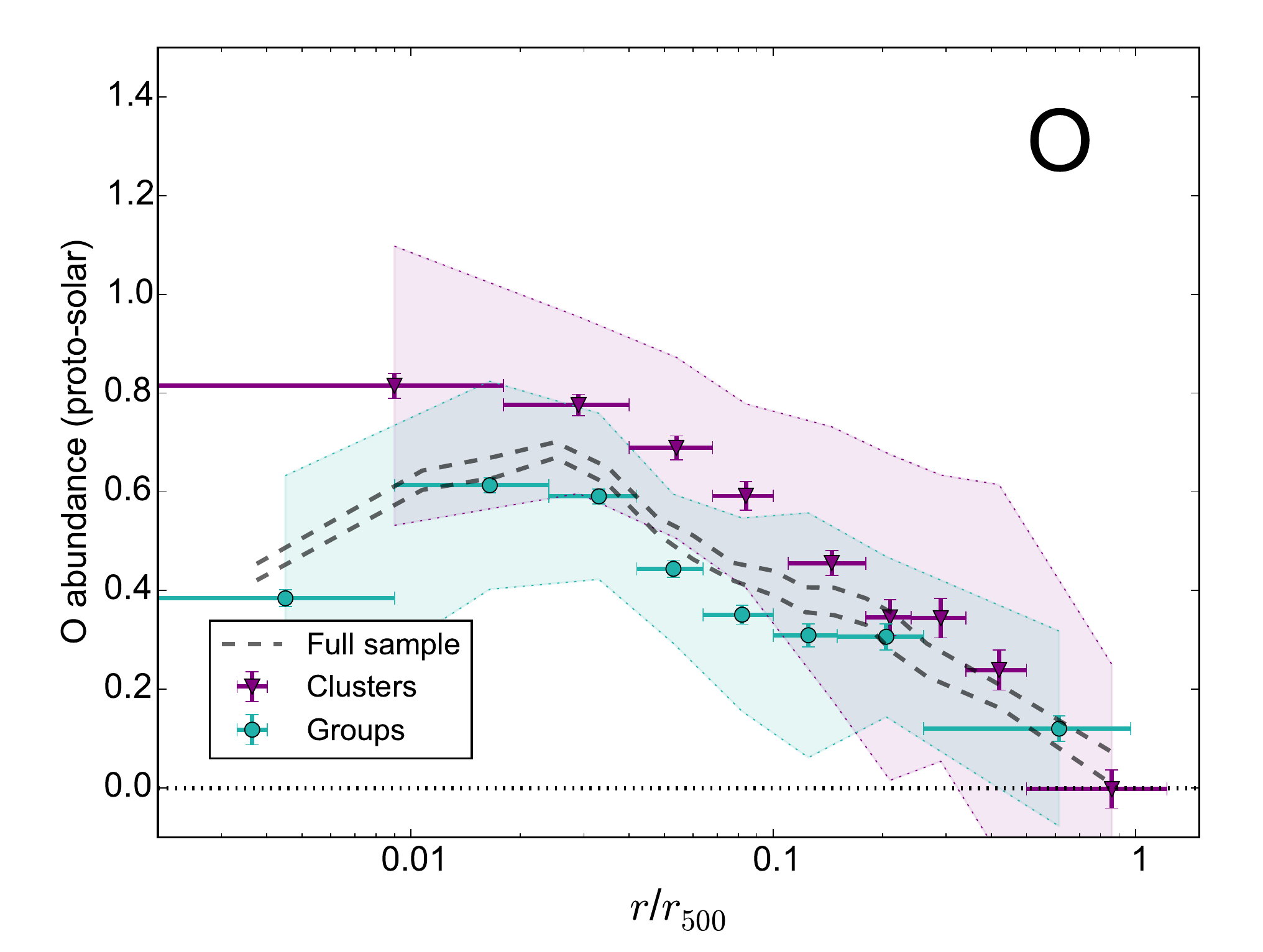}
                \includegraphics[width=0.41\textwidth]{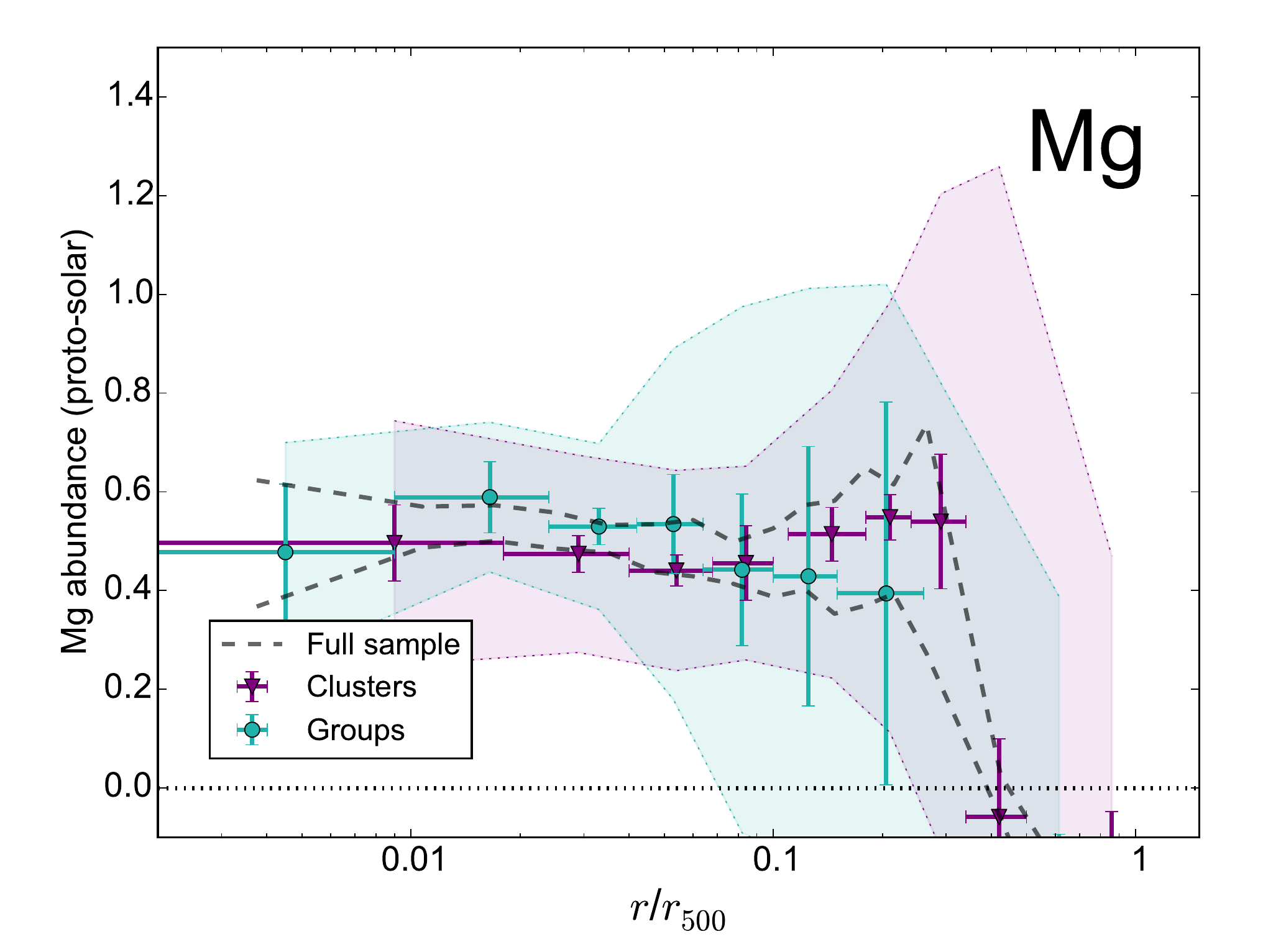}
\\
                \includegraphics[width=0.41\textwidth]{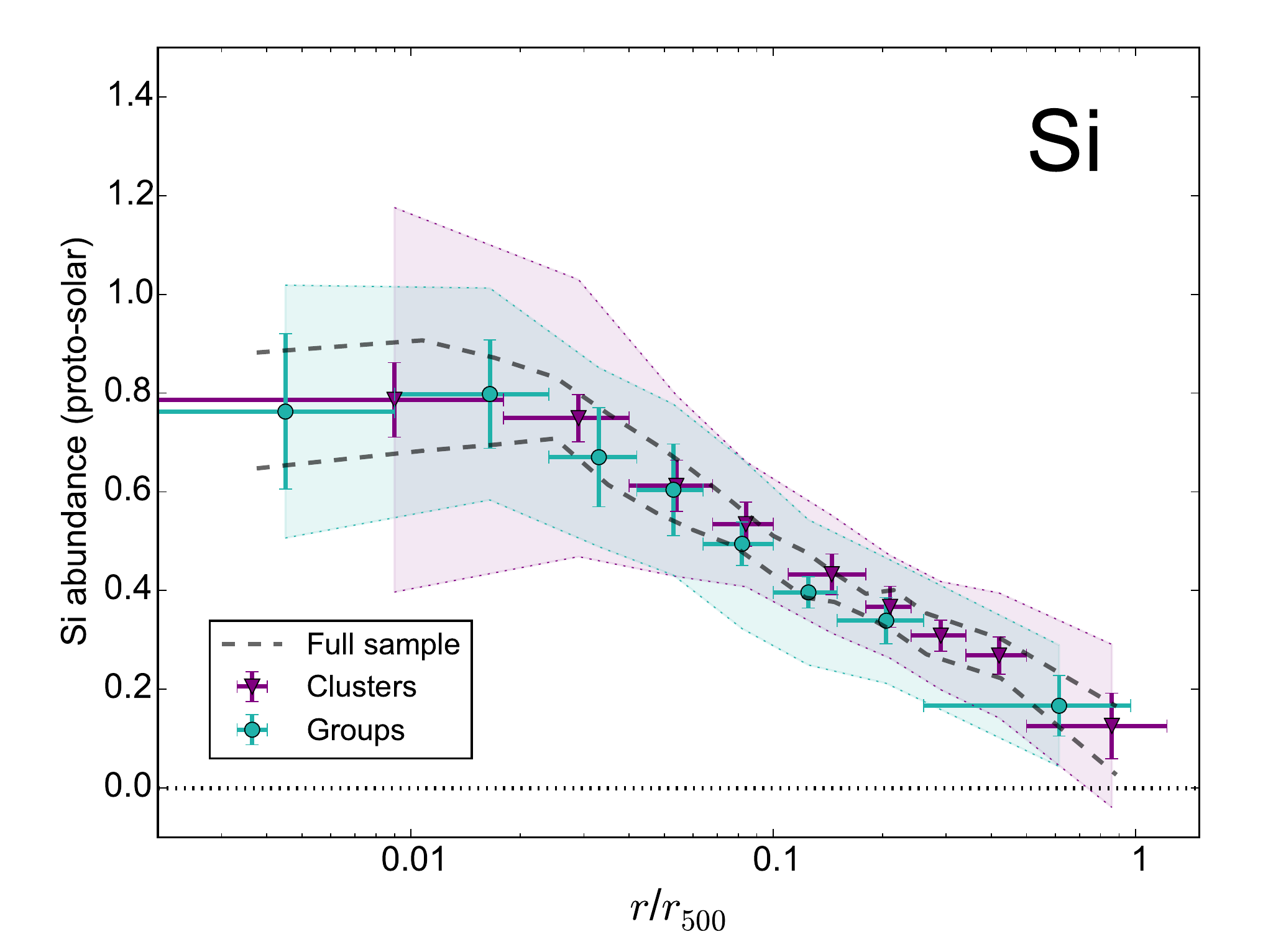}
                \includegraphics[width=0.41\textwidth]{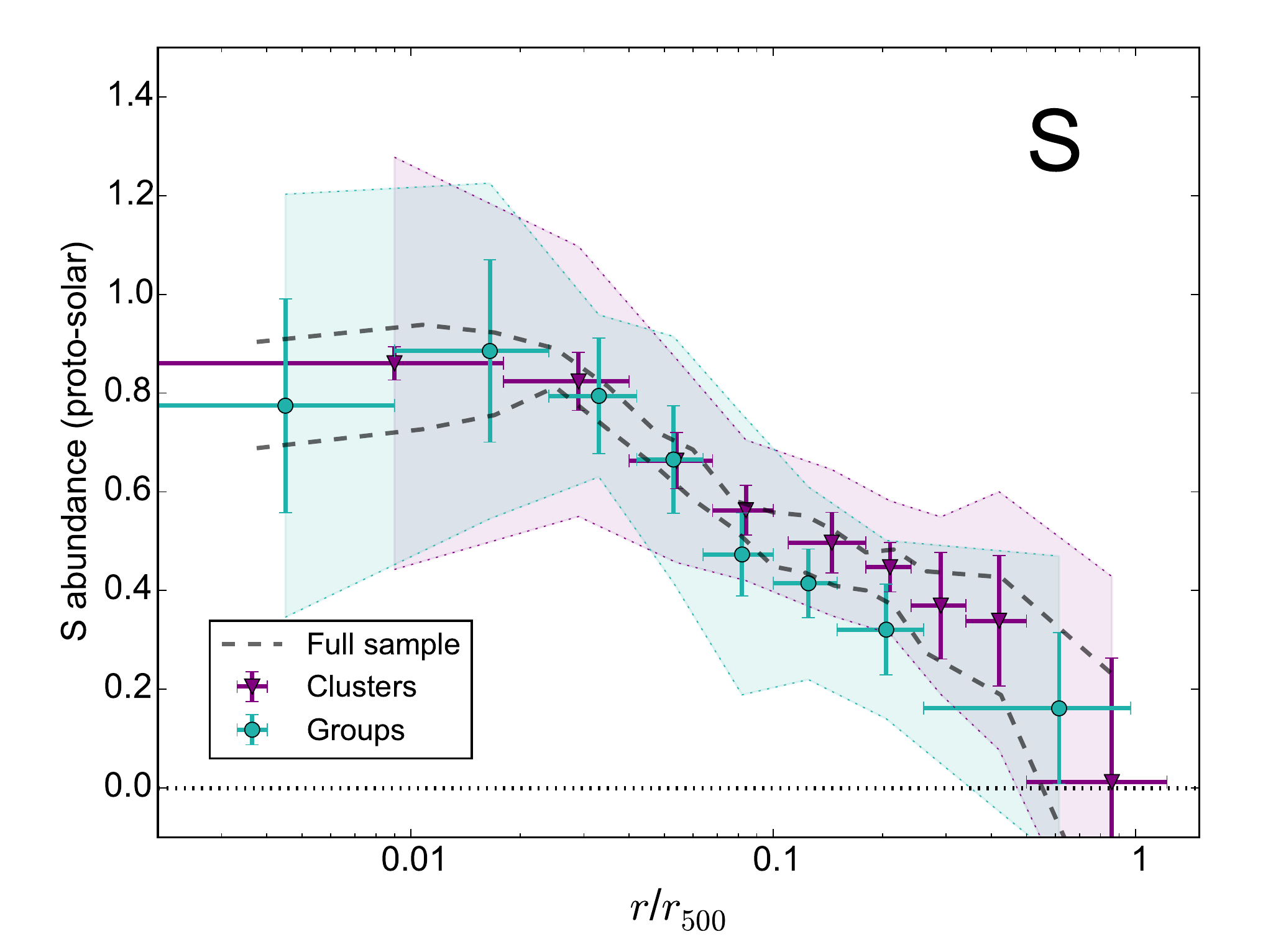}
\\
                \includegraphics[width=0.41\textwidth]{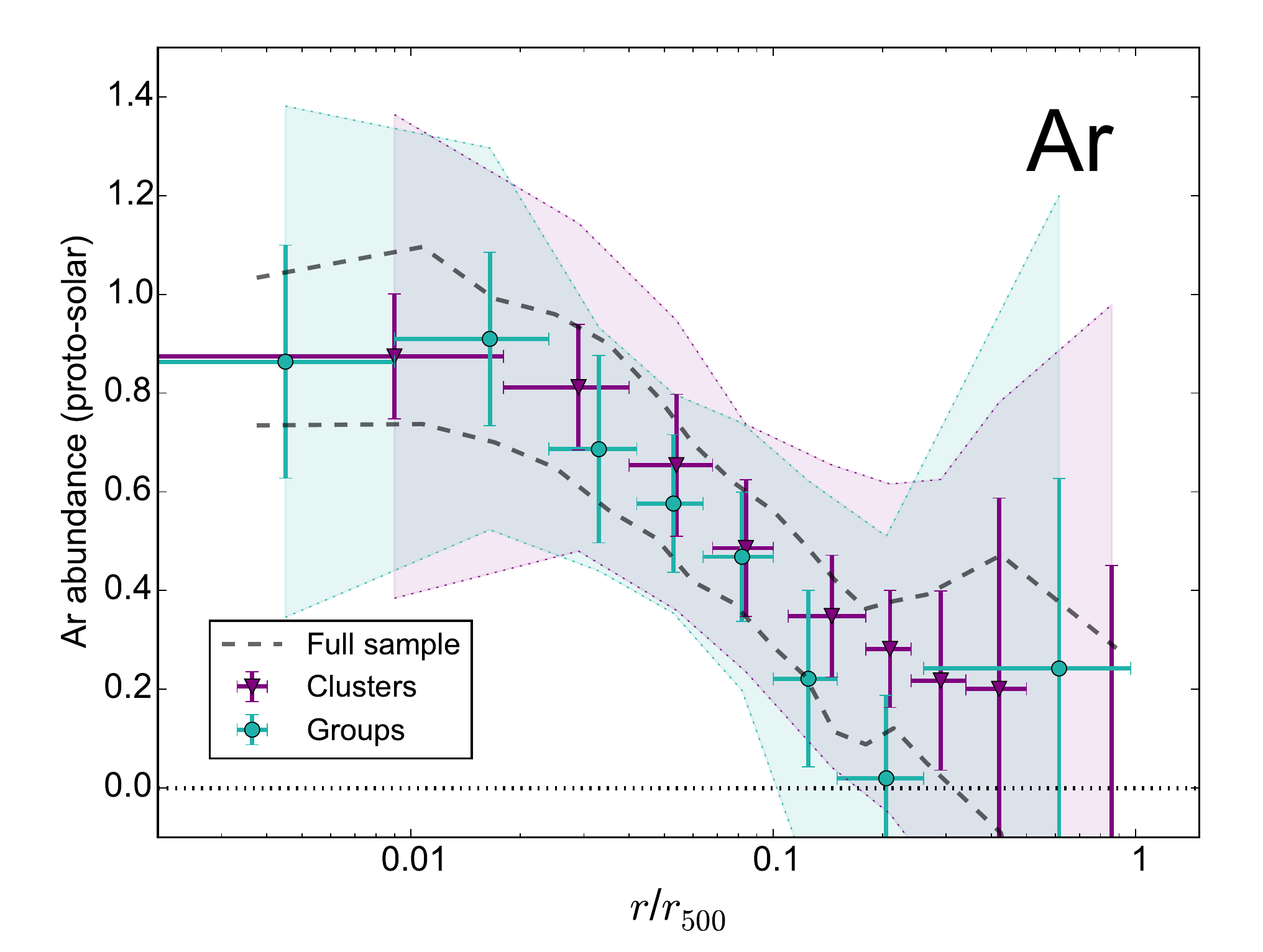}
                \includegraphics[width=0.41\textwidth]{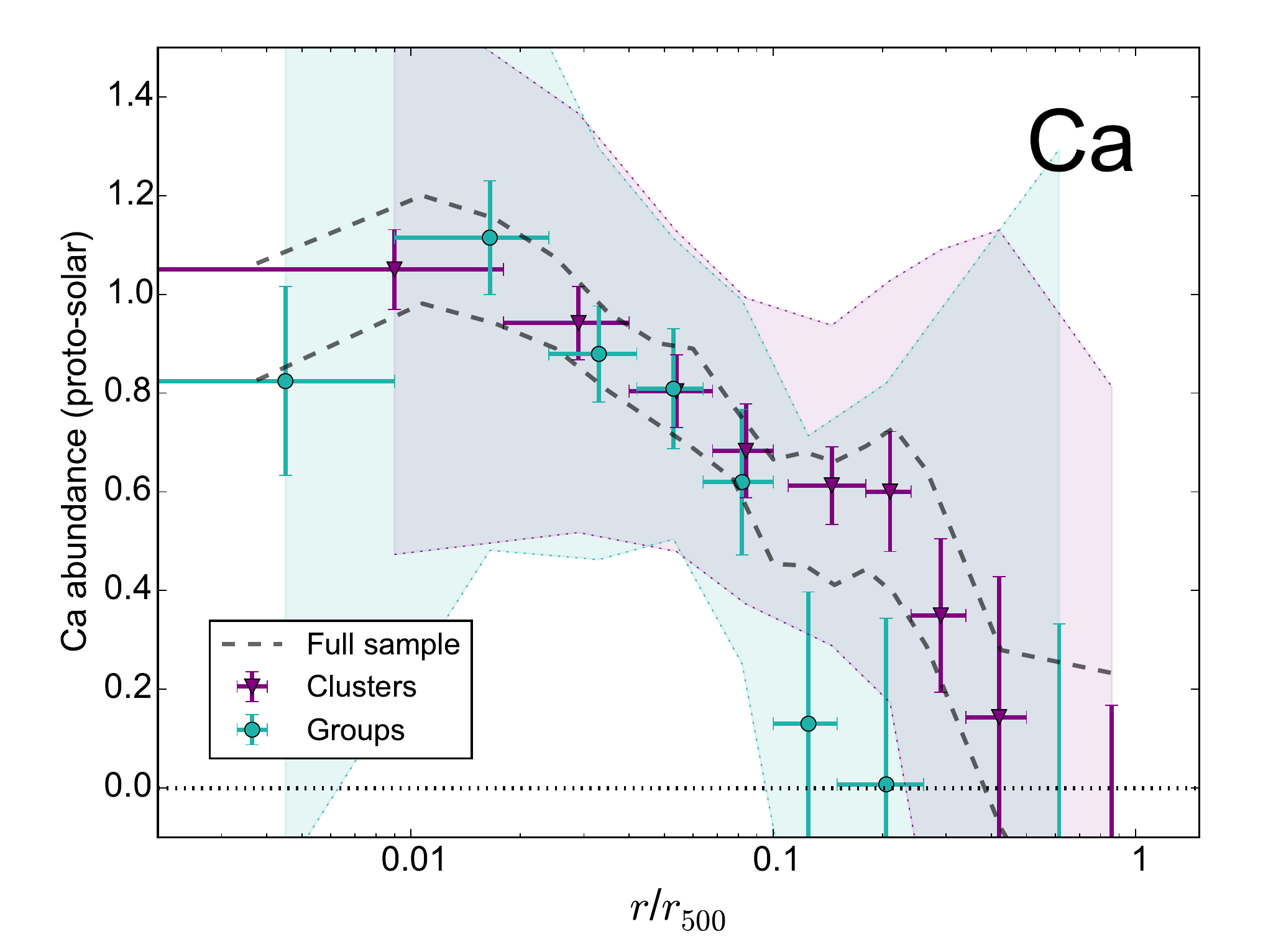}

        \caption{Comparison of the average abundance radial profiles between clusters (>1.7 keV) and groups/ellipticals (<1.7 keV). The error bars contain the statistical uncertainties and MOS-pn uncertainties (Sect. \ref{sect:MOS-pn_uncertainties}) except for the O abundance profiles, which are only measured with MOS. The corresponding shaded areas show the scatter of the measurements. The two dashed lines indicate the upper and lower error bars of the corresponding profiles over the full sample (Fig. \ref{fig:abundance_profiles}), without scatter for clarity.}
\label{fig:abundance_profiles_clgr}
\end{figure*}

Second, and similarly to Fig. \ref{fig:Fe_radial_stacked_profile_clgr}, we compute the average O, Mg, Si, S, Ar, and Ca abundance profiles (and their respective scatters) for clusters, on the one hand, and for groups, on the other hand. These profiles are shown in Fig. \ref{fig:abundance_profiles_clgr} and Table \ref{table:abundance_profiles_clgr}. For comparison, the average profiles using the full sample (Fig. \ref{fig:abundance_profiles}, without scatter) are also shown (dashed grey lines). All the profiles (groups and clusters) show an abundance decrease towards the outskirts. Globally, the clusters and groups abundance profiles are very similar for a given element. We note, however, the exception of the O profiles, for which the groups show on average a lower level of enrichment (similar to the case of Fe). A drop in the innermost bin for groups is also clearly visible for O (however, see Sect. \ref{sect:systematics_weight}). Moreover, the Ca profile for groups also suggests a drop in the innermost bin, followed by a more rapidly declining profile towards the outskirts. While these global trends are discussed further in Sect. \ref{sect:discussion_clgr}, we must recall that the large scatter of our measurements (shaded areas) prevents us from deriving any firm conclusion regarding possible differences in the cluster versus group profiles presented here.

\begin{figure*}[!]
        \centering
                \includegraphics[width=0.42\textwidth]{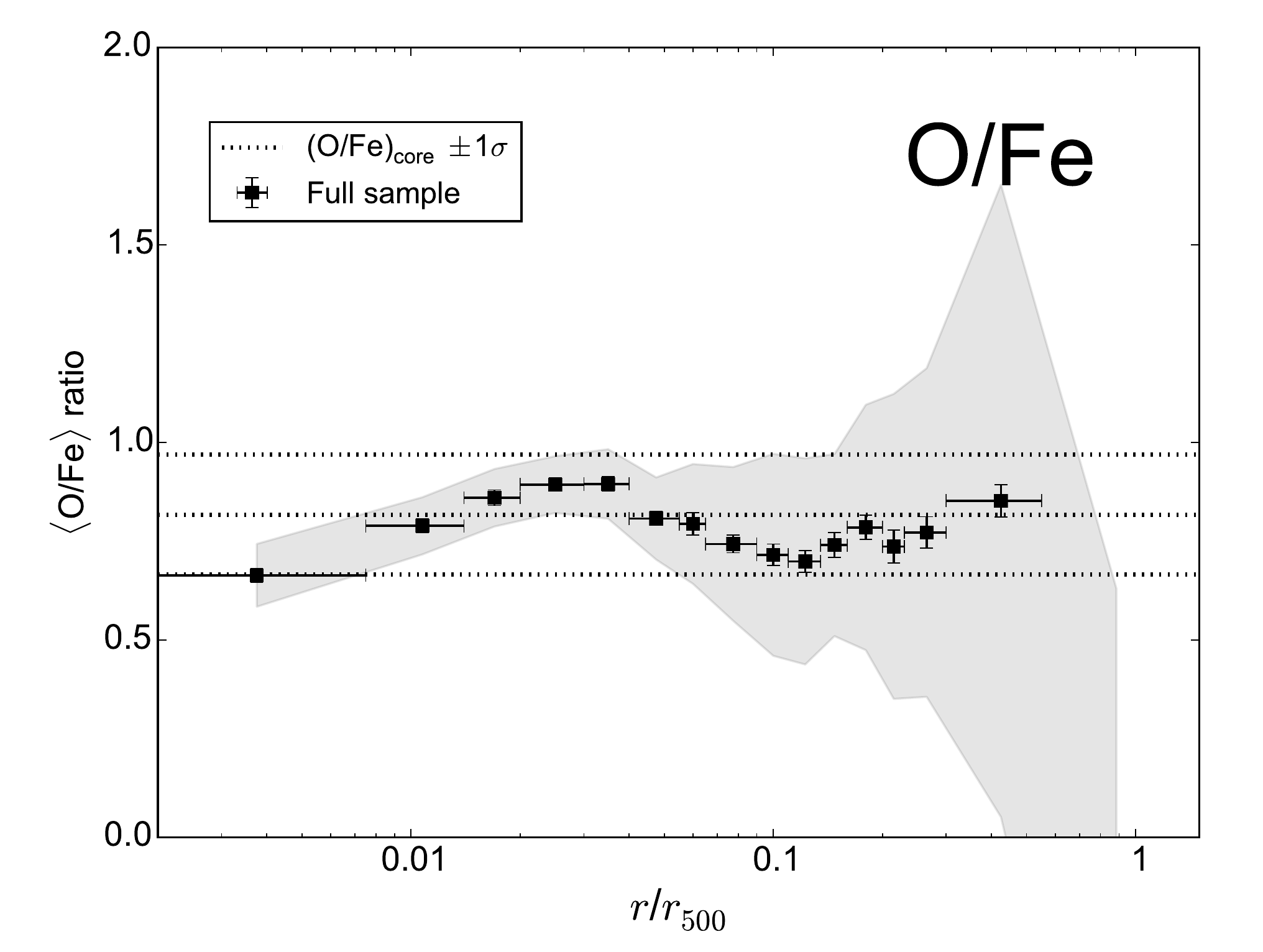}
                \includegraphics[width=0.42\textwidth]{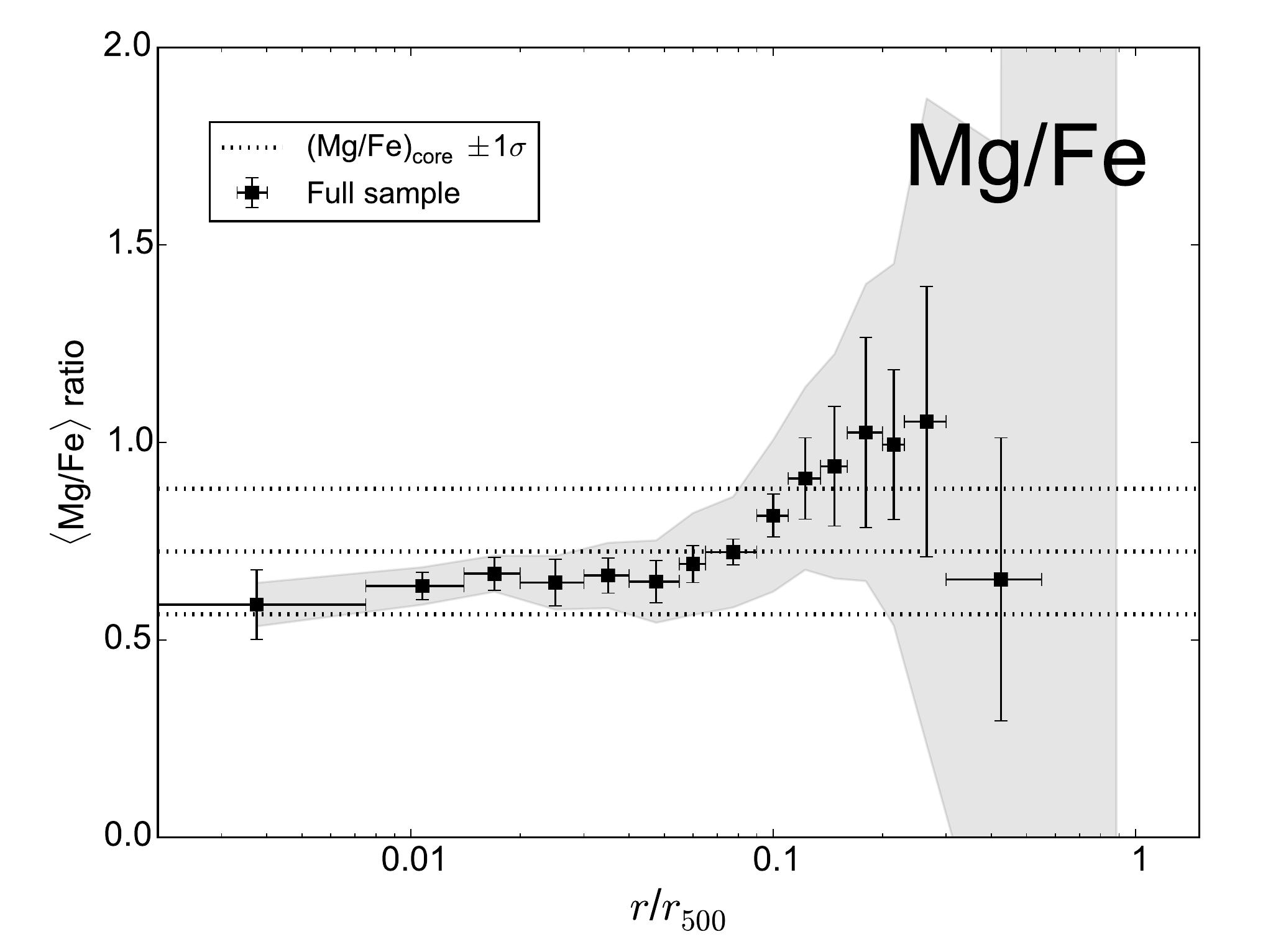}
\\
                \includegraphics[width=0.42\textwidth]{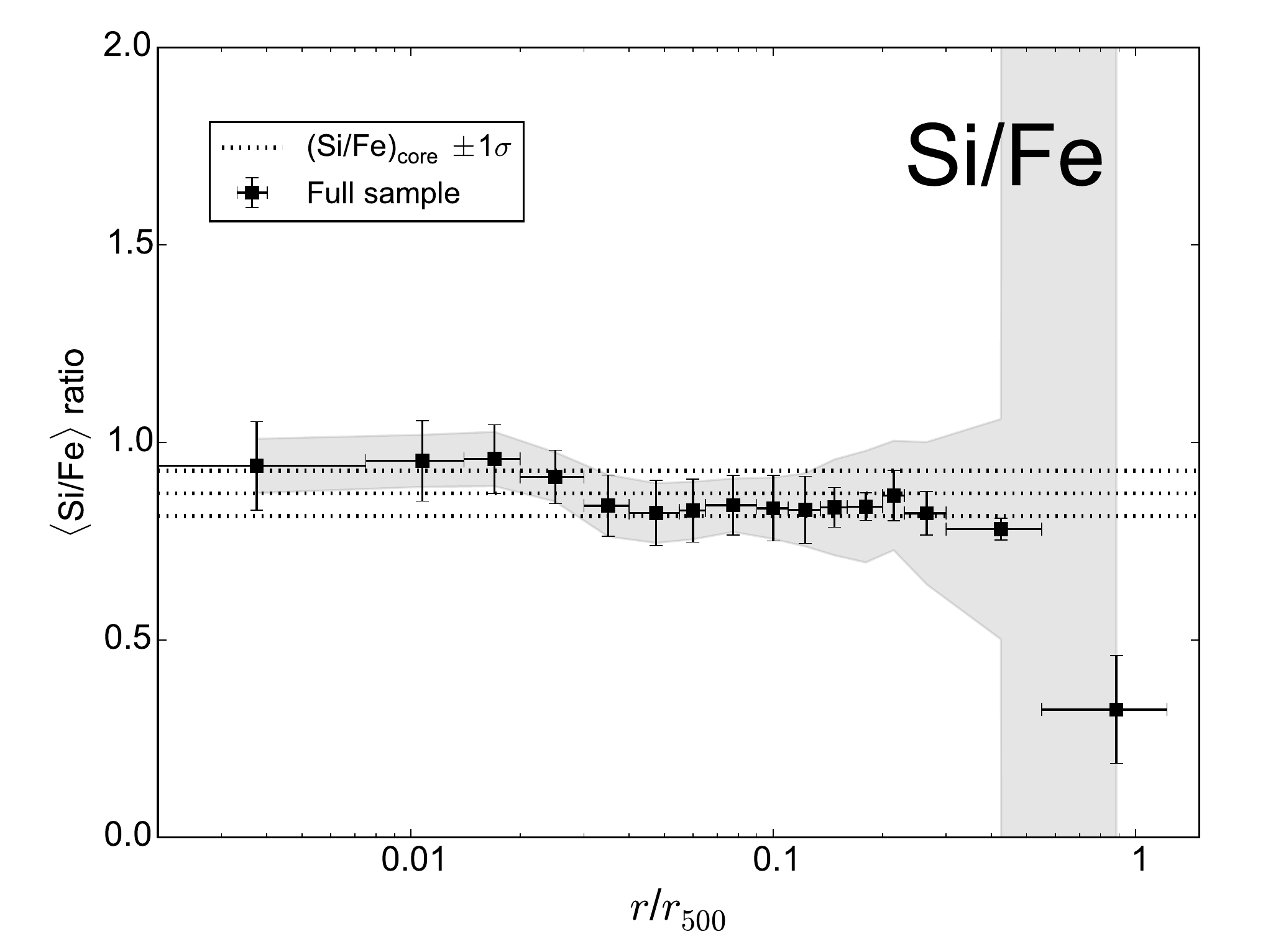}
                \includegraphics[width=0.42\textwidth]{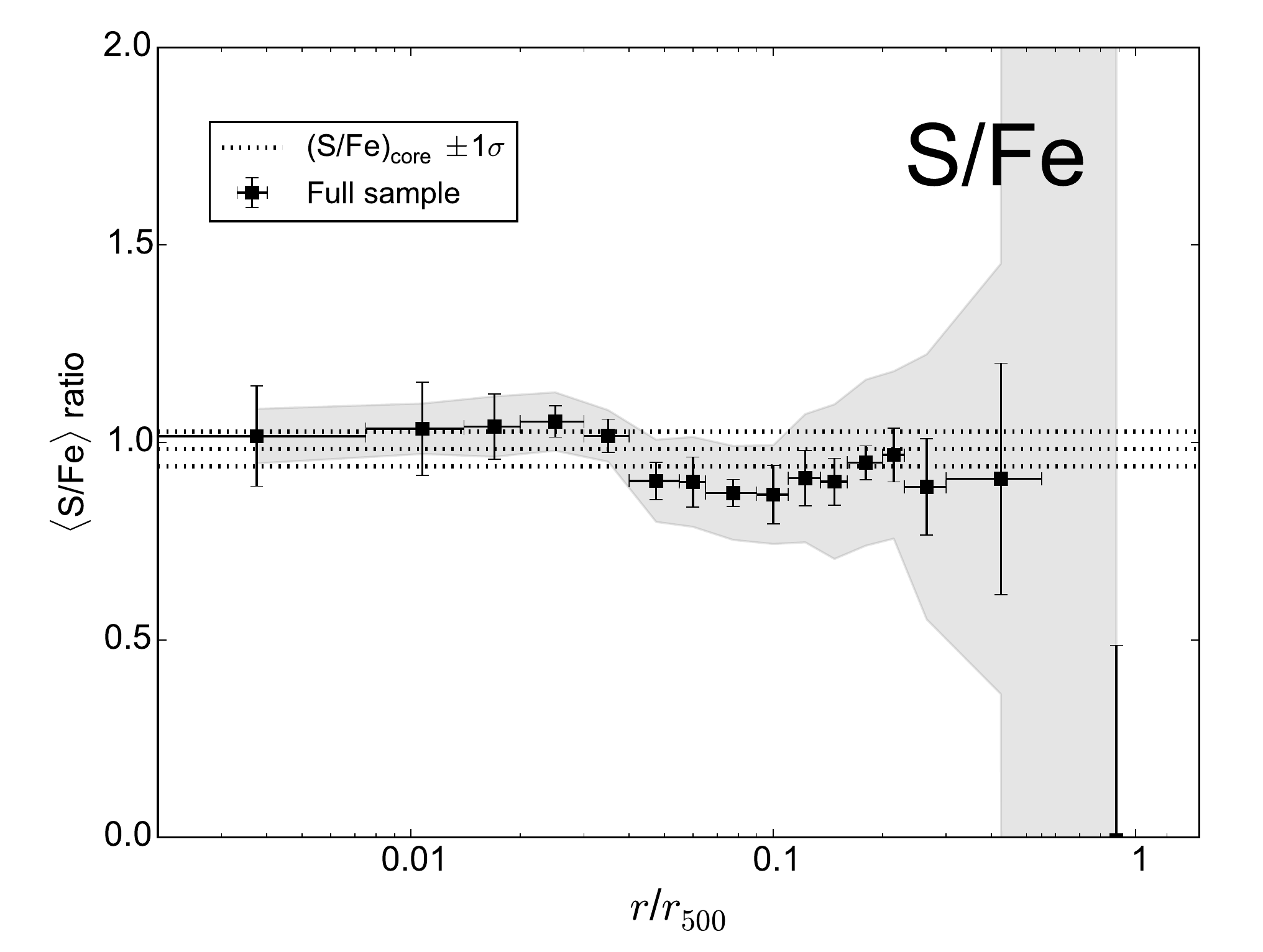}
\\
                \includegraphics[width=0.42\textwidth]{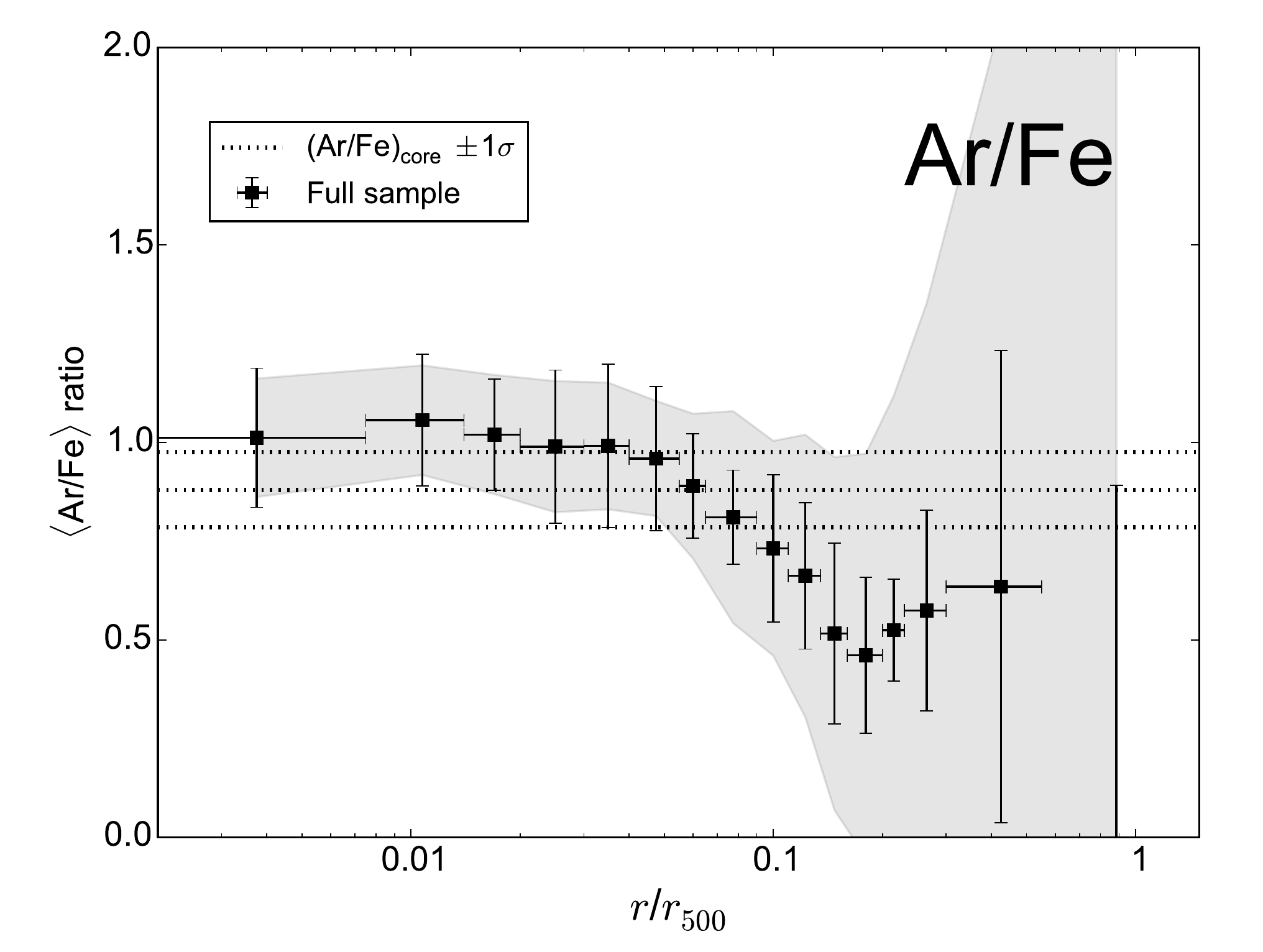}
                \includegraphics[width=0.42\textwidth]{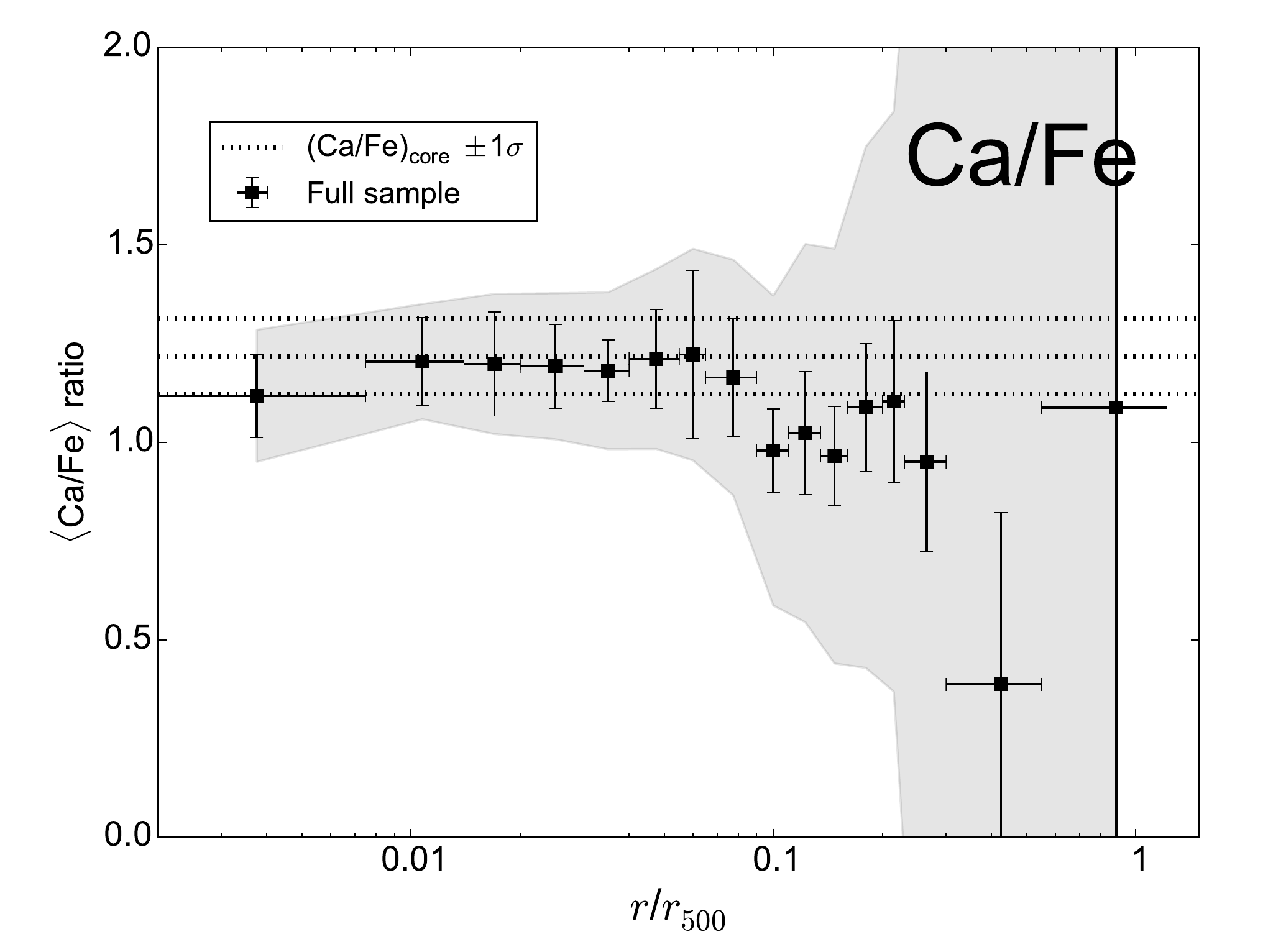}
\\
                \includegraphics[width=0.42\textwidth]{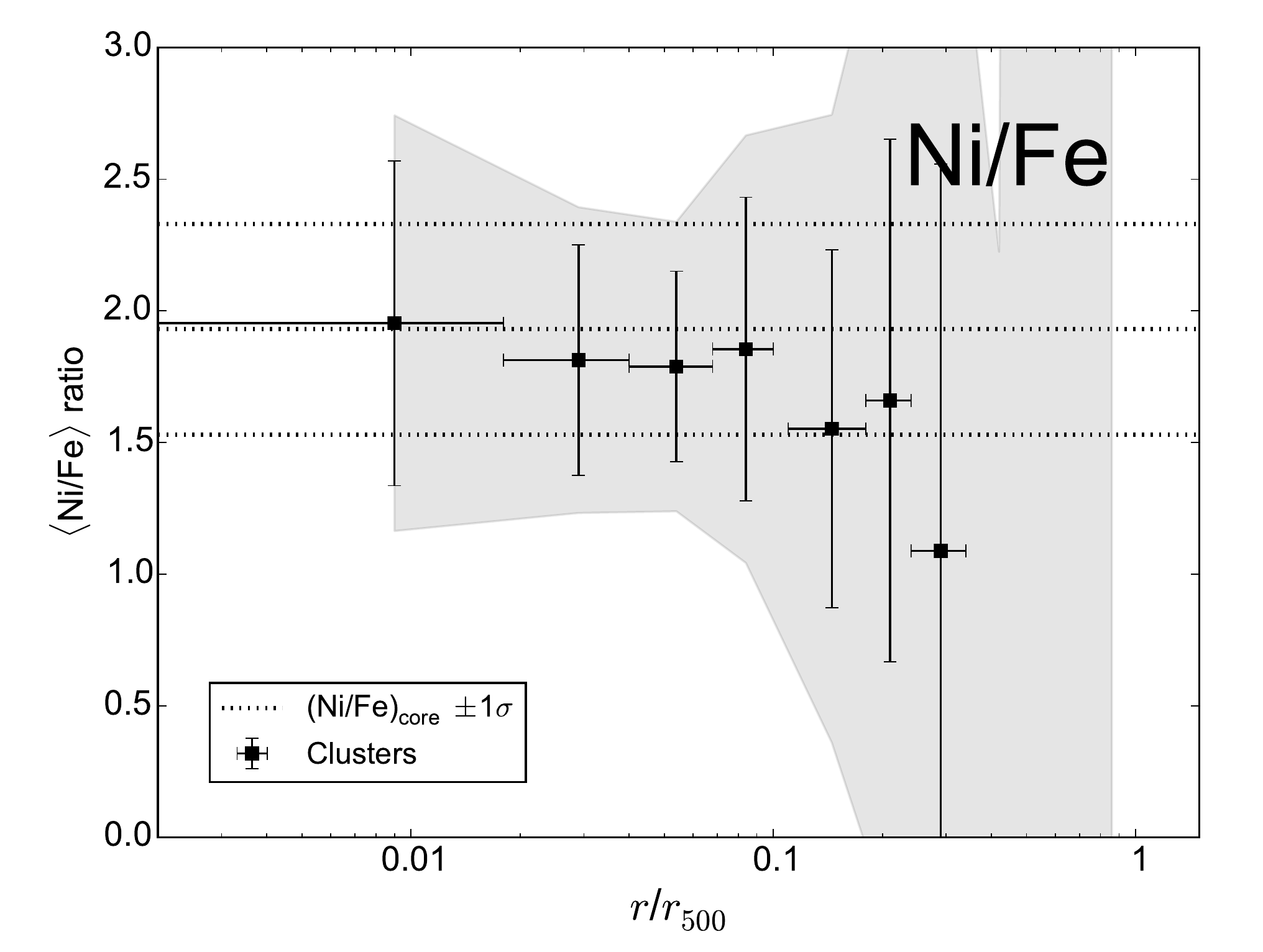}

        \caption{Individual radial X/Fe ratio measurements averaged over the full sample. The error bars contain the statistical uncertainties and MOS-pn uncertainties (Sect. \ref{sect:MOS-pn_uncertainties}) except for the O/Fe abundance profiles, which are only measured with MOS. The corresponding shaded areas show the scatter of the measurements. The average X/Fe abundance ratios (and their uncertainties) measured in the ICM core by \citet{2016A&A...592A.157M}, namely (X/Fe)$_\text{core}$, are also plotted.}
\label{fig:abundance_profiles_normFe}
\end{figure*}

Another method for comparing the Fe abundance profile with the abundance profiles of other elements is to compute the X/Fe abundance ratios in each annulus of each individual observation. We stack all these measurements over the full sample as described in Sect. \ref{sect:building_radial_profiles} to build average X/Fe profiles. These Fe-normalised profiles are shown in Fig. \ref{fig:abundance_profiles_normFe}. In each panel, we also indicate (X/Fe)$_\text{core}$, the average X/Fe ratio measured within the ICM core (i.e. $\le$0.05$r_{500}$ when possible, $\le$0.2$r_{500}$ otherwise) adopted from \citet{2016A&A...592A.157M}, and their total uncertainties (dotted horizontal lines; including the statistical errors, intrinsic scatter, and MOS-pn uncertainties). As mentioned earlier, the Ni/Fe profile could only be reasonably derived for clusters. Despite a usually large scatter (in particular in the outskirts), the X/Fe profiles are all in agreement with being flat, hence following the Fe average profile, and are globally consistent with their respective average (X/Fe)$_\text{core}$ values. Despite this global agreement, we note the clear drop of Ar/Fe beyond $\sim$0.064$r_{500}$. This outer drop corresponds to the steeper Ar profile seen in Fig. \ref{fig:abundance_profiles} and reported above. Finally, and similarly to Fig \ref{fig:abundance_profiles}, most of the outermost average X/Fe values are biased low with respect to their (X/Fe)$_\text{core}$ counterparts (often coupled with very large scatters), perhaps indicating the observational limits of measuring these ratios.

\section{Systematic uncertainties}\label{sect:systematics}


In the previous section, we presented the average abundance profiles measured for our full sample (CHEERS) and for the clusters and groups subsamples. Before discussing their implications on the ICM enrichment, we must ensure that our results are robust and do not (strongly) depend on the assumptions we invoke throughout this paper.
In this section, we explore the systematic uncertainties that could potentially affect our results. They can arise from: (i) the intrinsic scatter in the radial profiles of the different objects of our sample; (ii) MOS-pn discrepancies in the abundance measurements due to residual EPIC cross-calibration issues; (iii) projection effects on the plane of the sky; (iv) uncertainties in the thermal structure of the ICM; (v) uncertainties in the background modelling; and (vi) the weight of a few individual highest quality observations, which might dominate the average measurements.

We already took items (i) and (ii) taken into account in our analysis (Sect. \ref{sect:results_Fe} and \ref{sect:MOS-pn_uncertainties}, respectively), and here we focus on items (iii), (iv), (v), and (vi).

\subsection{Projection effects}\label{sect:systematics_proj}

Throughout this paper, we report the average abundance profiles of the ICM as observed by \textit{XMM-Newton}/EPIC, i.e. projected on the plane of the sky. Several models are currently available to deproject cluster data and estimate the radial metal distribution contained in concentric spherical shells \citep[e.g.][]{2003ApJ...590..225C,2004A&A...413..415K,2005MNRAS.356..237J,2008MNRAS.390.1207R}. However, all of them assume a spherical symmetry in the ICM distribution, which may not always be true. Moreover, some methods are known for introducing artefacts in the deprojected measurements \citep[for a comparison, see][]{2008MNRAS.390.1207R}, as deprojection methods assume a dependency between all the fitted annuli. We thus prefer to work with projected results to keep a statistical independence in the radial bins.

Several past works investigated the effects of deprojection on the abundance estimates at different radii. The general outcome is that these effects have a very limited impact on the abundance measurements \citep[e.g.][]{2007MNRAS.380.1554R,2008MNRAS.390.1207R}. Therefore, we do not expect them to be a source of significant systematic uncertainty for the purpose of this work.

\subsection{Thermal modelling}\label{sect:systematics_GDEM}

As explained in Sect. \ref{sect:thermal_mod}, the abundance determination is very sensitive to the assumed thermal structure of the cluster/group. Therefore, it is crucial to fit our spectra with a thermal model that reproduces the projected temperature structure as realistically as possible. In particular, a \texttt{cie} (single-temperature) model is clearly not optimal for our analysis. The thermal model used in this work (\texttt{gdem}) has been used in many previous studies and is thought to be rather successful at reproducing the true temperature structure of some clusters \citep[e.g.][]{2009A&A...493..409S,2013ApJ...764...46F}, as it represents one of the simplest way of accounting for a continuous mixing of temperatures in the ICM (coming from either projection effects or a locally intrinsic multi-phase plasma). The precise temperature distribution is however difficult to determine with the current spectrometers and may somewhat differ from the \texttt{gdem} assumption. Alternatively, some previous works suggest that the temperature distribution in cool-core clusters may be reasonably approximated by a truncated power law \citep[typically between $0.2 \text{ keV} \lesssim kT \lesssim 3 \text{ keV}$, with more emission towards higher temperatures; see e.g.][]{2004A&A...413..415K,2008MNRAS.385.1186S}. Such a distribution can be modelled in SPEX via the \texttt{wdem} model \citep[for more details, see e.g.][]{2004A&A...413..415K}.

\begin{figure}[!]
        \centering
                \includegraphics[width=0.49\textwidth]{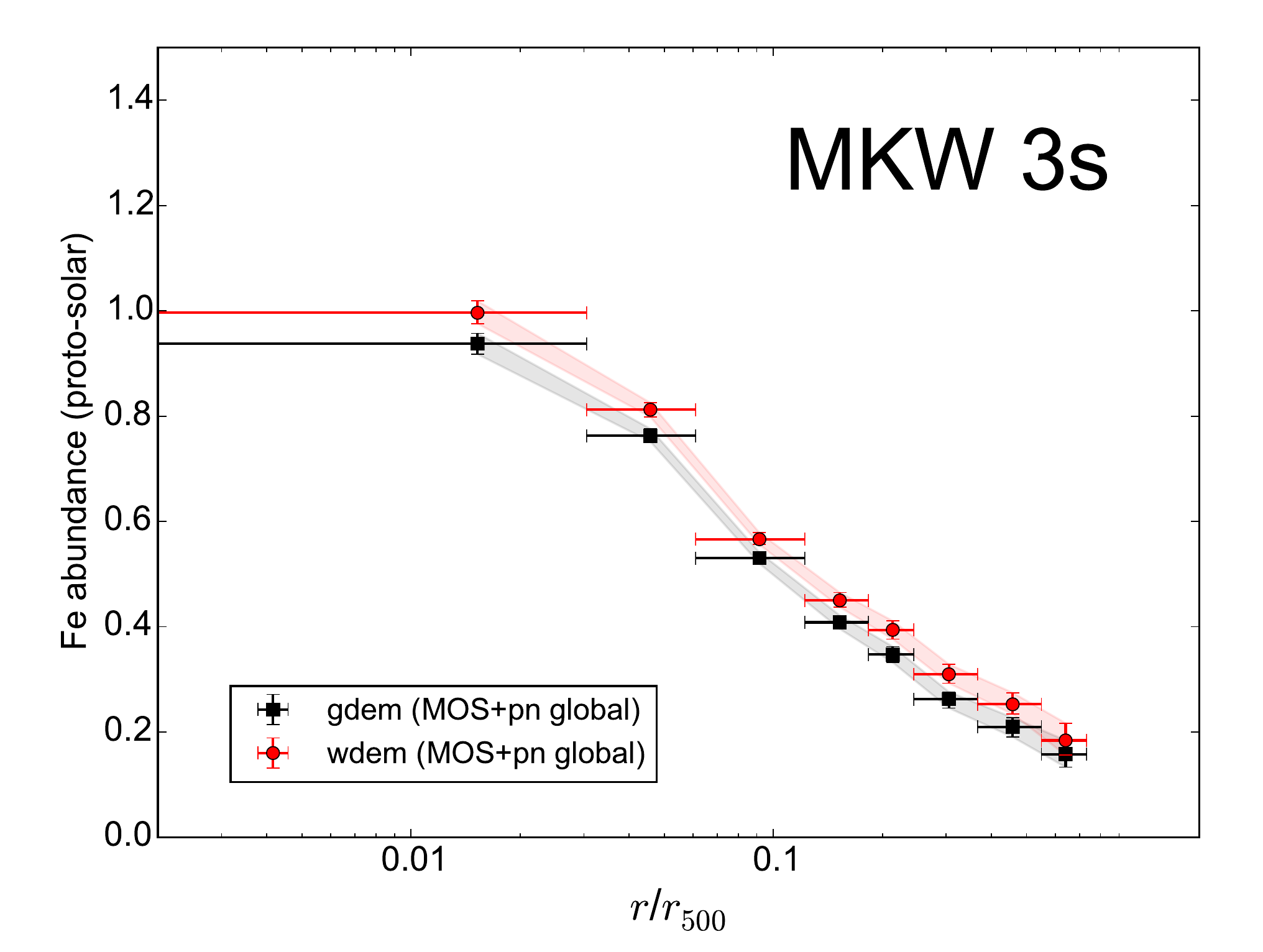}

        \caption{Comparison of the radial Fe profiles derived in MKW\,3s by assuming successively a Gaussian (\texttt{gdem}, black) and a power law (\texttt{wdem}, red) temperature distribution (see text for more details).}
\label{fig:MKW3s_gdem_wdem}
\end{figure}

Using a \texttt{wdem} model instead of a \texttt{gdem} model can potentially lead to differences in the measured abundances, hence contributing to add further systematic uncertainties to the derived profiles \citep[for a RGS comparison, see][]{dePlaa2017}. Unfortunately, the large computing time required by the \texttt{wdem} model in the fits does not allow us to perform a full comparison between the two models over the whole sample. We thus select one cluster, MKW\,3s, and we explore how the use of a \texttt{wdem} model affects its Fe profile. MKW\,3s has the advantage of emitting a moderate ICM temperature ($\sim$3.4 keV) inside 0.05$r_{500}$, which is very close to the mean temperature of the clusters in the sample ($\sim$3.2 keV) within this radius. Moreover, the Fe radial profile of MKW\,3s (Fig. \ref{fig:Fe_profiles_indiv_clusters}) is rather similar to the average Fe profile presented in Fig. \ref{fig:Fe_radial_stacked_profile}. The \texttt{gdem}-\texttt{wdem} comparison on the Fe radial profile of MKW\,3s is presented in Fig. \ref{fig:MKW3s_gdem_wdem}.
The use of a \texttt{wdem} model in MKW\,3s systematically predicts higher Fe abundances than using a \texttt{gdem model}, where the increase may vary from +$6\%$ (core) up to +$20\%$ (outskirts). Since there is a difference of temperature between the core ($kT_\text{mean} \simeq 3.5$ keV) and the outskirts ($kT_\text{mean} \simeq 1$ keV), this may suggest a temperature dependence \citep[see also][]{dePlaa2017}. However, there is no substantial change in the slope of the overall profile. The same trend is also found for the abundance profiles of the other elements. For comparison, we also check that we obtain similar results for NGC\,507, i.e. a cooler group. In conclusion, we do not expect any variation in the shape of the average abundance profiles owing to the use of another temperature distribution in our modelling. The normalisation of these profiles, which might slightly be revised upwards in the case of a \texttt{wdem} model, still lies within the scatter of our measurements and does not affect our results.

Nevertheless, as said above, it is worth keeping in mind that the current spectral resolution offered by CCDs does not allow us to resolve the precise temperature structure in the ICM. Further improvements on the thermal assumptions invoked here are expected with X-ray micro-calorimeter spectrometers on board future missions.

\subsection{Background uncertainties}\label{sect:systematics_bg}

As mentioned in Sect. \ref{sect:bg_mod}, a proper modelling of the background is crucial for a correct determination of the abundances in the ICM. This is especially true in the outskirts, where the background contribution is significant and may easily introduce systematic biases when deriving spectral properties. Presumably, the Si, S, Ar, Ca, Fe, and Ni abundances are more sensitive to the modelling of the non-X-ray background, as the HP and SP components dominate beyond $\sim$2 keV. On the other hand, the O abundance is more sensitive to the X-ray background, in particular the GTE and LHB components, which may have their greatest influence below $\sim$1 keV. We investigate the effects of background-related uncertainties on the abundance profiles using two different approaches.

\begin{figure*}[!]

                \includegraphics[width=0.5\textwidth]{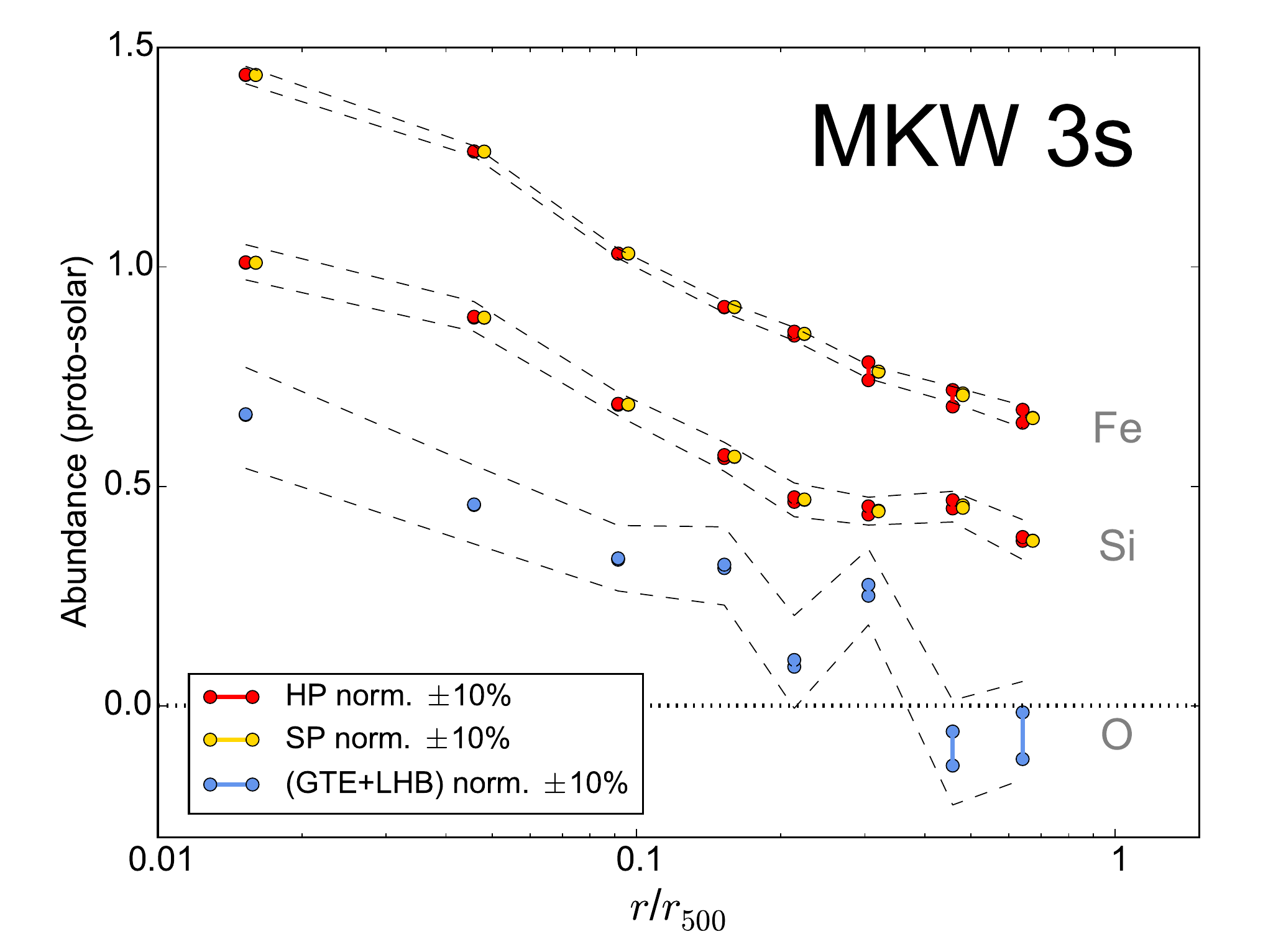}
                \includegraphics[width=0.5\textwidth]{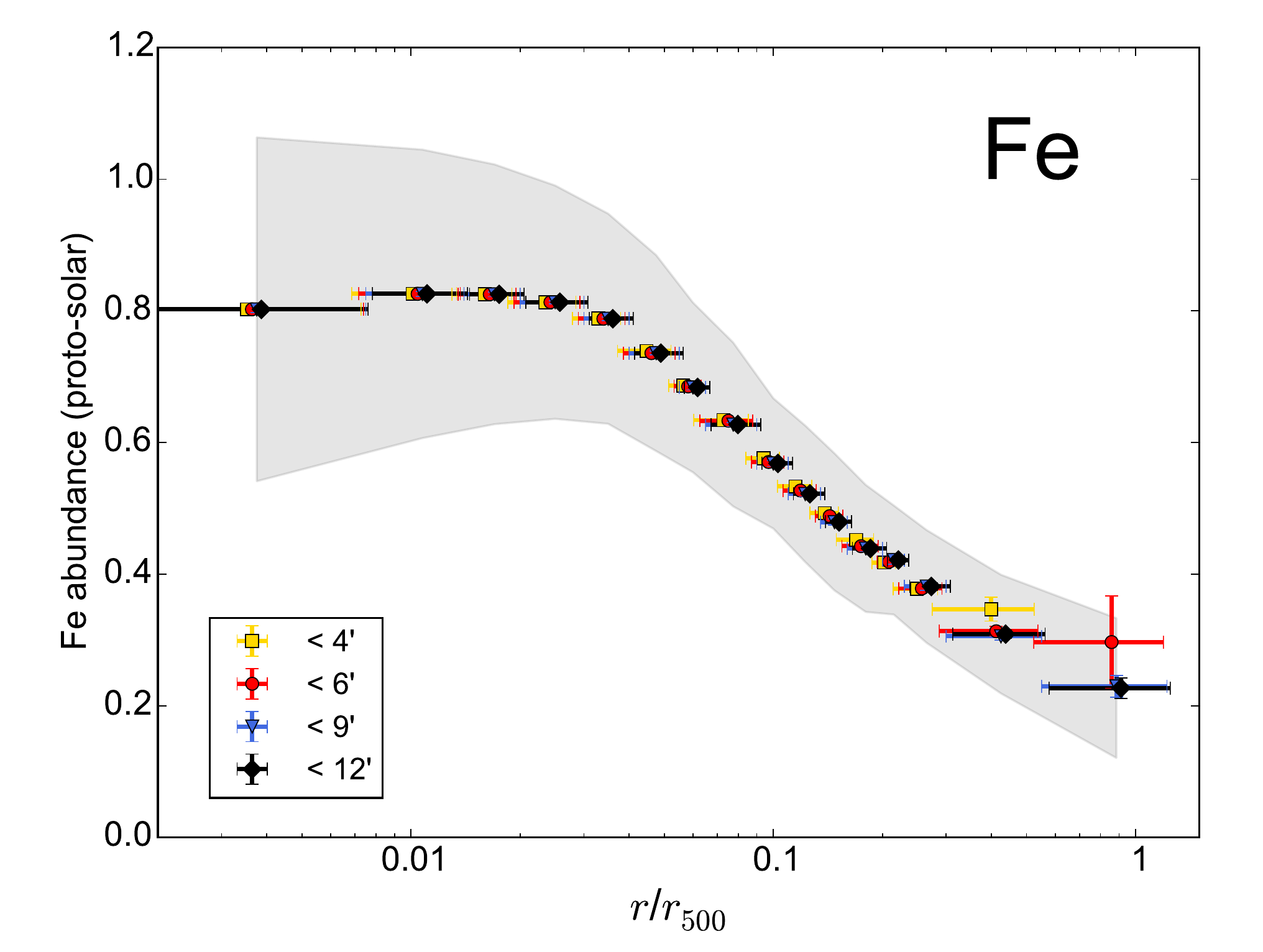}

        \caption{\textit{Left:} Effects of the background model uncertainties on the Fe, Si, and O radial profiles of MKW\,3s. The normalisation of the HP, SP, and GTE+LHB were successively fixed to $\pm$10\% of their best-fit values (see text). The dashed lines show the range constrained by the statistical uncertainties for each profile. For clarity, the Si and Fe profiles are shifted up by 0.25 and 0.5, respectively. \textit{Right:} Comparison of the average Fe profile for different truncated radii adopted in each observation. Data points with different colours are slightly shifted for clarity.}
\label{fig:systematics_bg}
\end{figure*}

First, similar to Sect. \ref{sect:systematics_GDEM}, we take MKW\,3s as an object representative of the whole sample. In each annulus and for all the EPIC instruments, we successively fix the normalisations of the HP and SP background components to $\pm$10\% of their best-fit values. We then refit the spectra and measure the changes in the best-fit Si and Fe profiles. We do the same for the O profile, this time by fixing the normalisations of the GTE and  LHB components together to $\pm$10\% of their best-fit values. The results are shown in the left panel of Fig. \ref{fig:systematics_bg}, where the Si and Fe profiles were shifted up for clarity. In all cases, the changes in the best-fit abundances are smaller than (or similar to) the statistical uncertainties from our initial fits. This clearly illustrates that a slightly ($\lesssim$10\%) incorrect scaling of the modelled background has a limited impact on our results, even at large radii. Moreover, we may reasonably expect that the possible deviations from the true normalisation of the background components average out when stacking all the objects. 

Second, and despite the encouraging previous indication that the background-related systematic uncertainties are under control, we still consider the possibility that the outer regions of every observation would be too contaminated and should be discarded from the analysis. In this respect, in the right panel of Fig. \ref{fig:systematics_bg} we rebuild the average Fe profile by successively ignoring the $\ge$9$'$, $\ge$6$'$, and $\ge$4$'$ regions (corresponding to keeping only the first seven, six, and five annuli, respectively) from each observation. Restricting our analysis to <6$'$ still allows us to derive a mean Fe abundance in the outermost average radial bin (0.55--1.22$r_{500}$). However, most of the area from the only two measurements that partly fall into this bin (A\,2597 and A\,1991) overlap the inner reference bin (0.3--0.55$r_{500}$). This spatial resolution issue may thus explain the slight ($\sim$30\%, albeit non-significant) increase of the average Fe value observed in outermost bin when truncating the $\ge$6$'$ regions. A similar explanation can be invoked for the <4$'$ case, in the second outermost bin (0.3--0.55$r_{500}$), where an average increase of $\sim$12\% is observed (though less than 2$\sigma$ significant). In any case, the changes related to the truncation of the profiles at different radii are always smaller than the scatter (grey area) even in the outskirts. Therefore, this scatter can reasonably be seen as a conservative limit encompassing all the background uncertainties mentioned here.

In summary, our results clearly suggest that our careful modelling of the background allows us to keep all its related systematic uncertainties on the abundances under control, even at larger radii. However, it is not impossible that the background dominates in the outermost radial bin ($\ge$ 0.55$r_{500}$) too much, thereby biasing low the average abundances of some elements (Sect. \ref{sect:results_abun}).

\subsection{Weight of individual observations}\label{sect:systematics_weight}

Among the 44 objects of our sample, the three brightest objects (A\,3526 a.k.a. Centaurus, M\,87, and Perseus) benefit from excellent data quality, leading to very small statistical uncertainties ($\sigma^2_{\text{X}(i)_j}$) of their measured abundances. Consequently, these observations may have an important contribution in shaping the average abundance profiles (as $1 / \sigma^2_{\text{X}(i)_j} \gg 1$). The consequences of this weighting selection effect is explored in this section.

\begin{figure*}[!]

                \includegraphics[width=0.5\textwidth]{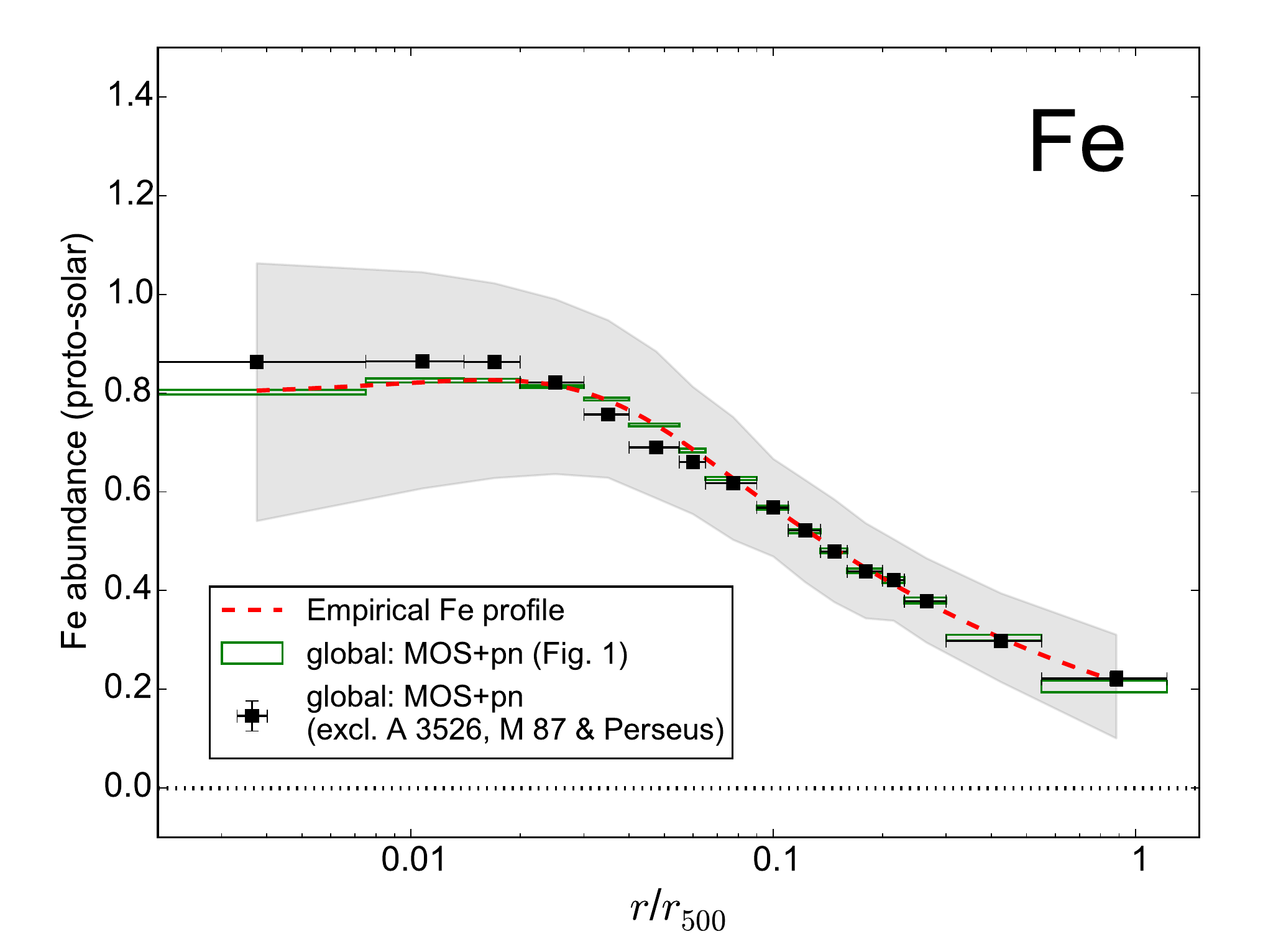}
                \includegraphics[width=0.5\textwidth]{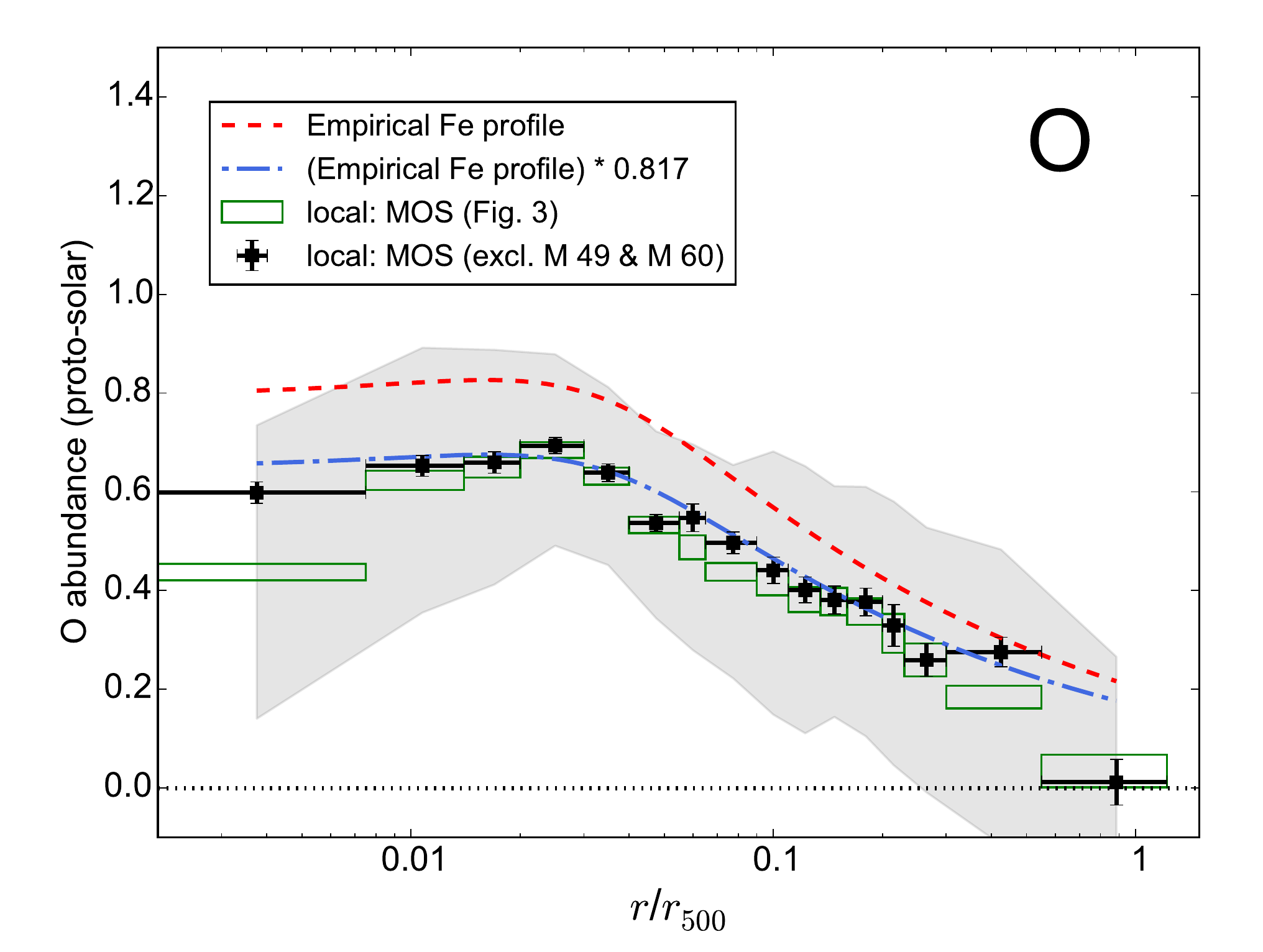} \\
                \includegraphics[width=0.5\textwidth]{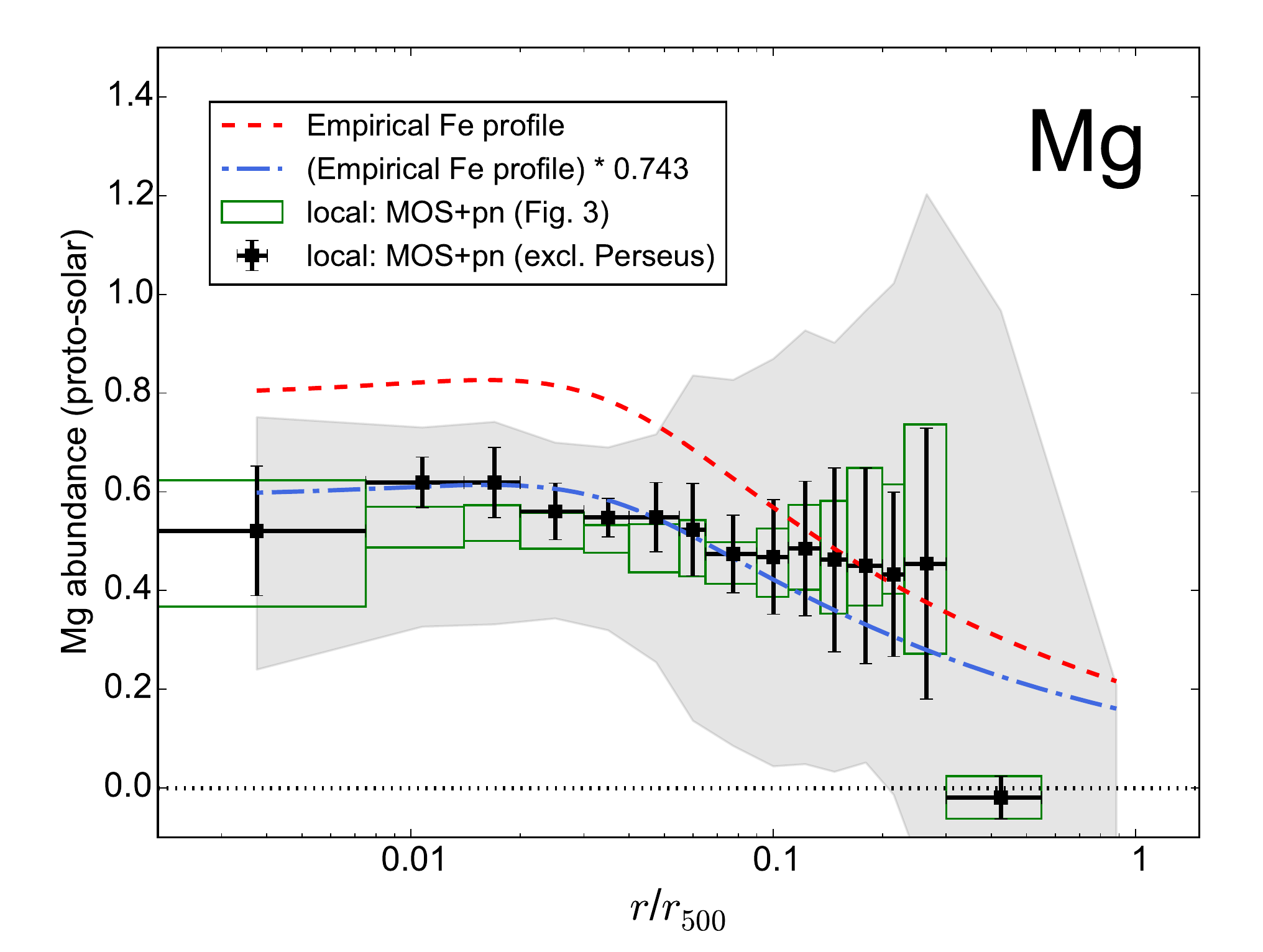}
                \includegraphics[width=0.5\textwidth]{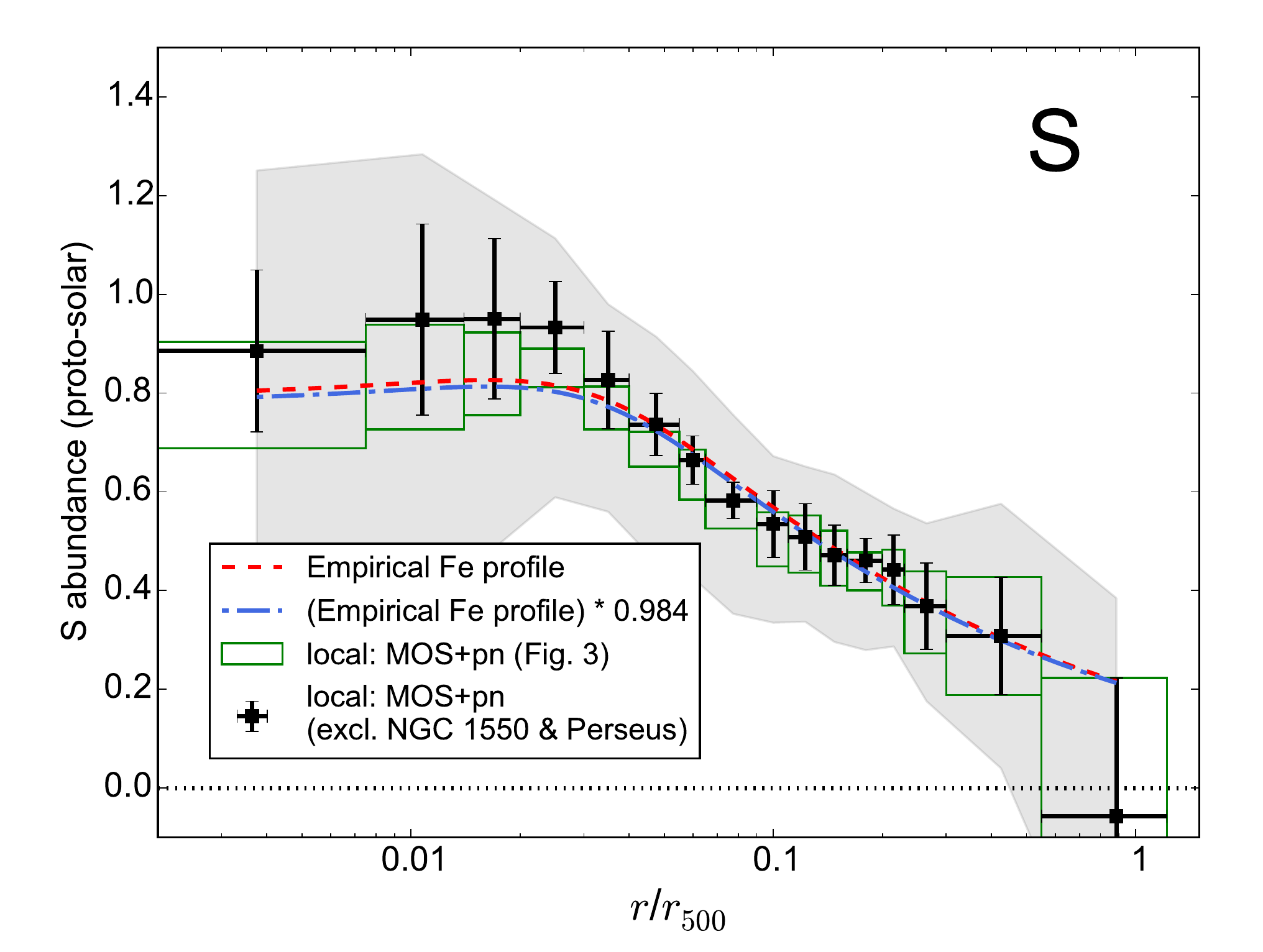}

        \caption{Same as Figs. \ref{fig:Fe_radial_stacked_profile} and \ref{fig:abundance_profiles} (Fe, O, Mg, and S, blue empty boxes), where we discard A\,3526, M\,87, and Perseus from the Fe average profile (\textit{top left}), M\,49, M\,60, and NGC\,4636 from the O average profile (\textit{top right}), Perseus from the Mg average profile (\textit{bottom left}), and NGC\,1550 and Perseus from the S average profile (\textit{bottom right}). These modified profiles are shown by the black squares.}
\label{fig:excl_profiles}
\end{figure*}

In Fig. \ref{fig:excl_profiles} (top left panel), we show how the average Fe profile changes when we exclude A\,3526, M\,87, and Perseus from the sample. Compared to the initial Fe profile (blue empty boxes; see also Fig \ref{fig:Fe_radial_stacked_profile}), the largest effect is an increase of $\sim$8\% in the innermost average radial bin ($\le 7.5 \times 10^{-3} r_{500}$), while the rest of the radial profile varies a few per cent at most.

Similarly, this weighting effect may affect the other abundance profiles. In Fig. \ref{fig:abundance_profiles}, we showed that the average Si, Ca, and Ni radial profiles follow very well the fitted average Fe radial profile normalised by the average X/Fe ratio found in the core. However, the innermost region ($\le$0.01$r_{500}$) shows an O drop about $\sim$20\% lower than predicted by our empirical profile, while the Mg profile looks significantly flatter than expected. Similarly, some deviations from the expected S profile are also observed within 0.04--0.1$r_{500}$. In this section, we show that these profiles are more affected by the weight of a few individual observations, and that the empirical O/Mg/S profiles can be very well reproduced when temporarily ignoring these peculiar measurements.

When we exclude M\,49, M\,60, and NGC\,4636 from the analysis, we find a much better agreement between the O abundance and its corresponding empirical prediction in the innermost bin (Fig. \ref{fig:excl_profiles}, top right panel). Indeed, these three ellipticals/groups are characterised by a suspiciously low O abundance within their respective <0.5$'$ annuli \citep[inconsistent with the values found within 0.8$'$ with RGS by][]{dePlaa2017}, which, together with very small errors bars, contribute to substantially lower the average O abundance in the $\le 7.5 \times 10^{-3} r_{500}$ region.

When we exclude Perseus from the analysis, the average Mg measurements agree much better with the expected empirical profile, especially within $\sim$0.01--0.05$r_{500}$ (Fig. \ref{fig:excl_profiles}, bottom left panel). The significant MOS-pn discrepancies measured in the Perseus spectra make the Mg abundance somewhat uncertain over the region considered above. However, and coincidentally, combining these (discrepant) MOS/pn measurements from Perseus with those from the rest of the sample brings the average MOS and pn estimates of Mg at very similar levels, thereby dramatically reducing the total MOS-pn uncertainties that we consider in Sect. \ref{sect:MOS-pn_uncertainties}. This case is thus a good illustration that care must be taken when combining individual systematic uncertainties over a large data sample.
Finally, the exclusion of NGC\,1550 and Perseus from the sample contributes to a better agreement of the measured S profile with its empirical expectation (Fig. \ref{fig:excl_profiles}, bottom right panel).

To sum up, in addition to showing that the average measured radial abundance profiles for all elements can reproduce very well their empirical counterparts, this section illustrates that care must be taken when strictly interpreting the error bars shown in the figures of this paper, as only one or two individual observations may slightly (usually, within a few per cent) but significantly raise or lower our measurements. That said, in the rest of the paper we consider our full sample, including the peculiar measurements discussed here.

\subsection{Atomic code uncertainties}\label{sect:systematics_atomic}

The CIE model employed to fit our EPIC spectra is based on the MEKAL model \citep[][also present in XSPEC]{1985A&AS...62..197M,1986A&AS...65..511M}, with important updates up to now. The atomic database and routines on which this model relies is called SPEXACT\footnote{SPEX Atomic Code and Tables}. Whereas the initial version of SPEXACT can be simply attributed to the original MEKAL model, the version used in this work (corresponding to the atomic code that was regularly updated between 1996 and 2016) is referred to SPEXACT v2. In recent months, substantial efforts have been devoted towards a major update of the code (SPEXACT v3), followed by a newly released version of SPEX \citep[see also][]{dePlaa2017}. For example, this new version includes a more precise parametrisation of the radiative recombination rates \citep{2016A&A...587A..84M}, updated collisional ionisation coefficients \citep{2017arXiv170206007U}, and the calculation of many more transitions. Following \citet{2016A&A...592A.157M}, we included the correction of this latest update on our O abundance measurements (Sect. \ref{sect:thermal_mod}). However the abundances of the other elements may also be affected by such improved calculations.

\begin{figure}[!]

                \includegraphics[width=0.5\textwidth]{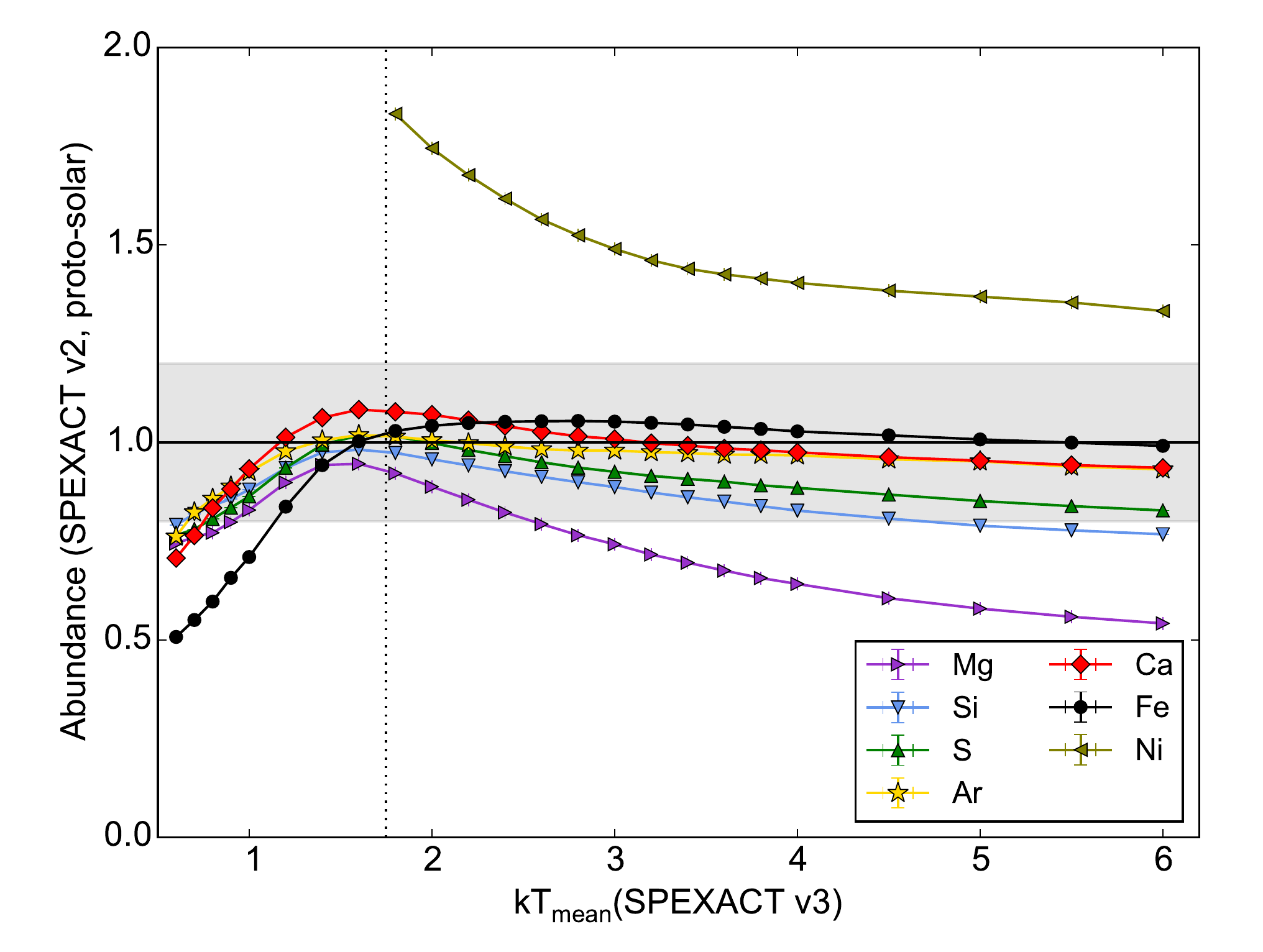}

        \caption{Abundance results from (\texttt{gdem}) local fits with SPEXACT v2 to simulated SPEXACT v3 spectra for a range of temperatures. The measured abundances are shown and compared to their input value of 1 proto-solar. The grey shaded area shows the $\pm$20\% level of uncertainty. The vertical dotted line indicates our (arbitrary) separation between clusters and groups.}
\label{fig:SPEX2vs3}
\end{figure}

Unfortunately, fitting all our EPIC spectra using SPEXACT v3 would require unrealistic amounts of computing time and resources. Therefore, we evaluate the impact of these atomic code differences on the EPIC abundances by following a similar approach as carried out by \citet{dePlaa2017} for RGS. Here, we simulate EPIC spectra assuming a \texttt{gdem} distribution calculated from SPEXACT v3, for a range of mean temperatures from 0.6 keV to 6.0 keV and by setting all the abundances to 1. We then fit these mock spectra locally with a \texttt{gdem} model calculated from SPEXACT v2 (i.e. the version used in this work), and we measure the changes in the best-fit abundances. The result is shown in Fig. \ref{fig:SPEX2vs3}.

For temperatures hotter than $\sim$1.5 keV, most of the abundances do not change by more than $\sim$20\%. The two exceptions are Mg and Ni, which can change by almost a factor of 2 at high and low temperatures, respectively. For temperatures cooler than $\sim$1.5 keV, we see a dramatic decrease (by more than a factor of 2) of the measured Fe abundance. The main difference between the spectral models generated by SPEXACT v2 and SPEXACT v3 resides in the Fe-L complex, which is foremost used by the fits to determine the Fe abundance in cool ($kT \lesssim 2$ keV) plasmas.

Since most of the computed abundances remain fairly constant within the typical temperature range ($\sim$1--5 keV) of all the spectra of our sample, such atomic code uncertainties are not expected to affect our results. Nevertheless, we note that these changes between SPEXACT v2 and SPEXACT v3 may have a non-negligible impact on the integrated abundances (and X/Fe abundance ratios) reported in previous works. For instance, if updated atomic calculations indeed revise the average Ni/Fe abundance downwards \citep[so far measured to be surprisingly high; e.g.][]{2016A&A...592A.157M,2016A&A...595A.126M}, a more simple agreement than previously assumed between the ICM abundance pattern and SN yield models may be expected. This issue (and further use of SPEXACT v3 on real cluster data) will be discussed extensively in a forthcoming paper.

\subsection{Instrumental limitations for O and Mg abundances}\label{sect:systematics_EPIC}

Finally, we must warn that the EPIC instruments have limitations in deriving accurate O and Mg abundances. 

The main K-shell transitions of O ($\sim$0.6 keV rest-frame) are situated close to the oxygen absorption edge, and the interstellar absorption may affect the O abundance determination, as the EPIC spectral resolution cannot resolve the emission and absorption features within this band \citep[see e.g.][]{2004A&A...423...49D}. Moreover, and despite our considerations from Sect. \ref{sect:systematics_bg}, the Galactic foreground may play a more important role than expected, which can potentially bias the O abundance, especially when the background dominates. Although affecting on average 3--4\% of the \textit{XMM-Newton} observations, solar wind charge exchange might also be a source of (limited) bias for the O abundance, at it may affect the \ion{O}{vii} and \ion{O}{viii} lines in the contaminated spectra \citep[e.g.][]{2004ApJ...610.1182S,2011A&A...527A.115C}.

On the other hand, the main K-shell emission line of Mg ($\sim$1.5 keV rest-frame) falls partly into the Fe-L complex, which is unresolved by the EPIC instruments. Moreover, measuring the Mg abundance in clusters outskirts is challenging because the EPIC hard particle background is contaminated by the Al K$\alpha$ fluorescence line, which is also situated at $\sim$1.5 keV both in MOS and pn instruments \citep[e.g.][]{2015A&A...575A..37M}, and thus impossible to disentangle from the Mg K-shell ICM emission lines using the EPIC spectrometers.

Despite all these limitations, the good agreement of our average O and Mg profiles with their respective empirical predictions (at least out to $\sim$0.3$r_{500}$, and after discarding specific observations from the sample, see Sect. \ref{sect:systematics_weight}) is very encouraging, and makes us confident about the results presented in this work.


\section{Discussion}\label{sect:discussion}


We derived the average radial abundance profiles of 44 galaxy clusters, groups, and elliptical galaxies. In Sect. \ref{sect:results}, we were able to provide constraints on the radial ICM distributions of Fe, but also O, Mg, Si, S, Ar, Ca, and Ni, comparing them within clusters (>1.7 keV) and groups (<1.7 keV). In the previous section, we also showed that the major systematic uncertainties are kept under control. We now discuss our results and we compare them with measurements and predictions from previous studies.

\subsection{Enrichment in clusters and groups}\label{sect:discussion_clgr}

In Fig. \ref{fig:Fe_radial_stacked_profile_clgr}, we compared the radial Fe abundance profile averaged 
over clusters, on the one hand, and groups, on the other hand. Although the scatter in each profile is large, the average enrichment level in clusters is slightly higher than in groups. This result is not surprising, as an increase of the ICM metallicity with the cluster/group temperature (at least up to $kT \simeq 3$ keV) has been commonly observed in previous studies \citep[e.g.][]{2009MNRAS.399..239R,2017MNRAS.464.3169Y}. This trend is also consistent with the results of \citet{2016A&A...592A.157M}, who analysed the same sample with the same data and same definition for clusters versus groups/ellipticals. They found that, within $0.05 r_{500}$, the Fe abundance in clusters is on average $\sim$22\% higher than in groups. We find a similar Fe enhancement (on average $\sim$21\%) in our profiles for clusters and groups at all radii, except in their respective innermost and outermost radial bins (see also Sect. \ref{sect:results_Fe}). The absence of difference of Fe abundance in the innermost bin of clusters and groups can be explained by the important weight of a few individual clusters, as already discussed in Sect. \ref{sect:systematics_weight}. In particular, the cores of Perseus and A\,3526 show deep and significant Fe drops (see also Sect. \ref{sect:Fe_drop}), which tend to lower the innermost average Fe abundance for clusters. Removing these two objects from the sample increases this innermost Fe abundance (Fig. \ref{fig:excl_profiles}, top left panel), and, therefore, should contribute towards keeping a similar enhancement between clusters and groups within $\sim$0.01$r_{500}$. On the other hand, among the 11 measurements in the outermost radial bin of the groups profile, only 2 ($\sim$18\%) are located beyond $0.5 r_{500}$, i.e. covering the outermost bin of the clusters profile. The Fe abundance averaged over this outermost bin of the groups profile is thus weighted towards the measurements at smaller radii, roughly at the location of the third (<0.34$r_{500}$) and second (0.34--0.5$r_{500}$) outermost bins of the clusters profile. This explains the illusion of a Fe enhancement in the outskirts of groups with respect to those of clusters.

In summary, the average Fe profile of clusters is consistent with being more enhanced in a similar way not only in the core, but also at all radial distances at least out to $0.5 r_{500}$. The origin of such a difference of ICM enrichment between cooler and hotter objects is still unclear, and has been already debated in the literature \citep[e.g.][]{2009MNRAS.399..239R,2016MNRAS.456.4266L,2017MNRAS.464.3169Y}. For example, in contrast to clusters, galaxy groups may not be closed boxes \citep[e.g.][]{2011MNRAS.412.1965M} and AGN feedback may contribute to remove enriched material out of the groups. It may even be possible that part of this apparent difference of enrichment could be due to underestimated Fe abundances in low temperature plasmas, as mentioned in Sect. \ref{sect:systematics_atomic}. A thorough discussion of these aspects is somewhat beyond the scope of this paper. However, our radial profiles may provide useful constraints on the dominant mechanisms that are responsible for such a difference.

Interestingly, the same trend between clusters and groups is not clearly observed in the average profiles of other elements. Instead, these abundance profiles are consistent within clusters and groups (Fig. \ref{fig:abundance_profiles_clgr}). In fact, we report a slight (but not significant) enhancement in the X/Fe ratio profiles of groups compared to those of clusters, up to $0.03 r_{500}$, whose effect is visible in the Si/Fe and S/Fe profiles of the full sample (Fig. \ref{fig:abundance_profiles_normFe}). However, the large error bars (including systematic uncertainties from the MOS-pn cross-calibration) prevent us from firmly confirming this trend. We also note the exception of the O profiles, which clearly show an enhancement in the case of clusters with respect to that of groups. However, we must recall that O measurements using EPIC may be still uncertain (Sect. \ref{sect:systematics_EPIC}). Moreover, the measured O abundance in hotter systems may be biased high compared to its true value, essentially owing to issues in determining the correct continuum coupled to the weak emissivity of the \ion{O}{viii} line at these temperatures \citep{2008ApJ...674..728R}.

\subsection{Central metallicity drop}\label{sect:Fe_drop}

As seen in Figs. \ref{fig:Fe_profiles_indiv_clusters} and \ref{fig:Fe_profiles_indiv_groups}, some clusters and groups clearly exhibit a central drop in their Fe abundances. The presence of drops in these systems also appear in Fig. \ref{fig:Fe_radial_stacked_profile}, where a slight central decrease is observed in the average Fe abundance profile. Figures \ref{fig:abundance_profiles} and \ref{fig:abundance_profiles_normFe} suggest that these drops are not exclusive to Fe, as the other metals seem to be concerned. In this section, we attempt to quantify these abundance drops (focussing mainly on Fe) and then discuss their possible origins.

One way of quantifying the Fe drops is to measure their "depths". We choose arbitrarily the quantity Fe($r_\text{max}$)/Fe$_\text{drop}$: we divide the Fe abundance at its off-centre peak (or the Fe abundance at its second innermost bin, if the profile is monotonically decreasing) by the Fe abundance at the first innermost bin. With this definition, all the objects with Fe($r_\text{max}$)/Fe$_\text{drop}$ significantly greater than 1 are considered to host a significant drop. We find that 14 objects ($\sim$32\%) of our sample show a decrease of Fe abundance in their very core. Three of these objects (2A\,0335+096, A\,3526, and Perseus) are classified as clusters (i.e. $\sim$13\% of the subsample), while the remaining 11 (A\,189, A\,3581, Fornax, HCG\,62, M\,49, M\,86, NGC\,4325, NGC\,4636, NGC\,5044, NGC\,5813, and NGC\,5846) are classified as groups (i.e. $\sim$52\% of the subsample). This apparent larger proportion of groups hosting a central metallicity drop should be treated with caution because the larger distance of many clusters does not allow us to investigate their very core with the same spatial resolution as for nearer groups and ellipticals. Similarly, the drop seen in the average Fe profile (Fig. \ref{fig:Fe_radial_stacked_profile}) is smoothed by the lower spatial resolution of more distant systems, and thus appears less pronounced than in individual nearby objects.

\begin{figure}[!]

                \includegraphics[width=0.5\textwidth]{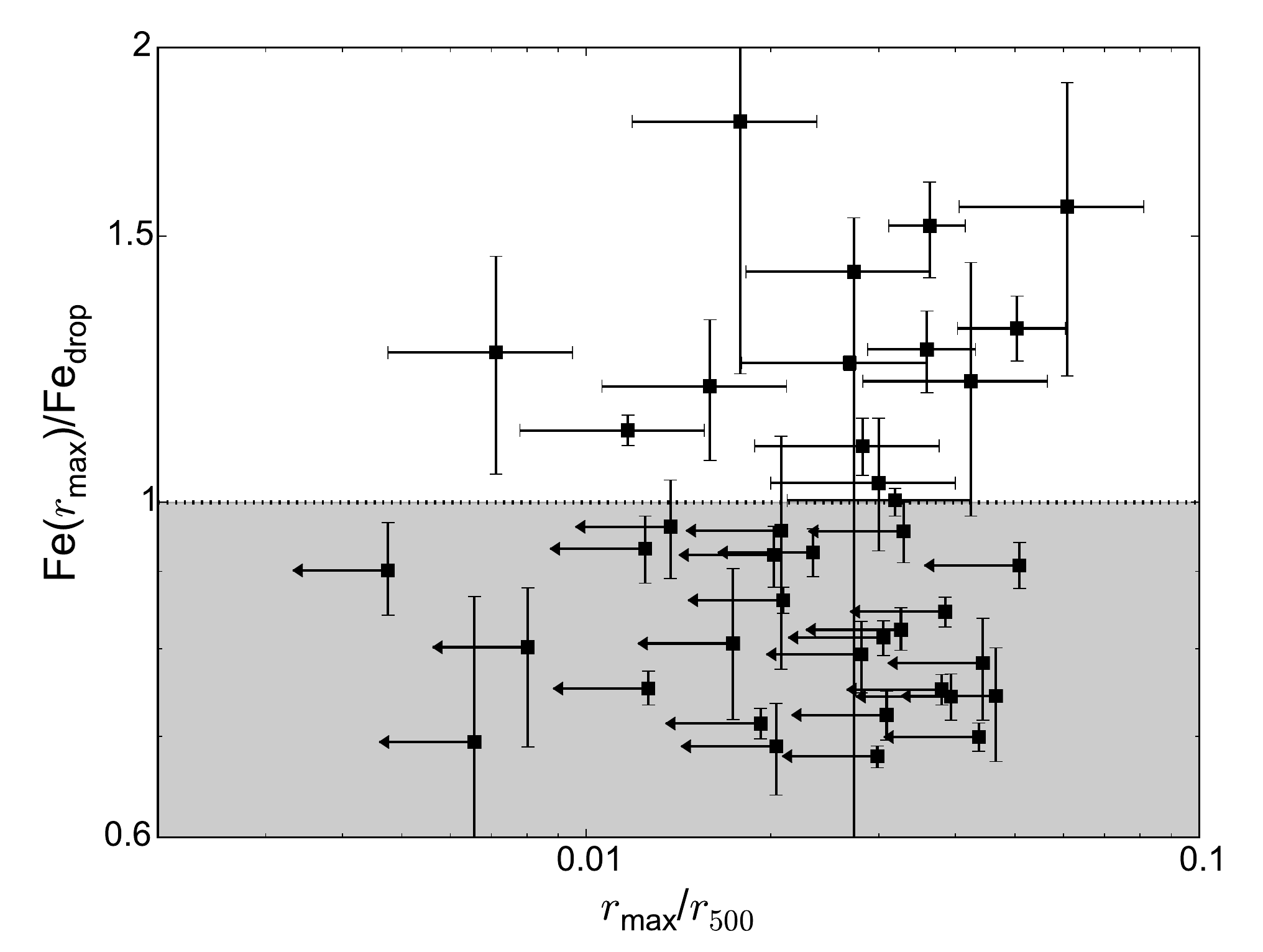}

        \caption{Depth of the central Fe drop (Fe($r_\text{max}$)/Fe$_\text{drop}$) vs. location of the Fe off-centre maximum ($r_\text{max}/r_{500}$) for all the objects of our sample. A value Fe($r_\text{max}$)/Fe$_\text{drop} \lesssim 1$ (grey area below the dotted horizontal line) means that no Fe drop could be significantly detected and only upper limits of $r_\text{max}/r_{500}$ could be estimated.}
\label{fig:Fe_drop_correlations}
\end{figure}

In most cases, the Fe drop is only seen in the innermost bin. However, some objects (e.g. Perseus, Fornax, M\,49, and NGC\,5044) clearly exhibit a drop extending within several radial bins. Therefore, for each object we also evaluate $r_\text{max}/r_{500}$, i.e. the location of the (off-centre) Fe peak, in units of $r_{500}$. For objects not showing any apparent drop, we adopt the extent of the innermost bin, which only provides an upper limit. Figure \ref{fig:Fe_drop_correlations} shows a diagram of Fe($r_\text{max}$)/Fe$_\text{drop}$ versus $r_\text{max}/r_{500}$ (i.e. the depth of the drops versus the location of the Fe off-centre peaks). The grey shaded area corresponds to the objects with no apparent drop (Fe($r_\text{max}$)/Fe$_\text{drop} \le 1$), where only an upper limit of $r_\text{max}/r_{500}$ could be constrained. When restricting ourselves to the objects exhibiting a drop (white area), we do not find evidence for a clear correlation ($\rho \simeq 0.19$) between the depth and radial extent of the drops. In fact, the error bars and scatter of the measurements are quite large and prevent us from deriving any firm conclusion on this assessment. The ACIS instrument on board \textit{Chandra} could help to reduce the error bars and to confirm (or rule out) this correlation. Such a detailed study, however, is beyond the scope of this present paper, and we leave it for future work.

This is not the first time that central metallicity drops have been found in the core of the ICM \citep[e.g.][]{2002MNRAS.331..273S,2002MNRAS.336..299J,2007MNRAS.381.1381S,2013MNRAS.428...58R}. However, their interpretation is not yet established. Below we discuss several possibilities that could explain the metallicity drops found in this work.

First, these apparent drops in metallicity could be the result of an artefact when fitting the spectra of the central regions. For example, an inappropriate modelling of the X-ray emission of the central AGN (or cumulated X-ray binaries in the BCG) could potentially introduce an incorrect estimate of the continuum of the ICM emission and underestimate the abundances in the very core. However, the abundance decrease extends sometimes outside the innermost region (Fig. \ref{fig:Fe_profiles_indiv_groups}, see e.g. Fornax, M\,49, and NGC\,5044), where no contamination from AGN emission is expected. Similarly (and perhaps more interestingly), the presence of non-thermal electrons in X-ray cavities could produce an additional power law component, which would underestimate the abundances if not properly modelled. However, we would then expect a good match between cavities extents (and morphologies) and abundance drops, which is not actually observed \citep{2015MNRAS.447..417P,2016MNRAS.457...82S}. Another possibility would be that the abundances measured in the very core suffer from the Fe/Si/S-bias \citep[e.g.][]{2000MNRAS.311..176B} owing to too simple assumptions concerning the thermal modelling. While accounting for a multi-temperature structure may sometimes help to remove the abundance drop \citep[e.g. in 2A\,0335+096;][]{2006A&A...449..475W}, this is not necessarily true for all the sources \citep[e.g.][]{2004MNRAS.349..952S,2013MNRAS.433.3290P}. Moreover, we must recall that all our spectra are fitted with a \texttt{gdem} model, which already assumes a multi-phase plasma. As a test, we also checked that the use of a \texttt{wdem} model does not remove the central drop in A\,3526. We admit, however, that a better knowledge of the true temperature distribution in cooling cores would be required to investigate in detail its impact on the very central abundance measurements. We can also discard other artificial effects such as projection on the plane of the sky \citep{2007MNRAS.381.1381S} or resonant scattering \citep{2006MNRAS.370...63S}, since they are not efficient enough to fully remove the drops. Finally, the underestimate of the Fe abundance at CCD resolution for low temperatures (Sect. \ref{sect:systematics_atomic}) could be an alternative fitting bias to explain the abundance drops. Although this could explain some abundance drops found in very cool group cores (e.g. NGC\,5813), this bias can hardly be invoked in the case of core temperatures above $\sim$1 keV still exhibiting a drop (e.g. A\,3526). 

Second, assuming that the drops are real, it may be reasonable to speculate that a fraction of the central metal mass has been redistributed from the core, by either AGN feedback, or sloshing motions. Whereas it is now well established that AGN feedback may play a key role in transporting the metals outside of the very core via jets and/or buoyant bubbles, as already observed in M\,87 \citep{2008A&A...482...97S} and in Hydra A \citep{2009A&A...493..409S}, simulations do not favour any clear formation of inner drops \citep[e.g.][]{2010ApJ...717..937G}. Furthermore, we do not find any correlation between AGN radio luminosities ($L_\text{1.4 GHz}$) reported in the literature \citep[e.g.][]{2012MNRAS.427.3468B} and the depths (Fe($r_\text{max}$)/Fe$_\text{drop}$) or the radial extent ($r_\text{max}/r_{500}$) of the drops in our sample. Similarly, while the extended drop seen in NGC\,5044 might be partly explained by its peculiar metal distribution in the sloshed gas \citep{2014MNRAS.437..730O}, sloshing process can probably not explain the (narrower) drops seen in other objects \citep{2011MNRAS.413.2057R,2012MNRAS.420.3632R}.

\begin{figure}[!]

                \includegraphics[width=0.5\textwidth]{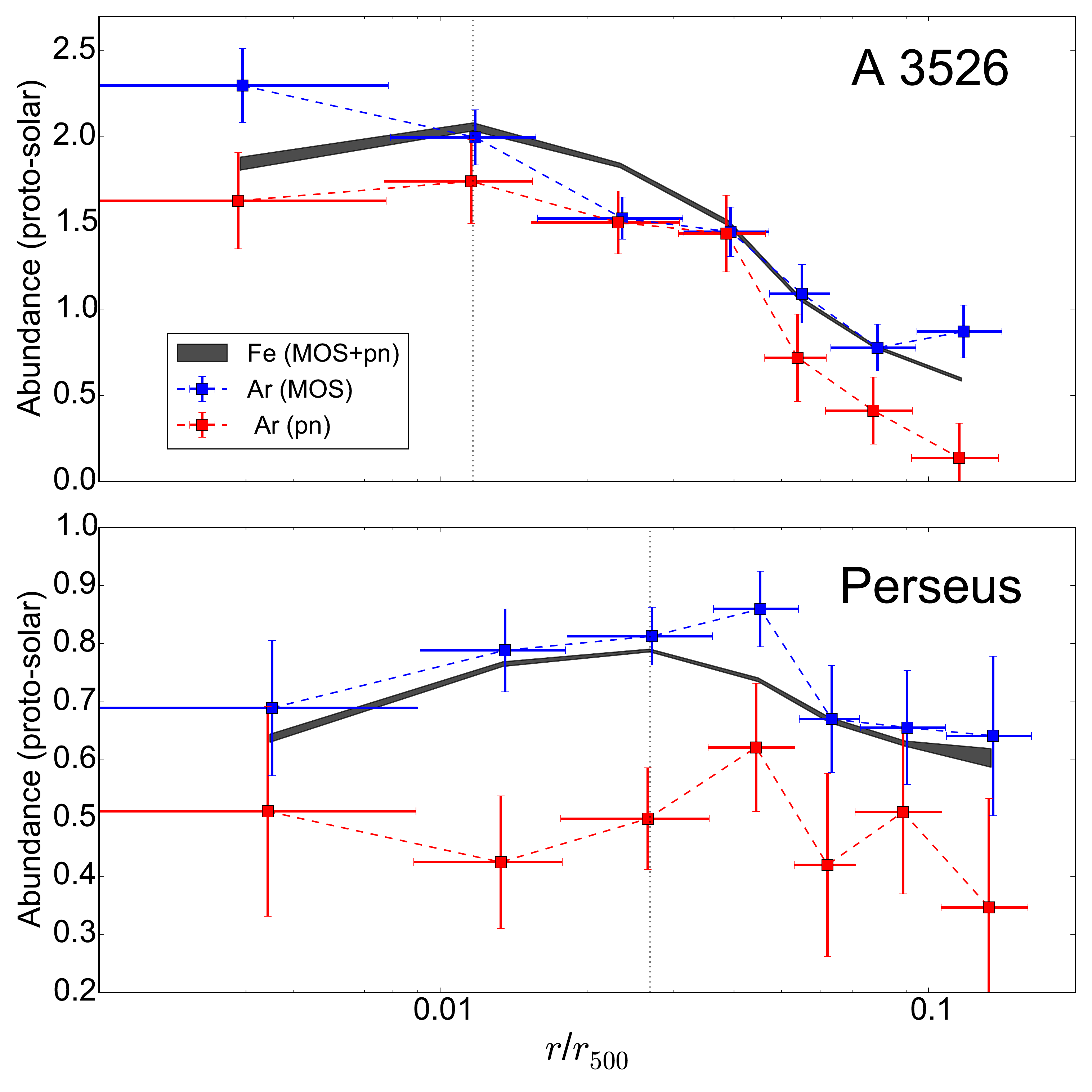}

        \caption{Ar radial profiles in A\,3526 (upper panel) and Perseus (lower panel) for independent MOS and pn measurements. The shaded areas show the (combined MOS+pn) Fe radial profiles. The grey vertical dotted lines indicate the Fe peaks.}
\label{fig:Ar_drop}
\end{figure}

Third, and alternatively, the drops could be the result of the depletion of a part of the ICM-phase metals into dust grains. In the scenario proposed by \citet{2013MNRAS.433.3290P,2015MNRAS.447..417P}, a significant part of the metals released by SNe within the BCG remain in the form of cold dust grains \citep{2011ApJ...738L..24V} and become incorporated into the central dusty filaments. These dust grains are then dragged out by buoyant bubbles caused by the AGN activity and are released back in the hot ICM phase out of the very core, thereby forming the off-centre Fe peak. This idea is supported by the presence of dust in most of the objects studied by \citet{2015MNRAS.447..417P} and showing a metallicity drop. The authors emphasise that such a scenario can be tested by the behaviour of the Ne and Ar radial profile in the very core of clusters and groups. Indeed, while elements like Fe, Si, and S are known to be easily embedded in dust grains, Ne and Ar are noble gases and are not expected to be incorporated into dust\footnote{Dusty Ar might appear in the form of cold molecular gas $^{36}$ArH$^+$ \citep{2013Sci...342.1343B}, but the presence of such a gas in cluster cores still remains highly uncertain.}. Consequently, their radial abundance profiles should not show any sign of drop or flattening in the innermost regions. As mentioned in Sect. \ref{sect:thermal_mod}, the EPIC spectral resolution does not allow us to investigate the Ne radial distribution. Interestingly, the radial Ar distribution does not follow well the (rescaled) Fe distribution as it shows a sharper gradient than expected by the empirical Fe profile (Figs. \ref{fig:abundance_profiles} and \ref{fig:abundance_profiles_normFe}). This sharper gradient is consistent with the different average Ar/Fe ratios measured by \citet{2016A&A...595A.126M} in the $\le$0.05$r_{500}$ and the $\le$0.2$r_{500}$ regions. Similarly, the central ($\lesssim 0.02 r_{500}$) measured Ar abundances lie somewhat higher than expected. As an (speculative) explanation for this particular feature seen only in the Ar profile, dust depletion in the cool-core ICM (presumably affecting all the considered elements, except Ar) might play a substantial role in shaping the abundance profiles of depleted elements, in particular within $\sim$0.1$r_{500}$. However, our average Ar profile points towards the presence of a flattening (if not a drop) in the innermost bin (Fig. \ref{fig:abundance_profiles}), suggesting that dust depletion only may not be sufficient to explain the innermost metal drops. That said, the very large scatter of the Ar abundance prevents us from claiming any firm evidence for/against this scenario. When investigating the individual abundance profiles of Perseus and A\,3526 (i.e. the two objects hosting an abundance drop and providing the best statistics), as shown in Fig. \ref{fig:Ar_drop}, we find that the MOS measurements in A\,3526 suggest a monotonic increase of Ar towards the centre. The other measurements (pn in A\,3526, and MOS and pn in Perseus) instead suggest that Ar follows the Fe drop. To summarise, although we are not able to firmly favour or rule out this dust depletion scenario, our results might suggest a non-negligible effect of dust depletion of gas-phase metals in clusters, but do not confirm that metals that are embedded in dust in the very core of clusters/groups would be the unique origin of the abundance drops.

\begin{figure}[!]

                \includegraphics[width=0.5\textwidth]{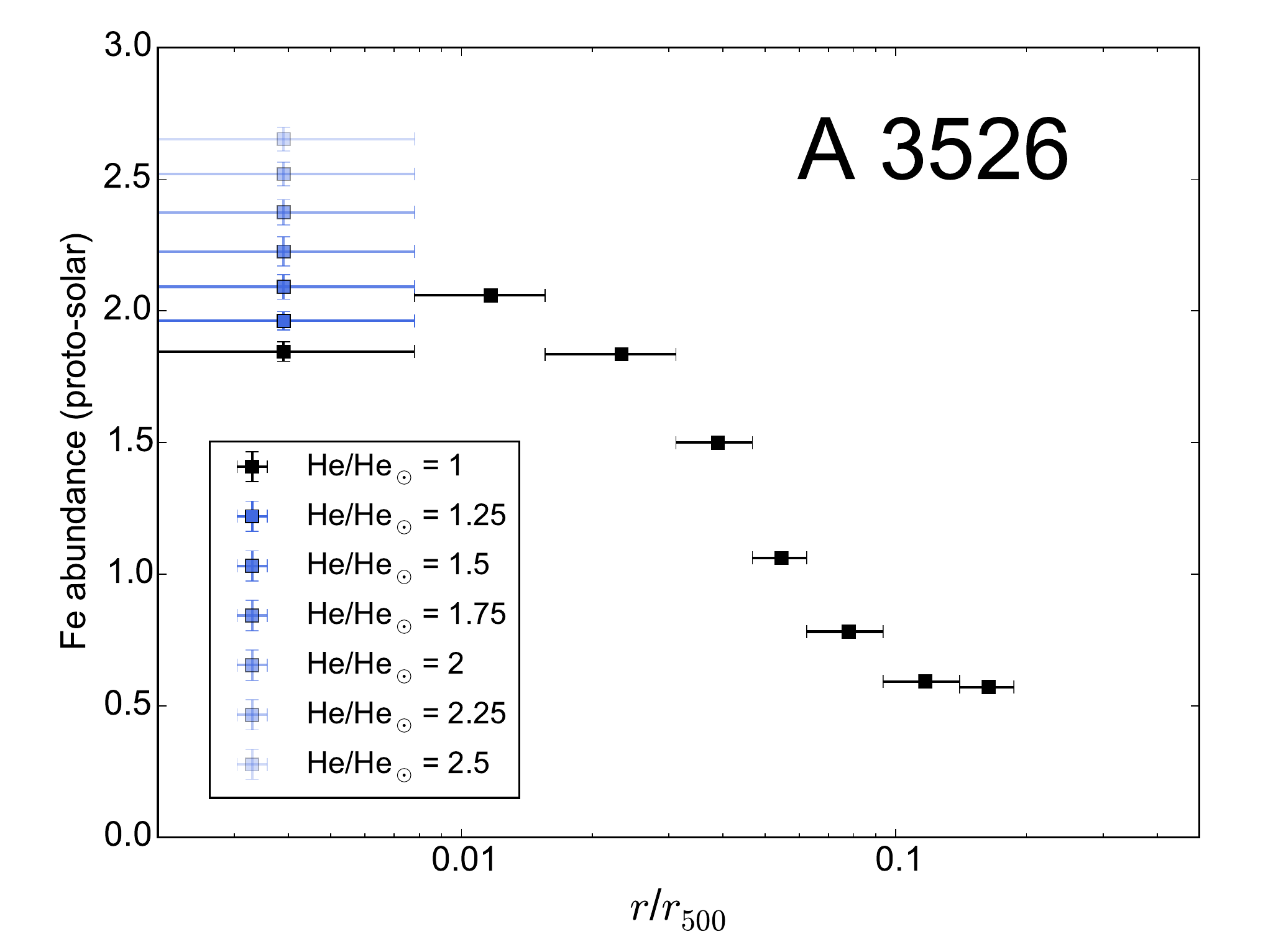}

        \caption{Effects of a hypothetical underestimate of the He fraction on the measured Fe abundance in the innermost bin of A\,3526.}
\label{fig:Fe_A3526_He}
\end{figure}

Fourth, the apparent drops may be the result of an underestimate of the helium content in the very core of such objects. Because He transitions do not occur at X-ray energies, it is impossible to provide any direct constraint on the He abundance in the ICM.
In all our fits (as in the majority of the similar studies found in the literature), we assume that He follows the primordial abundance \citep[$\sim$25\% of mass fraction; e.g.][]{2016arXiv160802062P}. However, the large gravitational potential in the core of clusters and groups may be efficient in retaining He, which could be more centrally peaked than H \citep{1977MNRAS.181P...5F,1981ApJ...248..429A}. If we effectively underestimate the He abundance in our fits of the core region, the net continuum would be overestimated, resulting in a bias of all our metal abundances towards lower values \citep[e.g.][]{2006MNRAS.369L..42E}. We illustrate this effect in Fig. \ref{fig:Fe_A3526_He}, where we assume the He abundance in our fits of the innermost bin of A\,3526 to be successively 1.25, 1.5, 1.75, 2, 2.25, and 2.50 times the primordial value. As can be seen, a He abundance that is 1.5 higher than previously assumed in the ICM core is sufficient to remove the inner Fe drop significantly. 
However, recent models point towards a less important He sedimentation in the very centre of cool-core clusters than in their surroundings \citep[$\sim$0.4--0.8$r_{500}$;][]{2009ApJ...693..839P}. Moreover, as already noted by \citet{2015MNRAS.447..417P}, thermal diffusion may also play an important role in counteracting He sedimentation and in removing He and other metals (including Fe) out of the very core of clusters \citep[Medvedev et al. \citeyear{2014MNRAS.440.2464M}; see also][]{2016ApJ...824...32B,2016ApJ...833..164B}. Nevertheless, the relative importance of thermal diffusion is also expected to be significantly weaker than the importance of AGN feedbacks, especially in galaxy groups, where most of the Fe drops are found.

Finally, and interestingly, some hydrodynamical simulations \citep{2005A&A...435L..25S,2009A&A...504..719K} predict a drop of central abundances when assuming galactic winds as the dominant mechanism transporting the metals from galaxies to the ICM. However, the typical extent of such a drop is $\sim$400 kpc, which is always much larger than the typical extents derived from our observations (a few tens of kpc at most). Moreover, this suppression of metal enrichment by galactic winds should preferentially happen in hot and massive clusters, where the ICM pressure is high enough. Instead, we find metal drops for a large portion of less massive objects.

\subsection{The overall Fe profile}\label{sect:discussion_Fe}

\subsubsection{Comparison with previous measurements}\label{sect:Fe_literature}

\begin{figure*}[!]

                \includegraphics[width=0.5\textwidth]{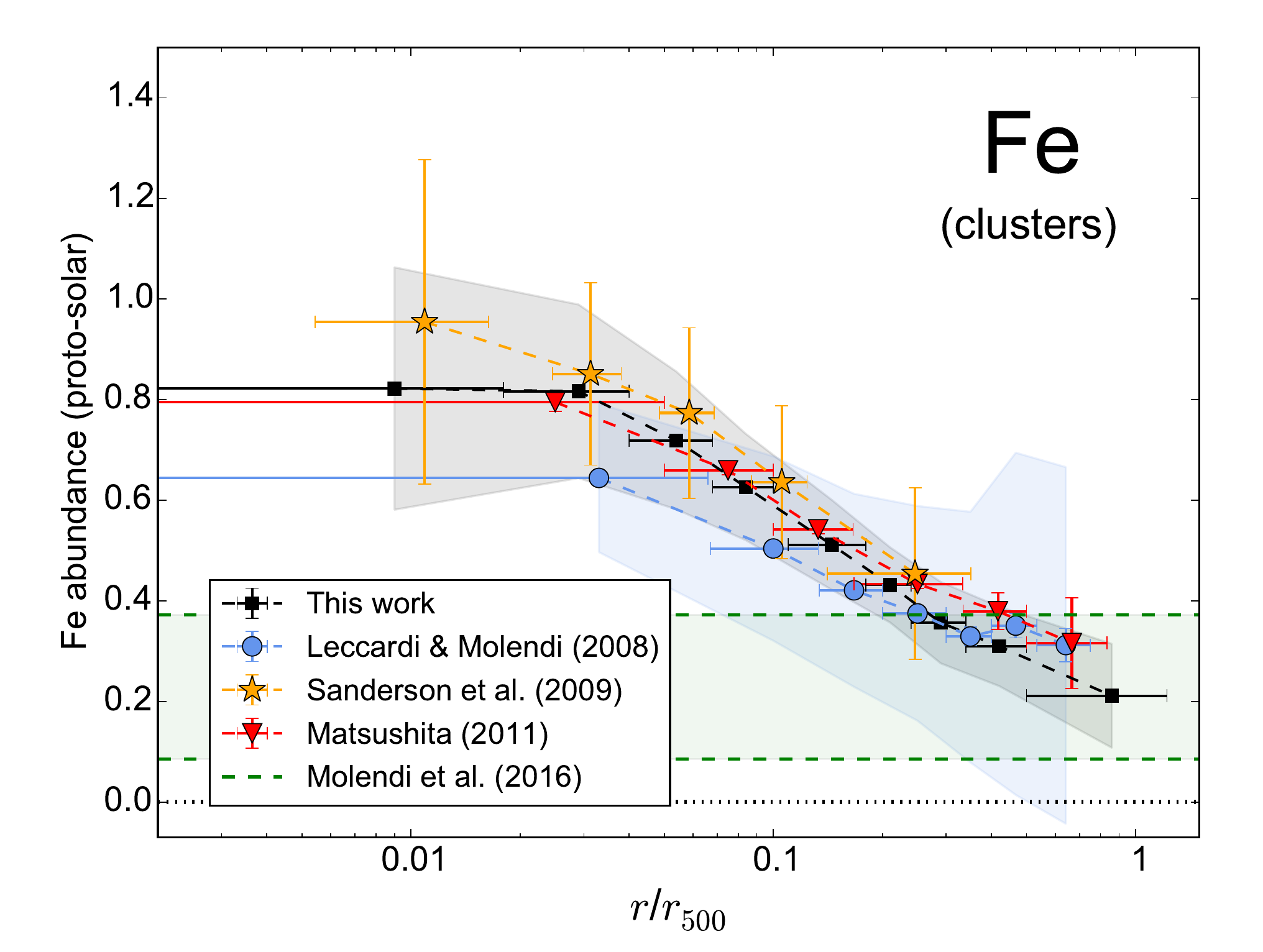}
                \includegraphics[width=0.5\textwidth]{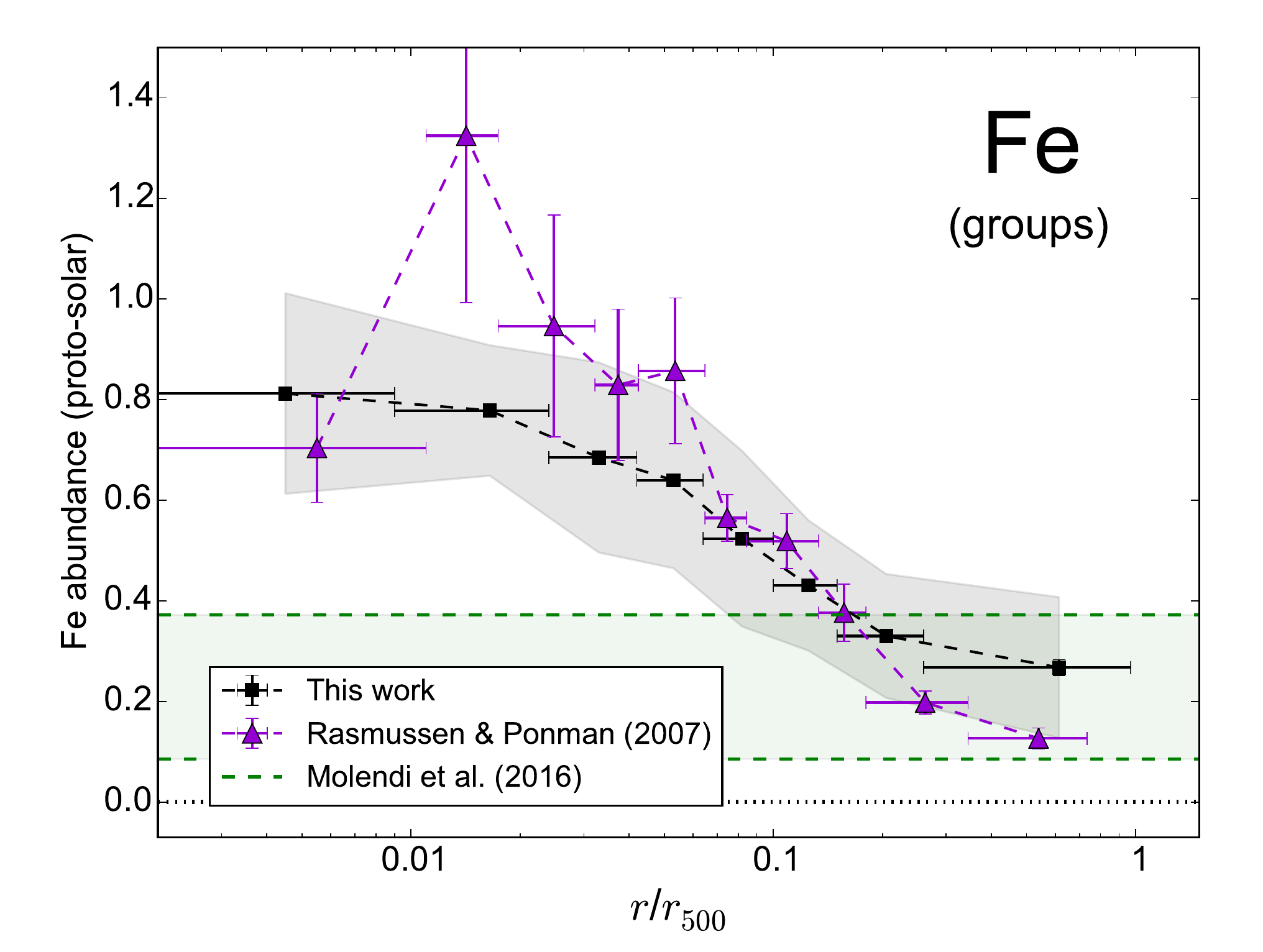}

        \caption{Comparison of our average radial Fe profiles (Fig. \ref{fig:Fe_radial_stacked_profile_clgr}) with estimations from previous works for clusters (\textit{left}) and groups (\textit{right}). Green dashed lines (and the corresponding shaded area) show the best constrained limits of the Fe abundance at $r_{180}$ ($\simeq 1.7 r_{500}$) derived by \citet{2016A&A...586A..32M}.}
\label{fig:Fe_profiles_literature}
\end{figure*}

The average Fe radial profile of our full sample (Fig. \ref{fig:Fe_radial_stacked_profile}) can be compared to other average profiles reported in the literature. \citet{2008A&A...487..461L} measured radial metallicity profiles for a sample of 48 hot ($\gtrsim$3.3 keV) intermediate redshift ($0.1 \lesssim z \lesssim 0.3$) clusters using \textit{XMM-Newton}. Similar studies for nearby cool-core clusters have been carried out by \citet[][\textit{Chandra}, $z < 0.1$]{2009MNRAS.395..764S} and \citet[][\textit{XMM-Newton}, $z < 0.08$]{2011A&A...527A.134M}. Finally, \citet{2007MNRAS.380.1554R} measured radial metallicity profiles for a sample of 15 nearby galaxy groups using \textit{Chandra}. 
Figure \ref{fig:Fe_profiles_literature} illustrates the comparison between our measurements and the three sample-based studies mentioned above. The choice of the reference (solar or proto-solar) abundance tables often varies in the literature; the most commonly used is \citet{1989GeCoA..53..197A}. Before comparing the profiles, all the abundances were rescaled to the proto-solar values of \citet{2009LanB...4B...44L} used in this work.

As seen in the left panel of Fig. \ref{fig:Fe_profiles_literature} (clusters), our Fe abundance profile is in excellent agreement with the measured profiles of \citet{2009MNRAS.395..764S} and \citet{2011A&A...527A.134M}. Only the second outermost bin of the profile of \citet{2011A&A...527A.134M} deviates from our values by <2$\sigma$, while all the other radial bins of these two profiles are 1$\sigma$ consistent with our measurements. The two innermost bins of the average profile of \citet{2008A&A...487..461L}, however, have significantly lower Fe abundances than this study. This can be easily explained, as the sample of \citet{2008A&A...487..461L} contains both cool-core and non-cool-core clusters. Because of their substantially less steep abundance decrease (Sect. \ref{sect:intro}), including non-cool-core clusters in a sample naturally flattens its average metallicity profile. Interestingly (and encouragingly), the four compared profiles agree very well beyond $\sim$$0.15r_{500}$ up to their respective outermost bins. This, together with the limited Fe scatter in the outermost radial bins of this work, may suggest a universal metallicity distribution outside cluster cores. We note that, however, from the 17 cool-core objects of the sample of \citet{2011A&A...527A.134M}, 13 are present in our sample as well (including M\,87). Very similar abundance profiles were thus expected, even at the cluster outskirts. Nevertheless, none of the clusters from the sample of \citet{2008A&A...487..461L} are also present in our sample, and the very similar average abundance ($\sim$0.2--0.3) found beyond $\sim$0.5$r_{500}$ for both nearby and intermediate redshift clusters is clearly an interesting result. Finally, the average Fe abundance measured in this work is fully consistent with the (large but conservative) limits at $r_{180}$ ($\simeq 1.7 r_{500}$) established by \citet{2016A&A...586A..32M}.

The right panel of Fig. \ref{fig:Fe_profiles_literature} (groups) shows a comparison between our average Fe abundance profile for groups and the average profile derived by \citet{2007MNRAS.380.1554R}. There is an overlap of six groups between the two samples. While the results agree below $0.01 r_{500}$ and within $0.07$--$0.2r_{500}$, disagreements can be seen elsewhere. Within $0.01$--$0.07r_{500}$, the \citet{2007MNRAS.380.1554R} abundances are <2$\sigma$ consistent with our average groups profile. However, the authors detect a deep average central abundance drop, which does not appear in our stacked profile. This difference may be explained by the large variety of metallicity profiles within the very core of groups, as seen in Fig. \ref{fig:Fe_profiles_indiv_groups} and in \citet[][see their fig. 3]{2007MNRAS.380.1554R}, and by the different groups selected in each respective sample. In particular, \citet{2007MNRAS.380.1554R} consider MKW\,4 part of their group sample, and using the ACIS instrument, they detect an off-centre Fe peak reaching $\sim$5--10 times the proto-solar value, which is more than two times the Fe abundance in its centre. This extreme measured metallicity should partly explain the high value of their second innermost average bin (Fig. \ref{fig:Fe_profiles_literature}, right).
On the other hand, mismatch is also found beyond $\sim$0.2$r_{500}$, where the average metallicity of \citet{2007MNRAS.380.1554R} in the outskirts is measured $\sim$2 times lower than in this work (although still within our inferred scatter). This issue is important to point out since \citet{2007MNRAS.380.1554R} interpret the lower enrichment in the group outskirts as a different groups enrichment history compared to more massive clusters. While uncertainties in the respective background treatments of the studies might explain the disagreement with our results, we must point out that an updated \textit{Chandra} calibration may revise upwards the Fe abundance in the outermost bins \citep[e.g. $\sim$+25\% for NGC\,4325][]{2009MNRAS.399..239R}. Moreover (and perhaps more importantly), \citet{2007MNRAS.380.1554R} measured the Fe abundances via only the Fe-L complex, and they assumed a single-temperature model in the spectra of each of their outermost bins. This may significantly underestimate the Fe abundance in case of a multi-phase plasma in the group outskirts.

\subsubsection{Comparison with simulations}\label{sect:Fe_simulations}

\begin{figure}[!]

                \includegraphics[width=0.5\textwidth]{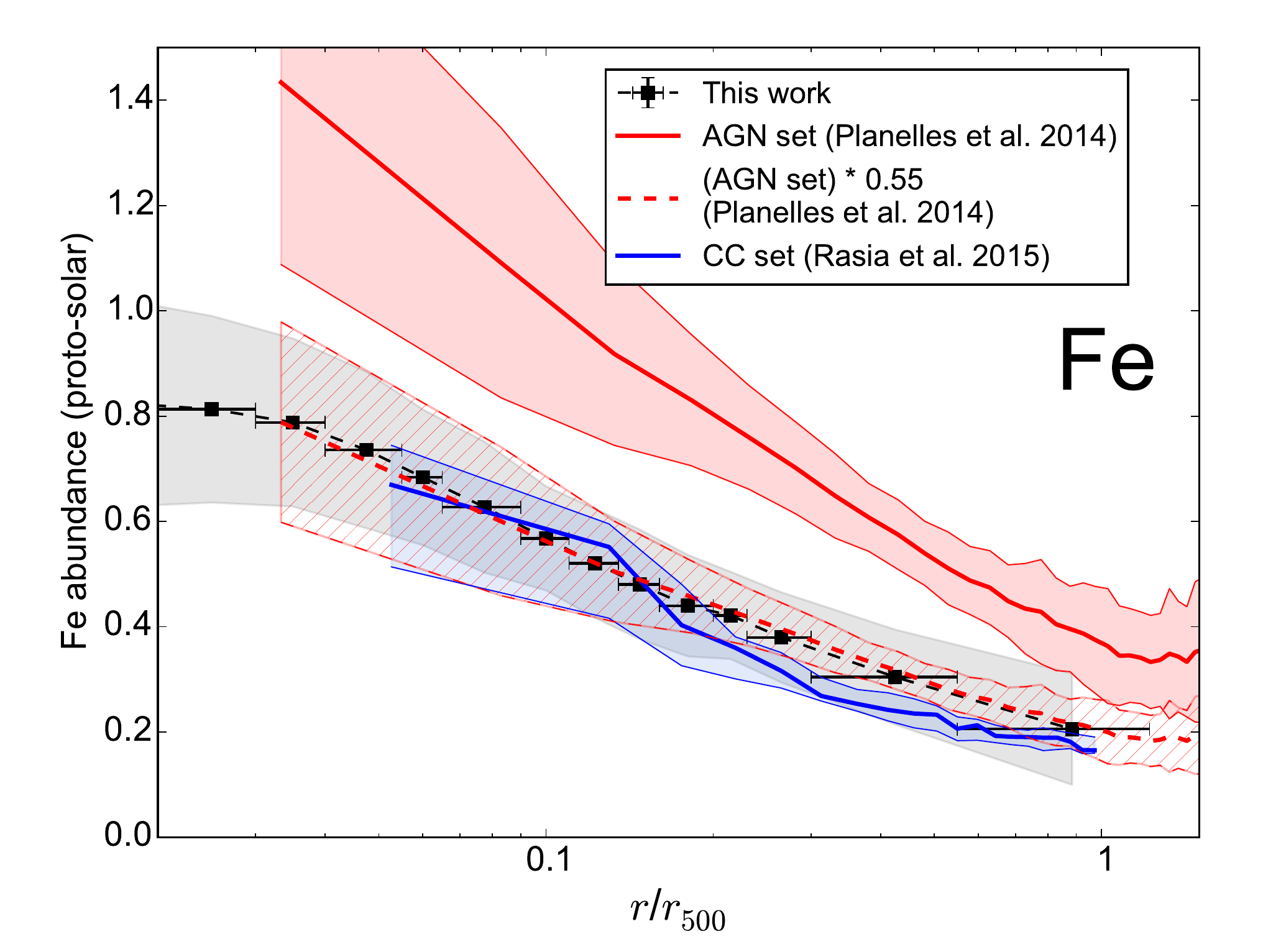}

        \caption{Comparison between our average Fe measured radial profile (Fig. \ref{fig:Fe_radial_stacked_profile}) and predictions from hydrodynamical simulations from \citet[][solid red lines]{2014MNRAS.438..195P} and \citet[][solid blue lines]{2015ApJ...813L..17R}, both modelling AGN feedback effects on the chemical enrichment. The red dashed lines show the same simulation set from \citet[][]{2014MNRAS.438..195P} with a normalisation rescaled by a factor of 0.55.}
\label{fig:Fe_profiles_simulations}
\end{figure}

The average Fe radial profile derived in this work (Fig. \ref{fig:Fe_radial_stacked_profile}) can also be compared with the average Fe profile predicted by hydrodynamical simulations. Two of the most recent simulation sets of the ICM including metal enrichment were performed by \citet{2014MNRAS.438..195P} and \citet{2015ApJ...813L..17R}. Both sets use the smooth particle hydrodynamics code GADGET-3, assume a Chabrier initial mass function \citep[IMF;][]{2003PASP..115..763C}, and incorporate the chemical evolution model (including metal production by SNIa, SNcc, and AGB stars) of \citet{2007MNRAS.382.1050T}, taking SN-powered galactic winds and AGN feedback into account. The comparison of our average Fe profile with these two simulation sets is shown in Fig. \ref{fig:Fe_profiles_simulations}.

The mean emission-weighted Fe profile from the "AGN set" of \citet[][derived from a sample of 36 hot nearby systems within 29 simulated regions]{2014MNRAS.438..195P}, shown in solid red lines (with its scatter in the shaded red area) in Fig. \ref{fig:Fe_profiles_simulations}, does not agree with our observations. In fact, a similar result was already discussed by the authors when comparing their predictions with the observations of \citet{2008A&A...487..461L}. However, as explained by \citet[][]{2014MNRAS.438..195P}, this significantly higher normalisation can be easily explained by outdated assumptions on the SN yields, the assumed IMF, the fraction of binary systems (eventually resulting in SNIa), and/or the SN efficiency to release metals into the ICM. The overall shape of the AGN set profile, however, is more crucial to confront with observational data, since AGN feedbacks presumably have a strong influence on (i) displacing metals from star-forming regions, (ii) suppressing star formation, and (iii) preventing cooling of the hot gas to temperatures emitting outside of the X-ray energy band. Interestingly, when applying a factor of $\sim$0.55 to the normalisation of this predicted Fe profile (dashed red lines in Fig. \ref{fig:Fe_profiles_simulations}), we find an excellent agreement with our measurements. In other words, the simulations of \citet[][]{2014MNRAS.438..195P} are remarkably successful at reproducing the measured chemical properties of the ICM, as long as the overall metal content produced and released in the gas phase is $\sim$1.8 times lower than originally assumed. This is not impossible, as both SN yields and SNIa rates are still uncertain within a factor of $\sim$2 \citep{2009MNRAS.399..574W}. However, a direct comparison between our results and the simulations of \citet[][]{2014MNRAS.438..195P} should be treated with caution. In fact, the simulation sets of \citet[][]{2014MNRAS.438..195P} contain both relaxed and non-relaxed systems (and fail to recover the cool-core versus non-cool-core dichotomy), while our observation are only based on cool-core clusters. Moreover, the simulated profiles are extracted from three-dimensional spherical shells, whereas our results are projected on the plane of the sky. This latter difference, however, is not expected to strongly affect the present comparison (Sect. \ref{sect:systematics_proj}).

A significant improvement of the simulation sets of \citet[][]{2014MNRAS.438..195P} has been achieved by \citet[][]{2015ApJ...813L..17R}, shown by the solid blue lines (with its scatter in the shaded blue area) in Fig. \ref{fig:Fe_profiles_simulations}. This more recent set of simulations, also including AGN feedback effects, constitutes the first success of disentangling cool-core (shown in Fig. \ref{fig:Fe_profiles_simulations}) and non-cool-core clusters. We find a reasonable agreement between the simulated profile of \citet[][]{2015ApJ...813L..17R} and our observed profile within $\sim$0.05--0.2$r_{500}$. Beyond $\sim$0.2$r_{500}$, the simulated profile slightly underestimates our observations ($\sim$20--25\%), but still lies within the scatter, which also includes possible systematic uncertainties (see Sect. \ref{sect:systematics}). Here as well, care must be taken when directly comparing observations and simulations. Similar to \citet[][]{2014MNRAS.438..195P}, the simulated profile of \citet[][]{2015ApJ...813L..17R} is also unprojected. Moreover, this profile is also mass weighted, while our measurements are directly derived from spectroscopy and are thus emission weighted. The conversion of mass weighted to emission weighted Fe profiles may result in a $\sim$30\% increase of the normalisation within $r_{500}$ \citep[][]{2014MNRAS.438..195P}. Such a change in the profile normalisation would lead to an excellent agreement with our results outside $\sim$0.2$r_{500}$, but to predictions that are slightly too high below this radius.

Furthermore, from a numerical point of view, simulations of the chemo-dynamical state of the very core ($\lesssim$0.05$r_{500}$) of the ICM are extremely challenging. Nevertheless, the good overall agreement between theoretical models and observations presented in this paper must emphasise the remarkable progress achieved by simulation groups in recent years. Future and more complete simulations will surely help to further improve the current picture of metal distributions in the ICM \citep[e.g.][]{2017arXiv170108164B}.

\subsection{Radial contribution of SNIa and SNcc products}\label{sect:discussion_SNe}

From Figs. \ref{fig:abundance_profiles} and \ref{fig:abundance_profiles_normFe} and the discussion above (e.g. Sect. \ref{sect:systematics_weight}), it clearly appears that the radial abundance profiles of O, Mg, Si, S, Ar, Ca, and Ni decrease with radius. Except Ar (see Sect. \ref{sect:Fe_drop}), all these profiles also scale quite remarkably with the Fe radial distribution, keeping a constant X/Fe ratio out to (and sometimes even beyond) $0.2 r_{500}$. In particular, the uniform radial O/Fe ratio is an important result. It is in contradiction with the flat O profiles found in, for example A\,496 \citep{2001A&A...379..107T}, M\,87 \citep{2001A&A...365L.181B,2003A&A...401..443M,2006A&A...459..353W}, NGC\,5044 \citep{2003ApJ...595..151B}, AWM\,7 \citep{2008PASJ...60S.333S}, and a sample of 19 clusters \citep{2004A&A...420..135T}. On the contrary, this trend is consistent with the peaked O profiles found in, for example A\,S1101 \citep{2006A&A...452..397D}, A\,3526 \citep{2006MNRAS.371.1483S}, Hydra A \citep{2009A&A...493..409S}, A\,3112 \citep{2012ApJ...747...32B}, A\,4059 \citep{2015A&A...575A..37M}, and 5 cool-core clusters \citet{2011A&A...528A..60L}.

\begin{figure}[!]

                \includegraphics[width=0.48\textwidth]{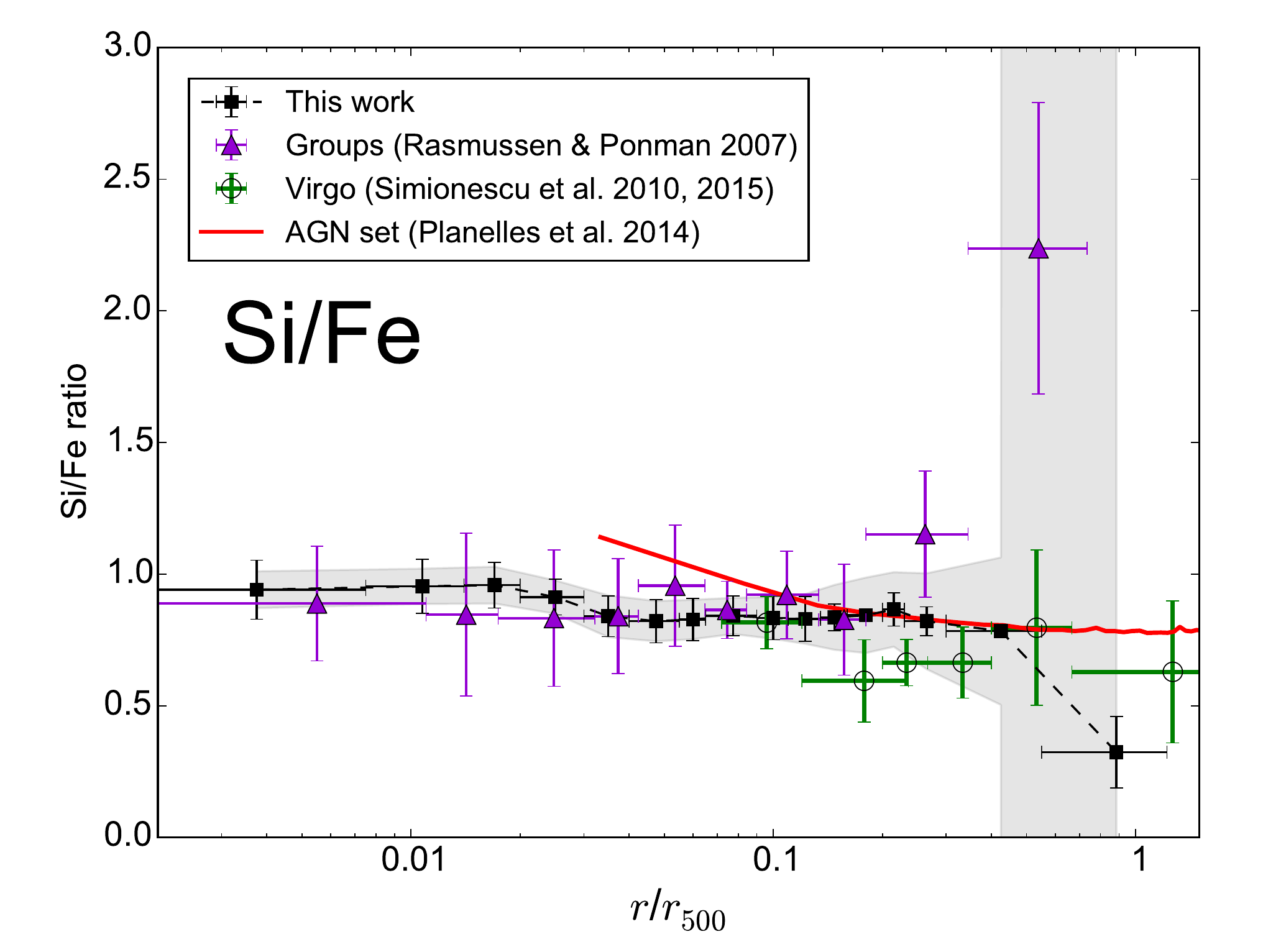}

        \caption{Comparison of the measured average radial Si/Fe profile (Fig. \ref{fig:abundance_profiles_normFe}) with previous observations for galaxy groups \citep{2007MNRAS.380.1554R} and Virgo \citep{2010MNRAS.405...91S,2015ApJ...811L..25S}, and with the AGN simulation set of \citet{2014MNRAS.438..195P}.}
\label{fig:SiFe_profiles}
\end{figure}

In Fig. \ref{fig:SiFe_profiles}, we show a comparison of our measured Si/Fe profile (from Fig. \ref{fig:abundance_profiles_normFe}) with two equivalent profiles reported from the literature. In their sample of 15 nearby galaxy groups, \citet[][purple triangles]{2007MNRAS.380.1554R} measured a flat Si/Fe profile up to $0.5 r_{500}$, followed by a dramatic increase in the outskirts (although observed with rather large error bars in two radial bins only). In a companion paper \citep{2009MNRAS.399..239R}, the same authors interpret this increase as a dominant enriching fraction of SNcc products in group outskirts, in agreement with the increasing O/Fe and/or Mg/Fe profiles observed in other studies (see above). Taking advantage of the low instrumental background of \textit{Suzaku} XIS, \citet[][four outermost green circles]{2015ApJ...811L..25S} reported a flat Si/Fe radial distribution in the outskirts of the Virgo cluster, in agreement with the Si/Fe ratios measured at smaller radii \citep[][two innermost green circles]{2010MNRAS.405...91S}. Our flat Si/Fe profile is in agreement with the results of \citet{2010MNRAS.405...91S,2015ApJ...811L..25S} and contradicts the results of \citet{2007MNRAS.380.1554R}. Furthermore, our results are consistent with the Si/Fe predictions from the simulation sets of \citet[][solid red line]{2014MNRAS.438..195P}, but we do not observe the slight predicted increase of Si relative to Fe towards the core below $0.1 r_{500}$, expected from the suppression of cooling (predominantly processed by SNcc products) due to the AGN feedback \citep[see also][]{2010MNRAS.401.1670F}. This issue was already discussed by \citet{2014MNRAS.438..195P}, and could be due to efficient diffusion or transport mechanisms that were not yet implemented in the simulations.

\begin{figure}[!]

                \includegraphics[width=0.48\textwidth]{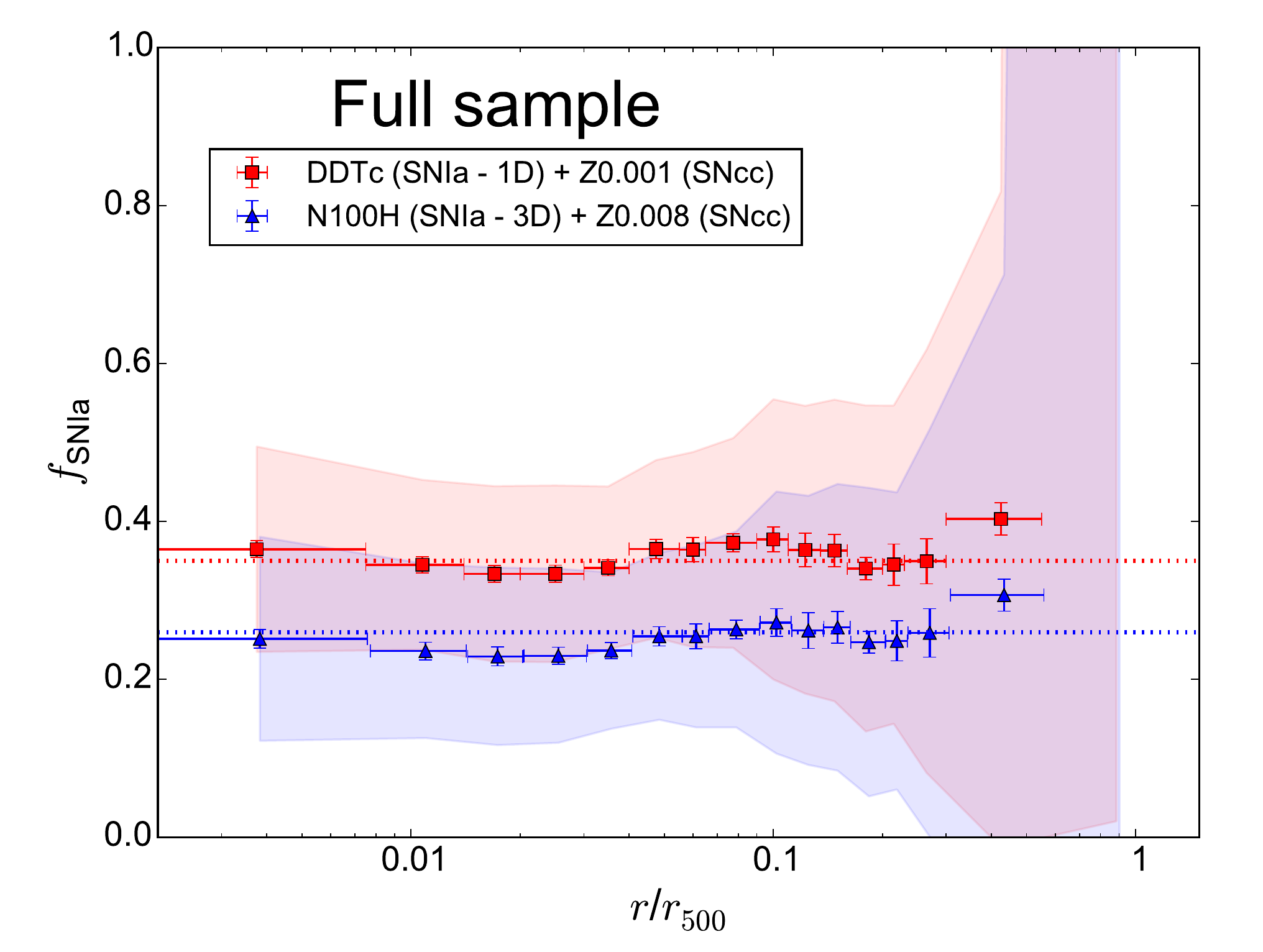}

        \caption{Radial dependency of the SNIa fraction contributing to the ICM enrichment ($f_\text{SNIa}$). Two combinations of SN yield models were adopted successively (see text). The corresponding shaded areas show the uncertainties when accounting for the scatter of the measurements. For each combination, the dotted line corresponds to $f_\text{SNIa}$ estimated within the core (0.2$r_{500}$ or 0.05$r_{500}$), averaged over the full sample \citep[see][]{2016A&A...595A.126M}.}
\label{fig:SNe_ratios}
\end{figure}

\begin{figure*}[!]

                \includegraphics[width=0.48\textwidth]{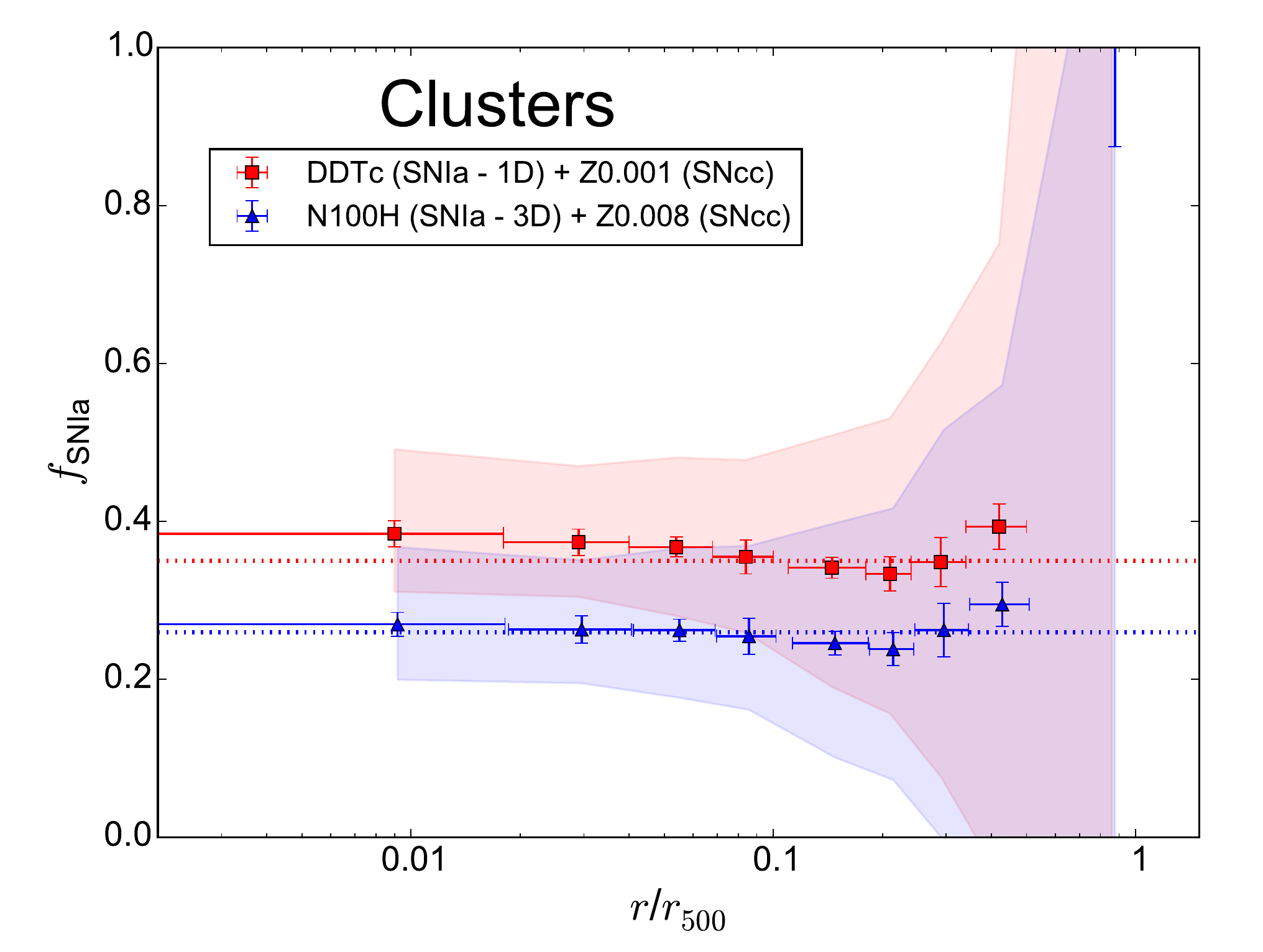}
                \includegraphics[width=0.48\textwidth]{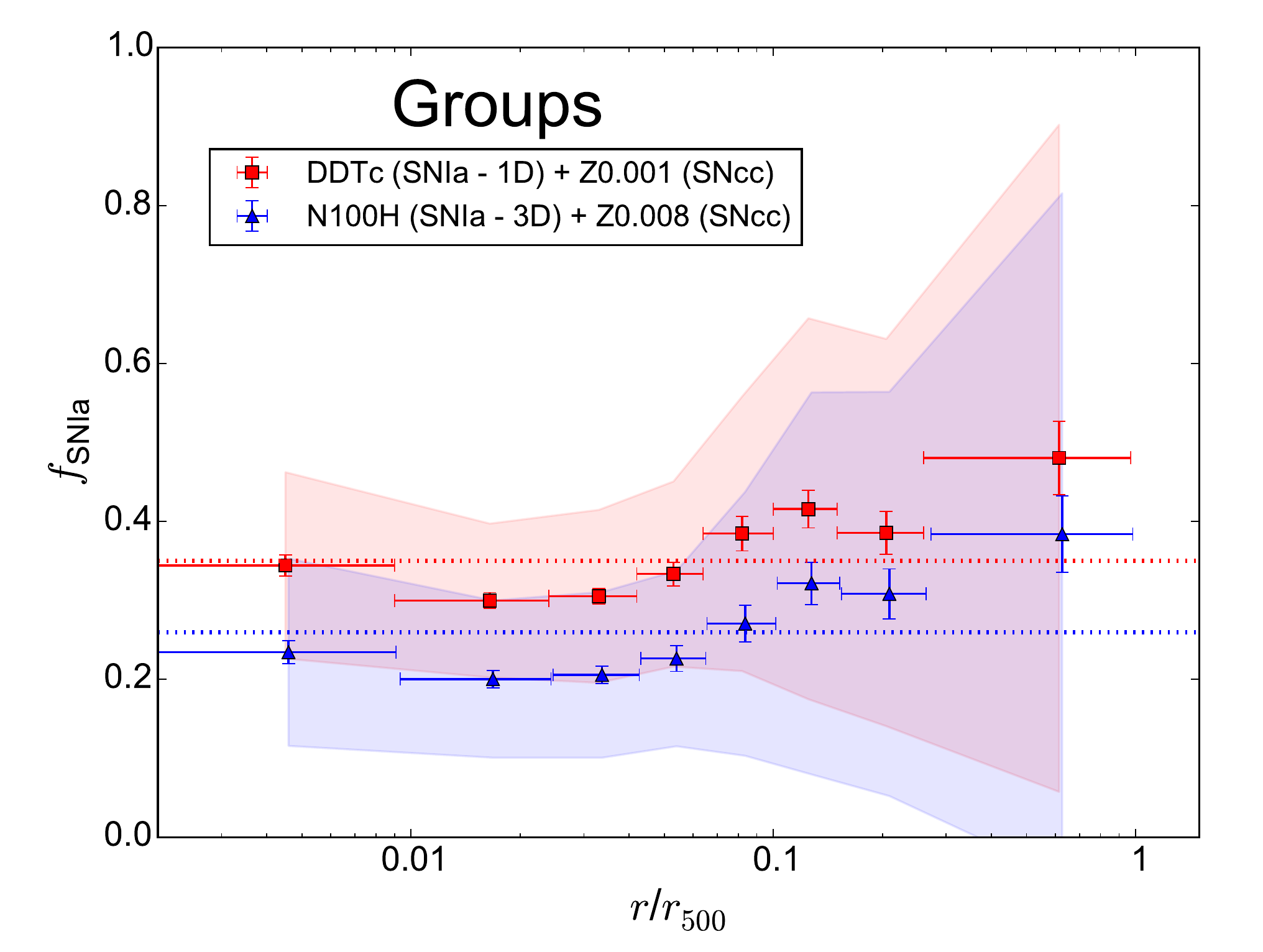}

        \caption{Same as Fig. \ref{fig:SNe_ratios}, with a differentiation for clusters (>1.7 keV, \textit{left}) and groups (<1.7 keV, \textit{right}).}
\label{fig:SNe_ratios_clgr}
\end{figure*}

In order to better quantify the radial contribution of SNIa and SNcc products, we fit the X/Fe abundance ratios in each radial bin with a combination of SNIa and SNcc yield models as described in \citet{2016A&A...595A.126M}. Based on their results, and because the large uncertainties of the measured abundances in individual bins do not allow us to favour any yield model in particular, we select the following two combinations of one SNIa and one SNcc model that reproduce equally well the average abundance pattern within the ICM core \citep[$0.2 r_{500}$ or $0.05 r_{500}$;][]{2016A&A...595A.126M}:
\begin{enumerate}

\item The one-dimensional delayed-detonation SNIa yield model ("DDTc") introduced in \citet{2005ApJ...624..198B} that reproduces the spectral features of the Tycho supernova remnant \citep{2006ApJ...645.1373B}, combined with the SNcc yield model from \citet{2013ARA&A..51..457N} assuming an initial metallicity of stellar progenitors of $Z_\text{init} = 0.001$, and averaged over a Salpeter IMF \citep{1955ApJ...121..161S} between 10 and 40 $M_\sun$ ("Z0.001");

\item The three-dimensional delayed-detonation SNIa yield model ("N100H") from \citet{2013MNRAS.429.1156S}, combined with the SNcc yield model from \citet{2013ARA&A..51..457N}, assuming an initial metallicity of stellar progenitors of $Z_\text{init} = 0.008$ and IMF-averaged similarly as for the Z0.001 model ("Z0.008").

\end{enumerate}
We fit the X/Fe abundance pattern measured in each radial bin (Fig. \ref{fig:abundance_profiles_normFe}) successively with these two combinations of models. This allows us to estimate $f_\text{SNIa}$, the fraction of SNIa over the total number of SNe (i.e. SNIa+SNcc) contributing to the enrichment, as a function of the radial distance. This is shown in Fig. \ref{fig:SNe_ratios} (full sample) and Fig. \ref{fig:SNe_ratios_clgr} (clusters, left panel; groups, right panel). In all the (sub)samples, $f_\text{SNIa}$ is fully consistent with being uniform up to $\sim$0.5$r_{500}$, and agrees very well with the average values found in the ICM core \citep[][dotted horizontal lines in the figures]{2016A&A...595A.126M}. In some radial bins, we observe slight but significant (>1$\sigma$) deviations from these core-averaged values. For example, we cannot exclude a slight increase of $f_\text{SNIa}$ in groups, at least from $\sim$0.01$r_{500}$ to $\sim$0.1$r_{500}$. However, these deviations completely vanish when we account for the scatters of Fig. \ref{fig:abundance_profiles_normFe} in the estimation of $f_\text{SNIa}$ (shaded areas). Such a radially uniform fraction has also been recently measured in A\,3112 \citep{2016arXiv160903581E}.

As discussed in Sect. \ref{sect:systematics}, the average values may be affected by systematic uncertainties and accounting for the scatters is conservative enough to keep all the systematic effects under control. Consequently, and although the flat radial behaviour of $f_\text{SNIa}$ based on the average X/Fe ratios is quite remarkable (at least in clusters), we cannot fully exclude a changing SNIa/SNcc contribution to the enrichment beyond $\sim$0.2$r_{500}$ in clusters and/or groups. Finally, and unsurprisingly, we find that a different choice of SN yield models only affects the absolute average $f_\text{SNIa}$ value and not its relative radial distribution.

\subsubsection*{Implications for the enrichment history of the ICM}

As discussed throughout this paper, our results are fully consistent with a uniform contribution of SNIa (SNcc) products in the ICM from its very centre up to (at least) $\sim$0.5$r_{500}$. Although, accounting for various systematic uncertainties (including the population scatter, which dominates over the other uncertainties even at large radius), we cannot fully exclude an increase/decrease in the SNIa contribution to the enrichment outside $\sim$0.2$r_{500}$, we do not observe a clear trend supporting that scenario. If true, the uniform radial contribution of SNIa products in the ICM has interesting consequences, as it provides valuable constraints on the enrichment history of galaxy clusters/groups.

One of the main pictures (Sect. \ref{sect:intro}) that had been proposed to explain the results showing a flat O profile in the previous literature, is that the bulk of SNcc events would have exploded early on, during or shortly before the formation of clusters/groups ($\sim$10 Gyr ago), and their products would have efficiently diffused within the entire cluster. The Fe central excess, tracing the SNIa products, would then mostly originate from the BCG at later cosmic time, hence supporting the idea that SNIa explode significantly later than the time required for more massive stars to release (mostly via SNcc explosions) and diffuse their metals into the ICM. One issue with this scenario was that, whereas one should expect a shallower Si profile than the Fe profile (since Si is synthesised by both SNIa and SNcc), many previous studies reported a constant \citep[e.g.][]{2006MNRAS.371.1483S,2008PASJ...60S.333S} or sometimes even decreasing \citep{2011MNRAS.418.2744M} Si/Fe ratio across radius. To solve this paradox, \citet{2002A&A...381...21F} propose a diversity of SNIa to contribute to the ICM enrichment: promptly exploding SNIa (whose products are supposed to be efficiently mixed over the whole cluster) produce less Si than SNIa with longer delay times (mostly enriching the cluster core). 
Since our results suggest a uniform contribution of SNIa (SNcc) products in the core and in the outskirts, invoking a diversity in SNIa (as well as in their delay times) is not required anymore, and alternative scenarii should be considered.

In their study of the chemical enrichment in Hydra A, \citet{2009A&A...493..409S} found that the central O excess can be explained either if stellar winds are 3 to 8 times more efficient in releasing metals than previously predicted, or if 3--$8 \times 10^{8}$ SNcc had exploded in the cluster core over the last $\sim$10 Gyr. Alternatively, ram-pressure stripping may help to build a central peak of SNcc (and SNIa) products from infalling cluster galaxy members \citep{2006A&A...452..795D}; however such a process should also occur at rather large distances ($\sim$1 Mpc), while the O excess is only observed in Hydra A within $\sim$120 kpc. Similarly, \citet{2011MNRAS.418.2744M} found centrally peaked profiles for eight elements in the core of M\,87. In addition to the peaked Mg profile, they measured a steeper gradient for Si and S than for Fe, and interpret their findings as the result of efficient enriching winds from a central pre-enriched stellar population and/or intermittent formation of massive stars in the BCG.

If the central O (and/or Mg) excess is indeed due to a significant amount of concentrated SNcc explosions in the cluster core, one relevant question is whether this SNcc peak was produced prior to the formation of the BCG, or by the BCG itself at a later stage of the cluster assembly. 
Recent simulations \citep{2007MNRAS.382.1050T,2010MNRAS.401.1670F} suggest that the enrichment time of both O and Fe in the inner  $\sim$0.4$r_{500}$ is significantly shorter than outside this radius, which may imply that the BCG is indeed responsible for the central excess in the ICM observed for both SNIa and SNcc products. Moreover, the recent analysis of WARPJ1415.1+3612 ($z \simeq 1$) by \citet{2014A&A...567A.102D} shows that the bulk of the central Fe excess was already present $\sim$8 Gyr ago and that its slope is steeper than at present times. This suggests in turn that the BCG is the dominant source responsible for the enrichment in the ICM core, and that the metals released by the BCG spread out of the core with time via diffusive/mixing processes.

If the Fe peak indeed comes from the BCG \citep[as the Fe mass in the ICM could suggest;][]{2004A&A...416L..21B,2004A&A...419....7D} and has a similar (scaled) radial distribution as SNcc products, as our results suggest, this central SNcc enrichment may also originate from the BCG. Although most BCGs appear red and dead at present times \citep[with typical star formation rates of a few $M_\sun$/yr at most; e.g.][]{2011ApJ...734...95M}, their star formation was dramatically higher over the last $\sim$9 Gyr \citep{2016ApJ...817...86M}, and in some cases, can still reach a few tens to hundreds $M_\sun$/yr at $z \simeq 0$ \citep{2008ApJ...681.1035O}. This past (and, sometimes, present) high star formation in BCGs could thus be responsible  for the central excess of SNcc products seen in the ICM. In this case, and assuming that some mechanisms diffuse out the metals from the cluster core (see above), the consistency between the slopes of the radial SNIa and SNcc distributions suggests that the bulk of SNIa exploded quite shortly after the period of star formation in the BCG. More precisely, the typical delay time of SNIa should not be larger than the timescale of metal mixing/diffusing processes in the ICM. 

More generally, and regardless of whether the central excess of SNcc products reported in this study originates from the BCG or not, the (lack of) radial dependence translates into a time dependence of the chemical enrichment patterns that we can infer. Specifically, the consistent radial profiles for all the measured abundances may suggest that the SNIa and SNcc components of the enrichment originate from the same astrophysical source(s) and have been occurring at similar epochs. 
Such a reasoning can be applied to the case of the intra-cluster stellar population. Both observations \citep[e.g.][]{2006AJ....131..168K,2007AJ....134..466K} and simulations \citep{2004MNRAS.355..159W} provide increasing evidence for a significant fraction (10--50\%) of stars that are unbound to any cluster galaxy and could potentially contribute to the ICM enrichment \citep{2004A&A...425L..21D}. As it takes a substantial time for these stars to be ejected and travel away from their galaxy hosts, the intra-cluster population should essentially contain low-mass stars, and thus enrich the ICM predominantly with SNIa, likely providing a different radial distribution of SNIa products than that of SNcc products (coming from other sources). This picture disagrees with our present results. Therefore, under these assumptions, intra-cluster stars may not be the dominant source of the ICM enrichment. A similar conclusion is reached by \citet{2010A&A...516A..41K} on the basis of hydrodynamical simulations and SNIa expected rates.

In summary, while it was commonly thought from previous studies that the bulk of the SNcc (SNIa) enrichment would contribute only at early (late) times, recent works --- including our present study --- have provided increasing evidence that the SNIa versus SNcc dichotomy is not pronounced since the chemical composition does not evolve dramatically with radius.

The astrophysical implications discussed here hold only if further and definitive confirmation of the uniform distribution of $f_\text{SNIa}$ is achieved with more accurate instruments on board future missions. In particular, the high spectral resolution and effective area of \textit{Athena} will be required to investigate the distribution of key elements, like O or Mg, with unprecedented accuracy from the core to the outskirts. Moreover, a complete discussion would be required to fully quantify the speculative arguments used here, and therefore, to pursue the extensive use of realistic hydrodynamical simulations, preferably including all the potential sources of (SNIa and SNcc) enrichment and all the mixing and diffusion mechanisms known so far.


\section{Conclusions}\label{sect:conclusion}


In this work, we used deep \textit{XMM-Newton}/EPIC observations of 44 nearby cool-core galaxy clusters, groups, and ellipticals (all taken from the CHEERS catalogue, i.e. $\sim$4.5 Ms of total net exposure) to derive the average projected radial abundance profiles of eight elements in the ICM. Whereas average Fe and Si abundance profiles had been previously reported in the literature (though over limited samples only), the O, Mg, S, Ar, Ca, and Ni profiles are measured and averaged over a large sample for the first time. This allows an unprecedented estimation of the average radial contribution of SNIa and SNcc products in the ICM. Our results can be summarised as follows.

\begin{itemize}

\item The Fe abundance can be robustly constrained out to $\sim 0.9 r_{500}$ and $\sim 0.6 r_{500}$ in clusters and groups, respectively, while most of the other abundances are uncertain beyond $\sim$0.5$r_{500}$. Owing to a robust and conservative modelling of the EPIC background, the systematic background uncertainties are limited typically to a few per cent, which are usually smaller than (or comparable to) the statistical uncertainties for each object. The other systematic uncertainties (related to MOS-pn discrepancies, projection effects, an uncertain temperature distribution, or selection effects) are always smaller than the population scatter derived in each average profile. Therefore, the latter can be considered as a conservative limit for our measurements. 

\item The average radial profiles of all the considered elements exhibit a centrally peaked distribution, and seem to converge at large radii consistently towards the limits (0.09--0.37 times proto-solar) assessed at $r_{180}$ by \citet{2016A&A...586A..32M}. When rescaled by the X/Fe ratios measured previously in the ICM core \citep{2016A&A...592A.157M}, the average profiles of all the elements (except perhaps Ar) follow the average Fe profile very well out to at least $\sim$0.5$r_{500}$. Similarly, the average radial X/Fe profiles (again, with the possible exception of Ar) are remarkably uniform out to this radius.

\item Subdividing our sample into clusters (>1.7 keV) and groups (<1.7 keV) subsamples, we find that groups are on average $\sim$21\% less enriched in Fe than clusters. From $0.01 r_{500}$ to $0.5 r_{500}$, this fraction is rather constant and no significant change is observed in the slopes of the two subsamples. Below and beyond this radial range, the similar enrichment level found in clusters and groups can be explained by selection and binning effects. Interestingly, no sign of metal enhancement towards more massive objects could be significantly detected in the other profiles (with the possible exception of the O profile).

\item The average Fe profile for clusters reported here agrees remarkably well with previous observations \citep{2008A&A...487..461L,2009MNRAS.395..764S,2011A&A...527A.134M}. The agreement of our average Fe profile for groups with the previous observations of \citet{2007MNRAS.380.1554R} is less good, but still comparable within uncertainties. Although it should be treated with caution, the comparison of our measured Fe profile with predictions from recent hydrodynamical simulations, taking AGN feedback and galactic winds effects into account \citep{2014MNRAS.438..195P,2015ApJ...813L..17R}, is also very encouraging. Future cluster simulations will be interesting to compare with our measurements.

\item In 14 systems ($\sim$32\% of our sample), we detect a significant central drop of the Fe abundance. This can also be observed in the average abundance profiles (both for Fe and the other elements) by an apparent flattening below $\sim$0.01$r_{500}$. We do not see a clear correlation between the depth of such metal drops and their radial extent. These drops are probably real and could be related to dust depletion of metals in the very core of the ICM, before they are dragged out by AGN feedback and released back in the hot gas phase. The slightly steeper profile of Ar (expected not to be incorporated in dust grains), compared to that of Fe, could (at least partly) witness dust depletion of the other elements within $\sim$0.1 $r_{500}$. However, the (statistical and systematic) uncertainties prevent us from firmly confirming or ruling out the presence of a central Ar drop.

\item Using the approach described in \citet{2016A&A...595A.126M}, we estimate the radial contribution of SNIa products to the ICM enrichment ($f_\text{SNIa}$). Although the scatter (and, by extension, the other systematic uncertainties) prevents us from excluding sudden changes in the outskirts, our observations suggest, on average, a remarkably uniform $f_\text{SNIa}$ distribution out to, at least, $0.5 r_{500}$. This result contrasts with the dramatic increase of SNcc contribution in the outskirts inferred by \citet{2009MNRAS.399..239R}, but is consistent with more recent measurements \citep{2015ApJ...811L..25S,2016arXiv160903581E} and simulations \citep{2010MNRAS.401.1670F,2014MNRAS.438..195P,2017arXiv170108164B}. This suggests that the major fraction of the SNIa and SNcc enriching the ICM may share the same origins and may have both exploded before mixing and diffusion processes played a significant role in spreading out the metals. In particular, since there is increasing evidence that the central Fe excess originates from the BCG, it is likely that a past intense period of star formation in the BCG had released SNcc products in the ICM core in a similar way.

\item Finally, we emphasise that, although the systematic uncertainties considered here are under control, the Ni abundance may be systematically overestimated when using SPEXACT v2. Whereas it should not have a significant impact on the shape of the Ni profile presented here, such a bias might affect the average Ni/Fe ratio \citep[e.g.][]{2016A&A...592A.157M} and the subsequent constraints inferred on the SNIa yield models \citep[e.g.][]{2016A&A...595A.126M}. We will devote a forthcoming paper to that specific issue.

\end{itemize}

While the abundance profiles of some elements (such as Fe or Si) could be remarkably constrained thanks to the large statistics of our sample, this paper clearly shows that, apart from the apparent scatter of the measurements, the most important limitations encountered so far are the systematic uncertainties, in particular related to MOS-pn cross-calibration imperfections \citep[see also][]{2015A&A...575A..30S,2016A&A...592A.157M}. Using the current X-ray facilities, a significant improvement of the accuracy of our results may only be achieved by improving the EPIC cross-calibration and better understanding all the systematic biases that could affect the EPIC instruments. Nevertheless, further improvement in interpreting these results could also come from studying a more representative sample, for example including non-cool-core systems as well.

Despite our current efforts and achievements, we must stress the considerable breakthrough that the next X-ray missions \citep[e.g. \textit{Athena};][]{2013arXiv1308.6784B} will be able to achieve. On the one hand, the very large effective area of future instruments will allow us to probe a detailed view of the chemical state of cluster outskirts, which is still challenging for \textit{XMM-Newton}, as demonstrated in this paper. On the other hand, the remarkable spectral resolution of micro-calorimeters on board these future missions will considerably reduce the uncertainties on both the thermal structure and the distribution of various metals within and outside cluster cores. Therefore, there is no doubt that the next generation of X-ray observatories will bring further light on this study and provide a valuable understanding of the full history of the ICM enrichment.

\begin{acknowledgements}
The authors are very thankful to the referee for valuable comments that helped to improve the paper. The authors would also like to thank Jesper Rasmussen and Alastair Sanderson for kindly providing their observational data, as well as Susana Planelles and Elena Rasia for kindly providing their simulation outputs and for useful discussions. This work is partly based on the \textit{XMM-Newton} AO-12 proposal "\emph{The XMM-Newton view of chemical enrichment in bright galaxy clusters and groups}" (PI: de Plaa), and is a part of the CHEERS (CHEmical Evolution Rgs cluster Sample) collaboration. H.A. acknowledges the support of NWO via a Veni grant. P.K. acknowledges financial support from STFC. C.P. acknowledges support from ERC Advanced Grant Feedback 340442. T.H.R. acknowledges support from the DFG through grant RE 1462/6 and through the Transregional Collaborative Research Centre TRR33 The Dark Universe, project B18. This project has been supported by the Lend\"ulet LP2016-11 grant awarded by the Hungarian Academy of Sciences. This work is based on observations obtained with \textit{XMM-Newton}, an ESA science mission with instruments and contributions directly funded by ESA member states and the USA (NASA). The SRON Netherlands Institute for Space Research is supported financially by NWO, the Netherlands Organisation for Scientific Research.
\end{acknowledgements}

\bibliography{Radial_abundances_hk}{}
\bibliographystyle{aa}

\newpage
\appendix

\section{Cluster properties and individual Fe profiles}

\begin{figure*}[!]

                \includegraphics[width=0.236\textwidth,trim={0 0 0 0},clip]{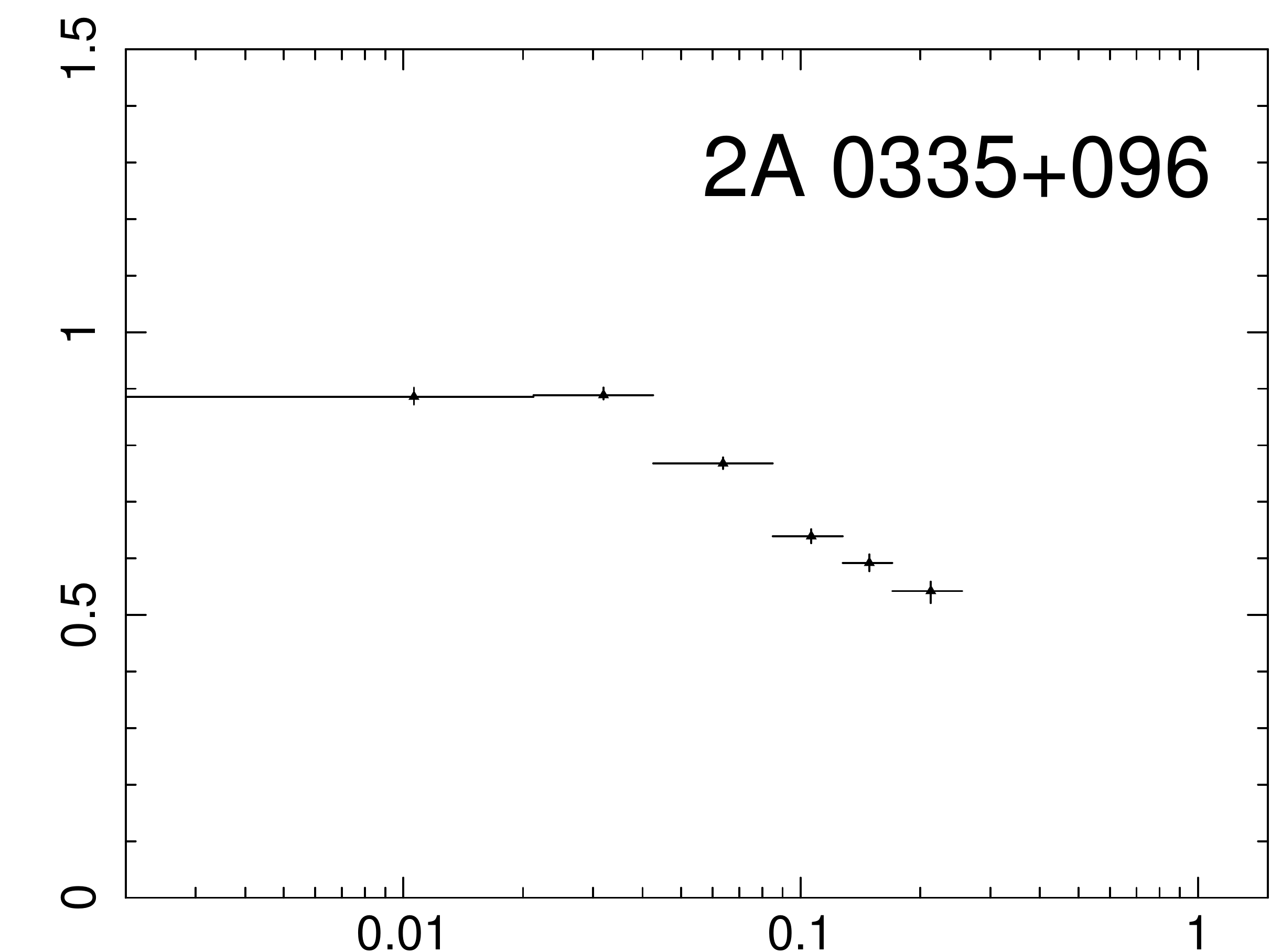}
                \includegraphics[width=0.236\textwidth,trim={0 0 0 0},clip]{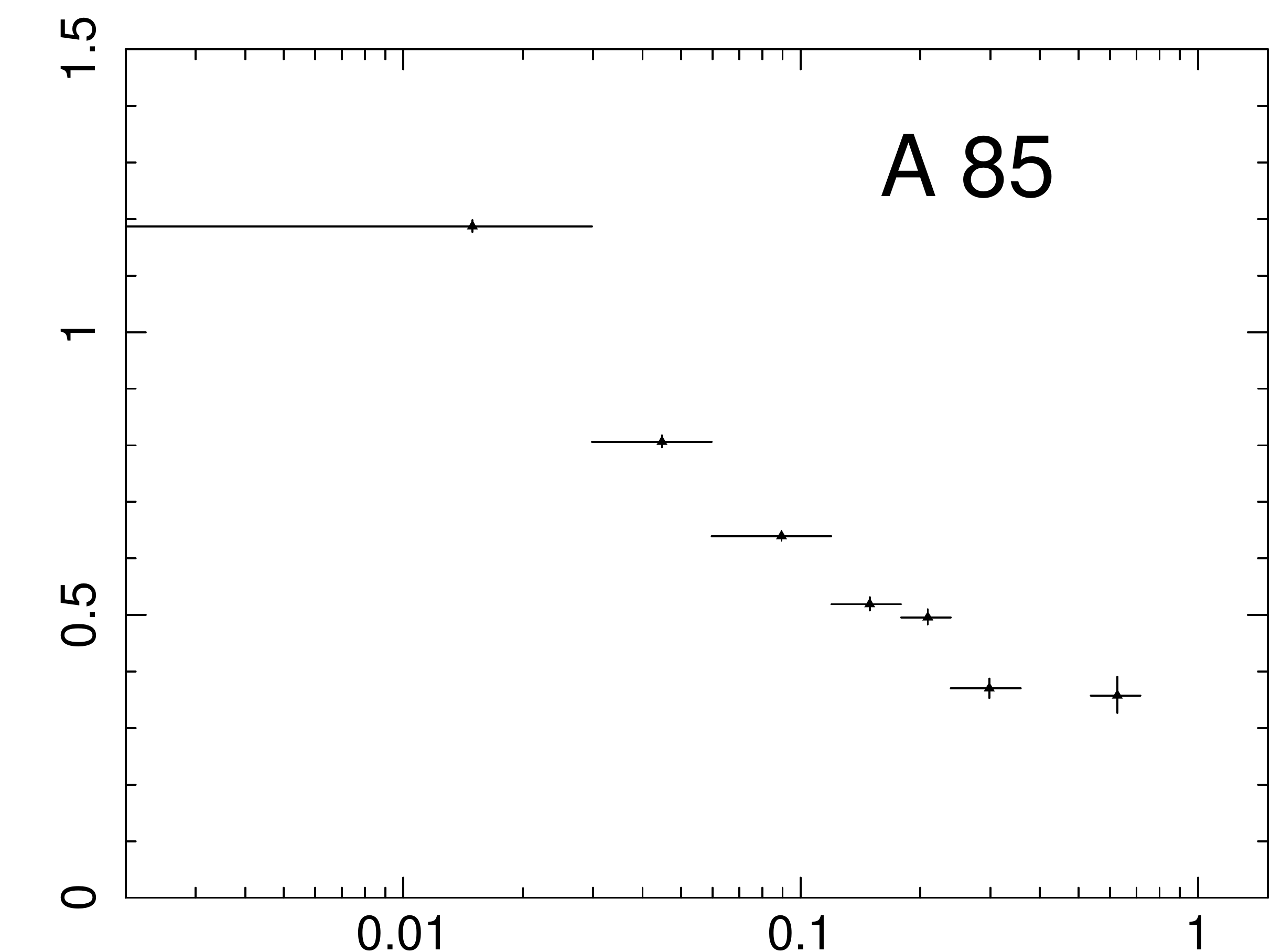}
                \includegraphics[width=0.236\textwidth,trim={0 0 0 0},clip]{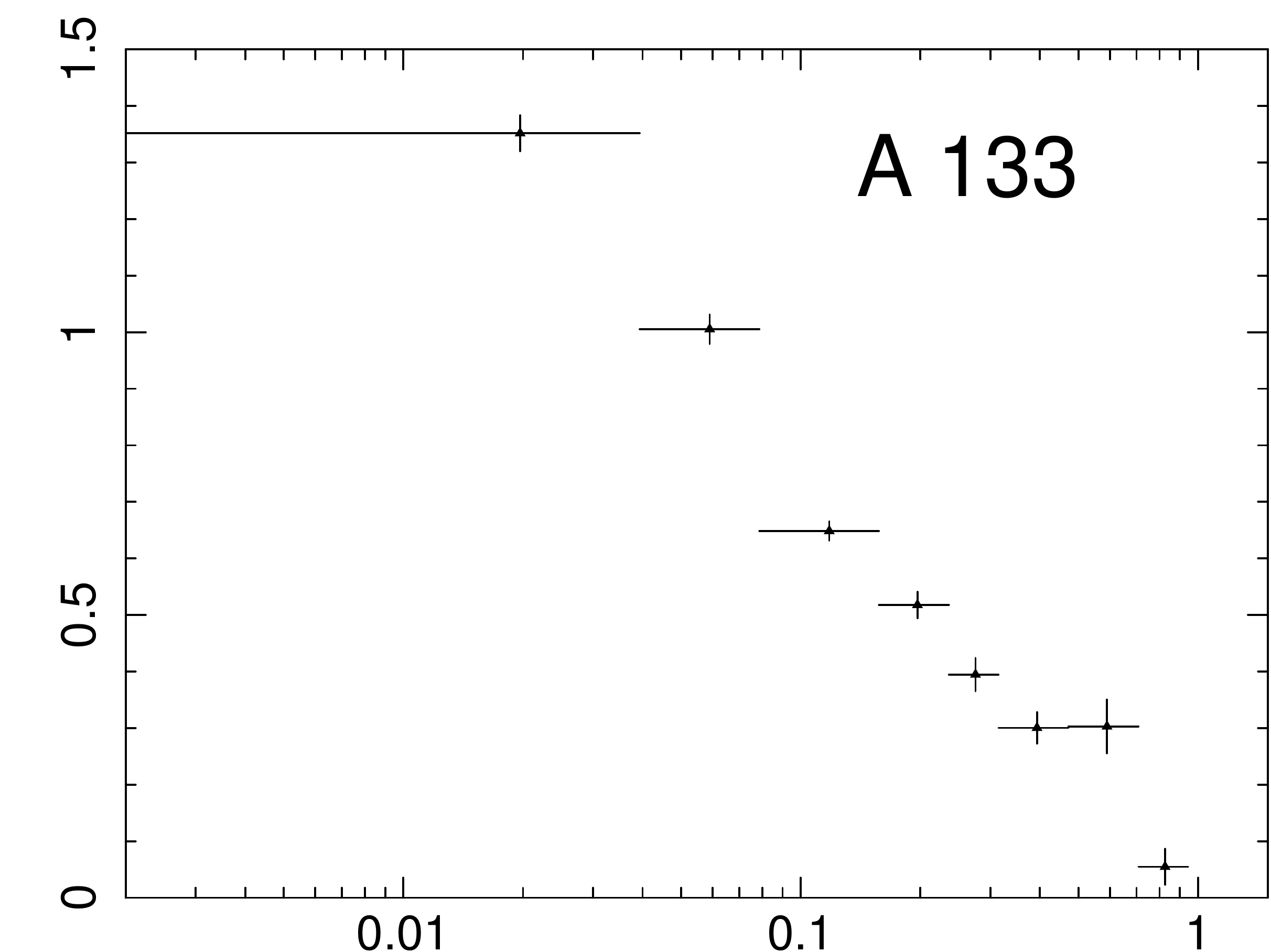}
                \includegraphics[width=0.236\textwidth,trim={0 0 0 0},clip]{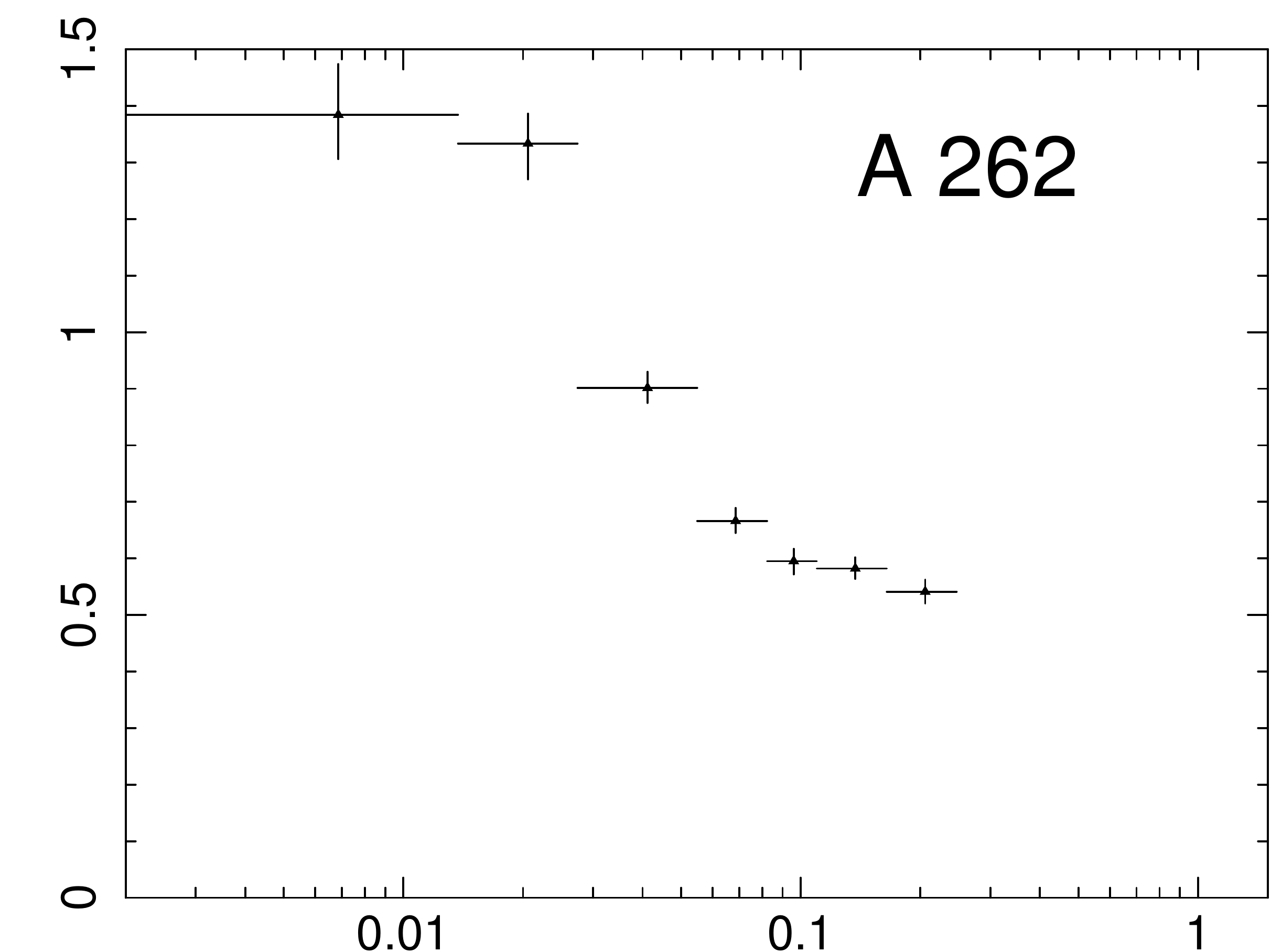} \\
                \includegraphics[width=0.236\textwidth,trim={0 0 0 0},clip]{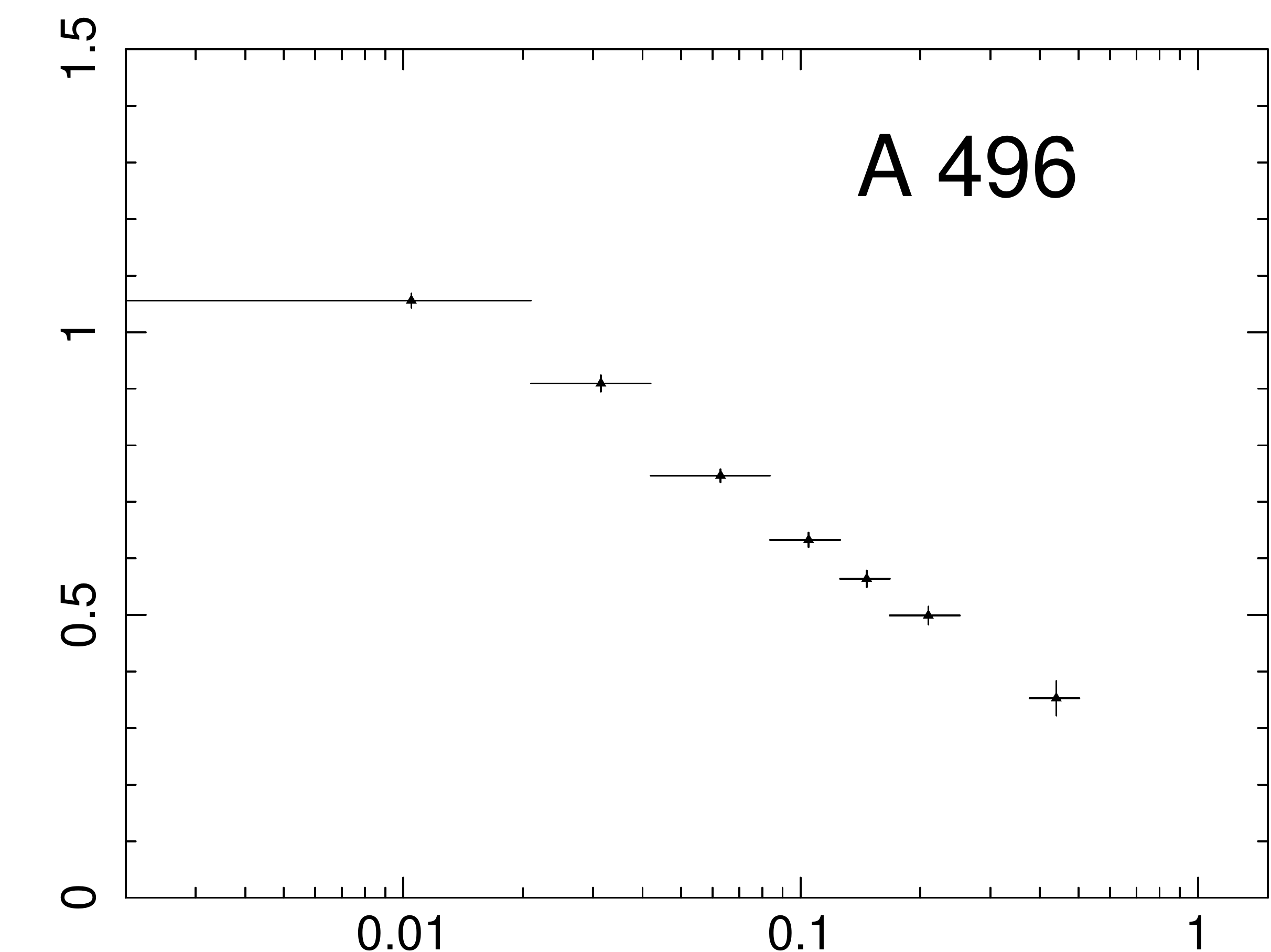}
                \includegraphics[width=0.236\textwidth,trim={0 0 0 0},clip]{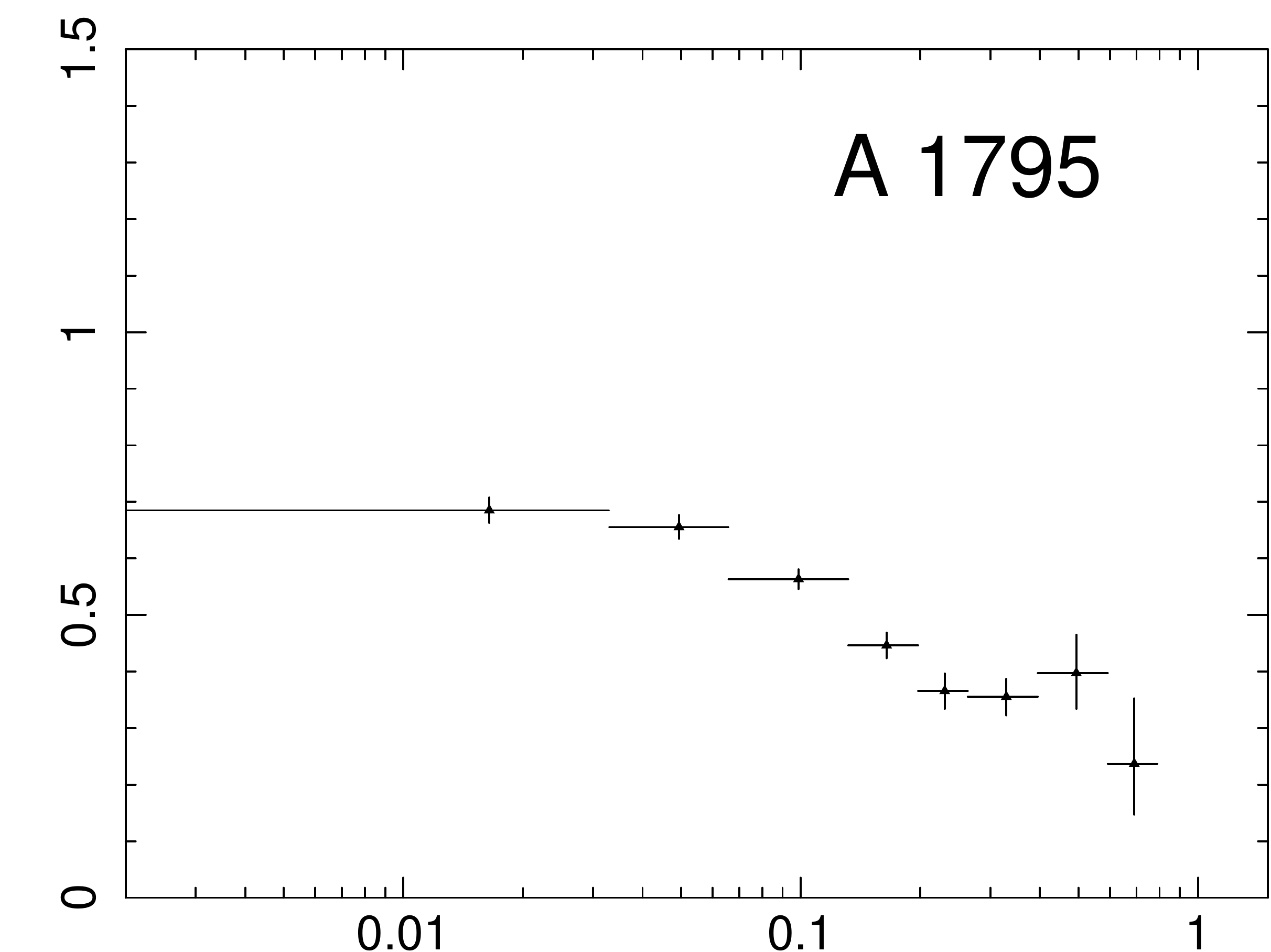}
                \includegraphics[width=0.236\textwidth,trim={0 0 0 0},clip]{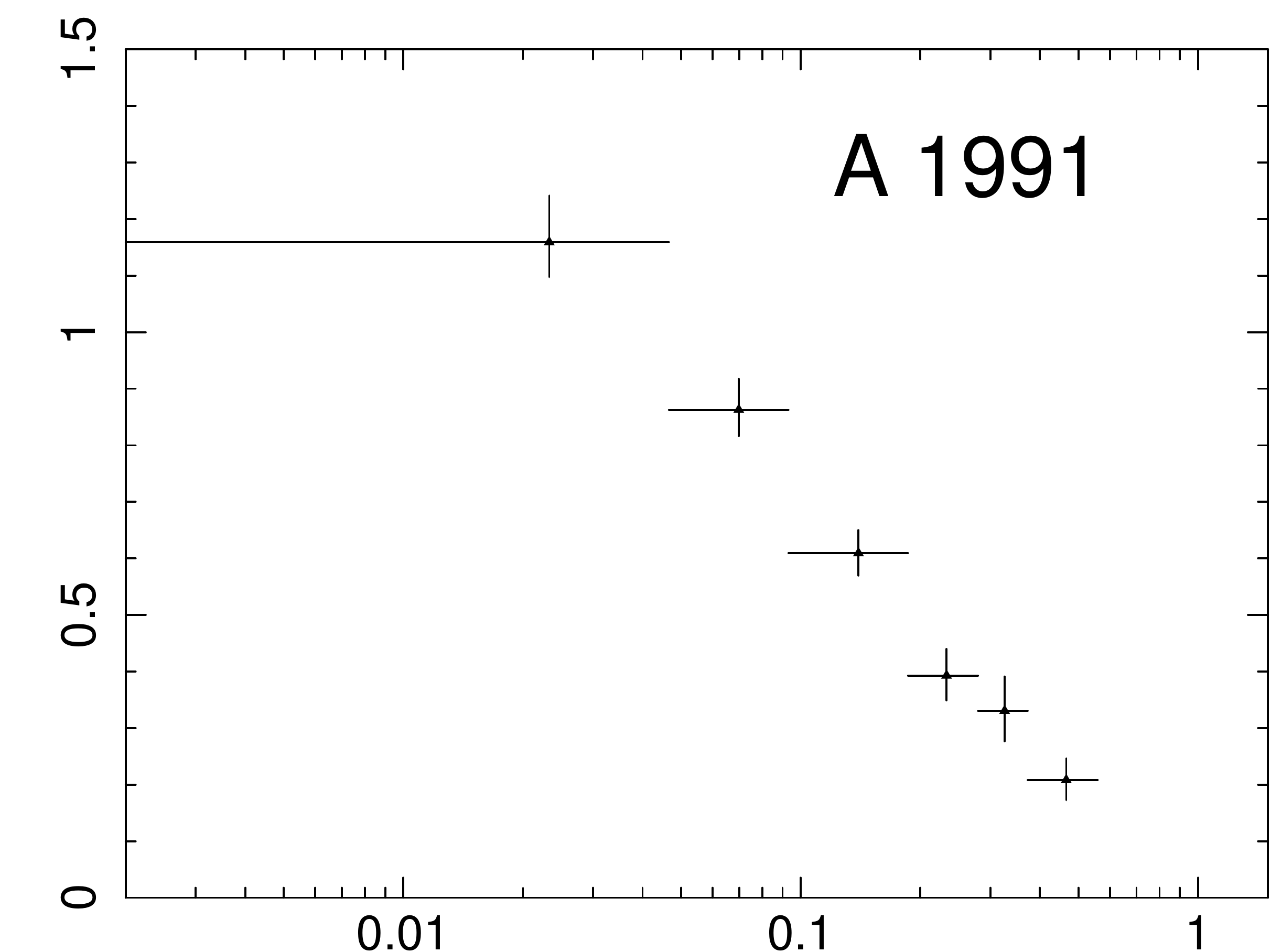}
                \includegraphics[width=0.236\textwidth,trim={0 0 0 0},clip]{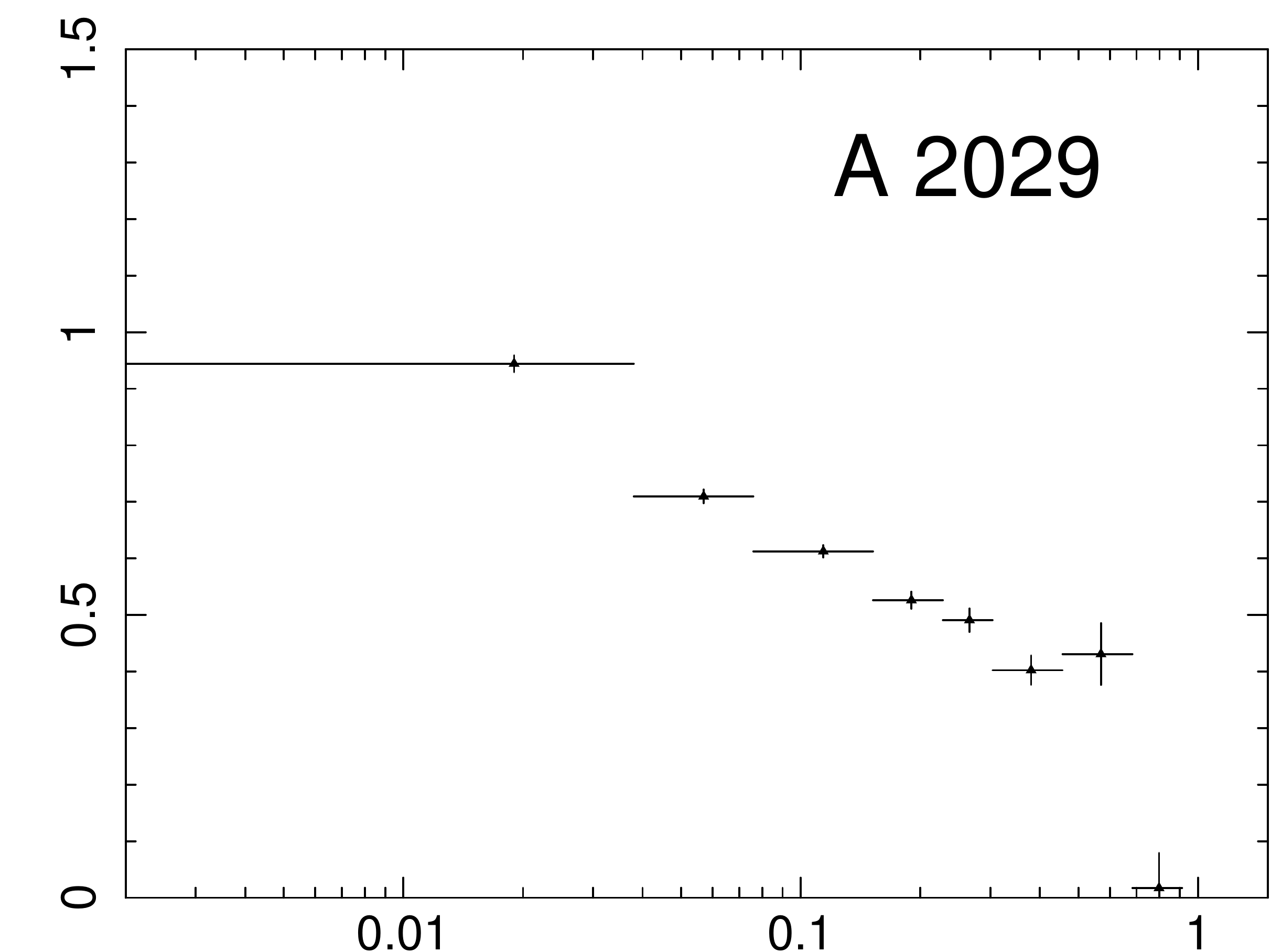} \\
                \includegraphics[width=0.236\textwidth,trim={0 0 0 0},clip]{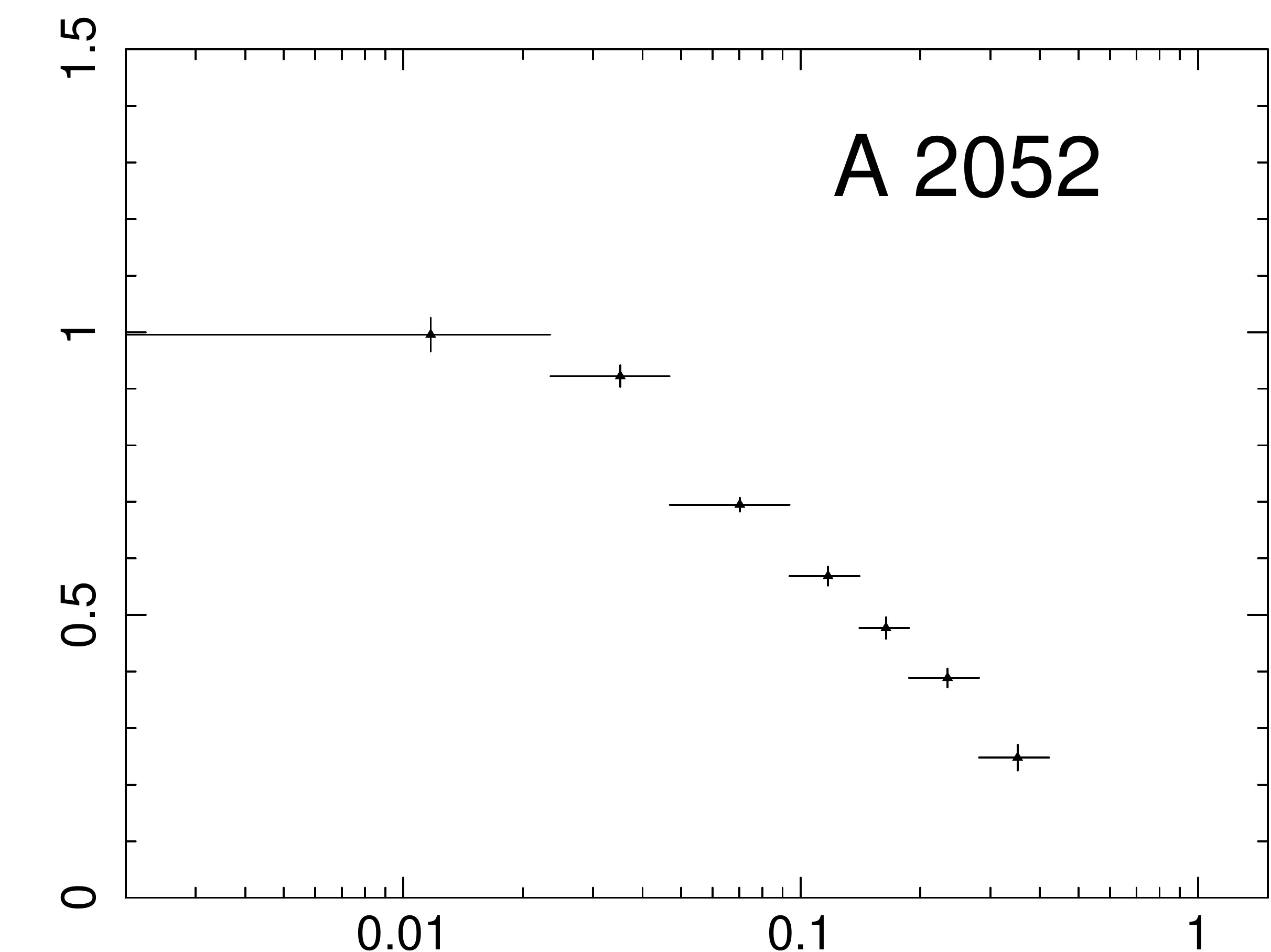}
                \includegraphics[width=0.236\textwidth,trim={0 0 0 0},clip]{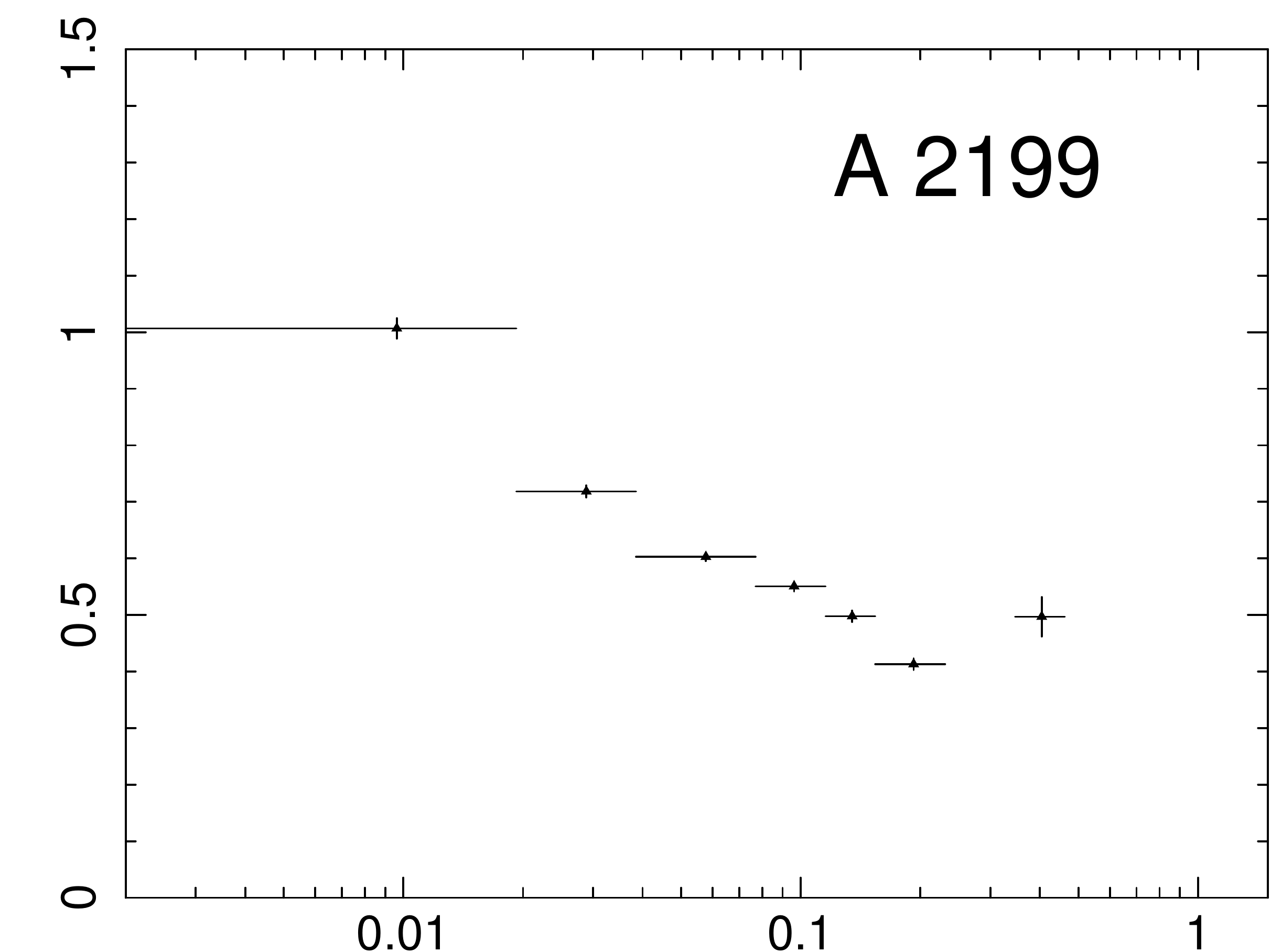}
                \includegraphics[width=0.236\textwidth,trim={0 0 0 0},clip]{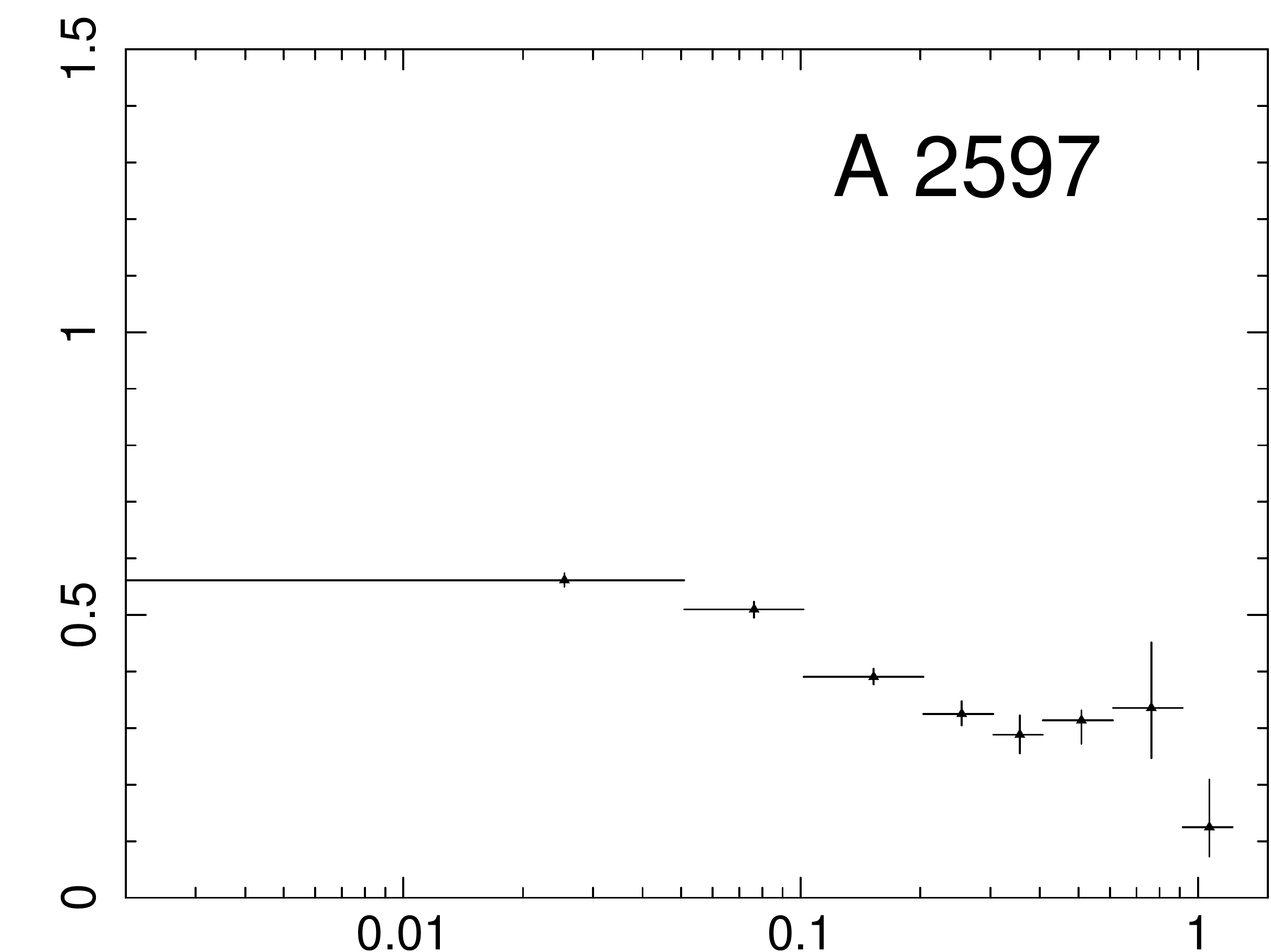}
                \includegraphics[width=0.236\textwidth,trim={0 0 0 0},clip]{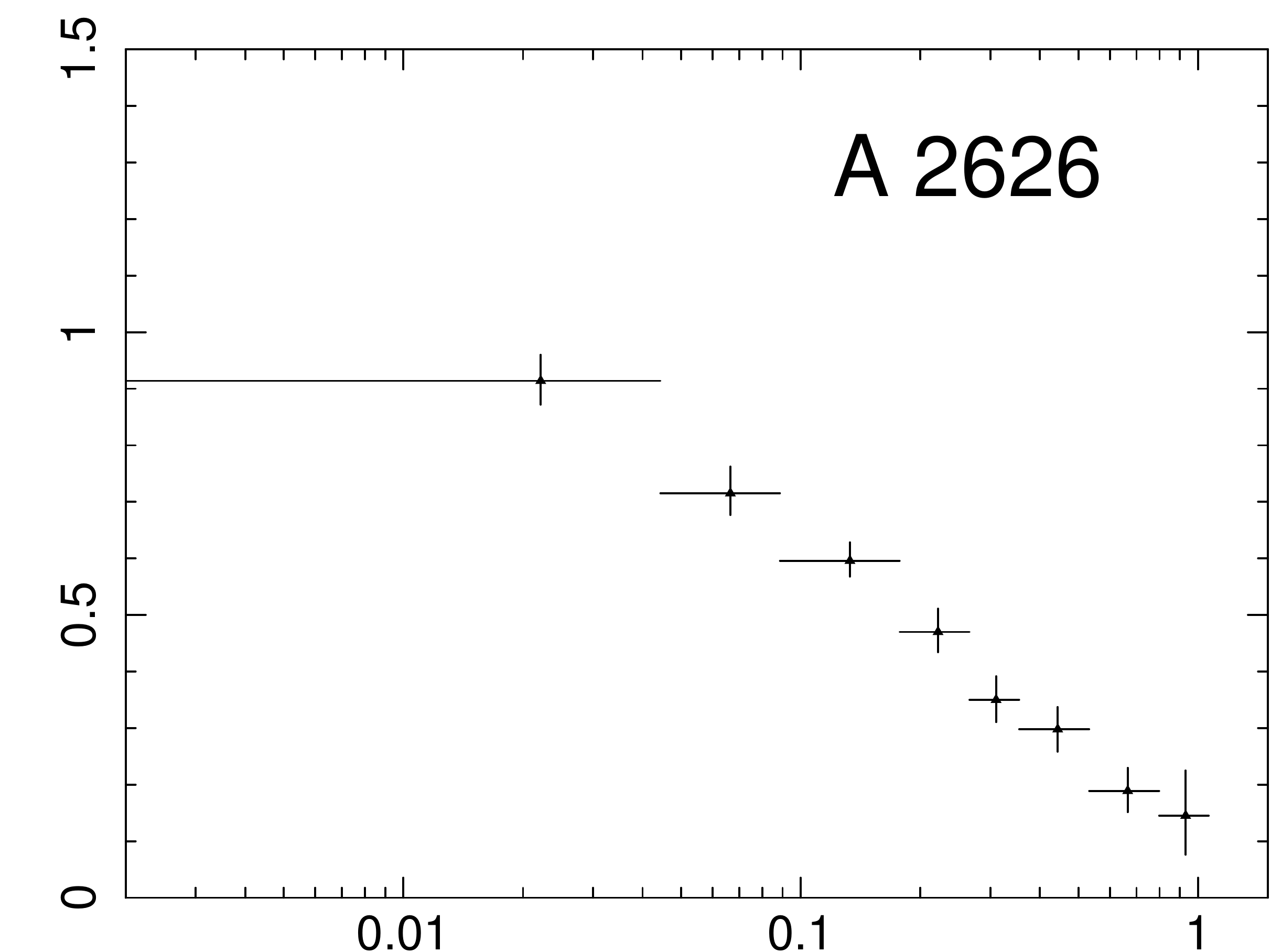} \\
                \includegraphics[width=0.236\textwidth,trim={0 0 0 0},clip]{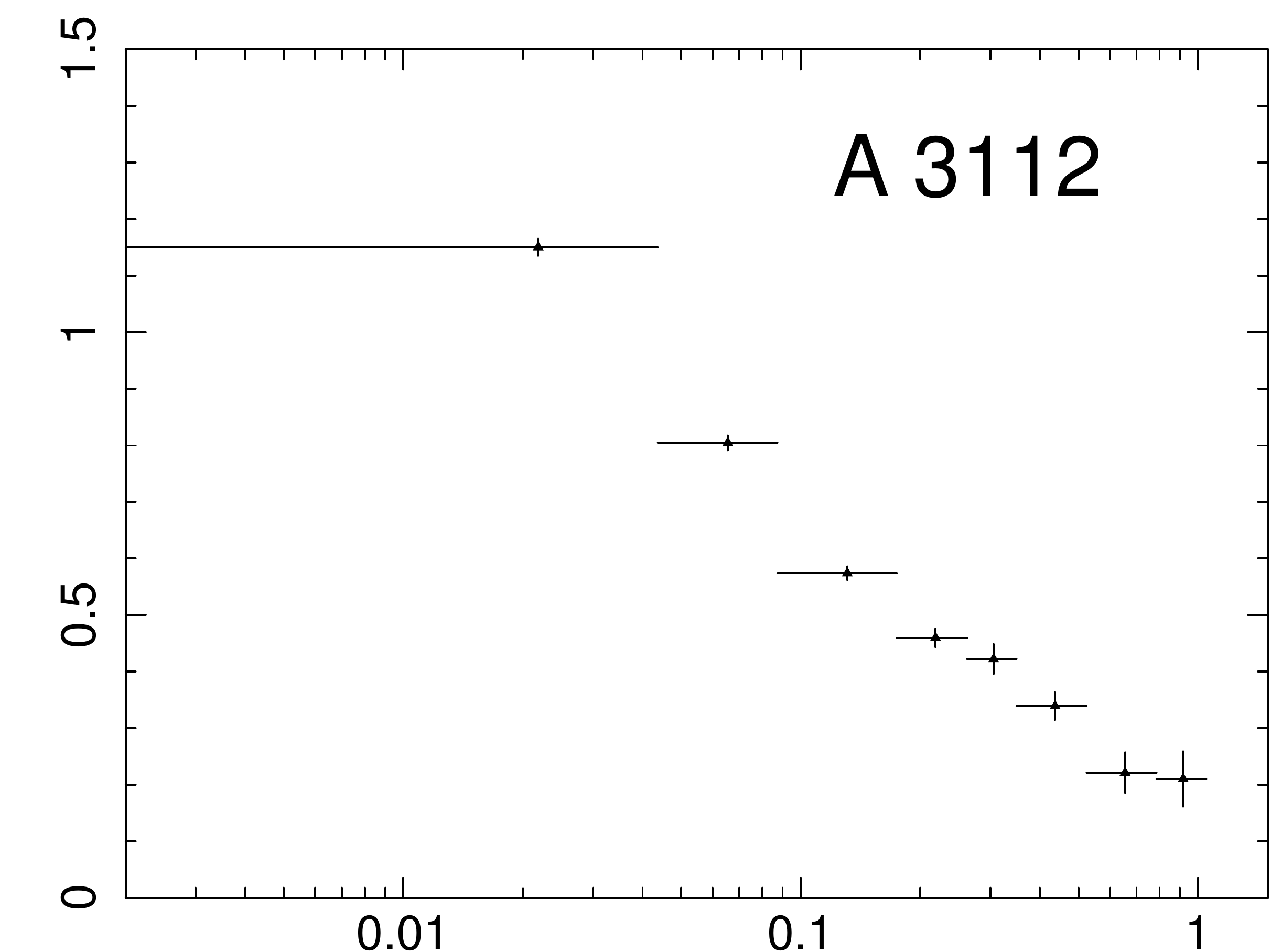}
                \includegraphics[width=0.236\textwidth,trim={0 0 0 0},clip]{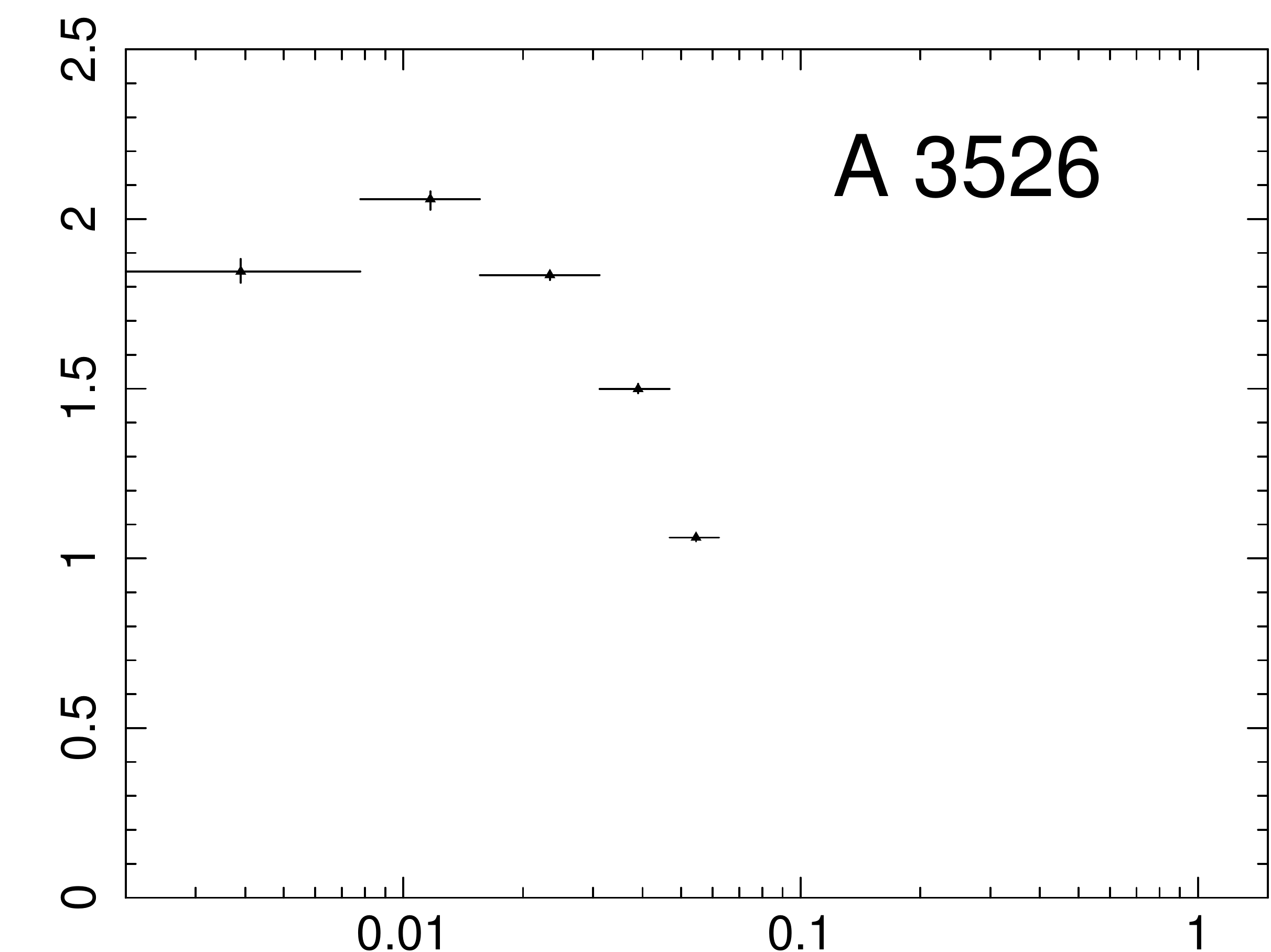}
                \includegraphics[width=0.236\textwidth,trim={0 0 0 0},clip]{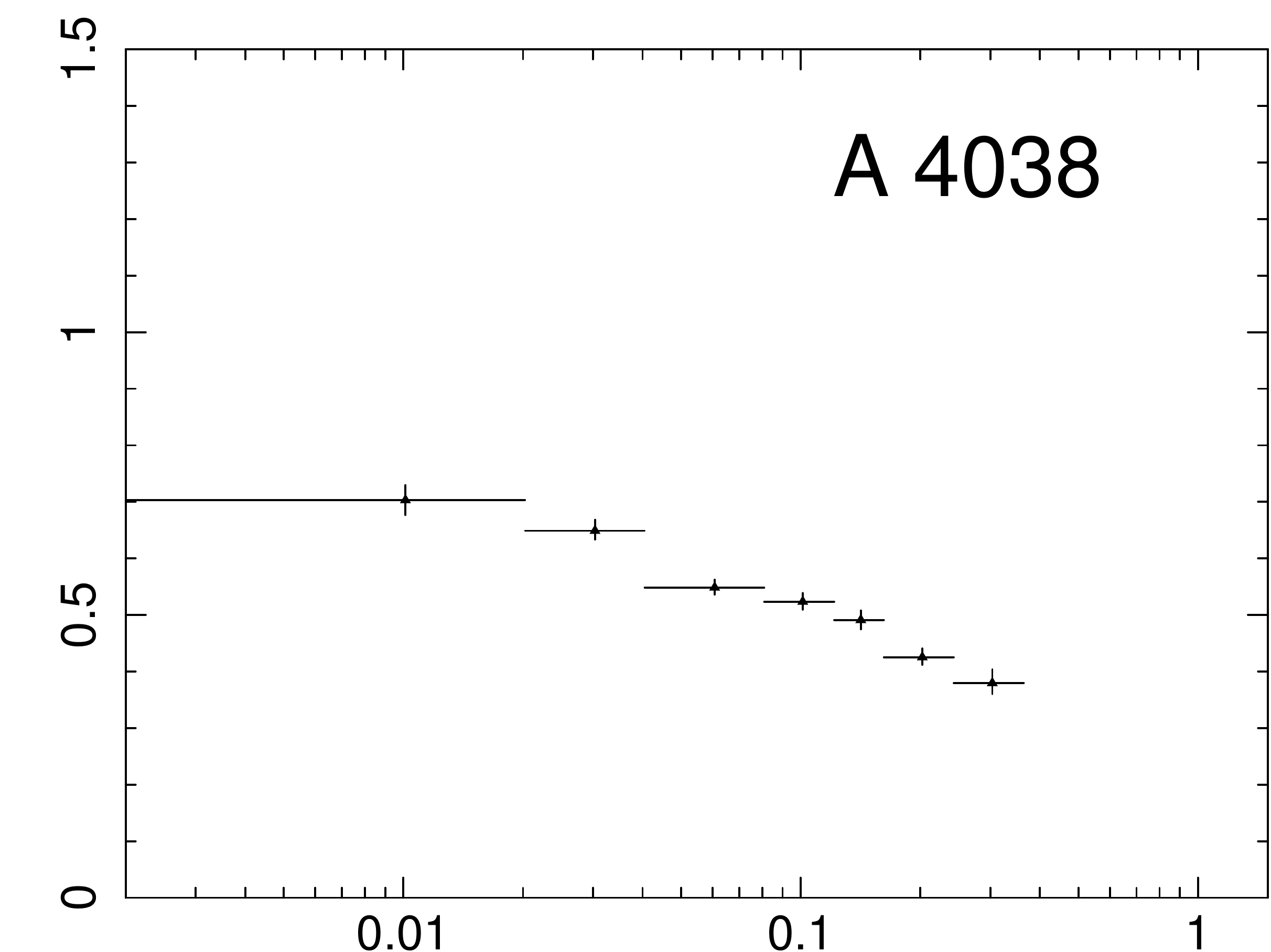}
                \includegraphics[width=0.236\textwidth,trim={0 0 0 0},clip]{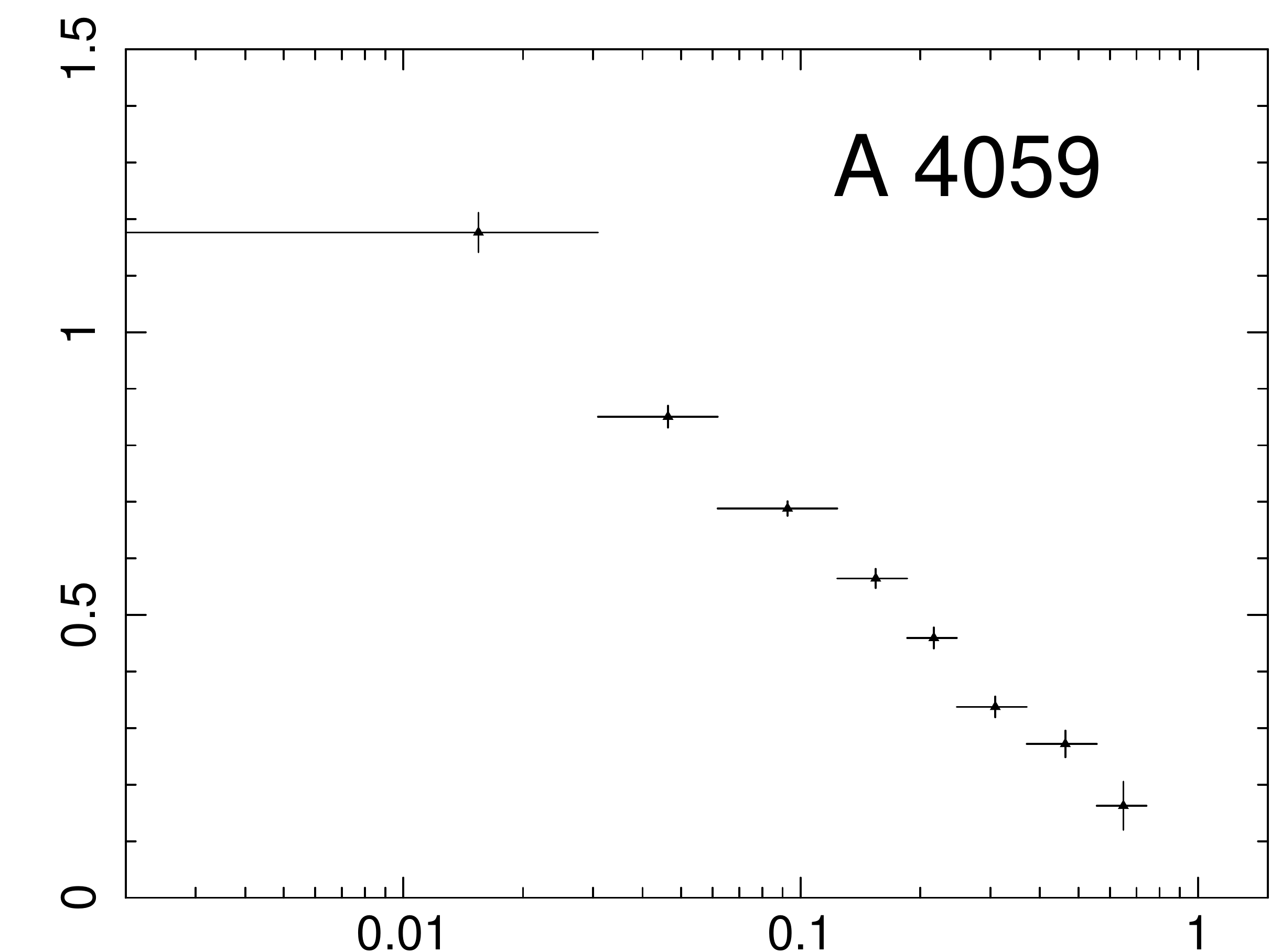} \\
                \includegraphics[width=0.236\textwidth,trim={0 0 0 0},clip]{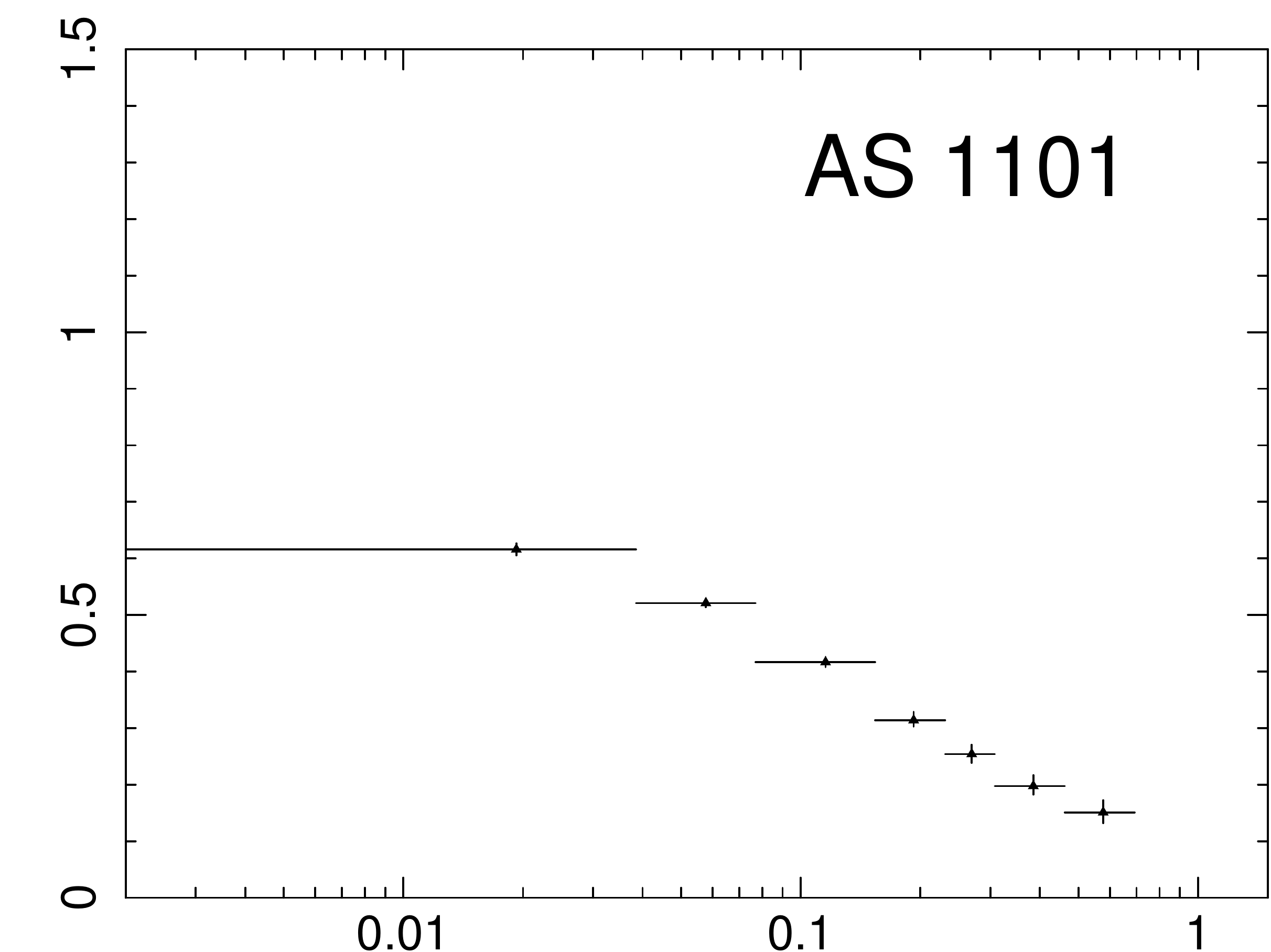}
                \includegraphics[width=0.236\textwidth,trim={0 0 0 0},clip]{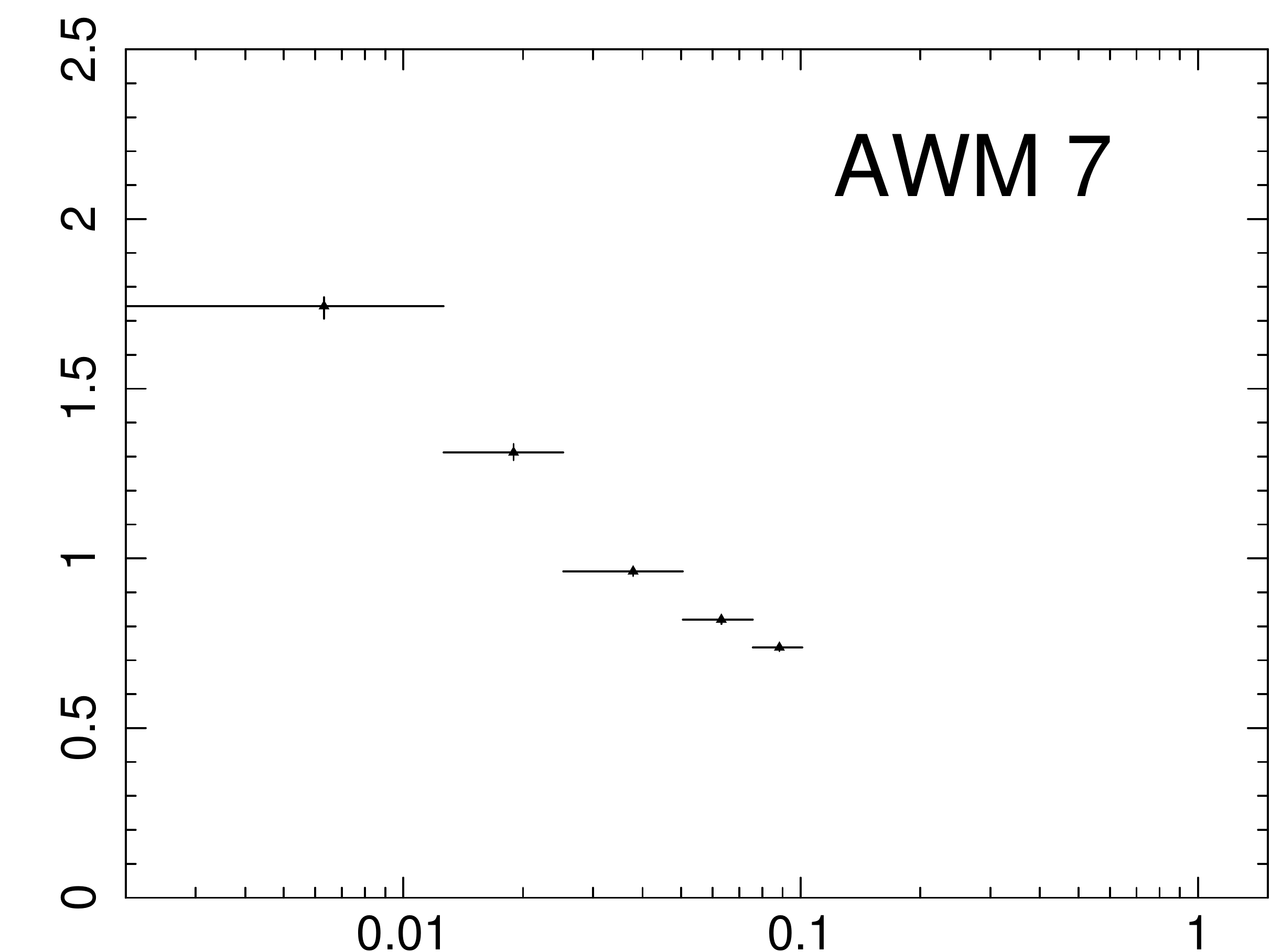}
                \includegraphics[width=0.236\textwidth,trim={0 0 0 0},clip]{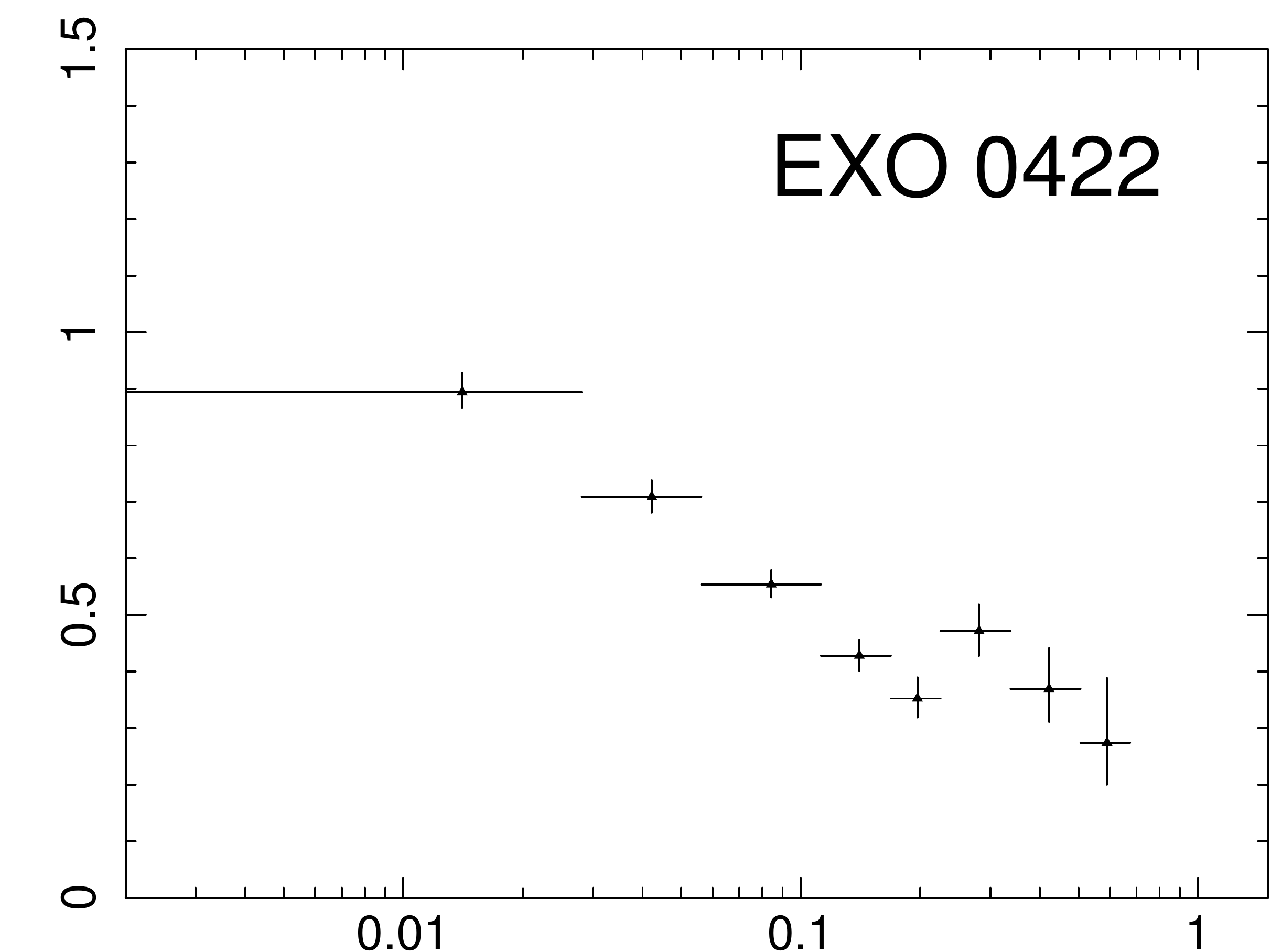}
                \includegraphics[width=0.236\textwidth,trim={0 0 0 0},clip]{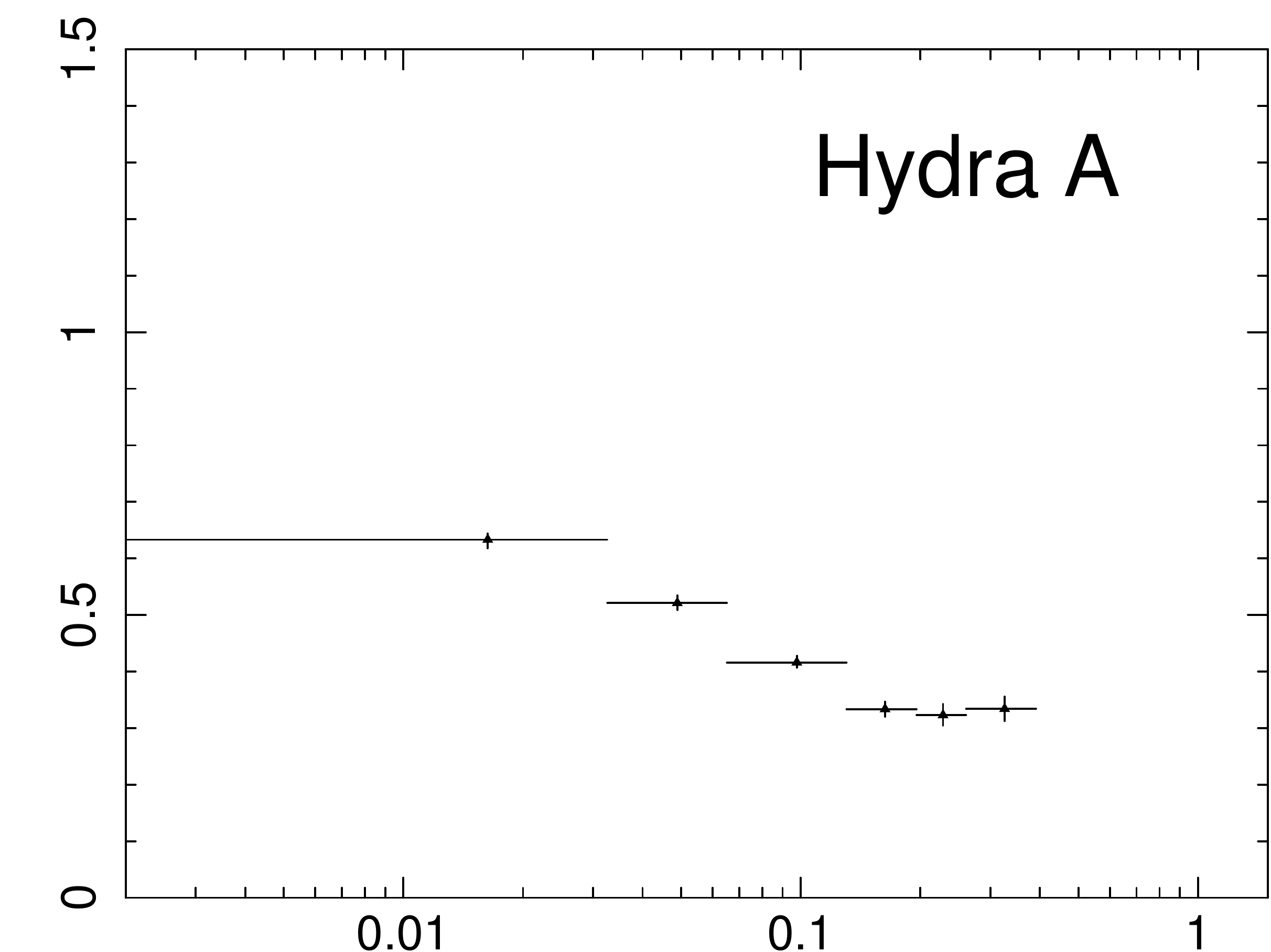} \\
                \includegraphics[width=0.236\textwidth,trim={0 0 0 0},clip]{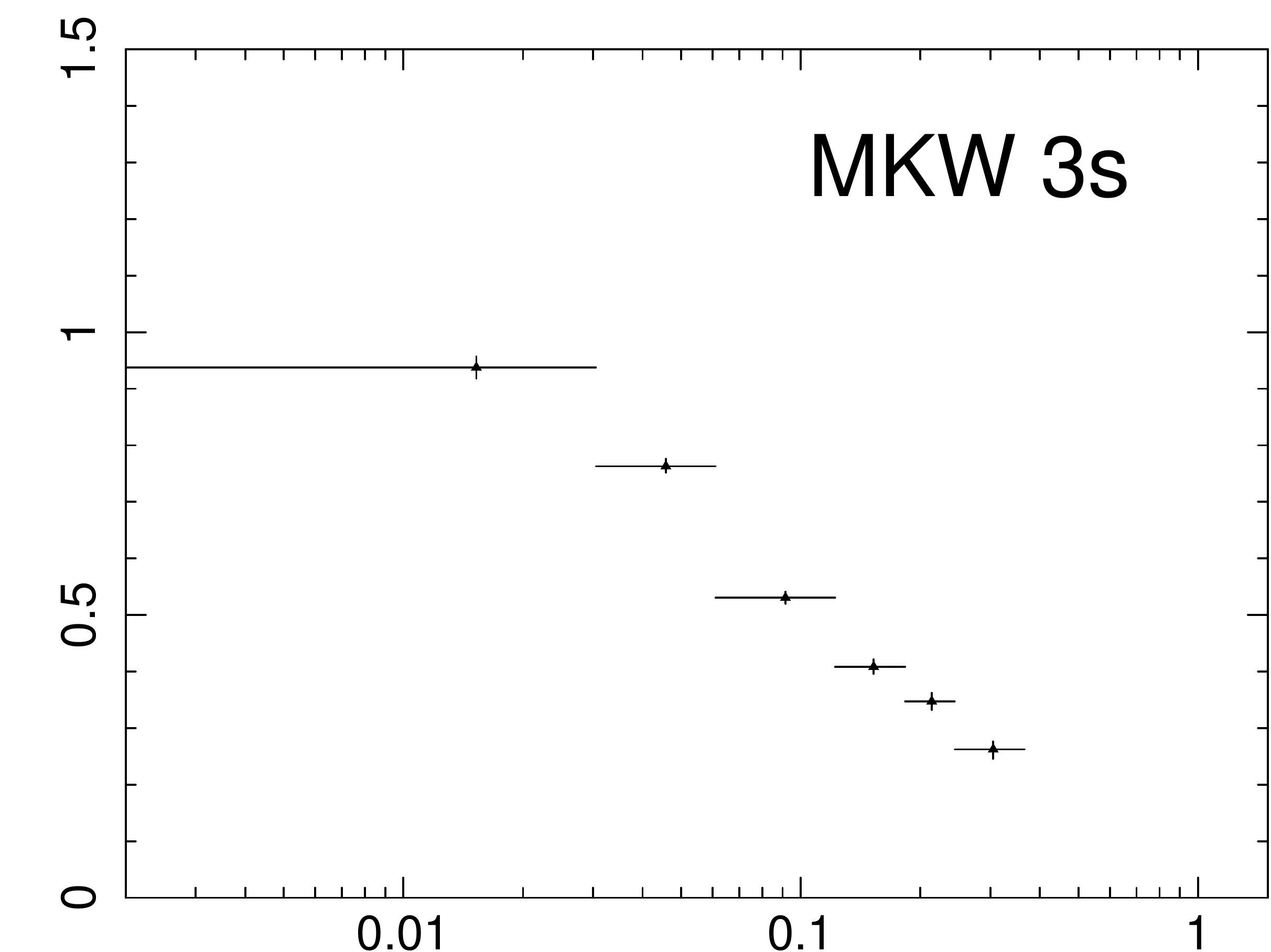}
                \includegraphics[width=0.236\textwidth,trim={0 0 0 0},clip]{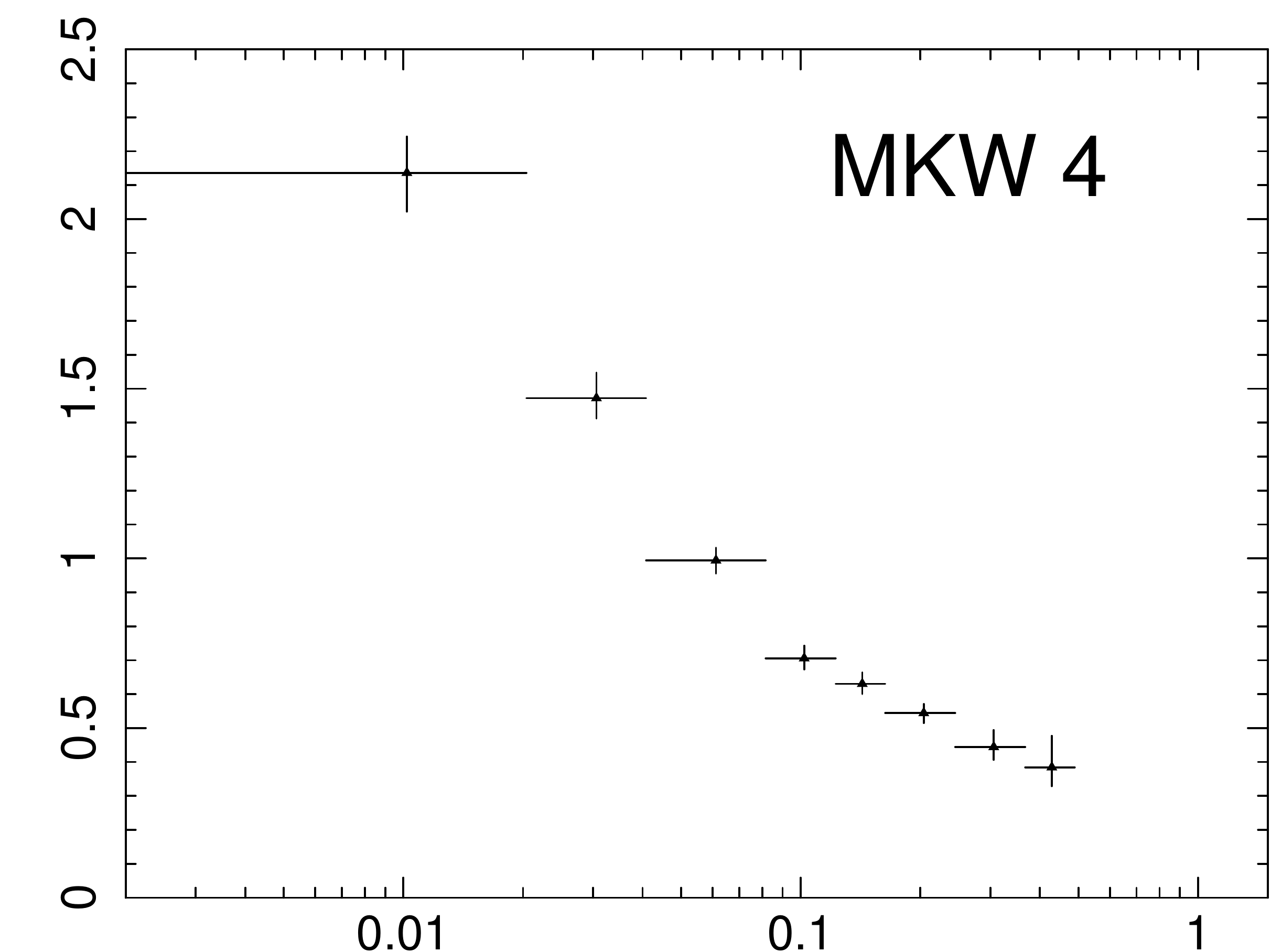}
                \includegraphics[width=0.236\textwidth,trim={0 0 0 0},clip]{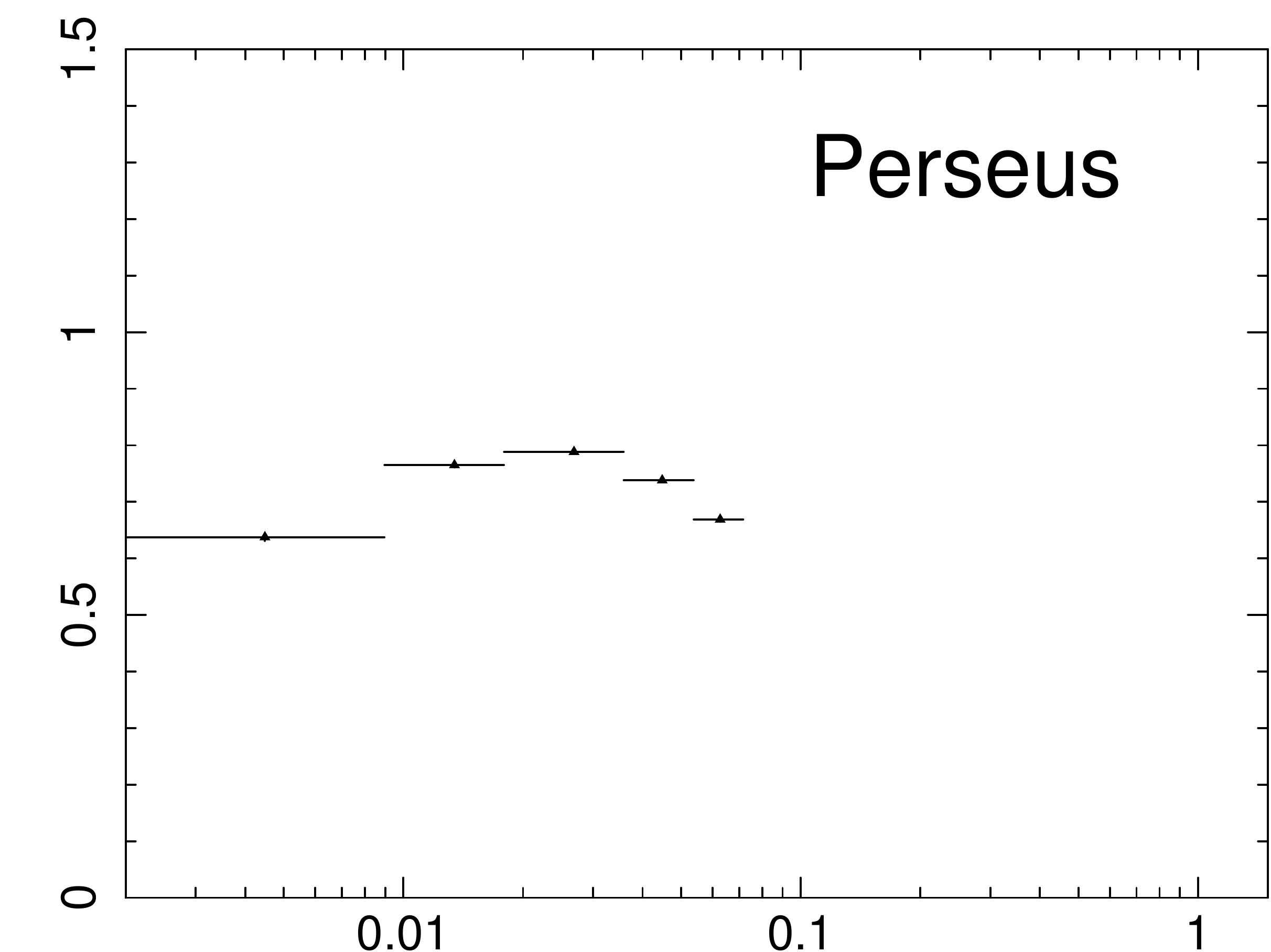}

        \caption{Radial Fe abundance profiles for all the clusters ($kT_\text{mean}$ > 1.7 keV) in our sample. The radial distances ($x$-axis) are expressed in fractions of $r_{500}$ while the Fe abundances ($y$-axis) are given with respect to their proto-solar values \citep{2009LanB...4B...44L}. Data points that were not included when computing the average profile were removed (Sect. \ref{sect:excl_artefacts}).}
\label{fig:Fe_profiles_indiv_clusters}
\end{figure*}

\begin{figure*}[!]

                \includegraphics[width=0.236\textwidth,trim={0 0 0 0},clip]{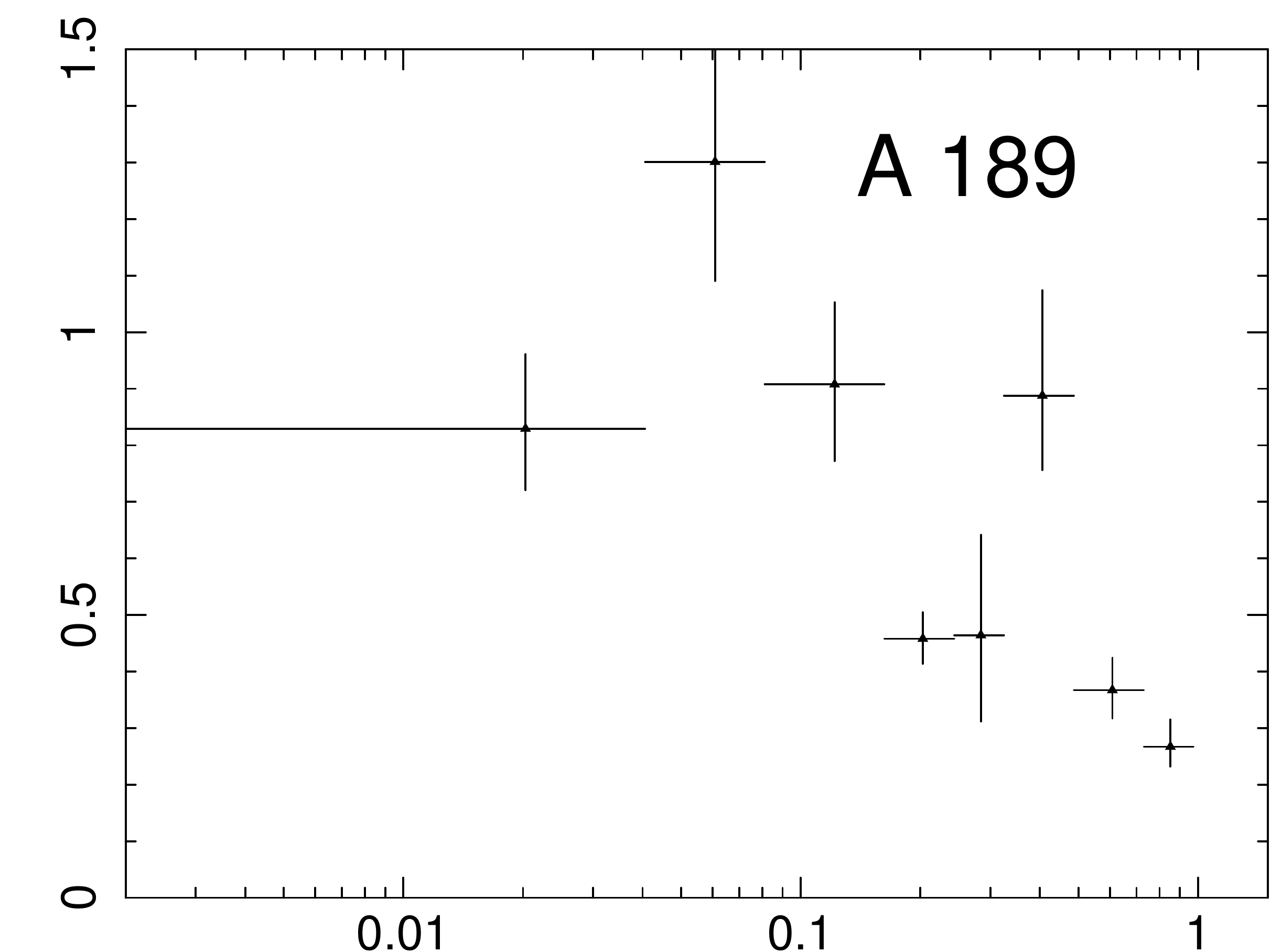}
                \includegraphics[width=0.236\textwidth,trim={0 0 0 0},clip]{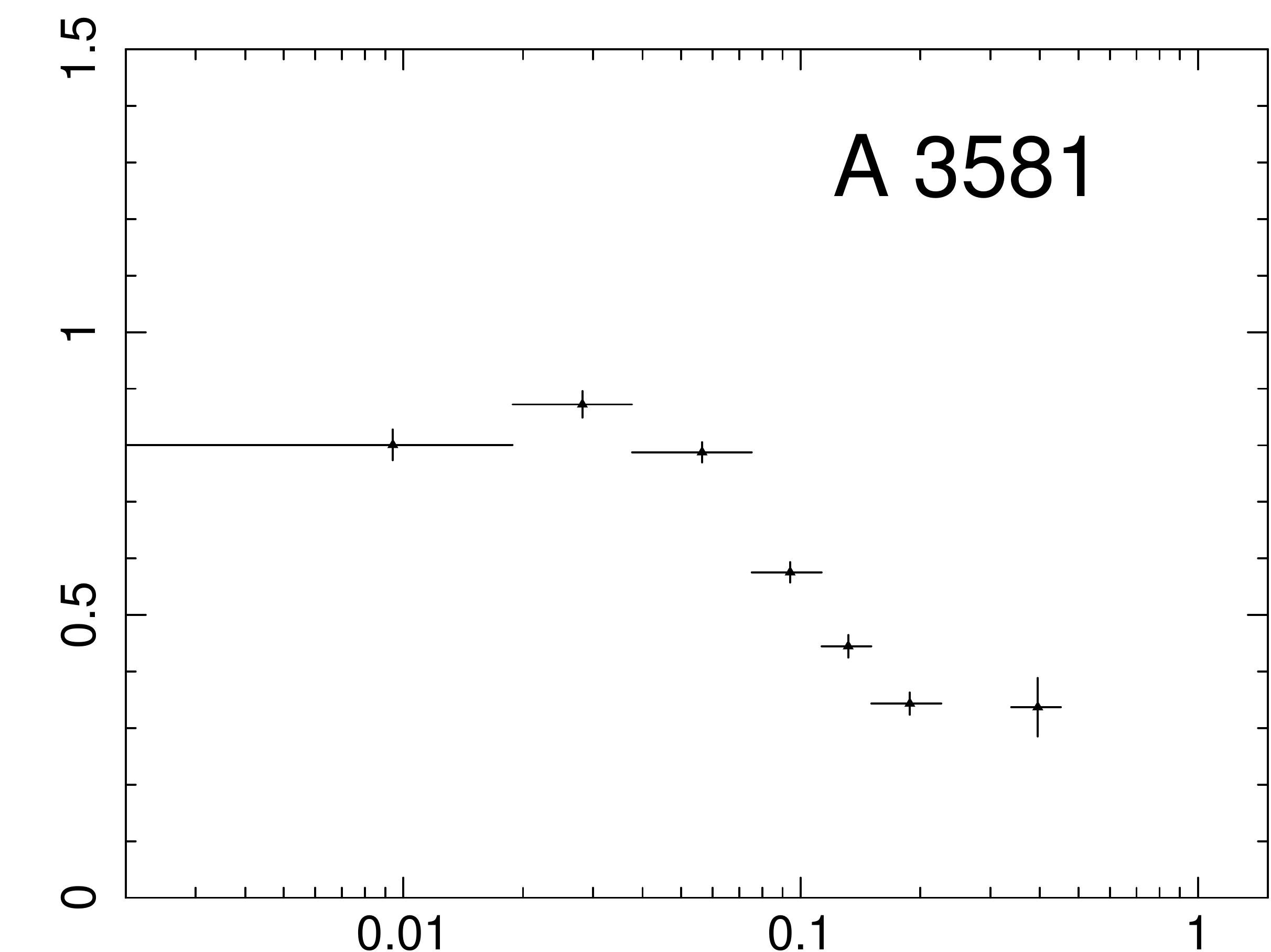}
                \includegraphics[width=0.236\textwidth,trim={0 0 0 0},clip]{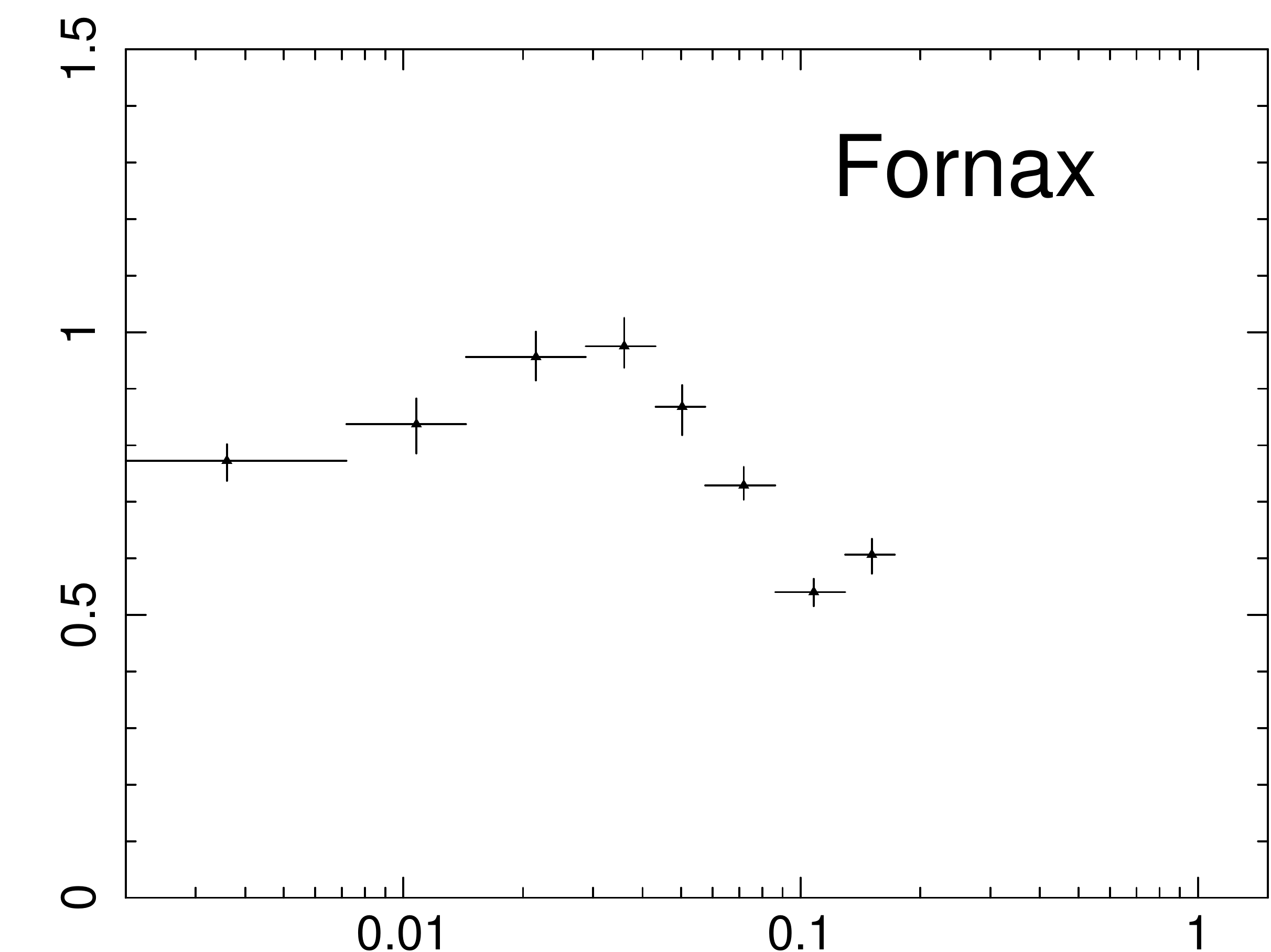}
                \includegraphics[width=0.236\textwidth,trim={0 0 0 0},clip]{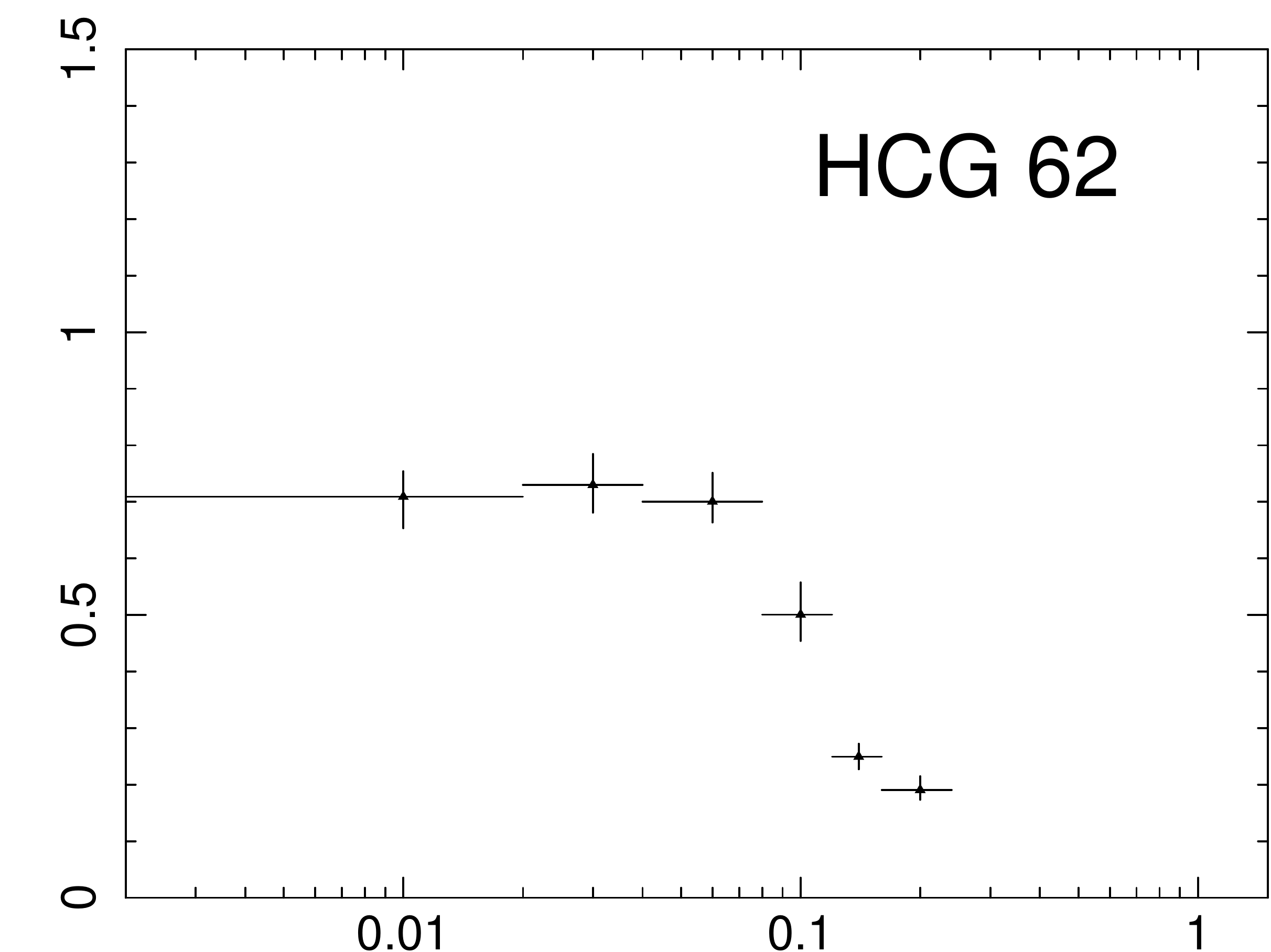} \\
                \includegraphics[width=0.236\textwidth,trim={0 0 0 0},clip]{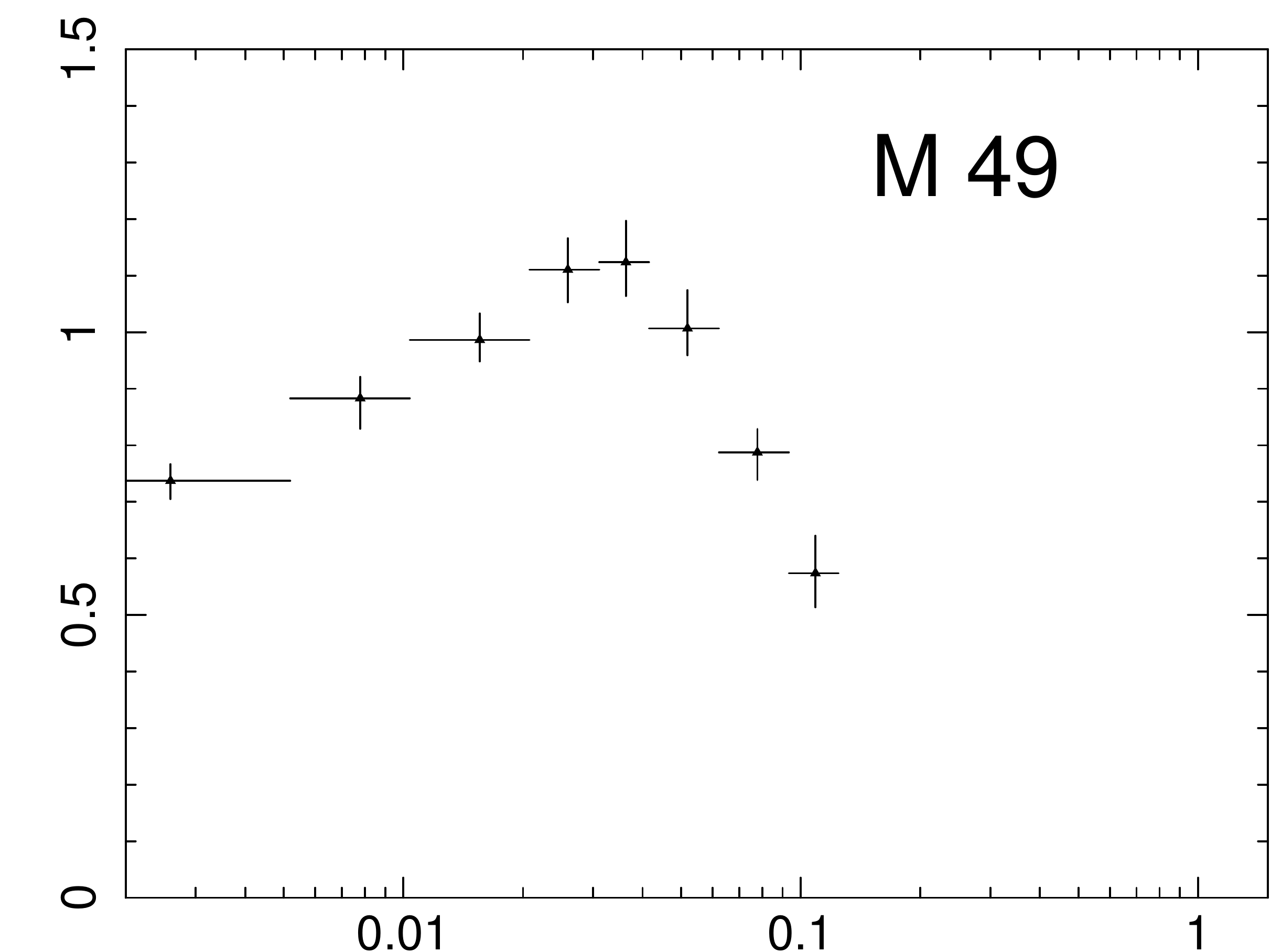}
                \includegraphics[width=0.236\textwidth,trim={0 0 0 0},clip]{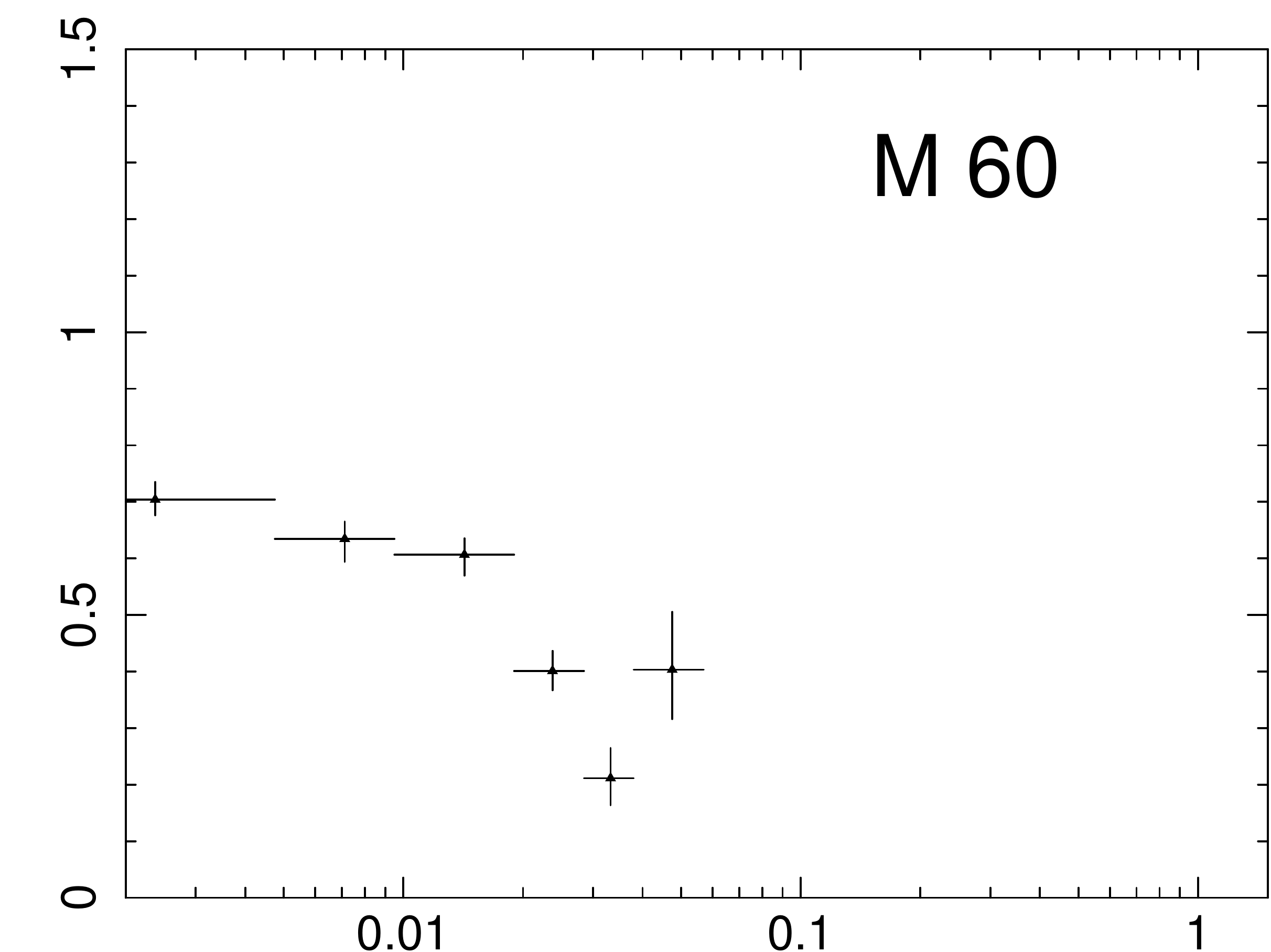}
                \includegraphics[width=0.236\textwidth,trim={0 0 0 0},clip]{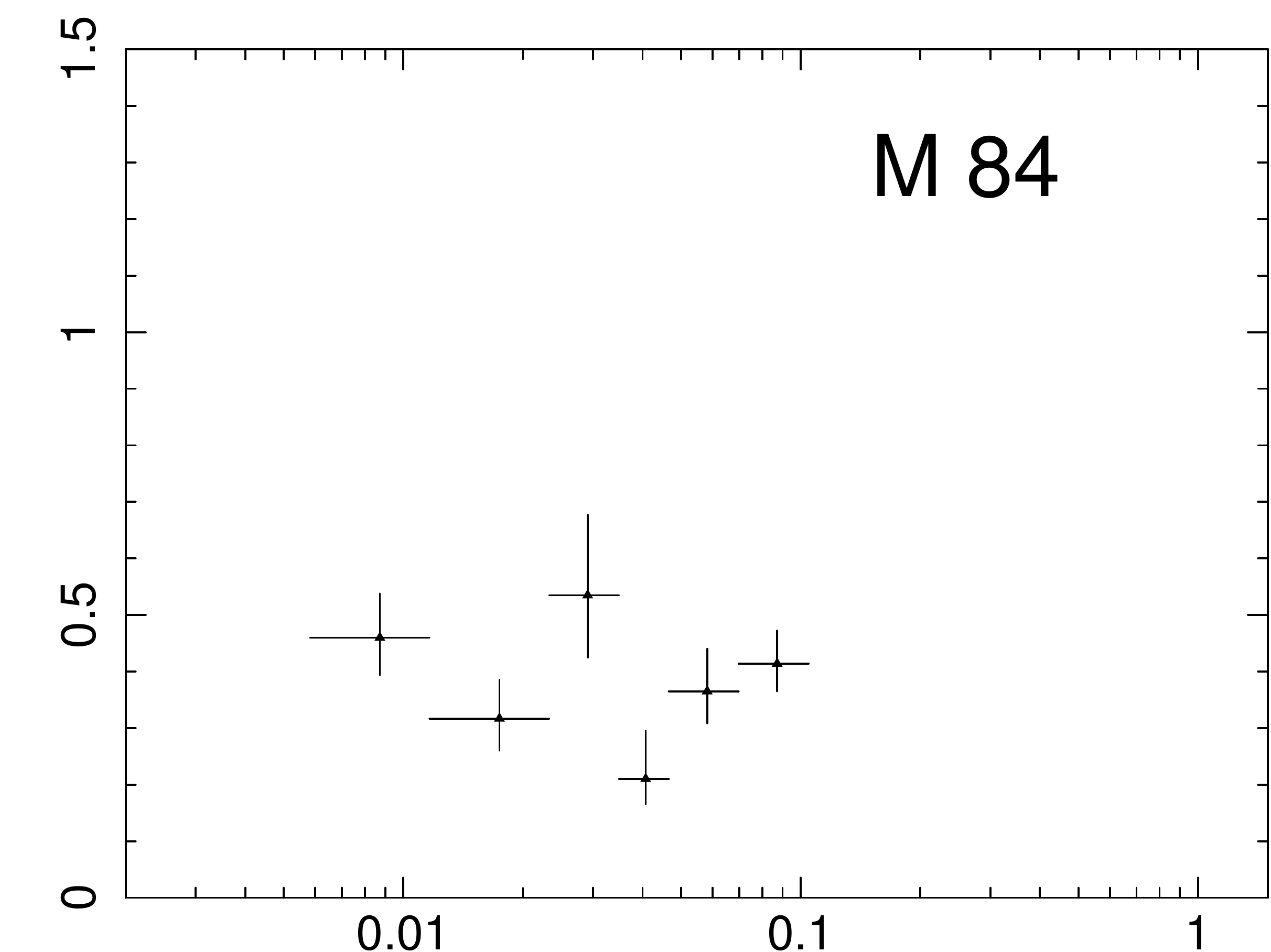}
                \includegraphics[width=0.236\textwidth,trim={0 0 0 0},clip]{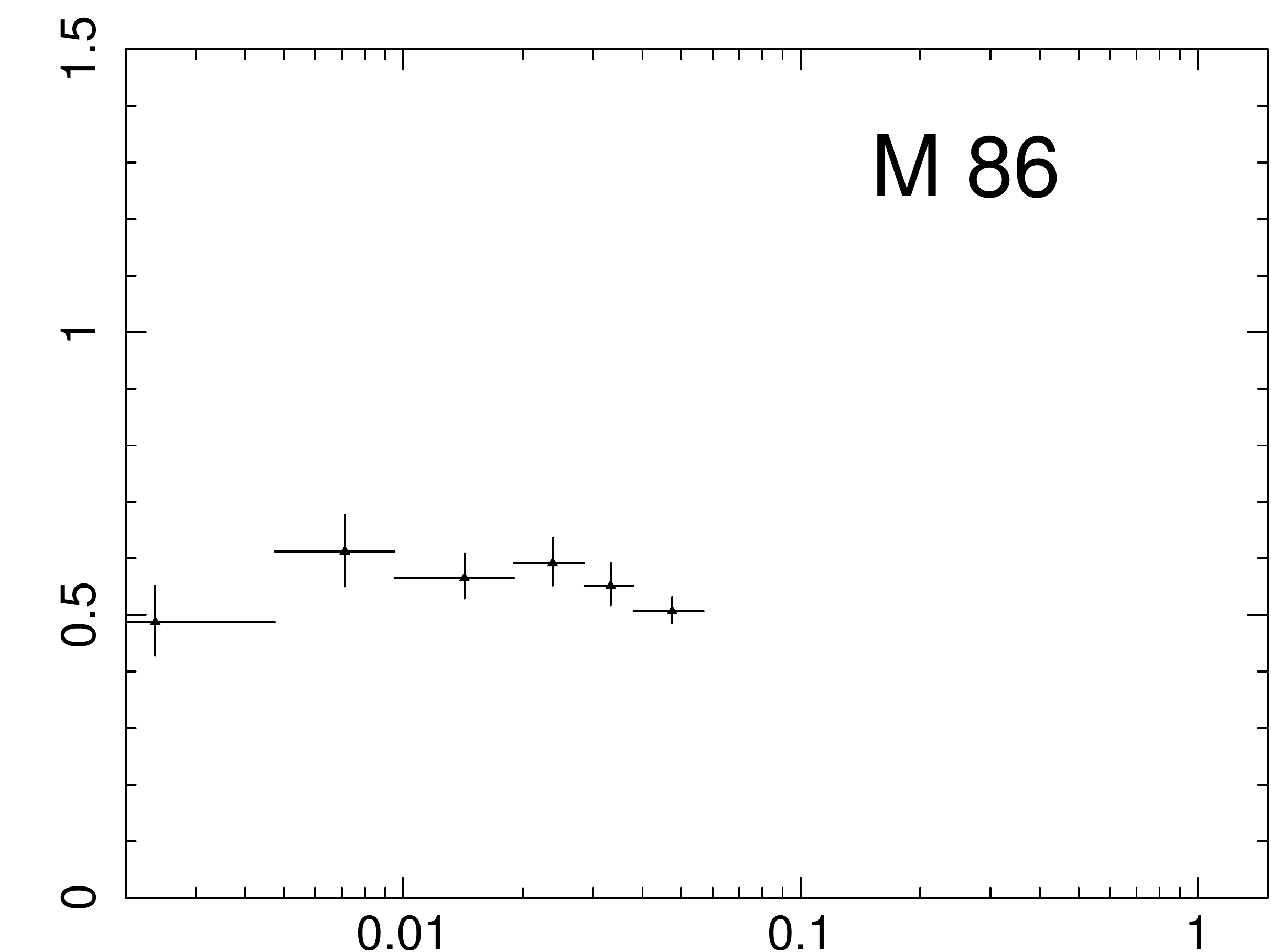} \\
                \includegraphics[width=0.236\textwidth,trim={0 0 0 0},clip]{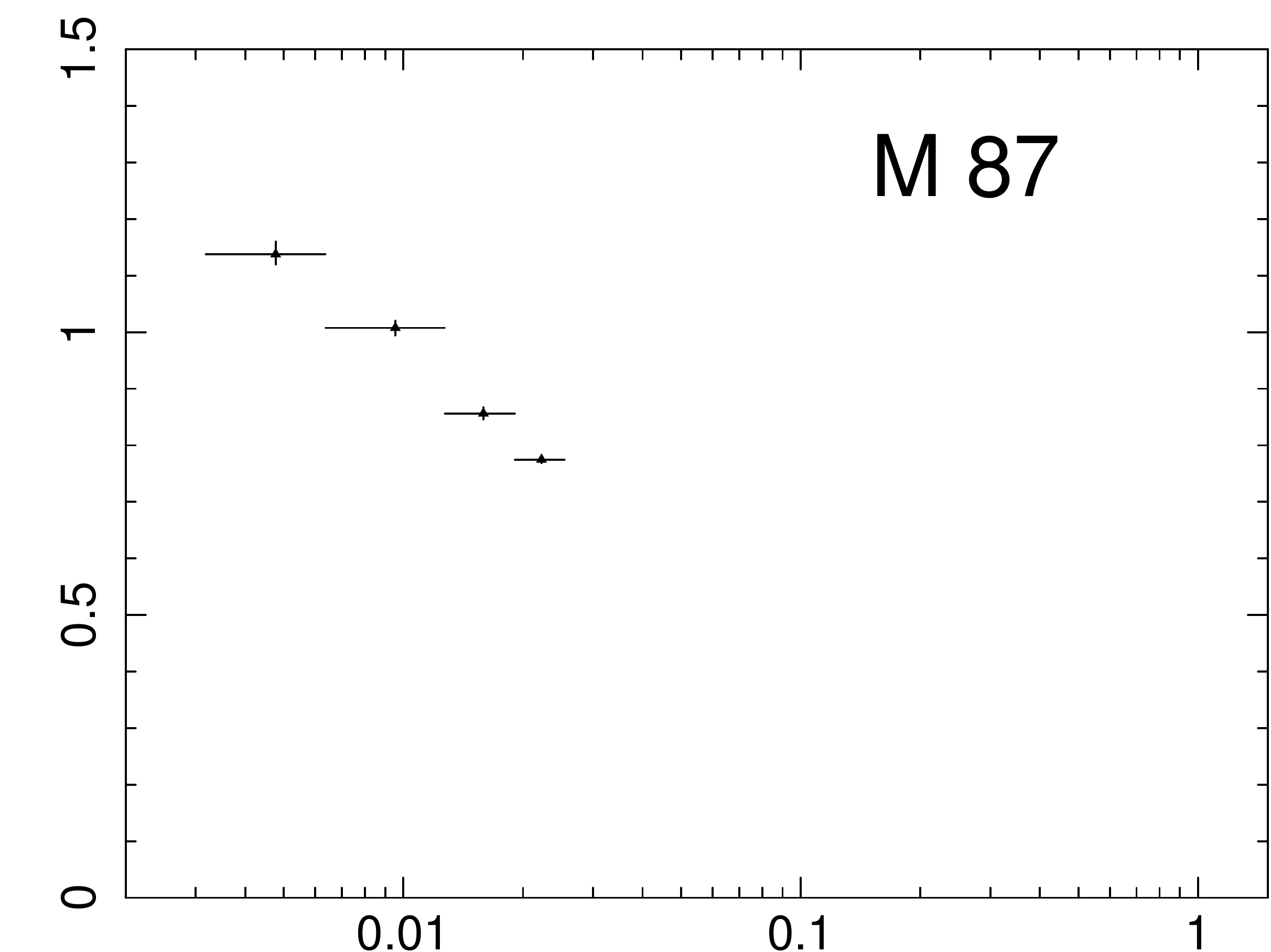}
                \includegraphics[width=0.236\textwidth,trim={0 0 0 0},clip]{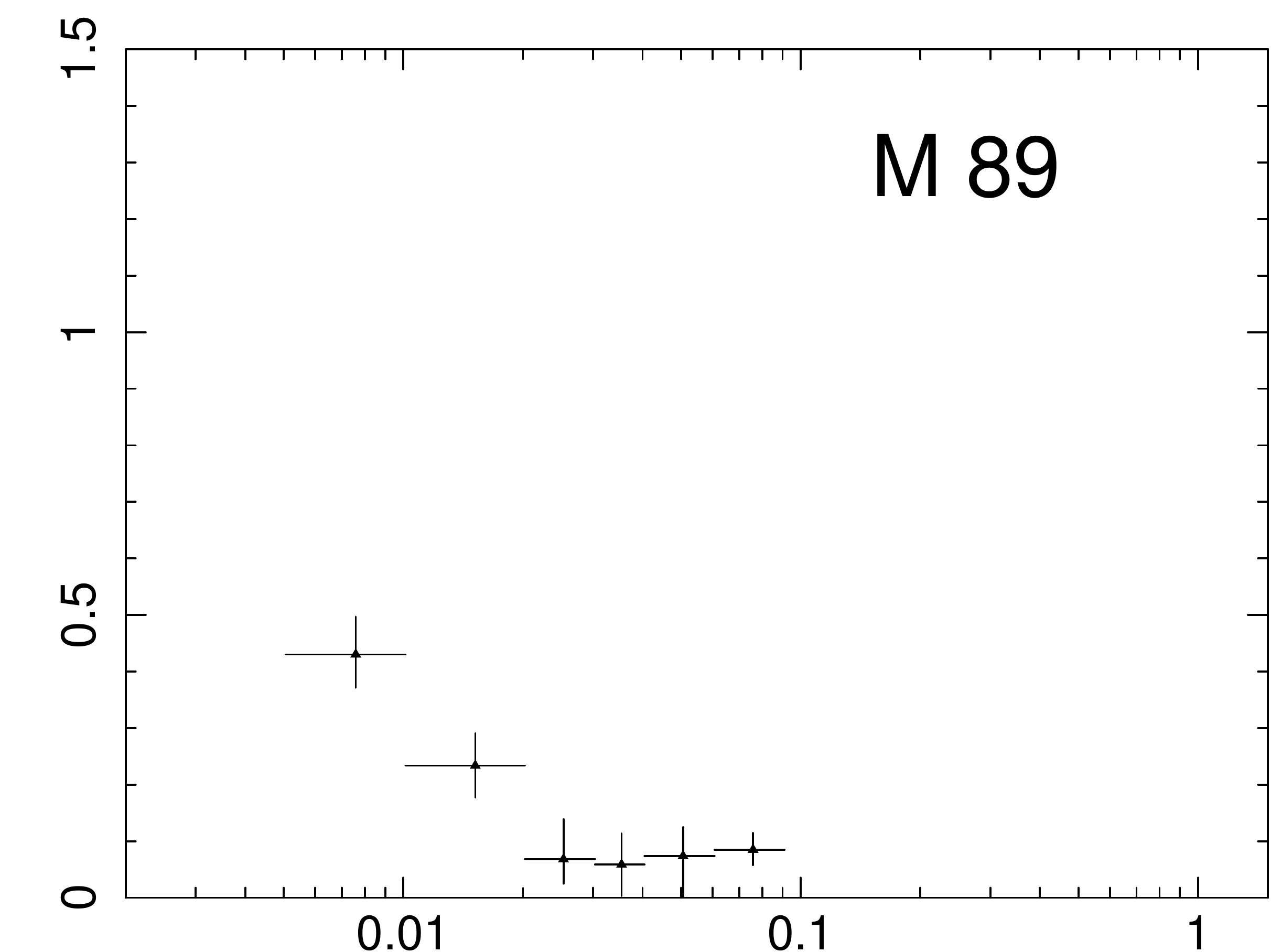}
                \includegraphics[width=0.236\textwidth,trim={0 0 0 0},clip]{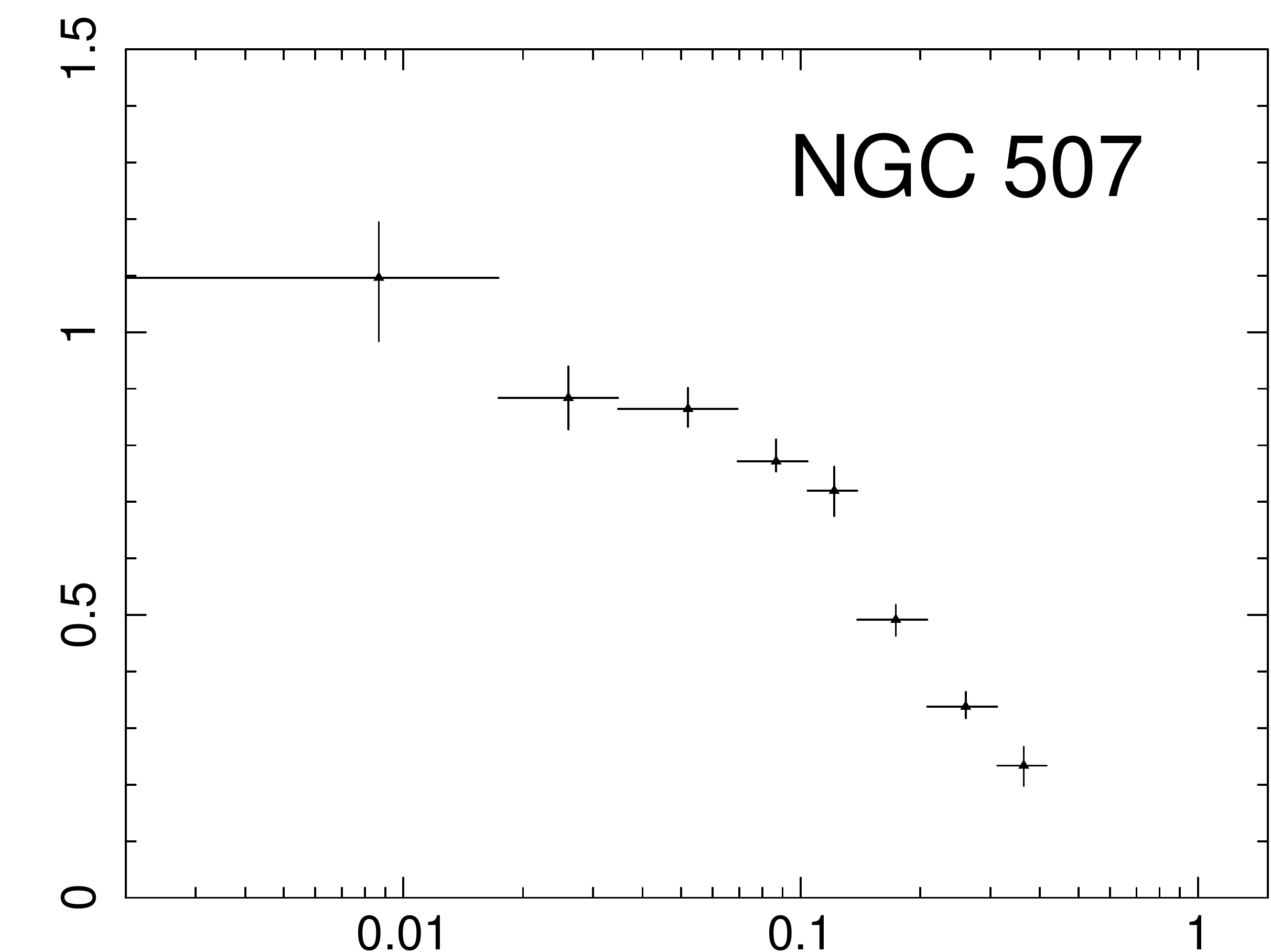}
                \includegraphics[width=0.236\textwidth,trim={0 0 0 0},clip]{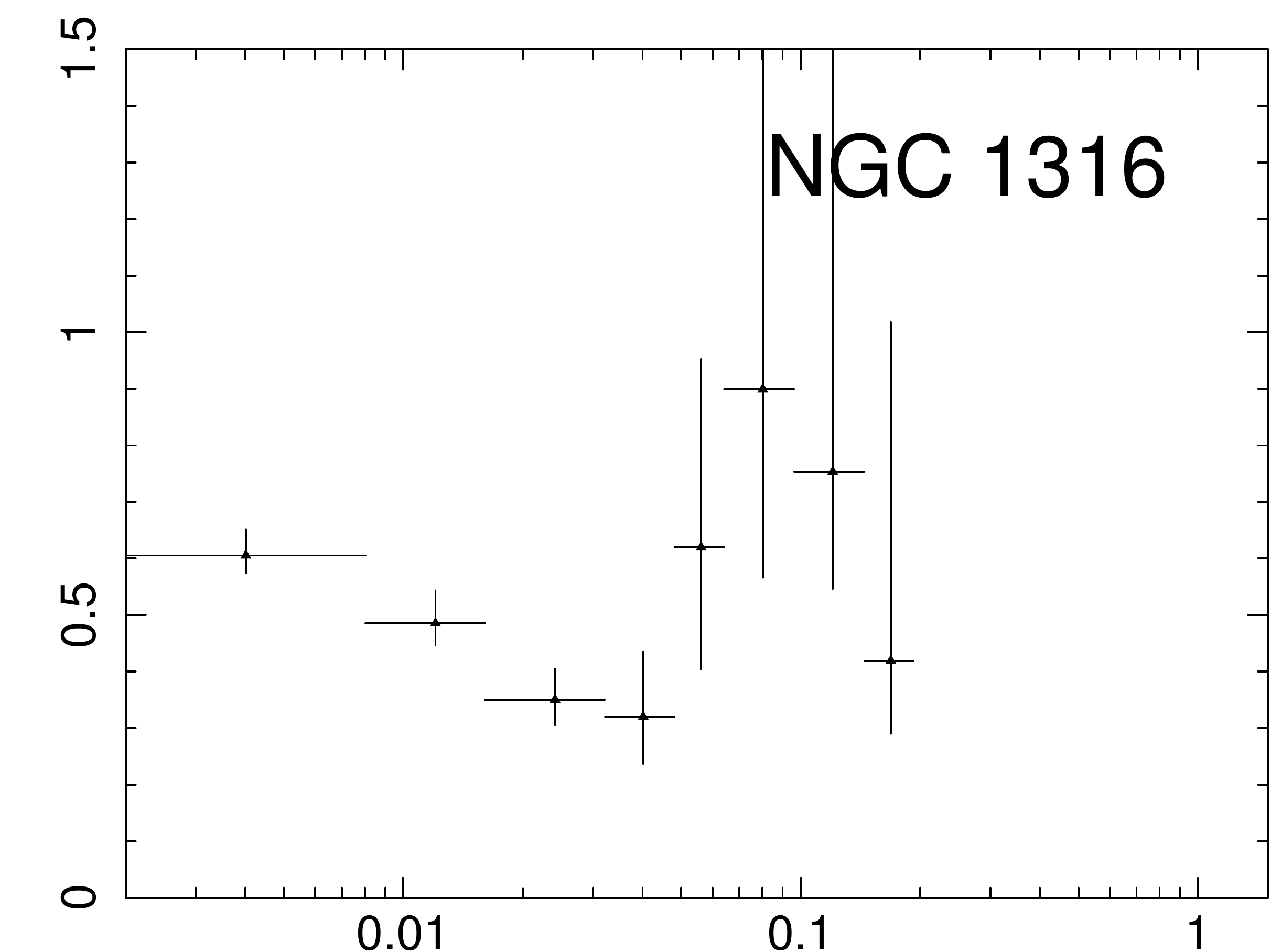} \\
                \includegraphics[width=0.236\textwidth,trim={0 0 0 0},clip]{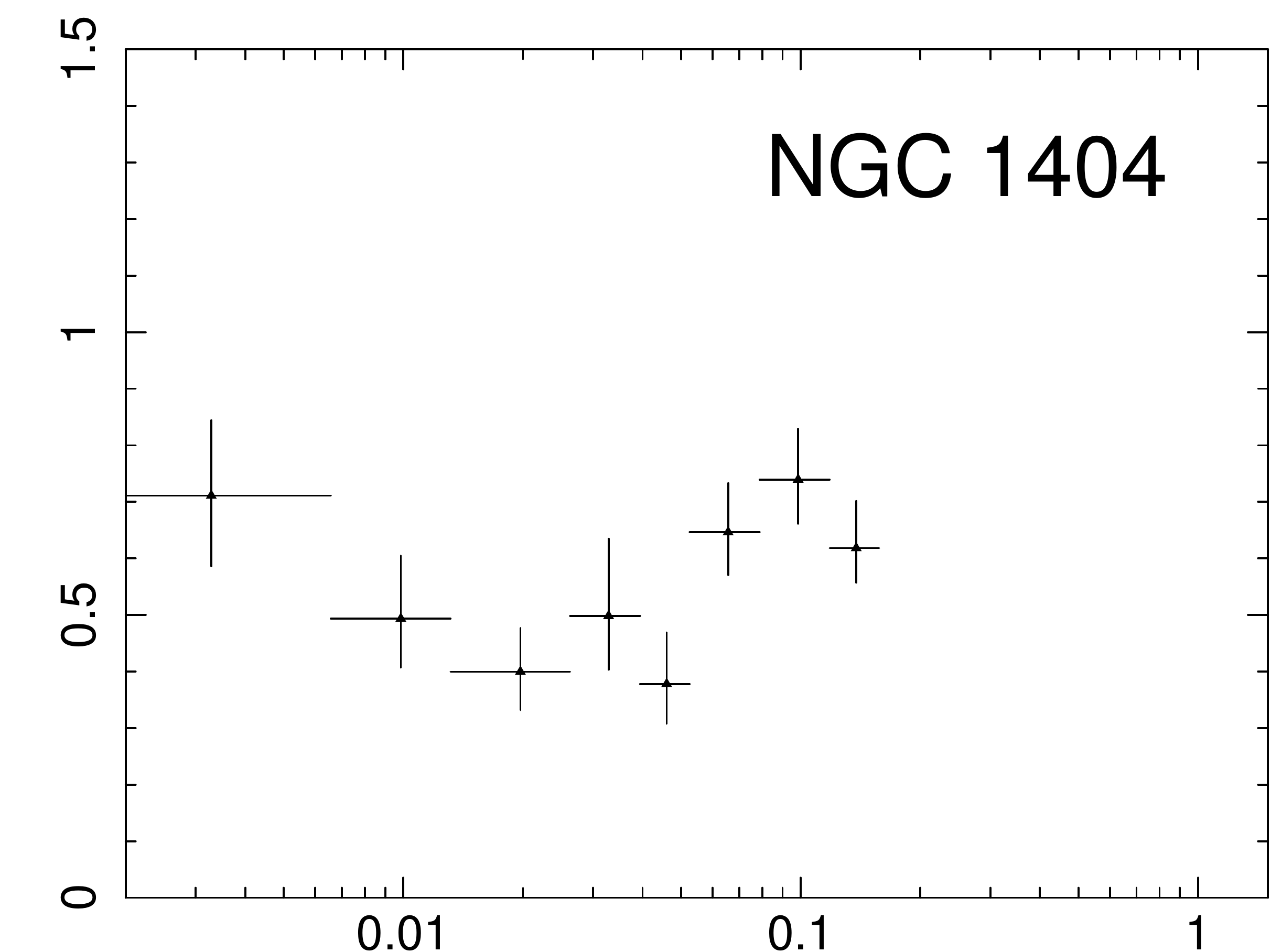}
                \includegraphics[width=0.236\textwidth,trim={0 0 0 0},clip]{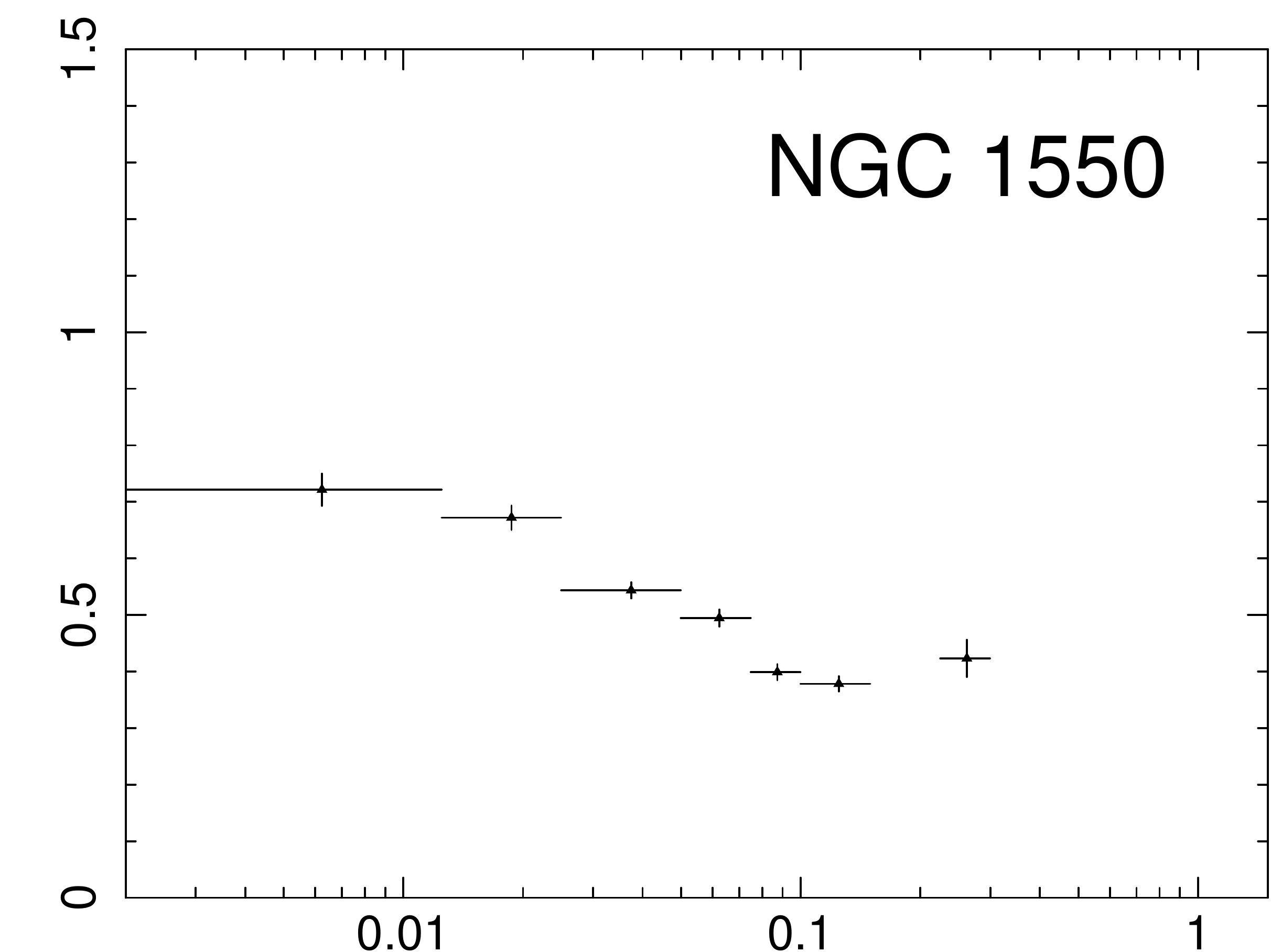}
                \includegraphics[width=0.236\textwidth,trim={0 0 0 0},clip]{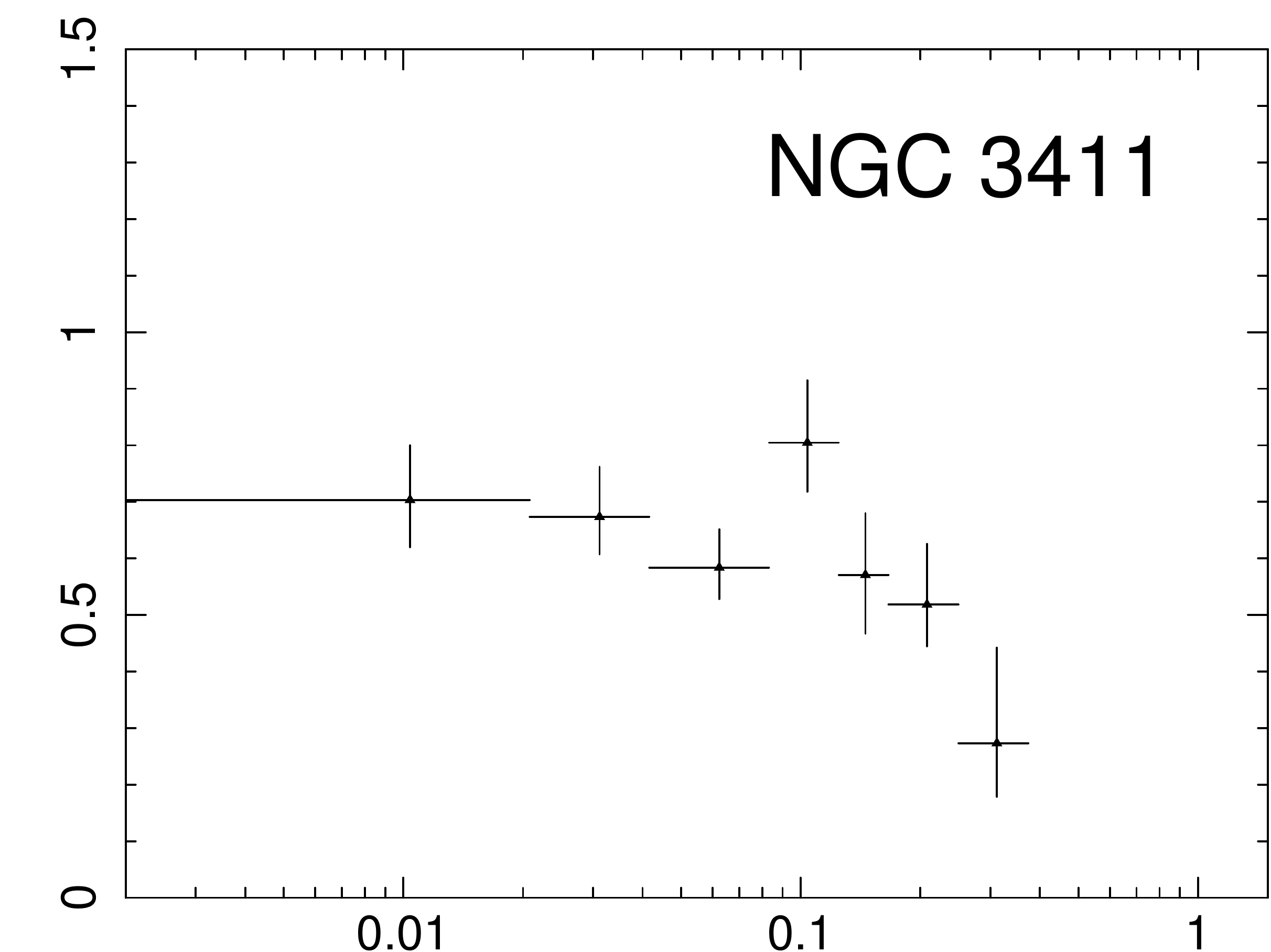}
                \includegraphics[width=0.236\textwidth,trim={0 0 0 0},clip]{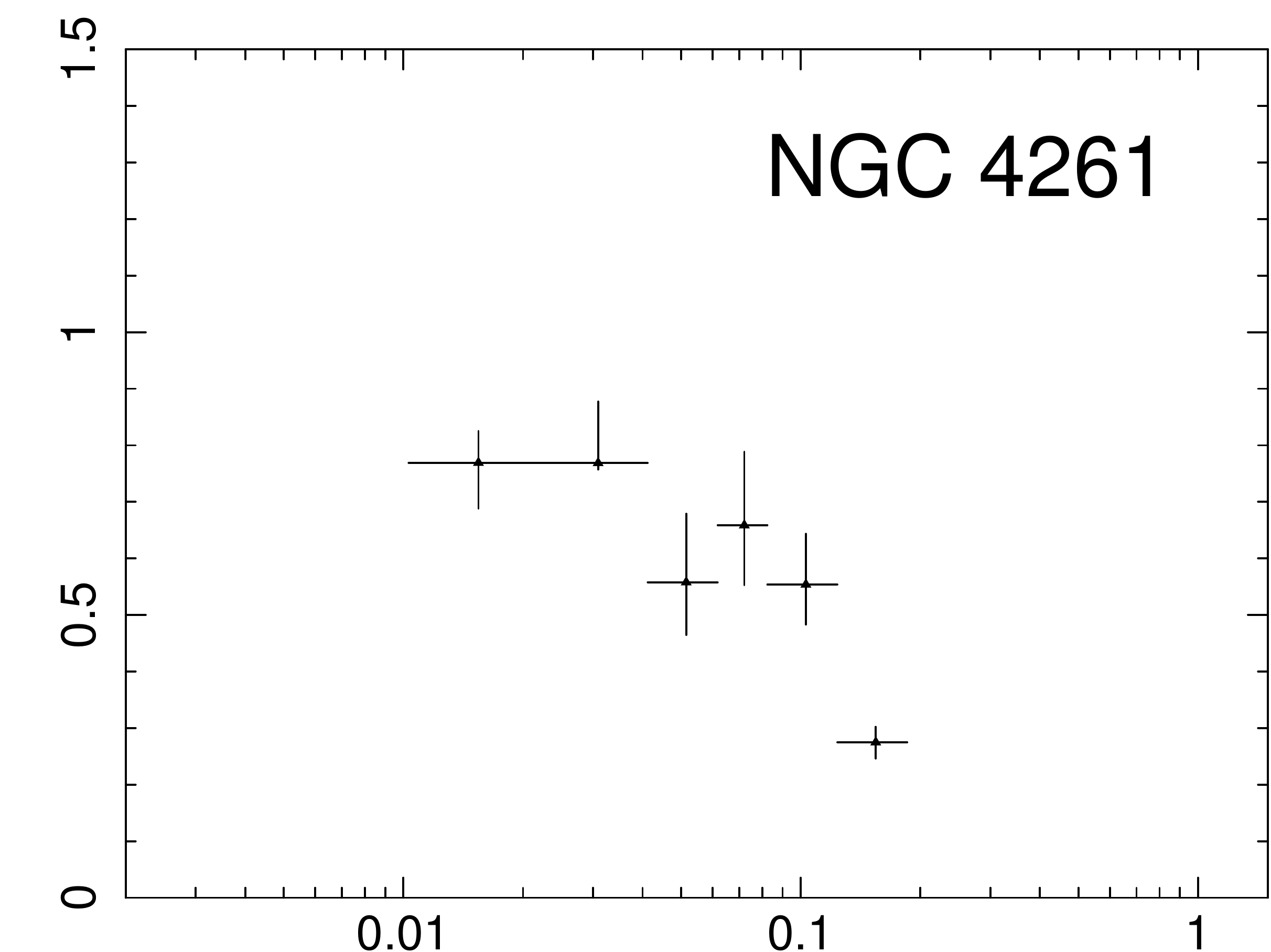} \\
                \includegraphics[width=0.236\textwidth,trim={0 0 0 0},clip]{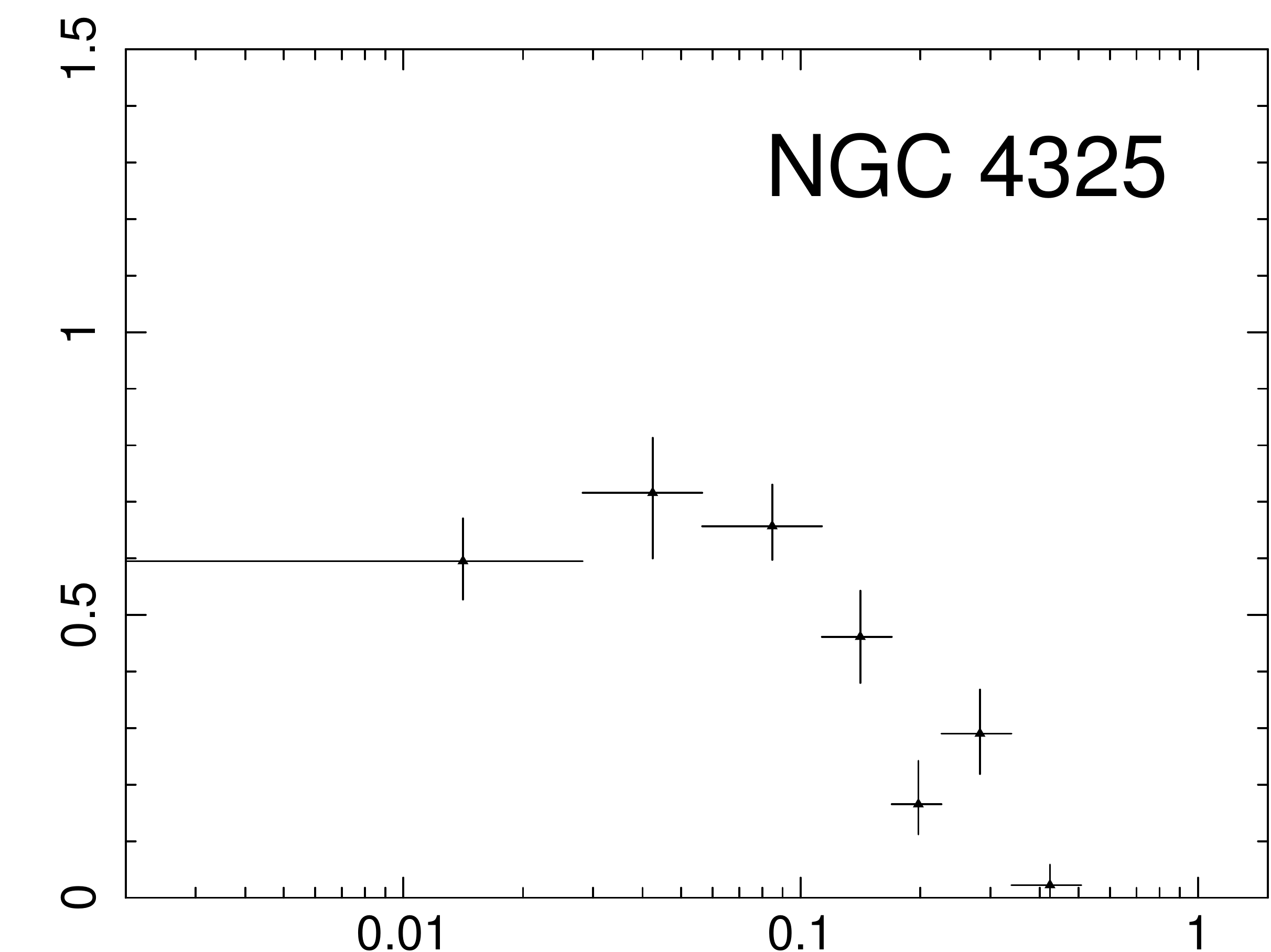}
                \includegraphics[width=0.236\textwidth,trim={0 0 0 0},clip]{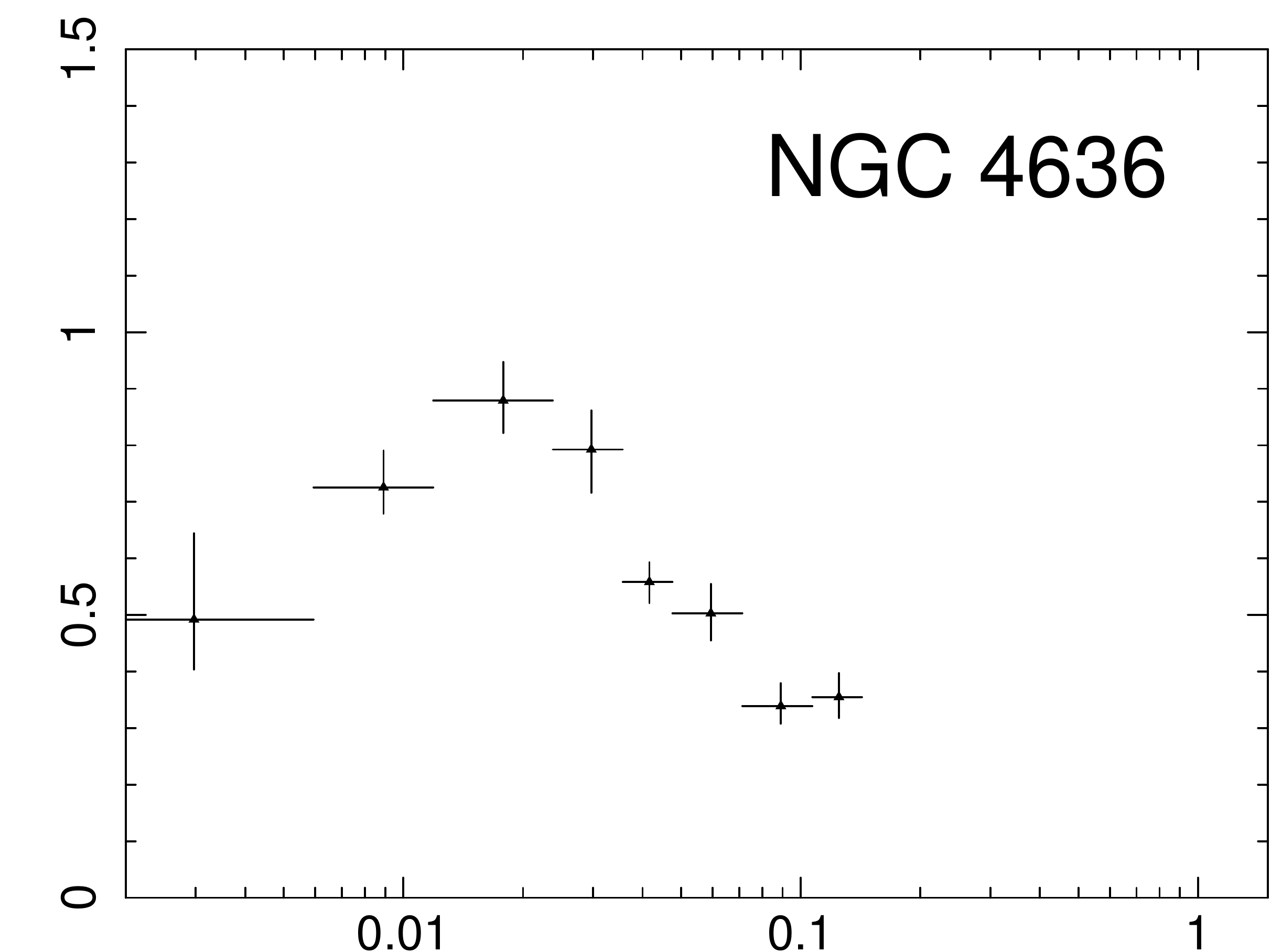}
                \includegraphics[width=0.236\textwidth,trim={0 0 0 0},clip]{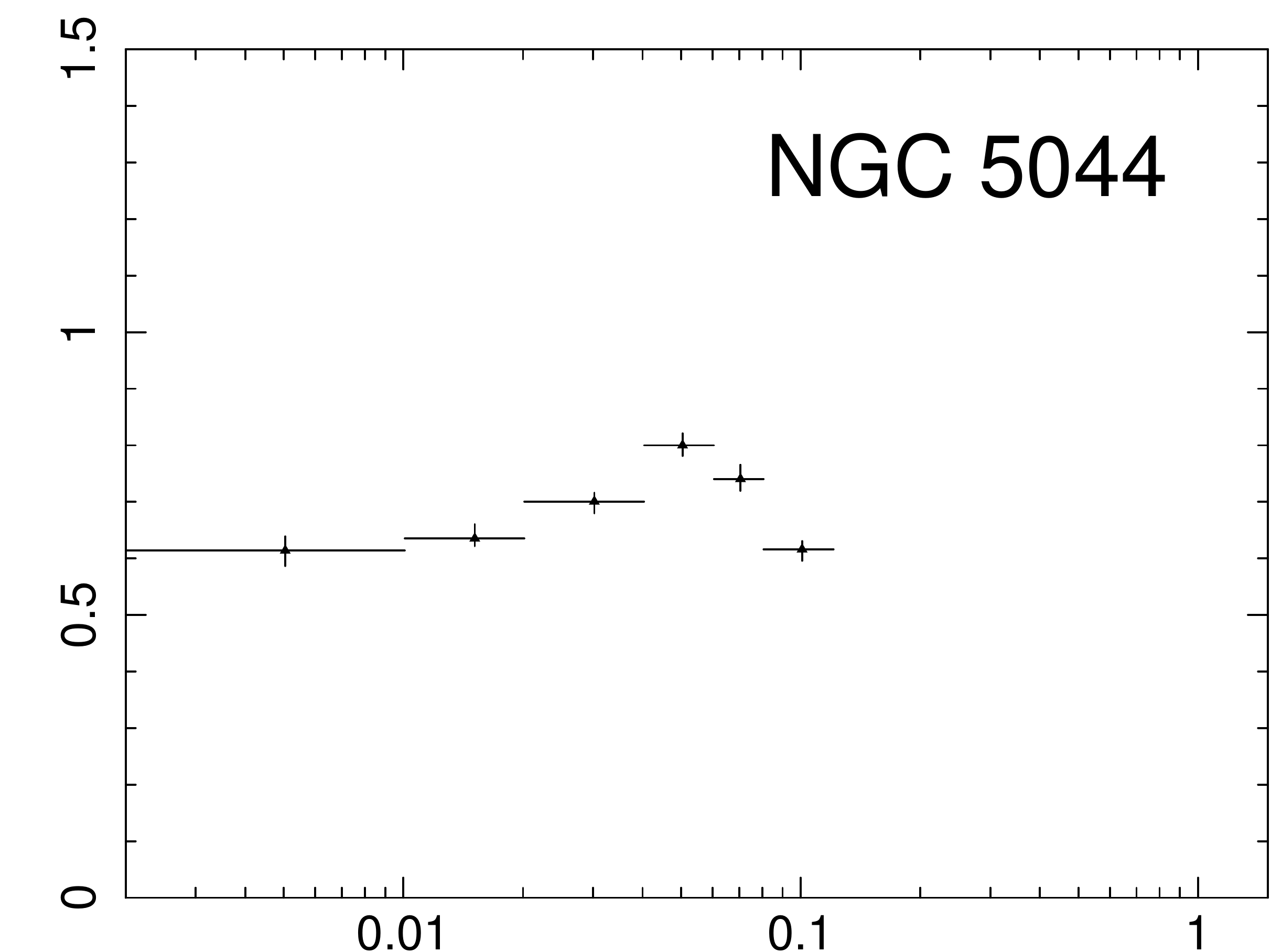}
                \includegraphics[width=0.236\textwidth,trim={0 0 0 0},clip]{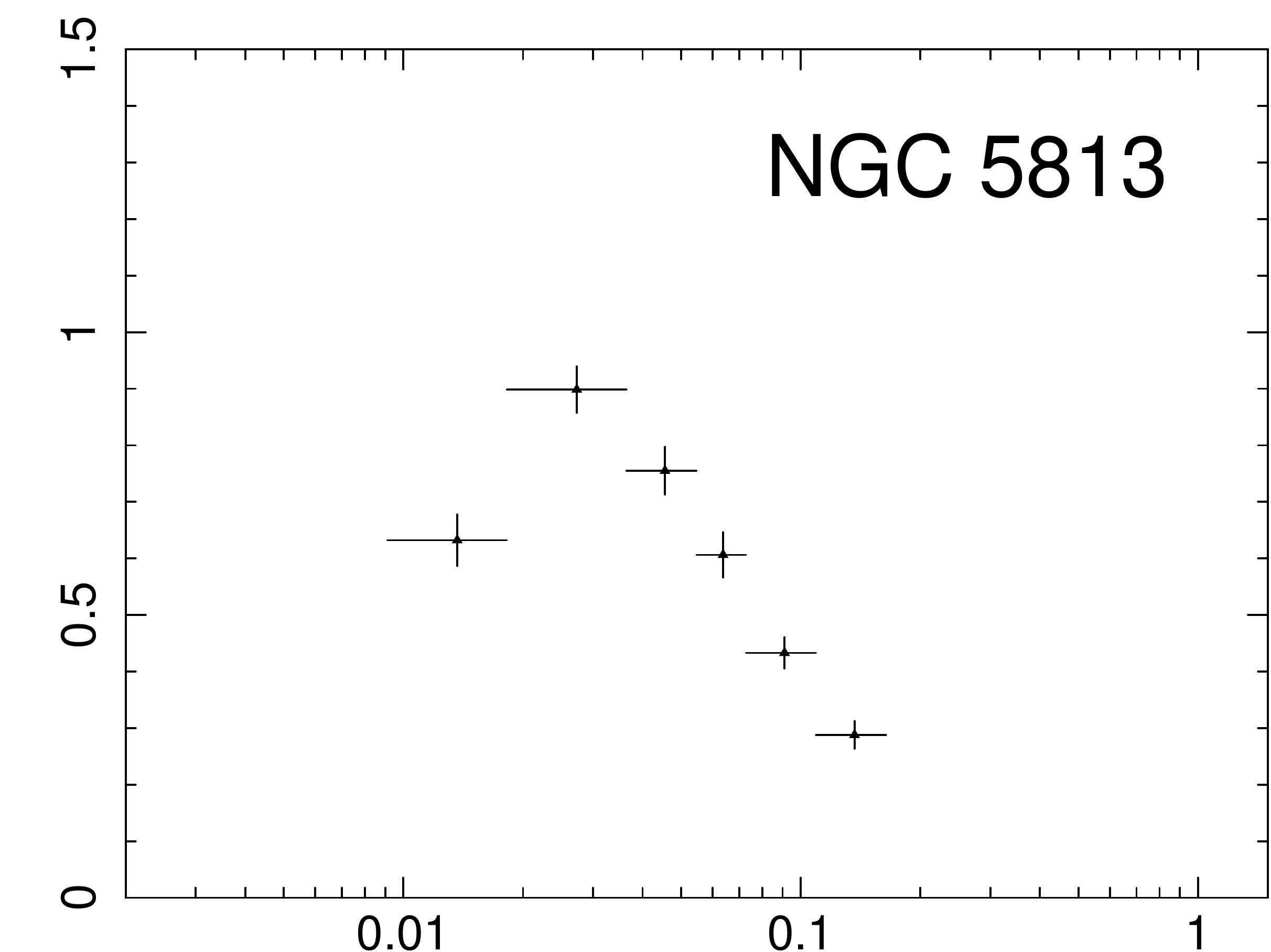} \\
                \includegraphics[width=0.236\textwidth,trim={0 0 0 0},clip]{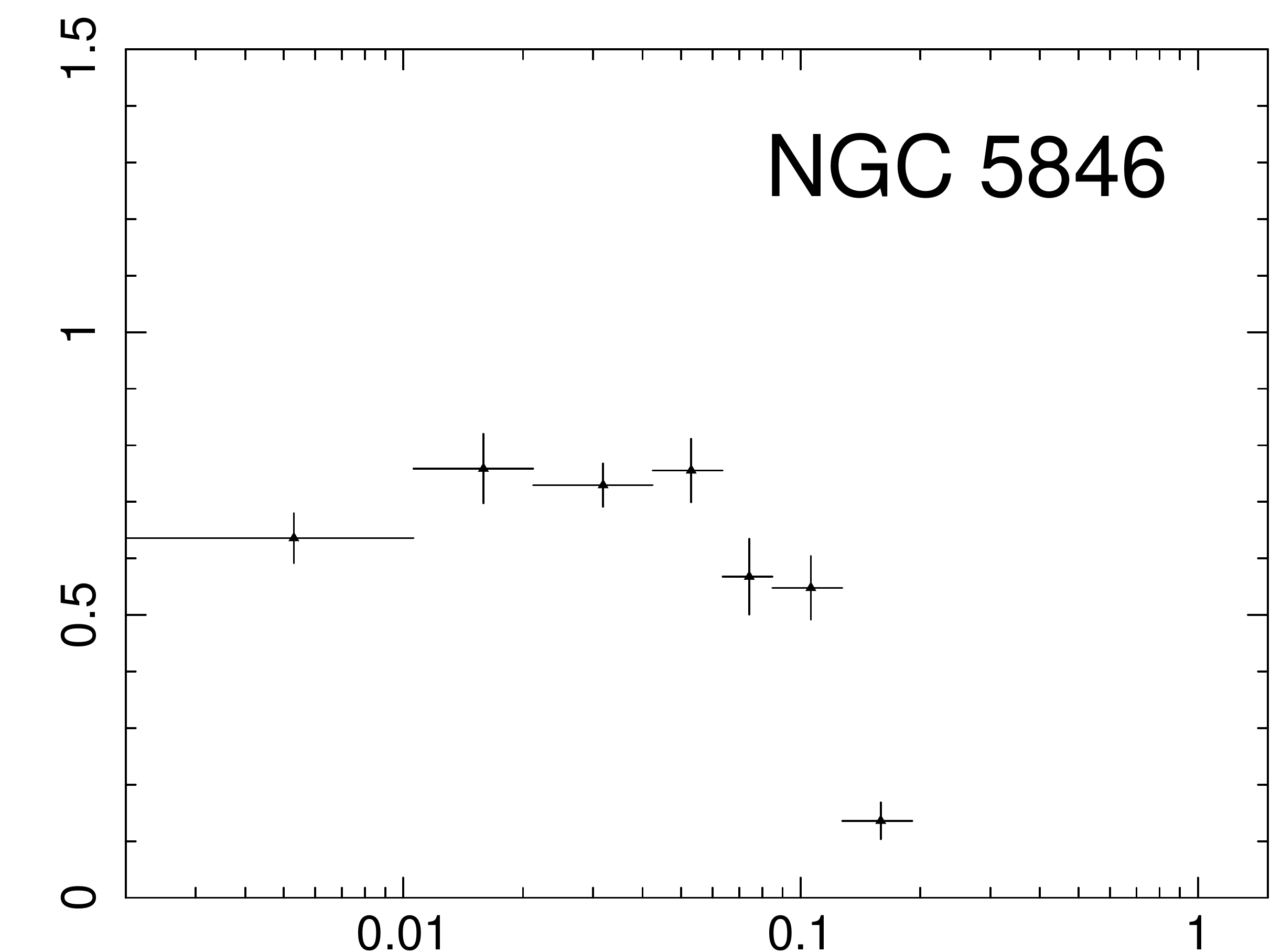}

        \caption{Radial Fe abundance profiles for all the groups/ellipticals ($kT_\text{mean}$ < 1.7 keV) in our sample. The radial distances ($x$-axis) are expressed in fractions of $r_{500}$ while the Fe abundances ($y$-axis) are given with respect to their proto-solar values \citep{2009LanB...4B...44L}. Data points that were not included when computing the average profile were removed (Sect. \ref{sect:excl_artefacts}). 
}
\label{fig:Fe_profiles_indiv_groups}
\end{figure*}

This section enumerates the objects of our sample (CHEERS) and provides supplementary information on their individual Fe profiles and radial extents. Table \ref{table:observations_radii} lists all the sources considered in this paper and their $r_{500}$ values \citep[adapted from][and references therein]{2015A&A...575A..38P}. For each element X, we also provide $r_\text{out,X}$, the maximum radius at which we evaluate the corresponding abundance (see Sect. \ref{sect:excl_artefacts} for further details). The Fe radial profiles of each source of our sample are shown in Figs. \ref{fig:Fe_profiles_indiv_clusters} (clusters) and \ref{fig:Fe_profiles_indiv_groups} (groups).

\begin{table*}[!]
\begin{centering}
\setlength{\tabcolsep}{2.5pt}
\caption{Properties of the observations used in this paper \citep[see][for further details]{2016A&A...592A.157M}.} 
\label{table:observations_radii}
\begin{tabular}{l c c c c c c c c c c c c}        
\hline \hline
Source   & $z$$^{(a)}$ & $r_{500}$$^{(b)}$ & $r_\text{out,O}$$^{(c)}$  & $r_\text{out,Mg}$$^{(c)}$  & $r_\text{out,Si}$$^{(c)}$  & $r_\text{out,S}$$^{(c)}$  & $r_\text{out,Ar}$$^{(c)}$  & $r_\text{out,Ca}$$^{(c)}$ & $r_\text{out,Fe}$$^{(c)}$ & $r_\text{out,Ni}$$^{(c)}$ & Cluster & Group\\
  &  & & & & & & &  &  & & & \\
   &  & (Mpc) & ($r_{500}$) & ($r_{500}$) & ($r_{500}$) & ($r_{500}$) & ($r_{500}$) & ($r_{500}$) & ($r_{500}$) & ($r_{500}$) &  & \\

\hline 

2A 0335+096 &     0.0349 & 1.05 & 0.26 & 0.26 & 0.26 & 0.26 & 0.26 & 0.26 & 0.26 & 0.26 & $\surd$ & $-$ \\ 

A\,85 &     0.0556 & 1.21 & 0.72 & 0.72 & 0.72 & 0.72 & 0.72 & 0.72 & 0.72 & 0.72 & $\surd$ & $-$ \\ 

A\,133 &     0.0569  & 0.94 & 0.94 & 0.94 & 0.94 & 0.94 & 0.94 & 0.94 & 0.94 & 0.94 & $\surd$ & $-$ \\ 

A\,189 &     0.0318 & 0.50 & 0.97 & 0.97 & 0.97 & 0.97 & 0.97 & 0.97 & 0.97 & $-$ & $-$ & $\surd$ \\ 

A\,262 &     0.0161 &  0.74 & 0.33 & 0.33 & 0.33 & 0.33 & 0.33 & 0.33 & 0.25 & 0.33 & $\surd$ & $-$ \\ 

A\,496 &     0.0328 & 1.00 & 0.50 & 0.50 & 0.50 & 0.50 & 0.50 & 0.50 & 0.50 & 0.50 & $\surd$ & $-$ \\ 

A\,1795 &     0.0616 & 1.22 & 0.79 & 0.79 & 0.79 & 0.79 & 0.79 & 0.79 & 0.79 & 0.79 & $\surd$ & $-$ \\

A\,1991 &     0.0587 & 0.82 & 0.84 & 1.12 & 1.12 & 1.12 & 1.12 & 1.12 & 0.56 & 1.12 & $\surd$ & $-$ \\

A\,2029 &     0.0767 & 1.33 & 0.91 & 0.91 & 0.91 & 0.91 & 0.91 & 0.91 & 0.91 & 0.91 & $\surd$ & $-$ \\

A\,2052 &     0.0348 & 0.95 & 0.42 & 0.56 & 0.56 & 0.56 & 0.56 & 0.56 & 0.42 & 0.56 & $\surd$ & $-$ \\

A\,2199 &     0.0302 & 1.00 & 0.46 & 0.46 & 0.46 & 0.46 & 0.46 & 0.46 & 0.46 & 0.46 & $\surd$ & $-$ \\

A\,2597 &    0.0852 & 1.11 & 1.22 & 1.22 & 1.22 & 1.22 & 1.22 & 1.22 & 1.22 & 1.22 & $\surd$ & $-$ \\

A\,2626 &     0.0573 & 0.84 & 1.06 & 1.06 & 1.06 & 1.06 & 1.06 & 1.06 & 1.06 & 1.06 & $\surd$ & $-$ \\

A\,3112 &     0.0750 & 1.13 & 1.05 & 1.05 & 1.05 & 1.05 & 1.05 & 1.05 & 1.05 & 1.05 & $\surd$ & $-$ \\

A\,3526 / Centaurus   & 0.0103 & 0.83 & 0.19 & 0.14 & 0.09 & 0.19 & 0.19 & 0.19 & 0.06 & 0.19 & $\surd$ & $-$ \\

A\,3581 &    0.0214 & 0.72 & 0.45 & 0.45 & 0.45 & 0.45 & 0.45 & 0.45 & 0.45 & $-$ & $-$ & $\surd$ \\

A\,4038 / Klemola 44  & 0.0283 & 0.89 & 0.36 & 0.36 & 0.36 & 0.36 & 0.36 & 0.36 & 0.36 & 0.36 & $\surd$ & $-$ \\

A\,4059 &    0.0460 & 0.96 & 0.74 & 0.74 & 0.74 & 0.74 & 0.74 & 0.74 & 0.74 & 0.74 & $\surd$ & $-$ \\

AS\,1101 / S\'ersic 159-03 &    0.0580 & 0.98 & 0.69 & 0.69 & 0.92 & 0.92 & 0.92 & 0.92 & 0.69 & 0.92 & $\surd$ & $-$ \\

AWM\,7 &   0.0172 & 0.86 & 0.30 & 0.30 & 0.15 & 0.30 & 0.30 & 0.30 & 0.10 & 0.30 & $\surd$ & $-$ \\

EXO\,0422 &    0.0390 & 0.89 & 0.67 & 0.67 & 0.67 & 0.67 & 0.67 & 0.67 & 0.67 & 0.67 & $\surd$ & $-$ \\

Fornax / NGC\,1399 &    0.0046 & 0.40 & 0.17 & 0.17 & 0.17 & 0.17 & 0.17 & 0.17 & 0.17 & $-$ & $-$ & $\surd$ \\

HCG\,62 &    0.0146 & 0.46 & 0.36 & 0.48 & 0.48 & 0.48 & 0.48 & 0.48 & 0.24 & $-$ & $-$ & $\surd$ \\

Hydra\,A &    0.0538 & 1.07 & 0.39 & 0.39 & 0.39 & 0.39 & 0.39 & 0.39 & 0.39 & 0.39 & $\surd$ & $-$ \\

M\,49 / NGC\,4472 &    0.0044 & 0.53 & 0.12 & 0.12 & 0.12 & 0.12 & 0.12 & 0.12 & 0.12 & $-$ &  $-$ & $\surd$ \\

M\,60 / NGC\,4649 &   0.0037 & 0.53 & 0.11 & 0.11 & 0.11 & 0.11 & 0.11 & 0.11 & 0.06 & $-$ & $-$ & $\surd$ \\

M\,84 / NGC\,4374 &   0.0034 & 0.46 & 0.14 & 0.14 & 0.14 & 0.14 & 0.14 & 0.14 & 0.14 & $-$ & $-$ & $\surd$ \\

M\,86 / NGC\,4406 &    -0.0009 & 0.49 & 0.06 & 0.06 & 0.06 & 0.06 & 0.06 & 0.06 & 0.06 & $-$ & $-$ & $\surd$ \\

M\,87 / NGC\,4486 &    0.0044 & 0.75 & 0.06 & 0.08 & 0.03 & 0.08 & 0.10 & 0.10 & 0.03 & $-$ & $-$ & $\surd$ \\

M\,89 / NGC\,4552 &    0.0010 & 0.44 & $-$ & $-$ & 0.12 & $-$ & $-$ & $-$ & 0.09 & $-$ & $-$ & $\surd$ \\

MKW\,3s &    0.0450 & 0.95 & 0.73 & 0.73 & 0.73 & 0.73 & 0.73 & 0.73 & 0.37 & 0.73 & $\surd$ & $-$ \\

MKW\,4 &   0.0200 & 0.62 & 0.49 & 0.49 & 0.49 & 0.49 & 0.49 & 0.49 & 0.49 & 0.49 & $\surd$ & $-$ \\

NGC\,507 &   0.0165 & 0.60 & 0.42 & 0.42 & 0.42 & 0.42 & 0.42 & 0.42 & 0.42 & $-$ & $-$ & $\surd$ \\

NGC\,1316 / Fornax A &   0.0059 & 0.46 & 0.19 & 0.19 & 0.19 & 0.19 & 0.19 & 0.19 & 0.19 & $-$ & $-$ & $\surd$ \\

NGC\,1404 &   0.0064 & 0.61 & 0.16 & 0.16 & 0.16 & 0.16 & 0.16 & 0.16 & 0.16 & $-$ & $-$ & $\surd$ \\

NGC\,1550 &   0.0123 & 0.62 & 0.30 & 0.30 & 0.22 & 0.30 & 0.30 & 0.30 & 0.30 & $-$ & $-$ & $\surd$ \\

NGC\,3411 &   0.0155 & 0.47 & 0.37 & 0.50 & 0.50 & 0.50 & 0.50 & 0.50 & 0.37 & $-$ & $-$ & $\surd$ \\

NGC\,4261 &   0.0074 & 0.45 & 0.25 & 0.25 & 0.25 & 0.25 & 0.25 & 0.25 & 0.19 & $-$ & $-$ & $\surd$ \\

NGC\,4325 &   0.0258 & 0.58 & 0.51 & 0.68 & 0.68 & 0.68 & 0.68 & 0.68 & 0.51 & $-$ & $-$ & $\surd$ \\

NGC\,4636 &   0.0037 & 0.35 &0.14 & 0.14 & 0.14 & 0.14 & 0.14 & 0.14 & 0.14 & $-$ & $-$ & $\surd$ \\

NGC\,5044 &   0.0090 & 0.56 & 0.18 & 0.18 & 0.12 & 0.18 & 0.18 & 0.18 & 0.12 & $-$ & $-$ & $\surd$ \\

NGC\,5813 &   0.0064 & 0.44 & 0.16 & 0.22 & 0.22 & 0.22 & 0.22 & 0.22 & 0.16 & $-$ & $-$ & $\surd$ \\

NGC\,5846 &   0.0061 & 0.36 & 0.19 & 0.25 & 0.25 & 0.25 & 0.25 & 0.25 & 0.19 & $-$ & $-$ & $\surd$ \\

Perseus &   0.0183 & 1.29 & 0.26 & 0.26 & 0.11 & 0.26 & 0.26 & 0.26 & 0.07 & 0.26 & $\surd$ & $-$ \\

\hline                                   
\end{tabular}
\par\end{centering}
\tablefoot{$^{(a)}$ Redshifts were taken from \citet[][and references therein]{2015A&A...575A..38P}.
$^{(b)}$ Values of $r_{500}$ (in Mpc) were taken from \citet[][and references therein]{2015A&A...575A..38P}.
$^{(c)}$ $r_\text{out,X}$ (in units of $r_{500}$) corresponds to the maximum radial extent of the abundance measurements of element X (see text).}

\end{table*}

\section{Average abundance profiles of O, Mg, Si, S, Ar, Ca, and Ni}

In Sect. \ref{sect:results_Fe} we provided numerical values of the radial Fe profile in the full sample (Table \ref{table:Fe_radial_stacked_profile}) and after subdividing it into clusters and groups (Table \ref{table:Fe_radial_stacked_profile_clgr}). In this Appendix we extend these numbers to the average O, Mg, Si, Ar, Ca, and Ni profiles that are shown in Figs \ref{fig:abundance_profiles} and \ref{fig:abundance_profiles_clgr} (see Sect. \ref{sect:results_abun} for further details). These values are listed in Table \ref{table:abundance_profiles} (full sample) and Table \ref{table:abundance_profiles_clgr} (comparison between clusters and groups).

\begin{table*}
\begin{centering}
\caption{Average radial abundance profiles for the full sample, as shown in Fig. \ref{fig:abundance_profiles}.}             
\label{table:abundance_profiles}
\setlength{\tabcolsep}{10pt}
\scalebox{1}{
\begin{tabular}{r@{ -- }l r@{$\pm$}l r@{$\pm$}l r@{$\pm$}l r@{$\pm$}l r@{$\pm$}l r@{$\pm$}l}        
\hline \hline                
\multicolumn{2}{c}{Radius} & \multicolumn{2}{c}{O} & \multicolumn{2}{c}{Mg} & \multicolumn{2}{c}{Si} & \multicolumn{2}{c}{S} & \multicolumn{2}{c}{Ar} & \multicolumn{2}{c}{Ca} \\    
\multicolumn{2}{c}{($/r_{500}$)} & \multicolumn{2}{c}{} & \multicolumn{2}{c}{} & \multicolumn{2}{c}{} & \multicolumn{2}{c}{} & \multicolumn{2}{c}{} & \multicolumn{2}{c}{} \\    

\hline                        
0 & 0.0075         & $0.437$ & $0.017$ & $0.50$ & $0.13$ & $0.76$ & $0.12$ & $0.80$ & $0.11$ & $0.88$ & $0.15$ & $0.95$ & $0.12$ \\
0.0075 & 0.014 & $0.624$ & $0.020$ & $0.53$ & $0.04$ & $0.79$ & $0.11$ & $0.83$ & $0.11$ & $0.92$ & $0.18$ & $1.09$ & $0.11$ \\
0.014 & 0.02      & $0.650$ & $0.021$ & $0.54$ & $0.04$ & $0.78$ & $0.09$ & $0.84$ & $0.08$ & $0.85$ & $0.15$ & $1.05$ & $0.11$ \\
0.02 & 0.03        & $0.685$ & $0.016$ & $0.52$ & $0.04$ & $0.77$ & $0.06$ & $0.85$ & $0.04$ & $0.80$ & $0.16$ & $0.98$ & $0.09$ \\
0.03 & 0.04        & $0.632$ & $0.017$ & $0.51$ & $0.03$ & $0.69$ & $0.07$ & $0.77$ & $0.04$ & $0.73$ & $0.17$ & $0.88$ & $0.08$ \\
0.04 & 0.055     & $0.533$ & $0.017$ & $0.49$ & $0.05$ & $0.63$ & $0.07$ & $0.69$ & $0.04$ & $0.65$ & $0.15$ & $0.82$ & $0.08$ \\
0.055 & 0.065   & $0.54$ & $0.03$ & $0.49$ & $0.06$ & $0.58$ & $0.06$ & $0.63$ & $0.05$ & $0.56$ & $0.14$ & $0.79$ & $0.10$ \\
0.065 & 0.09     & $0.480$ & $0.021$ & $0.46$ & $0.04$ & $0.53$ & $0.04$ & $0.55$ & $0.03$ & $0.50$ & $0.12$ & $0.70$ & $0.07$ \\
0.09 & 0.11       & $0.42$ & $0.03$ & $0.46$ & $0.07$ & $0.47$ & $0.04$ & $0.50$ & $0.05$ & $0.42$ & $0.14$ & $0.56$ & $0.11$ \\
0.11 & 0.135     & $0.38$ & $0.03$ & $0.49$ & $0.09$ & $0.43$ & $0.05$ & $0.49$ & $0.06$ & $0.36$ & $0.13$ & $0.57$ & $0.11$ \\
0.135 & 0.16     & $0.38$ & $0.03$ & $0.47$ & $0.11$ & $0.41$ & $0.03$ & $0.47$ & $0.06$ & $0.27$ & $0.15$ & $0.54$ & $0.12$ \\
0.16 & 0.2         & $0.38$ & $0.03$ & $0.51$ & $0.14$ & $0.371$ & $0.023$ & $0.44$ & $0.04$ & $0.23$ & $0.14$ & $0.57$ & $0.12$ \\
0.2 & 0.23         & $0.33$ & $0.04$ & $0.50$ & $0.11$ & $0.36$ & $0.04$ & $0.43$ & $0.06$ & $0.25$ & $0.13$ & $0.56$ & $0.17$ \\
0.23 & 0.3         & $0.26$ & $0.03$ & $0.50$ & $0.23$ & $0.31$ & $0.04$ & $0.36$ & $0.08$ & $0.22$ & $0.17$ & $0.47$ & $0.18$ \\
0.3 & 0.55         & $0.27$ & $0.03$ & $-0.02$ & $0.04$ & $0.26$ & $0.04$ & $0.31$ & $0.12$ & $0.2$ & $0.3$ & $0.10$ & $0.18$ \\
0.55 & 1.22        & $0.01$ & $0.05$ & $-0.49$ & $0.14$ & $0.10$ & $0.07$ & $-0.1$ & $0.3$ & $-0.4$ & $0.7$ & $-0.1$ & $0.4$ \\

\hline                                   
\end{tabular}}
\par\end{centering}
\tablefoot{The error bars contain the statistical uncertainties and MOS-pn uncertainties (Sect. \ref{sect:MOS-pn_uncertainties}), except for the O abundance profile, which is only measured with MOS.}
\end{table*}

\begin{table*}
\begin{centering}
\caption{Average radial abundance profiles for clusters (>1.7 keV) and groups (<1.7 keV), as shown in Fig. \ref{fig:abundance_profiles_clgr}.}             
\label{table:abundance_profiles_clgr}
\setlength{\tabcolsep}{7.7pt}
\scalebox{1}{
\begin{tabular}{r@{ -- }l r@{$\pm$}l r@{$\pm$}l r@{$\pm$}l r@{$\pm$}l r@{$\pm$}l r@{$\pm$}l r@{$\pm$}l}        
\hline \hline                
\multicolumn{2}{c}{Radius} & \multicolumn{2}{c}{O} & \multicolumn{2}{c}{Mg} & \multicolumn{2}{c}{Si} & \multicolumn{2}{c}{S} & \multicolumn{2}{c}{Ar} & \multicolumn{2}{c}{Ca} & \multicolumn{2}{c}{Ni} \\    
\multicolumn{2}{c}{($/r_{500}$)} & \multicolumn{2}{c}{} & \multicolumn{2}{c}{} & \multicolumn{2}{c}{} & \multicolumn{2}{c}{} & \multicolumn{2}{c}{} & \multicolumn{2}{c}{} & \multicolumn{2}{c}{} \\    

\hline                        
\multicolumn{16}{c}{Clusters} \\
\hline

0 & 0.018       & $0.815$ & $0.025$ & $0.50$ & $0.08$ & $0.79$ & $0.08$ & $0.86$ & $0.03$ & $0.87$ & $0.13$ & $1.05$ & $0.08$ & $1.6$ & $0.5$ \\
0.018 & 0.04 & $0.776$ & $0.021$ & $0.47$ & $0.04$ & $0.75$ & $0.05$ & $0.82$ & $0.06$ & $0.81$ & $0.13$ & $0.94$ & $0.07$ & $1.5$ & $0.4$ \\
0.04 & 0.068 & $0.689$ & $0.024$ & $0.44$ & $0.03$ & $0.61$ & $0.05$ & $0.66$ & $0.06$ & $0.65$ & $0.14$ & $0.80$ & $0.07$ & $1.3$ & $0.3$ \\
0.068 & 0.1   & $0.59$ & $0.03$ & $0.46$ & $0.08$ & $0.53$ & $0.04$ & $0.56$ & $0.05$ & $0.49$ & $0.14$ & $0.68$ & $0.09$ & $1.2$ & $0.4$ \\
0.1 & 0.18     & $0.46$ & $0.025$ & $0.51$ & $0.05$ & $0.43$ & $0.04$ & $0.50$ & $0.06$ & $0.35$ & $0.12$ & $0.61$ & $0.08$ & $0.9$ & $0.4$ \\
0.18 & 0.24   & $0.35$ & $0.04$ & $0.55$ & $0.03$ & $0.37$ & $0.04$ & $0.45$ & $0.05$ & $0.28$ & $0.12$ & $0.60$ & $0.12$ & $0.8$ & $0.6$ \\
0.24 & 0.34   & $0.34$ & $0.04$ & $0.54$ & $0.14$ & $0.31$ & $0.03$ & $0.37$ & $0.11$ & $0.22$ & $0.18$ & $0.35$ & $0.15$ & $0.5$ & $0.8$ \\
0.34 & 0.5     & $0.37$ & $0.05$ & $-0.06$ & $0.16$ & $0.27$ & $0.04$ & $0.34$ & $0.13$ & $0.20$ & $0.39$ & $0.1$ & $0.3$ & $-0.9$ & $0.4$ \\
0.5 & 1.22      & $-0.02$ & $0.05$ & $-0.27$ & $0.22$ & $0.13$ & $0.07$ & $0.01$ & $0.25$ & $-0.24$ & $0.69$ & $-0.1$ & $0.3$ & $-3.4$ & $2.1$ \\

\hline                        
\multicolumn{16}{c}{Groups} \\
\hline

0 & 0.009         & $0.384$ & $0.017$ & $0.48$ & $0.14$ & $0.76$ & $0.16$ & $0.77$ & $0.22$ & $0.86$ & $0.24$ & $0.82$ & $0.19$ & \multicolumn{2}{c}{--} \\
0.009 & 0.024 & $0.613$ & $0.015$ & $0.59$ & $0.07$ & $0.80$ & $0.11$ & $0.89$ & $0.18$ & $0.91$ & $0.18$ & $1.11$ & $0.11$ & \multicolumn{2}{c}{--} \\
0.024 & 0.042 & $0.591$ & $0.015$ & $0.53$ & $0.04$ & $0.67$ & $0.10$ & $0.79$ & $0.12$ & $0.69$ & $0.19$ & $0.88$ & $0.10$ & \multicolumn{2}{c}{--} \\
0.042 & 0.064 & $0.460$ & $0.018$ & $0.53$ & $0.10$ & $0.60$ & $0.09$ & $0.67$ & $0.11$ & $0.58$ & $0.14$ & $0.81$ & $0.12$ & \multicolumn{2}{c}{--} \\
0.064 & 0.1     & $0.366$ & $0.024$ & $0.44$ & $0.15$ & $0.49$ & $0.04$ & $0.47$ & $0.08$ & $0.47$ & $0.13$ & $0.62$ & $0.15$ & \multicolumn{2}{c}{--} \\
0.1 & 0.15       & $0.309$ & $0.023$ & $0.4$ & $0.3$ & $0.40$ & $0.03$ & $0.41$ & $0.07$ & $0.22$ & $0.18$ & $0.13$ & $0.27$ & \multicolumn{2}{c}{--} \\
0.15 & 0.26     & $0.327$ & $0.03$ & $0.4$ & $0.4$ & $0.34$ & $0.05$ & $0.32$ & $0.09$ & $0.01$ & $0.17$ & $0.01$ & $0.34$ & \multicolumn{2}{c}{--} \\
0.26 & 0.97      & $0.19$ & $0.04$ & $-0.23$ & $0.14$ & $0.17$ & $0.06$ & $0.16$ & $0.15$ & $0.24$ & $0.38$ & $-0.3$ & $0.7$ & \multicolumn{2}{c}{--} \\

\hline                                   
\end{tabular}}
\par\end{centering}
\tablefoot{The error bars contain the statistical uncertainties and MOS-pn uncertainties (Sect. \ref{sect:MOS-pn_uncertainties}), except for the O abundance profiles, which are only measured with MOS.}
\end{table*}




\end{document}